%% file: main.tex
\PassOptionsToPackage{unicode}{hyperref}
\PassOptionsToPackage{hyphens}{url}
\documentclass[12pt]{article}
\usepackage{amsmath,amssymb}

\usepackage{iftex}
\ifPDFTeX
\usepackage[T1]{fontenc}
\usepackage[utf8]{inputenc}
\usepackage{textcomp}
\else
\usepackage{unicode-math}
\defaultfontfeatures{Scale=MatchLowercase}
\defaultfontfeatures[\rmfamily]{Ligatures=TeX,Scale=1}
\fi
\usepackage{lmodern}
\usepackage{microtype,parskip}
\UseMicrotypeSet[protrusion]{basicmath}
\usepackage{booktabs}
\usepackage{graphicx}
\makeatletter
\def\maxwidth{\ifdim\Gin@nat@width>\linewidth\linewidth\else\Gin@nat@width\fi}
\def\maxheight{\ifdim\Gin@nat@height>\textheight\textheight\else\Gin@nat@height\fi}
\makeatother
\setkeys{Gin}{width=\maxwidth,height=\maxheight,keepaspectratio}

\usepackage{caption,subcaption}
\usepackage{natbib}
\usepackage[bookmarks=false]{hyperref}

\usepackage{xurl}
\hypersetup{
	pdftitle={High-Frequency Estimation of the Leverage Effect under Rough Volatility with Microstructure Noise},
	pdfauthor={},
	pdfkeywords={Fractional Brownian motion, Rough Volterra processes, Stable convergence, Microstructure noise},
    hidelinks,
    hypertexnames=false,
	pdfcreator={LaTeX}}

\usepackage{amsthm}
\usepackage{authblk}

\newtheorem{thm}{Theorem}[section]
\newtheorem{definition}{Definition}[section]
\newtheorem{ass}{Assumption}[section]
\newtheorem{pro}{Proposition}[section]

\newcommand{\convp}{\xrightarrow{p}}

\newcommand{\convst}{\xrightarrow{\rm st}}
\newcommand{\convucp}{\xrightarrow{\rm u.c.p.}}

\newtheorem{lemma}{Lemma}[section]
\theoremstyle{remark}
\newtheorem{remark}{Remark}[section]
\theoremstyle{plain}
\newcommand{\bbE}{\mathbb{E}}
\newcommand{\F}{\mathcal{F}}

\newcommand{\E}{\bbE}
\newcommand{\eps}{\varepsilon}
\newcommand{\1}{\mathbf 1}
\renewcommand{\P}{\mathbb{P}}

\begin{document}

	\newcommand{\spacingset}[1]{\renewcommand{\baselinestretch}{#1}\small\normalsize}
\spacingset{1}

\providecommand{\authororderstatement}{Authors are listed in alphabetical order.}
\author[1]{Zhao Chen}
\author[1]{Ming Cheng}
\author[2]{Christina Dan Wang}
\author[1]{Ziyang Xiong}
\affil[1]{School of Data Science, Fudan University}
\affil[2]{Business Division, New York University Shanghai}
\hypersetup{pdfauthor={Zhao Chen; Ming Cheng; Christina Dan Wang; Ziyang Xiong}}
\title{\normalfont\LARGE\bfseries High-Frequency Estimation of the Leverage Effect under Rough Volatility with Microstructure Noise\thanks{\authororderstatement\ We thank participants at the 2026 ICSA China Conference at Southern University of Science and Technology for helpful comments and suggestions.}}
\date{}
\maketitle

\bigskip
\begin{abstract}
This paper develops feasible high-frequency inference for the leverage effect when the spot variance follows rough Volterra dynamics, leverage loadings depend on endogenous rough states, and observed prices may be contaminated by microstructure noise. We define an integrated leverage parameter through a scaled covariation between log prices and spot variance that accounts for their different local fluctuation scales. The scaled realized covariation admits a stable functional central limit theorem with an explicit pathwise centering correction arising from within-interval variation in the leverage loading. A loading regularity condition renders this correction asymptotically negligible, and we verify the condition for direct Volterra loadings and for smooth functions of closed rough states for every Hurst index below one half. We construct a block-based estimator with an exact finite-window correction for the leading proportional bias, together with a nonnegative single-scale estimator of its asymptotic variance. Pre-averaging extends feasible inference to observations with independent and identically distributed microstructure noise. Simulations under the rough Heston model demonstrate accurate finite-sample calibration with and without noise. An empirical application to fifteen years of one-second data on nine U.S. equities and ETFs finds negative median leverage estimates for every asset at weekly and monthly horizons.
\end{abstract}

{\it Keywords:} Fractional Brownian Motion; Rough Volterra processes; Stable convergence; Microstructure noise

\vfill
\newpage

\spacingset{1.8}

\allowdisplaybreaks[4]

\input{sections/intro.tex}
\input{sections/method.tex}
\input{sections/theory.tex}
\input{sections/noise.tex}
\input{sections/sim.tex}
\input{sections/emp.tex}

\bibliography{references}
\newpage

\appendix
\setcounter{equation}{0}
\renewcommand{\tablename}{Appendix Table}
\renewcommand{\figurename}{Appendix Figure}
\spacingset{1.8}

\input{appendices/A_loc.tex}
\input{appendices/B_lr.tex}
\input{appendices/C_limit.tex}
\input{appendices/D_est.tex}
\input{appendices/E_noise.tex}
\input{appendices/F_sim.tex}

\input{appendices/G_emp.tex}

\end{document}

%% file: sections/intro.tex
\section{Introduction}

The leverage effect, the dependence between return shocks and contemporaneous or future changes in volatility, plays a fundamental role in asset pricing and volatility modeling (see, for example, \citet{Bollerslev2006}, \citet{BandiReno2012}, and \citet{AitSahaliaFanLi2013}). High-frequency data make it possible to estimate this dependence nonparametrically over short horizons. \citet{Wang2014} study integrated leverage through quadratic covariation between log prices and functions of spot variance and develop block-based estimators with feasible inference, both with and without microstructure noise, while \citet{Yacine2017} study the estimation of continuous and discontinuous leverage effects in a general It\^o semimartingale framework. Related high-frequency approaches include \citet{KalninaXiu2017}, \citet{YuanMuZhou2020}, \citet{CuratoSanfelici2022}, and \citet{Yang2023}; see also \citet{ChongTodorov2024} for inference on leverage and volatility-of-volatility effects from options. All of these methods rely on the spot variance being an It\^o semimartingale.

A large body of evidence indicates that volatility is much rougher than conventional semimartingale models allow. \citet{Gatheral2018} show that log-volatility behaves, over relevant time scales, like a fractional process with Hurst index well below $1/2$. This finding has stimulated a broad modeling literature, including the rough Heston framework of \citet{Eleuch2019}, the multifactor Markovian approximations of \citet{Abijaber2019}, and work on simulation and numerical methods such as \citet{Bennedsen2017,Heinrich2019,Richard2023,Bayer2024}. Processes with Hurst index below $1/2$ fall outside the It\^o semimartingale framework, which has led to a substantial literature on stochastic calculus and pathwise analysis for fractional models \citep{Alos2000,Bender2003,Donati2011,Banna2019}; related probabilistic structures arise in the literature on Brownian semistationary and Volterra-type processes \citep{Barndorffnielsen2009,Barndorffnielsen2014,Barndorffnielsen2018,Corcuera2013,Granelli2019,Passeggeri2019,Li2023,Murray2021}.

Statistical inference on roughness from high-frequency observations has also progressed rapidly. \citet{Chong2024a,Chong2024b} establish central limit theorems for estimating the roughness parameter and derive the corresponding minimax rates, \citet{Chong2025a} develop a nonparametric test that distinguishes rough volatility paths from semimartingale volatility paths using high-frequency spot-volatility increments, and \citet{Bolko2023} propose a GMM approach based on realized measures. These works show that roughness is not merely a modeling feature, but an object that can be inferred and tested in a rigorous high-frequency statistical framework. High-frequency inference on leverage under rough volatility remains relatively limited, particularly in the presence of endogenous rough loadings and microstructure noise.

Three features make this problem difficult. In the rough setting, return increments and spot-variance increments fluctuate on distinct local scales, of order $\Delta_n^{1/2}$ and $\Delta_n^{H}$ for a sampling interval $\Delta_n$, so the conventional realized covariation requires a different normalization to yield a finite, nondegenerate leverage parameter. Spot variance is latent, and the error of its local estimation interacts with the rough increments to produce a bias proportional to the target, which can be non-negligible at the central limit theorem scale. The leverage loading, the product of volatility, volatility of volatility, and correlation, may itself depend on the rough state, and its fluctuations within each sampling interval generate a centering error that requires explicit control beyond standard semimartingale arguments. Moreover, the sampling frequencies needed to resolve rough dynamics are precisely those at which microstructure noise cannot be ignored.

To address this problem, we work with a continuous-time log-price process whose instantaneous variance follows rough Volterra-type dynamics, and we allow the return shock and the variance shock to be locally correlated. This paper focuses on the integrated leverage parameter induced by the contemporaneous local correlation between return and variance shocks. Following the quadratic-covariation framework of \citet{Wang2014}, we define this parameter through a scaled covariation between log prices and spot variance that accounts for their different local scales; at $H=1/2$, it reduces to the quadratic covariation. The main contributions are as follows.

First, we identify the centering error generated by within-interval variation in endogenous rough leverage loadings. Under locally bounded c\`adl\`ag model coefficients, the scaled realized covariation admits a stable functional CLT with an explicit pathwise correction. A loading regularity condition (LR) makes this correction asymptotically negligible at the CLT scale. We verify LR for two classes of rough loadings: direct Volterra loadings with predictable random coefficients, which include the rough Heston model, and smooth functions of closed rough states, for every fixed $H\in(0,1/2)$; the second verification rests on Malliavin-derivative bounds for the conditional expectation kernel. A third route covers general adapted state coefficients for $H>1/6$.

Second, we develop single-scale feasible inference for both noise-free and noisy observations. A deterministic normalization at the actual block sizes removes the leading proportional bias from spot-variance estimation. The same nonnegative quadratic form, based on adjacent-window second differences of the cumulative estimator, directly estimates the total asymptotic variance arising from spot-variance estimation error and rough fluctuations, without requiring separate estimation of its constituent terms. Pre-averaging with exact finite-window correction extends this construction to noisy observations under the same latent loading condition, and the noisy-data results hold for independent and identically distributed observation errors, independent of the latent process, with finite eighth moment. At the balanced block sizes, the noise-free and noisy estimators attain rates $n^{-1/(4H+2)}$ and $n^{-1/[4(2H+1)]}$, respectively. We further establish the validity of plug-in inference with an estimated Hurst index under suitable conditions on the pilot estimation rate and a randomly shifted grid for bandwidth selection (Remark~\ref{rem:noisy-plugin}).

Third, simulations under the rough Heston model and two further loading specifications show accurate finite-sample coverage across $H\in[0.05,0.45]$, with and without noise. Using fifteen years of one-second observations for nine U.S.\ equities and ETFs with five-minute pre-averaging, the empirical analysis finds negative weekly and monthly median leverage estimates for every asset.

In concurrent and independent work, \citet{Guo2026} develop an It\^o-type formula for rough volatility and apply it to leverage estimation. Both studies characterize leverage between log prices and spot variance through the limit of the same scaled realized covariation. Our analysis focuses on centering errors under endogenous rough loadings and further establishes feasible inference in the presence of microstructure noise.

The remainder of the paper is organized as follows. Section~2 introduces the model, the leverage parameter, and the estimators. Section~3 develops the noise-free asymptotic theory, verifies the loading regularity condition, and establishes feasible inference and variance estimation. Section~4 extends the analysis to observations with microstructure noise and shows that a pilot estimate of the roughness index can be plugged in. Section~5 reports the simulation study, and Section~6 presents the empirical analysis. Proofs and additional numerical results are provided in the supplementary material.

%% file: sections/method.tex
\section{Methodology}
\subsection{Model Setup}
We work on a fixed time interval $[0,T]$. Let
$(\Omega,\mathcal F,(\mathcal F_t)_{t\in[0,T]},\mathbb P)$ be a filtered
probability space satisfying the usual conditions and supporting the log-price
process $X$ and the spot volatility process $\sigma$. The dynamics are
\begin{equation}\label{eq:model}
\begin{aligned}
&X_t=X_0+\int_0^t\mu_s\,ds+\int_0^t\sigma_s\,dW_s,\\
&\sigma_t^2=\sigma_0^2+\int_0^tf_s\,ds+\int_0^t(t-s)^{H-1/2}g_s\,ds
+K_H^{-1}\int_0^t(t-s)^{H-1/2}v_s\,dB_s.
\end{aligned}
\end{equation}
Here $W$ and $B$ are two correlated Brownian motions relative to
$(\mathcal F_t)$ satisfying
\[
d\langle W,B\rangle_t=\rho_t\,dt,\qquad |\rho_t|\le1.
\]
\begin{ass}\label{ass:model}
Equation~\eqref{eq:model} has a continuous, adapted, nonnegative variance
solution on $[0,T]$, with $\sigma_t=\sqrt{\sigma_t^2}$. The coefficient
processes $\mu,f,g,v,\rho$ are adapted, c\`adl\`ag and locally bounded.
\end{ass}
The Hurst parameter satisfies $H\in(0,1/2)$. We treat $H$ as known. The constant $K_H$ adopts a normalization matching the local increment scale of standard fractional Brownian motion (fBm), given by
\[
K_H=\frac{\Gamma(H+1/2)}{\sqrt{2H\sin(\pi H)\Gamma(2H)}}.
\]
For background on this representation, see, for example, \citep{Donati2011}.
We observe the log-price process $X$ at the regular times
\[
t_i=i\Delta_n,\qquad i=0,1,\ldots,n,\qquad \Delta_n=T/n.
\]
For any process $V$ and integers $i\ge k\ge1$, define
\[
\Delta_i^{n,k}V:=V_{t_i}-V_{t_{i-k}},\qquad
\Delta_i^nV:=\Delta_i^{n,1}V.
\]
\subsection{Scaled Realized Covariation and Leverage Parameter}
In the present model, $\Delta_i^nX$ and $\Delta_i^n\sigma^2$ are of order
$\Delta_n^{1/2}$ and $\Delta_n^H$, respectively, motivating the following scaled realized covariation.
\begin{definition}
We define the scaled realized covariation (scaled RCV) between $X$ and $\sigma^2$ by
\begin{equation}\label{eq:scaled-rcv}
T_n(t):=\sum_{i:t_i\le t}(\Delta_i^nX)
\frac{\Delta_i^n\sigma^2}{\Delta_n^{H-1/2}},\qquad 0\le t\le T.
\end{equation}
At the terminal time, write $T_n=T_n(T)$.
\end{definition}
Each summand in~\eqref{eq:scaled-rcv} is of order $\Delta_n$, so the statistic is of nondegenerate order over the fixed interval $[0,T]$. The scaled RCV uses latent spot volatility and provides a theoretical benchmark for identifying the continuous-time leverage parameter.
\begin{definition}
The parameter of the leverage effect is defined by
\begin{equation*}
\theta_t:=\frac{1}{K_H(H+1/2)}\int_0^t\sigma_sv_s\rho_s\,ds,
\qquad 0\le t\le T.
\end{equation*}
\end{definition}
\subsection{The Proposed Estimator}
In practice, the spot-volatility process is latent and must be estimated
from high-frequency returns. Let $\{k_n\}_{n\ge1}$ be a sequence of integers
satisfying
\[
k_n\to\infty,\qquad k_n\Delta_n\to0.
\]
For $j\ge k_n$, we use the following local realized-variation estimator
of spot variance:
\begin{equation}\label{eq:spot-rv}
\widehat\sigma_{t_j}^2=\frac1{k_n\Delta_n}
\sum_{l=-k_n+1}^{0}(\Delta_{j+l}^nX)^2.
\end{equation}
This estimator is evaluated at the coarse-grid
points $j=k_n,2k_n,\ldots$. The backward-looking block is used to avoid an
additional bias generated by the coupling between the Volterra memory in
$\Delta\widehat\sigma^2$ and the return increment.
\begin{definition}
Set $a_n:=1+2k_n^{-H-1/2}$. The proposed estimator is
\begin{equation}\label{eq:estimator}
\widehat\theta_n=\frac{2H+3}{2a_n}
\sum_{i=2}^{\lfloor T/(k_n\Delta_n)\rfloor}
(\Delta_{ik_n}^{n,k_n}X)
\frac{\Delta_{ik_n}^{n,k_n}\widehat\sigma^2}{(k_n\Delta_n)^{H-1/2}}.
\end{equation}
\end{definition}
The statistic $\widehat\theta_n$ is constructed from return increments
and estimated spot-volatility increments over blocks of length $k_n\Delta_n$.
The first coarse block provides the preceding spot-variance estimate, so the
sum starts at $i=2$. The factor $(2H+3)/2$ is the rough-scale analogue of the
factor $2$ in the classical semimartingale leverage estimator of
\citet{Wang2014}. At $H=1/2$, $K_H=1$, the fractional scaling disappears
and the block normalization becomes $2$; the finite-window factor $a_n$ tends to $1$.
Spot-volatility estimation error in~\eqref{eq:spot-rv} and rough-leverage fluctuations jointly determine the asymptotic variance.

%% file: sections/theory.tex
\section{Asymptotic Theory}
The local coefficient driving the leverage term is
\begin{equation*}
L_t:=\sigma_tv_t\rho_t.
\end{equation*}
\subsection{Asymptotics for the Scaled Realized Covariation}
Set
\begin{equation*}
U_i:=\Delta_i^nX\,\frac{\Delta_i^n\sigma^2}{\Delta_n^{H-1/2}}
-\frac{1}{K_H(H+1/2)}\int_{t_{i-1}}^{t_i}L_s\,ds.
\end{equation*}
Then the centered fluctuation can be written as
\[
\Delta_n^{-1/2}(T_n-\theta_T)=\sum_{i=1}^n\Delta_n^{-1/2}U_i.
\]
For $t_{i-1}\le s<t_i$, define
\begin{equation}\label{eq:cell-weight}
\Lambda_i(s)=\frac1{K_H}
\left\{\Delta_n^{1/2-H}(t_i-s)^{H-1/2}-\frac1{H+1/2}\right\}.
\end{equation}
Its integral over $[t_{i-1},t_i)$ is zero. Define the pathwise correction
\begin{equation*}
\widetilde P_n^L(t):=\Delta_n^{-1/2}\sum_{i:t_i\le t}
\int_{t_{i-1}}^{t_i}\Lambda_i(s)(L_s-L_{t_{i-1}})\,ds.
\end{equation*}
Process stable convergence is relative to $\mathcal F$ and is understood in $D([0,T])$ with the Skorokhod $J_1$ topology; u.c.p. denotes uniform convergence in probability on $[0,T]$.
\begin{thm}\label{thm:corrected-clt}
Under Assumption~\ref{ass:model},
\begin{equation*}
\begin{aligned}
&\Delta_n^{-1/2}(T_n(\cdot)-\theta_\cdot)-\widetilde P_n^L(\cdot)
\convst\int_0^\cdot\eta_s\,dZ_s,\\
&\eta_s^2=\sigma_s^2v_s^2+
\frac{L_s^2}{K_H^2(H+1/2)^2}.
\end{aligned}
\end{equation*}
Here $Z$ is a Brownian motion on an extension, independent of $\mathcal F$.
\end{thm}
To obtain the CLT centered at $\theta$, we impose a regularity
condition on $L$. Write $\|\cdot\|_p$ for the $L^p(\mathbb P)$ norm, using
Euclidean norms for vectors and Frobenius norms for matrices, and
$\operatorname{TV}(A;[0,T])$ for the total variation of a path $A$.
\begin{ass}\label{ass:lr}
There are stopping times $\tau_N$ with $\mathbb P(\tau_N\le T)\to0$ and
adapted c\`adl\`ag processes $\widetilde L^N$ such that
$\widetilde L_t^N=L_t$ for $t<\tau_N$ and
$\mathbb E\sup_{t\le T}|\widetilde L_t^N|^2<\infty$.
For every $N$, there is a finite-dimensional standard Brownian motion $\mathbf B^N$
with respect to $(\mathcal F_t)$ and a jointly measurable representation
\begin{equation}\label{eq:lr}
\widetilde L_t^N=A_t^{N,\mathrm{BV}}+A_t^{N,\mathrm{reg}}
+\int_0^t\psi_N(t,u)\,d\mathbf B_u^N.
\end{equation}
The processes $A^{N,\mathrm{BV}}$ and $A^{N,\mathrm{reg}}$ are adapted,
and $\psi_N(t,\cdot)$ is a predictable row-vector process for each $t$.
For some $H_N\in(0,1/2)$, $\zeta_N>1/2$ and $C_N<\infty$,
uniformly over the indicated times,
\begin{equation}\label{eq:lr-bounds}
\begin{aligned}
&\|\operatorname{TV}(A^{N,\mathrm{BV}};[0,T])\|_2\le C_N,\\
&\|A_t^{N,\mathrm{reg}}-A_s^{N,\mathrm{reg}}\|_2\le C_N|t-s|^{\zeta_N},
\qquad 0\le s<t\le T,\\
&\|\psi_N(t,u)\|_2\le C_N\{1+(t-u)^{H_N-1/2}\},\qquad 0\le u<t\le T,\\
&\|\psi_N(t,u)-\psi_N(s,u)\|_2
\le C_N\int_s^t\{1+(r-u)^{H_N-3/2}\}\,dr,
\quad 0\le u<s<t\le T.
\end{aligned}
\end{equation}
We refer to this condition as loading regularity, abbreviated \textbf{LR}.
\end{ass}
LR controls the accumulated variation of $L$ under the zero-integral weights
in~\eqref{eq:cell-weight}.
\begin{pro}\label{prop:lr}
Under Assumption~\ref{ass:model}, each of the following conditions implies LR.
\begin{enumerate}
\renewcommand{\theenumi}{\roman{enumi}}
\renewcommand{\labelenumi}{(\theenumi)}
\item\label{cond:volterra} Suppose
\begin{equation}\label{eq:direct-volterra}
L_t=L_0+A_t^L+K_H^{-1}
\int_0^t(t-u)^{H-1/2}\gamma_u^L\,d\mathbf B_u.
\end{equation}
Here $\mathbf B$ is an $m$-dimensional standard Brownian motion
with respect to $(\mathcal F_t)$, $\gamma^L$ is a predictable, locally bounded
$\mathbb R^{1\times m}$-valued process, and $A^L$ is continuous and adapted with $A_0^L=0$.
For some $\zeta\in(1/2,1]$, assume
\begin{equation}\label{eq:direct-regularity}
\sup_{0\le s<t\le T}\frac{|A_t^L-A_s^L|}{|t-s|^\zeta}<\infty
\quad\text{a.s.}
\end{equation}
\item\label{cond:closed-state} Let $\mathbf V\in\mathbb R^d$ and
$\mathbf b\in\mathbb R^\ell$ be driven by a finite-dimensional standard
Brownian motion $\mathbf B$ with respect to $(\mathcal F_t)$, with
\begin{equation}\label{eq:closed-state}
\begin{aligned}
&\begin{aligned}[t]
\mathbf V_t=&\mathbf V_0+\int_0^t\beta_0(\mathbf V_s)\,ds
+K_H^{-1}\int_0^t(t-s)^{H-1/2}\beta_H(\mathbf V_s)\,ds\\
&+K_H^{-1}\int_0^t(t-s)^{H-1/2}\Gamma_V(\mathbf V_s)\,d\mathbf B_s,
\end{aligned}\\
&\mathbf b_t=\mathbf b_0+\int_0^t\mu_b(\mathbf V_s,\mathbf b_s)\,ds
+\int_0^t\gamma_b(\mathbf V_s,\mathbf b_s)\,d\mathbf B_s,\\
&L_t=F(\mathbf V_t,\mathbf b_t).
\end{aligned}
\end{equation}
The deterministic coefficients and $F$ are locally $C^3$ on their open
domains. The state has finite, continuous paths remaining in these domains
throughout $[0,T]$. On exhausting compact sets, the fields admit $C_b^3$
extensions preserving the displayed dependence of the rough-state coefficients.
\item\label{cond:general-state} Let $H\in(1/6,1/2)$. There are stopping
times $\tau_N$ with $\mathbb P(\tau_N\le T)\to0$. For each $N$, let
$F_N\in C_b^4$ and define
\begin{equation*}
\begin{aligned}
&\mathbf V_t^N=\mathbf V_0^N+\mathbf A_t^{N,V}+K_H^{-1}\int_0^t(t-u)^{H-1/2}
\gamma_u^{N,V}\,d\mathbf B_u,\\
&\widetilde L_t^N=F_N(\mathbf V_t^N,\mathbf b_t^N).
\end{aligned}
\end{equation*}
Assume $\widetilde L_t^N=L_t$ for $t<\tau_N$. Without loss of generality, let $\mathbf A_0^{N,V}=0$, absorbing its initial value into $\mathbf V_0^N$.
Here $\mathbf B$ is an $m$-dimensional standard Brownian motion with respect
to $(\mathcal F_t)$, $\gamma^{N,V}$ is a predictable
$d\times m$ matrix process, and $\mathbf V^N,\mathbf A^{N,V}$ are adapted
$\mathbb R^d$-valued processes. The $\mathbb R^\ell$-valued continuous It\^o
process $\mathbf b^N$ is driven by $\mathbf B$. The process $\gamma^{N,V}$ and the
It\^o coefficients have uniformly bounded $L^p$ norms on $[0,T]$ for every
fixed finite $p\ge2$, with constants allowed to depend on $N,p$.
Define
\begin{equation*}
\Sigma_t^N:=\gamma_t^{N,V}(\gamma_t^{N,V})^\top.
\end{equation*}
For every fixed finite $p\ge2$ and $0\le s<t\le T$, assume
\begin{equation*}
\|\mathbf A_t^{N,V}-\mathbf A_s^{N,V}\|_p\le C_{N,p}|t-s|^{H+1/2},\qquad
\|\Sigma_t^N-\Sigma_s^N\|_p\le C_{N,p}|t-s|^H.
\end{equation*}
\end{enumerate}
\end{pro}
Conditions~(\ref{cond:volterra}) and~(\ref{cond:closed-state}) provide
complementary ways to verify LR for every $H\in(0,1/2)$.
Condition~(\ref{cond:volterra}) allows general predictable coefficients
in a direct Volterra representation of $L$ and covers rough Heston with nonnegative variance.
Condition~(\ref{cond:closed-state}) establishes LR for smooth functions of a closed rough state and It\^o coordinates, using the state equations to control the conditional expectation kernel.
Condition~(\ref{cond:general-state}) permits general adapted state coefficients
under the stated moment and increment bounds for $H>1/6$. All three conditions give $H_N=H$.

Under the basic model assumptions, correcting $\widetilde P_n^L$ requires
characterizing the within-cell dynamics of the leverage loading
$L_t=\sigma_tv_t\rho_t$. This may involve additional model information,
including the stochastic evolution of the volatility-of-volatility loading
$v_t$, making a general feasible correction difficult. Our conditional-projection
and kernel-cancellation arguments, together with the Malliavin analysis for smooth
functions of closed rough states, provide a framework for studying discretization
errors in other high-frequency statistics with endogenous rough loadings.
\begin{thm}\label{thm:rcv-clt}
Under Assumptions~\ref{ass:model} and~\ref{ass:lr},
\begin{equation*}
\begin{aligned}
&\widetilde P_n^L\convucp0,\\
&\Delta_n^{-1/2}(T_n(\cdot)-\theta_\cdot)\convst\int_0^\cdot\eta_s\,dZ_s.
\end{aligned}
\end{equation*}
\end{thm}
\subsection{Asymptotics for the Proposed Estimator}
We now consider the proposed estimator $\widehat\theta_n$ defined
in~\eqref{eq:estimator}. The local spot-volatility estimator uses
blocks of $k_n$ high-frequency increments, and the block size determines the
tradeoff between spot-volatility estimation error and the rough-leverage
fluctuation of $T_n$. We assume
\begin{equation}\label{eq:bandwidth}
\frac{k_n}{n^b}\longrightarrow c>0,\qquad 0<b<1,\qquad
\beta:=(1-2H+2Hb)\wedge(1-b).
\end{equation}
The unnormalized block sum has the finite-window bias
\begin{equation*}
R_n:=(a_n-1)\theta_T
=\frac{4k_n^{-H-1/2}}{K_H(2H+1)}\int_0^TL_s\,ds.
\end{equation*}
\begin{thm}\label{thm:feasible-clt}
Under Assumptions~\ref{ass:model} and~\ref{ass:lr} and the bandwidth
condition~\eqref{eq:bandwidth},
\begin{equation*}
n^{\beta/2}(\widehat\theta_n-\theta_T)
\convst\sqrt{\mathcal V_T}\,\mathcal Z,
\end{equation*}
where $\mathcal Z$ is standard normal and independent of $\mathcal F$
on an extension, and
\begin{equation}\label{eq:limit-variance}
\begin{aligned}
\mathcal V_T=&
c^{-2H}T^{1-2H}(2H+3)^2\int_0^T\sigma_s^6\,ds\,
\mathbf1_{\{b\le 2H/(2H+1)\}}\\
&+cT\frac{(2^{2H}-1)(2H+3)^2}{(2H+1)(2H+2)}
\int_0^T\sigma_s^2v_s^2\,ds\,
\mathbf1_{\{b\ge 2H/(2H+1)\}}\\
&+\frac{cT}{K_H^2(H+1/2)^2}\int_0^TL_s^2\,ds\,
\mathbf1_{\{b\ge 2H/(2H+1)\}}.
\end{aligned}
\end{equation}
\end{thm}
The deterministic factor $a_n$ in~\eqref{eq:estimator} depends only on the roughness parameter and block length and removes the proportional bias $R_n$, which can be non-negligible at the CLT scale. This bias arises from local spot-volatility smoothing, while the pathwise term $\widetilde P_n^L$ records within-cell loading variation.
The choice $b=2H/(2H+1)$ balances spot-volatility estimation error and
rough-leverage fluctuations, yielding the fastest convergence rate
$n^{-1/(4H+2)}$ within the bandwidth family considered.
\subsection{Estimation of Asymptotic Variance}
Following \citet[Definition~2 and Remark~9]{MyklandZhang2017}, we estimate
variance from the completed local contributions to $\widehat\theta_n$.
This single-scale construction captures spot-volatility estimation error and rough fluctuations jointly.

For $j=0,\ldots,n$, let
\begin{equation*}
\widehat\theta_{n,j}
=\frac{2H+3}{2a_n}(k_n\Delta_n)^{1/2-H}
\sum_{i=2}^{\lfloor j/k_n\rfloor}
\Delta_{ik_n}^{n,k_n}X\,
\Delta_{ik_n}^{n,k_n}\widehat\sigma^2.
\end{equation*}
An empty sum is zero, and $\widehat\theta_{n,n}=\widehat\theta_n$.
Extend these cumulative sums by
$\widehat\theta_{n,j+n}=\widehat\theta_{n,j}+\widehat\theta_n$.
For an integer $1\le B_n\le n/2$, define
\begin{equation*}
Q_n=\frac1{2B_n}\sum_{j=0}^{n-1}
\bigl(\widehat\theta_{n,j+2B_n}
      -2\widehat\theta_{n,j+B_n}+\widehat\theta_{n,j}\bigr)^2.
\end{equation*}
$Q_n$ is nonnegative and uses adjacent windows of $B_n$ observation intervals;
$k_n$ controls the point-estimation block size and $B_n$ the variance-estimation window.

For every fixed localization level $N$ in LR, assume
\begin{equation}\label{eq:q-bandwidth}
\frac{k_n}{B_n}\longrightarrow0,\qquad
B_n\Delta_n\longrightarrow0,\qquad
(B_n\Delta_n)^{1+2H_N}=o(n^{-\beta}).
\end{equation}
For example, $B_n=k_n\lceil\log(1+T/(k_n\Delta_n))\rceil$
satisfies these conditions for every fixed $H_N>0$.
\begin{pro}\label{prop:variance}
Under the assumptions of Theorem~\ref{thm:feasible-clt} and
\eqref{eq:q-bandwidth},
\begin{equation*}
n^\beta Q_n\convp\mathcal V_T,
\end{equation*}
where $\mathcal V_T$ is given in~\eqref{eq:limit-variance}.
If $\mathbb P(\mathcal V_T>0)=1$, then
\begin{equation*}
\frac{\widehat\theta_n-\theta_T}{\sqrt{Q_n}}\convst N(0,1).
\end{equation*}
For $0<\alpha<1$, an asymptotic $1-\alpha$ confidence interval for
$\theta_T$ is
\begin{equation*}
\left[\widehat\theta_n-z_{1-\alpha/2}\sqrt{Q_n},
      \widehat\theta_n+z_{1-\alpha/2}\sqrt{Q_n}\right],
\end{equation*}
where $z_q$ is the standard normal quantile.
\end{pro}
LR controls the contribution of the integrated target to adjacent-window
differences, allowing the quadratic form to estimate the variance of the centered fluctuation.
Set the ratio to zero on $\{Q_n=0\}$, whose probability tends to zero
under the stated positivity condition.

%% file: sections/noise.tex
\section{Microstructure Noise}
We extend the adjacent-block estimator to noisy observations through pre-averaging
and retain the single-scale variance construction.

\subsection{The Proposed Estimator}
Suppose that we observe $Y_i=X_{t_i}+\epsilon_i$, $i=0,\ldots,n$.
\begin{ass}\label{noisy:noise}
The observation errors are independent and identically distributed, and
independent of $\mathcal F$. Their common
law does not depend on $n$, and
\begin{align*}
\mathbb E\epsilon_0=0,\qquad
\mathbb E\epsilon_0^2=\nu\ge0,\qquad
\mathbb E|\epsilon_0|^8<\infty.
\end{align*}
\end{ass}
Define
\begin{equation}\label{noisy:pa}
\begin{aligned}
&R_j^n=\frac{c_{m_n}}{m_n}\sum_{a=0}^{m_n-1}
\{Y_{(j+1)m_n+a}-Y_{jm_n+a}\},\\
&c_{m_n}=\left(\frac{3m_n^2}{2m_n^2+1}\right)^{1/2},
\end{aligned}
\end{equation}
and the local quadratic averages used for spot volatility estimation
\begin{equation*}
\widehat\sigma_{t_{ik_nm_n}}^{2,\mathrm N}
=\frac1{k_nm_n\Delta_n}\sum_{j=(i-1)k_n}^{ik_n-1}(R_j^n)^2.
\end{equation*}
Each average contains the common noise mean
$2c_{m_n}^2\nu/(m_n^2\Delta_n)$, which cancels in adjacent differences.
Let
\begin{equation}\label{noisy:indices}
\mathcal I_n^{\mathrm N}=\{i\ge2:(ik_n+1)m_n-1\le n\}.
\end{equation}
For $x\in\mathbb R$, define
\begin{equation}\label{noisy:correction-kernels}
\begin{aligned}
&g_{m_n}(x)=\frac{c_{m_n}}{m_n}\sum_{a=0}^{m_n-1}
\mathbf1_{(a/m_n,1+a/m_n]}(x),\\
&p_{k_n}(x)=\sum_{r=0}^{k_n-1}g_{m_n}(x-r).
\end{aligned}
\end{equation}
The deterministic correction is
\begin{equation}\label{noisy:correction-factor}
\begin{aligned}
a_n^{\mathrm N}=&(H+1/2)(H+3/2)k_n^{-H-3/2}
\sum_{j=-k_n}^{k_n-1}(2\mathbf1_{\{j\ge0\}}-1)\\
&\times\int_{\mathbb R}\int_s^\infty
\{p_{k_n}(s)g_{m_n}(t-j)^2+2p_{k_n}(t)g_{m_n}(s-j)g_{m_n}(t-j)\}
(t-s)^{H-1/2}\,dt\,ds.
\end{aligned}
\end{equation}
It depends only on $H,k_n,m_n$ and satisfies
\begin{equation}\label{noisy:calibration-properties}
a_n^{\mathrm N}=c_{m_n}+O_H(k_n^{-H-1/2}),\qquad
a_n^{\mathrm N}\big|_{m_n=1}=1+2k_n^{-H-1/2}=a_n.
\end{equation}
In particular, $a_n^{\mathrm N}>0$ for all sufficiently large $n$.
\begin{definition}
The noisy estimator is
\begin{equation*}
\widehat\theta_n^{\mathrm N}
=\frac{2H+3}{2a_n^{\mathrm N}}(k_nm_n\Delta_n)^{1/2-H}
\sum_{i\in\mathcal I_n^{\mathrm N}}
\Bigl(\sum_{j=(i-1)k_n}^{ik_n-1}R_j^n\Bigr)
\bigl(\widehat\sigma_{t_{ik_nm_n}}^{2,\mathrm N}
      -\widehat\sigma_{t_{(i-1)k_nm_n}}^{2,\mathrm N}\bigr).
\end{equation*}
\end{definition}
Here $a_n^{\mathrm N}$ is the finite-window response of the leverage term to
pre-averaging and blockwise spot-volatility smoothing: the unnormalized block sum
has first-order asymptotic center $a_n^{\mathrm N}\theta_T$.
Dividing by the coefficient calculated for the actual $k_n,m_n$ corrects this
proportional shift; using only its limit can leave a non-negligible bias at the CLT scale.
The estimator attenuates observation noise through
pre-averaging and extracts leverage by multiplying block return sums and adjacent
differences of local quadratic averages.

\subsection{Asymptotic Distribution}
For $0<b<1$, assume
\begin{equation}\label{noisy:bandwidth}
\frac{m_n}{\sqrt n}\longrightarrow\lambda>0,\qquad
\frac{k_n}{(n/m_n)^b}\longrightarrow c>0,\qquad
\beta:=(1-2H+2Hb)\wedge(1-b).
\end{equation}
\begin{thm}\label{noisy:clt}
Under Assumptions~\ref{ass:model}, \ref{ass:lr} and~\ref{noisy:noise},
and~\eqref{noisy:bandwidth},
\begin{align*}
n^{\beta/4}(\widehat\theta_n^{\mathrm N}-\theta_T)
\convst\sqrt{\mathcal V_T^{\mathrm N}}\,\mathcal Z,
\end{align*}
where $\mathcal Z$ is standard normal and independent of $\mathcal F$
on an extension, and
\begin{equation}\label{noisy:limit-variance}
\begin{aligned}
\mathcal V_T^{\mathrm N}=&
c^{-2H}\lambda^{1-2H+2Hb}T^{1-2H}(2H+3)^2
\mathbf1_{\{b\le2H/(2H+1)\}}\\
&\times\int_0^T\left\{\frac98\sigma_s^6
 +\frac{9\nu}{2\lambda^2T}\sigma_s^4
 +\frac{27\nu^2}{2\lambda^4T^2}\sigma_s^2\right\}ds\\
&+c\lambda^{1-b}T
\frac{(2^{2H}-1)(2H+3)^2}{(2H+1)(2H+2)}
\int_0^T\sigma_s^2v_s^2\,ds\,
\mathbf1_{\{b\ge2H/(2H+1)\}}\\
&+\frac{c\lambda^{1-b}T}{K_H^2(H+1/2)^2}
\int_0^TL_s^2\,ds\,
\mathbf1_{\{b\ge2H/(2H+1)\}}.
\end{aligned}
\end{equation}
\end{thm}
The choice $b=2H/(2H+1)$ balances sampling error and rough-volatility
fluctuations, giving rate $n^{-1/[4(2H+1)]}$. Relative to noise-free
observations, pre-averaging reduces the effective number of observations
to order $\sqrt n$.

\subsection{Estimation of Asymptotic Variance}
Use the same cumulative-estimator construction as in Section~3.
Let $N_n=\lceil n/m_n\rceil$. For $j=0,\ldots,N_n$, define
\begin{equation*}
\widehat\theta_{n,j}^{\mathrm N}
=\frac{2H+3}{2a_n^{\mathrm N}}(k_nm_n\Delta_n)^{1/2-H}
\sum_{\substack{i\in\mathcal I_n^{\mathrm N}\\
                 (ik_n+1)m_n-1\le jm_n}}
\Bigl(\sum_{r=(i-1)k_n}^{ik_n-1}R_r^n\Bigr)
\bigl(\widehat\sigma_{t_{ik_nm_n}}^{2,\mathrm N}
      -\widehat\sigma_{t_{(i-1)k_nm_n}}^{2,\mathrm N}\bigr).
\end{equation*}
The summation includes exactly the blocks whose required observations are
available by $jm_n\Delta_n$.
Extend it additively by
$\widehat\theta_{n,j+N_n}^{\mathrm N}
=\widehat\theta_{n,j}^{\mathrm N}+\widehat\theta_n^{\mathrm N}$.
For an integer $1\le B_n\le N_n/2$, define
\begin{equation}\label{noisy:Q}
Q_n^{\mathrm N}
=\frac1{2B_n}\sum_{j=0}^{N_n-1}
\left(\widehat\theta_{n,j+2B_n}^{\mathrm N}
      -2\widehat\theta_{n,j+B_n}^{\mathrm N}
      +\widehat\theta_{n,j}^{\mathrm N}\right)^2.
\end{equation}
Here $B_n$ counts grid cells of length $m_n\Delta_n$. For every fixed
localization level $N$ in LR, assume
\begin{equation}\label{noisy:Q-bandwidth}
\frac{k_n}{B_n}\longrightarrow0,\qquad
B_nm_n\Delta_n\longrightarrow0,\qquad
(B_nm_n\Delta_n)^{1+2H_N}=o(n^{-\beta/2}).
\end{equation}
\begin{pro}\label{noisy:variance}
Under the assumptions of Theorem~\ref{noisy:clt} and
\eqref{noisy:Q-bandwidth},
\begin{align*}
n^{\beta/2}Q_n^{\mathrm N}\convp\mathcal V_T^{\mathrm N}.
\end{align*}
If $\mathbb P(\mathcal V_T^{\mathrm N}>0)=1$, then
\begin{align*}
\frac{\widehat\theta_n^{\mathrm N}-\theta_T}
     {\sqrt{Q_n^{\mathrm N}}}\convst N(0,1).
\end{align*}
An asymptotic $1-\alpha$ confidence interval for $\theta_T$ is
\begin{align*}
\left[\widehat\theta_n^{\mathrm N}-z_{1-\alpha/2}\sqrt{Q_n^{\mathrm N}},
      \widehat\theta_n^{\mathrm N}+z_{1-\alpha/2}\sqrt{Q_n^{\mathrm N}}\right].
\end{align*}
\end{pro}

\begin{remark}\label{rem:noisy-plugin}
Retain a deterministic pre-averaging sequence with $m_n/\sqrt n\to\lambda>0$.
Let $\widehat H_n\in(0,1/2)$ satisfy
$|\widehat H_n-H|=O_p(n^{-C})$ for some $C>0$.
For block selection, set $q_n=(\log n)^{-2}$ and let
$\overline H_n=q_n\{\lfloor\widehat H_n/q_n+U\rfloor-U\}$,
truncated to $[0,1/2]$, where a single $U$ is uniform on $[0,1]$ and
independent of the sigma-field generated by $\mathcal F$, all observation
errors and the pilot randomization.
For fixed $\delta\ge0$, take
\begin{equation*}
\begin{aligned}
&b(u)=\frac{2u}{1+2u},\qquad
\widehat k_n=\left\lceil c(n/m_n)^{b(\overline H_n+\delta)}\right\rceil,\\
&\beta_\delta=1-b(H+\delta).
\end{aligned}
\end{equation*}
Use a deterministic rule $B_n=B_n(\widehat k_n)$ satisfying
\eqref{noisy:Q-bandwidth} for bandwidth sequences of this asymptotic order,
with $\beta=\beta_\delta$.
Under the model, noise and LR assumptions of Proposition~\ref{noisy:variance},
replacing $H$ by $\widehat H_n$ throughout $\widehat\theta_n^{\mathrm N}$
and $Q_n^{\mathrm N}$, including the finite-window correction,
preserves the feasible CLT whenever
\begin{equation*}
C>\frac{\beta_\delta}{4},\qquad\text{equivalently}\qquad
\delta>\frac1{8C}-H-\frac12,
\end{equation*}
provided that the limiting variance at $b=b(H+\delta)$ is almost surely positive.
The grid value $\overline H_n$ and margin $\delta$ select the block length;
the estimator and its correction use $\widehat H_n$.
In particular, if $|\widehat H_n-H|=O_p(n^{-1/[4(2H+1)]})$,
any fixed $\delta>0$ suffices, at rate $n^{-\beta_\delta/4}$.
\end{remark}

Sections~3 and~4 combine finite-window normalization and single-scale variance
estimation to provide inference from observed prices in models with endogenous
rough loadings. This construction brings spot-volatility estimation error, rough
fluctuations, and observation noise into a common variance-estimation framework,
providing standard errors and confidence intervals computed from observed prices
for the same leverage parameter under both observation schemes.

%% file: sections/sim.tex
\section{Simulation Study}

In this section, we study the finite-sample behavior of the proposed estimator under the rough Heston model.
The simulations are carried out over a grid of Hurst parameters
\[
H\in \mathcal{H}:=\{0.05,\,0.10,\,0.15,\,0.20,\,0.25,\,0.30,\,0.35,\,0.40,\,0.45\}.
\]
Our focus is on three questions: whether observable standardization delivers stable calibration, whether the proposed variance estimator captures the relevant fluctuation scale, and how the approximation quality varies across different values of $H$.

\subsection{Feasible Inference}
We consider the rough Heston model
\begin{equation}\label{eq:rough-heston}
\begin{aligned}
&dX_t = \left(r-\frac 12\sigma^2_t\right)dt+ \sigma_t\,dW_t,\\
&\sigma_t^2
		=
		\sigma_0^2
		+
		\frac{1}{\Gamma(H+\frac12)}
		\int_0^t
		(t-s)^{H-\frac12}
		\Bigl(
		\kappa(\vartheta-\sigma_s^2)\,ds
		+
		v_s\,dB_s
		\Bigr),
\end{aligned}
\end{equation}
with
\[
d\langle W,B\rangle_t=\rho\,dt,\qquad v_t=\gamma\sigma_t.
\]
The normalization in \eqref{eq:rough-heston} is the common simulation convention.
It is equivalent to the theoretical normalization in \eqref{eq:model} under the deterministic rescaling
\[
v_t^{\mathrm{th}}=\frac{K_H}{\Gamma(H+1/2)}v_t^{\mathrm{sim}}.
\]
The targets, corrections, and variance constants are computed with this effective theoretical loading.
In the rough Heston model, $L_t$ is proportional to $\sigma_t^2$, so LR follows from
Proposition~\ref{prop:lr}(\ref{cond:volterra}).

In the baseline experiment, we use $5000$ paths with
\[
\begin{aligned}
&\sigma_0^2=\vartheta=0.02,\qquad \kappa=5,\qquad \rho=-0.7,\\
&T=\frac{1}{252},\qquad n=23400,\qquad u=10,\qquad n_{\mathrm{fine}}=un.
\end{aligned}
\]
Here $u$ denotes the upsampling factor used to generate the fine simulation grid.

The finite-sampling bias derived in our theory arises from the discrepancy
between local volatility constructed from discrete squared returns and its
continuous-time local average, together with the dependence of this discrepancy
on returns. To capture the within-interval dynamics underlying this bias, we use
upsampling: paths are generated on the fine grid and then sampled on the observation
grid to compute the feasible estimator
$\widehat\theta_n$ and the variance estimator $Q_n$.
The numerical approximation $\theta_T^{\mathrm{fine}}$ to the continuous-time
target is computed on the fine grid.
For each path, the feasible estimator is
\[
\begin{aligned}
&\widehat{\theta}_n
=
\frac{2H+3}{2a_n}
\sum_{i=2}^{\lfloor T/(k_n\Delta_n)\rfloor}
(\Delta_{ik_n}^{n,k_n}X)
\frac{\Delta_{ik_n}^{n,k_n}\widehat{\sigma}^2}{(k_n\Delta_n)^{H-\frac12}},\\
&\Delta_n=\frac{T}{n},
\end{aligned}
\]
where $\widehat{\sigma}^2$ is the rolling spot-volatility estimator and
$k_n\asymp n^{2H/(2H+1)}$. The resulting studentized statistic is reported as in
Proposition~\ref{prop:variance}.

For noisy observations, we use
\[
Y_i=X_{t_i}+\epsilon_i,\qquad i=0,\ldots,n,
\]
where
\[
\epsilon_i\overset{\mathrm{iid}}{\sim}N(0,\nu),\qquad
\{\epsilon_i\}_{i=0}^n\perp\mathcal F,\qquad \nu=(0.0005)^2.
\]
We take $m_n\asymp\sqrt n$ and $k_n\asymp(n/m_n)^{2H/(2H+1)}$;
the remaining model parameters and simulation grid are unchanged.
\begin{table}[!htbp]
\centering
\normalsize
\linespread{1.6}\selectfont
\setlength{\tabcolsep}{4pt}
\caption{Rough Heston: empirical means, variances, and 95\% coverage without noise and with observation noise.}
\label{tab:simulation}
\begin{tabular}{c|rrr|rrr}
\toprule
$H$ & \multicolumn{3}{c|}{Noise-free} & \multicolumn{3}{c}{Noisy} \\
\cmidrule(lr){2-4}\cmidrule(lr){5-7}
 & Mean $Z$ & Var. $Z$ & Cov. 95\% & Mean $Z$ & Var. $Z$ & Cov. 95\% \\
\midrule
0.05 & 0.053 & 0.986 & 0.9508 & 0.053 & 0.986 & 0.9506 \\
0.10 & 0.020 & 1.023 & 0.9464 & 0.013 & 0.997 & 0.9520 \\
0.15 & 0.010 & 0.983 & 0.9498 & 0.011 & 0.998 & 0.9528 \\
0.20 & 0.019 & 1.000 & 0.9508 & 0.034 & 1.046 & 0.9464 \\
0.25 & -0.002 & 1.003 & 0.9488 & 0.042 & 1.049 & 0.9442 \\
0.30 & 0.033 & 1.026 & 0.9458 & 0.026 & 1.027 & 0.9516 \\
0.35 & 0.018 & 0.999 & 0.9570 & 0.047 & 1.053 & 0.9454 \\
0.40 & 0.027 & 1.061 & 0.9422 & 0.052 & 1.085 & 0.9426 \\
0.45 & 0.001 & 1.002 & 0.9520 & 0.046 & 1.066 & 0.9496 \\
\bottomrule
\end{tabular}
\end{table}
\begin{figure}[!htbp]
\centering
\begin{subfigure}[t]{0.32\textwidth}
\includegraphics[width=\textwidth]{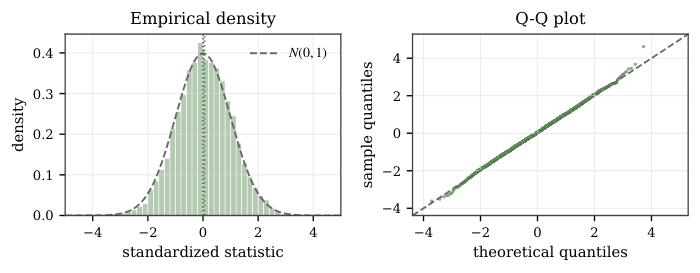}
\caption{$H=0.05$}
\end{subfigure}
\hfill
\begin{subfigure}[t]{0.32\textwidth}
\includegraphics[width=\textwidth]{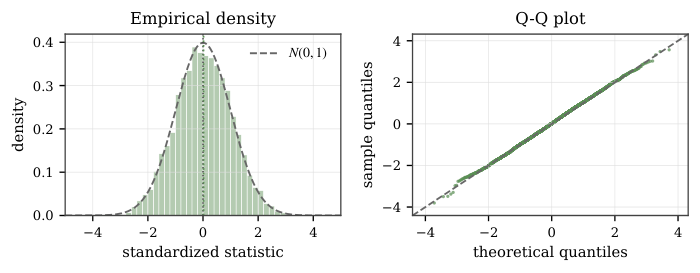}
\caption{$H=0.10$}
\end{subfigure}
\hfill
\begin{subfigure}[t]{0.32\textwidth}
\includegraphics[width=\textwidth]{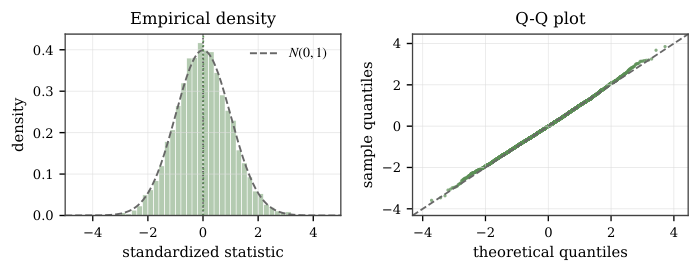}
\caption{$H=0.15$}
\end{subfigure}

\begin{subfigure}[t]{0.32\textwidth}
\includegraphics[width=\textwidth]{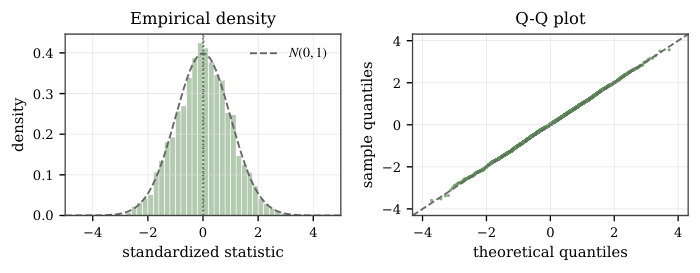}
\caption{$H=0.20$}
\end{subfigure}
\hfill
\begin{subfigure}[t]{0.32\textwidth}
\includegraphics[width=\textwidth]{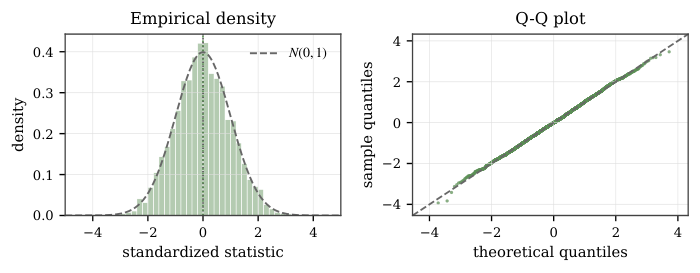}
\caption{$H=0.25$}
\end{subfigure}
\hfill
\begin{subfigure}[t]{0.32\textwidth}
\includegraphics[width=\textwidth]{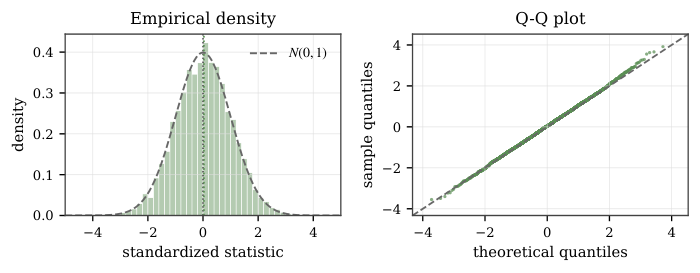}
\caption{$H=0.30$}
\end{subfigure}

\begin{subfigure}[t]{0.32\textwidth}
\includegraphics[width=\textwidth]{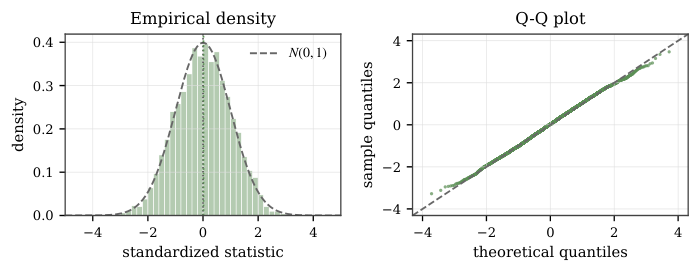}
\caption{$H=0.35$}
\end{subfigure}
\hfill
\begin{subfigure}[t]{0.32\textwidth}
\includegraphics[width=\textwidth]{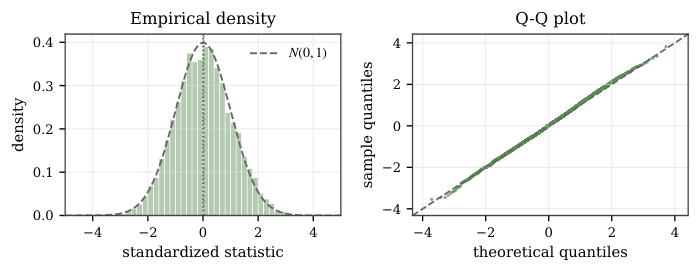}
\caption{$H=0.40$}
\end{subfigure}
\hfill
\begin{subfigure}[t]{0.32\textwidth}
\includegraphics[width=\textwidth]{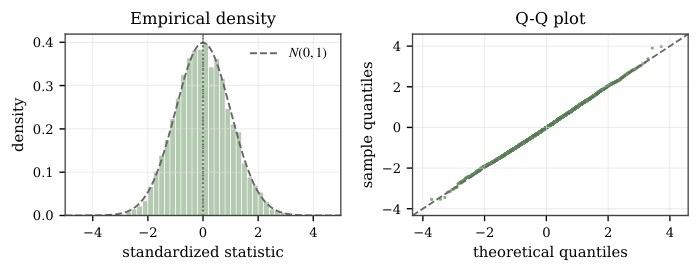}
\caption{$H=0.45$}
\end{subfigure}
\caption{Rough Heston without observation noise: densities of $Z$ and normal QQ plots at the nine values of $H$.}
\label{fig:simulation}
\end{figure}

\begin{figure}[!htbp]
\centering
\begin{subfigure}[t]{0.32\textwidth}
\includegraphics[width=\textwidth]{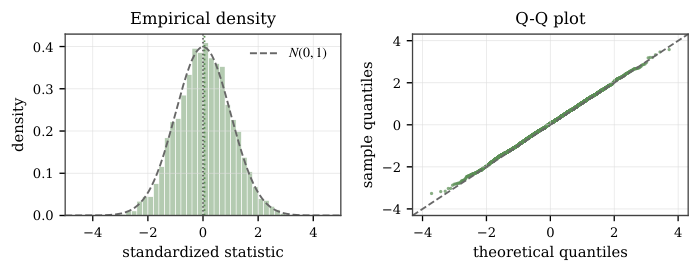}
\caption{$H=0.05$}
\end{subfigure}
\hfill
\begin{subfigure}[t]{0.32\textwidth}
\includegraphics[width=\textwidth]{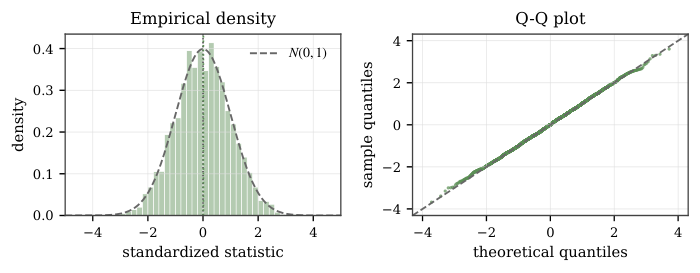}
\caption{$H=0.10$}
\end{subfigure}
\hfill
\begin{subfigure}[t]{0.32\textwidth}
\includegraphics[width=\textwidth]{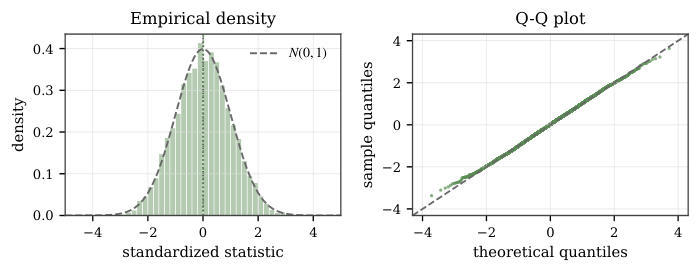}
\caption{$H=0.15$}
\end{subfigure}

\begin{subfigure}[t]{0.32\textwidth}
\includegraphics[width=\textwidth]{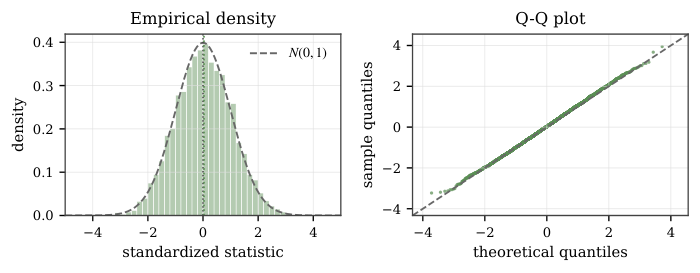}
\caption{$H=0.20$}
\end{subfigure}
\hfill
\begin{subfigure}[t]{0.32\textwidth}
\includegraphics[width=\textwidth]{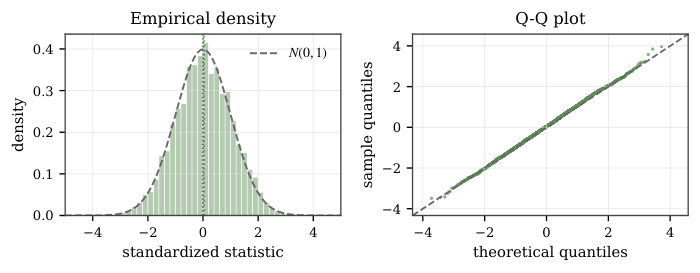}
\caption{$H=0.25$}
\end{subfigure}
\hfill
\begin{subfigure}[t]{0.32\textwidth}
\includegraphics[width=\textwidth]{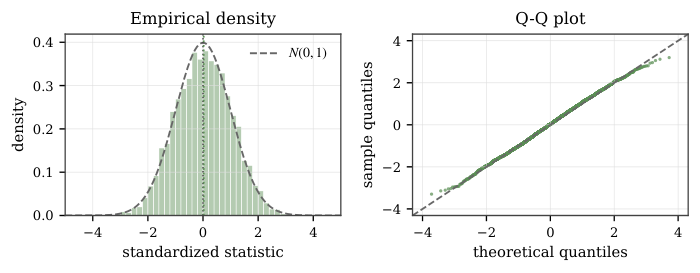}
\caption{$H=0.30$}
\end{subfigure}

\begin{subfigure}[t]{0.32\textwidth}
\includegraphics[width=\textwidth]{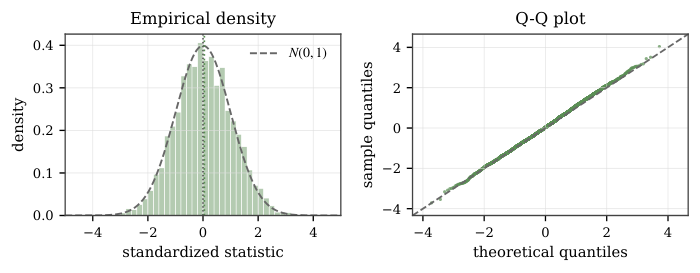}
\caption{$H=0.35$}
\end{subfigure}
\hfill
\begin{subfigure}[t]{0.32\textwidth}
\includegraphics[width=\textwidth]{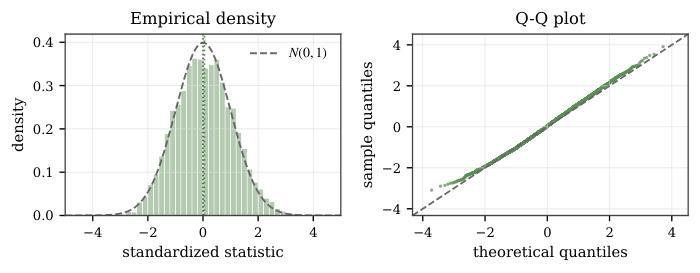}
\caption{$H=0.40$}
\end{subfigure}
\hfill
\begin{subfigure}[t]{0.32\textwidth}
\includegraphics[width=\textwidth]{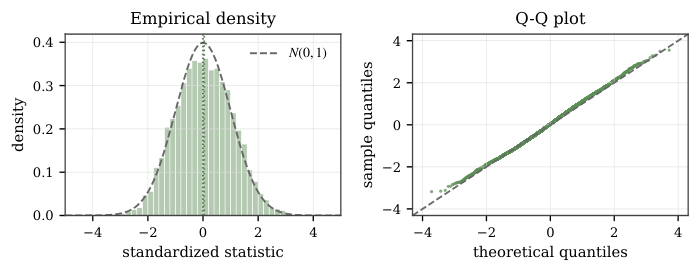}
\caption{$H=0.45$}
\end{subfigure}
\caption{Rough Heston with observation noise: densities of $Z$ and normal QQ plots at the nine values of $H$.}
\label{fig:simulation-noisy}
\end{figure}

Table~\ref{tab:simulation} reports the main numerical diagnostics for all
values of $H$. Figures~\ref{fig:simulation} and~\ref{fig:simulation-noisy} display
the corresponding noise-free and noisy studentized statistics.
The densities and Q--Q plots broadly follow the standard normal benchmark.
The empirical means remain close to zero, with modest variance inflation
in some noisy cases.
The 95\% coverages remain close to their nominal level, supporting the normal
approximation in Propositions~\ref{prop:variance} and~\ref{noisy:variance}
in this experimental setting.

\subsection{Validation of \texorpdfstring{$R_n$}{Rn}}
We finally examine the finite-sample role of the block-bias correction
\(R_n\) directly. To make the correction term visible, we run a separate
diagnostic experiment with a longer horizon and stronger negative leverage:
\[
H=0.05,\qquad
T=\frac{5}{252},\qquad
\rho=-0.95,\qquad
n=23400,\qquad
u=10,
\]
using \(5000\) independent paths. We compare
the uncorrected and corrected standardized statistics
\[
Z_n^{0}=\frac{a_n\widehat{\theta}_n-\theta_T^{\mathrm{fine}}}{a_n\sqrt{Q_n}},\qquad
Z_n^{R}=\frac{\widehat{\theta}_n-\theta_T^{\mathrm{fine}}}{\sqrt{Q_n}}.
\]
For noisy data, the corresponding statistics use $\widehat\theta_n^{\mathrm N}$,
$a_n^{\mathrm N}$, and $Q_n^{\mathrm N}$ in these definitions.
The empirical mean of $Z_n^0-Z_n^R$ measures the standardized correction;
without noise it equals $R_n/(a_n\sqrt{Q_n})$.

\begin{figure}[!htbp]
\centering
\begin{minipage}{\textwidth}
\centering
\normalsize
\linespread{1.6}\selectfont
\setlength{\tabcolsep}{4pt}
\captionof{table}{Finite-sample effect of \(R_n\) on the standardized feasible statistic.}
		\label{tab:bias}
		\begin{tabular}{lrrrrr}
			\toprule
				Data & Mean \(Z_n^{0}\)
				& Std. \(Z_n^{0}\)
				& Mean \(Z_n^0-Z_n^R\)
				& Mean \(Z_n^{R}\)
			& Std. \(Z_n^{R}\) \\
			\midrule
Noise-free & -0.497 & 0.985 & -0.592 & 0.095 & 0.988 \\
Noisy & -0.102 & 0.984 & -0.156 & 0.053 & 0.985 \\
			\bottomrule
		\end{tabular}
\end{minipage}

\begin{subfigure}[t]{0.48\textwidth}
\includegraphics[width=\textwidth]{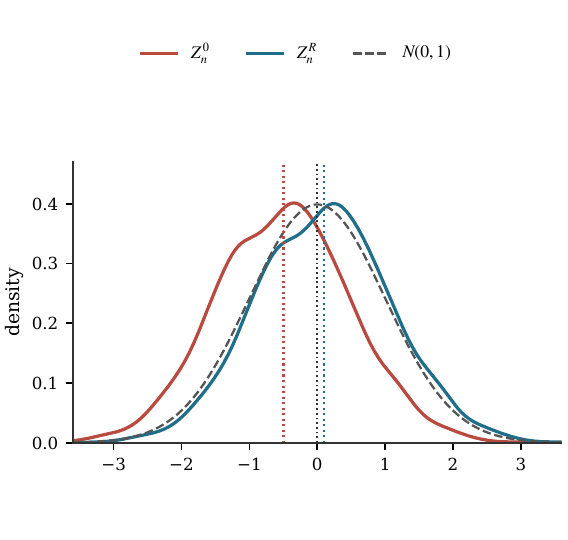}
\caption{Noise-free}
\end{subfigure}
\hfill
\begin{subfigure}[t]{0.48\textwidth}
\includegraphics[width=\textwidth]{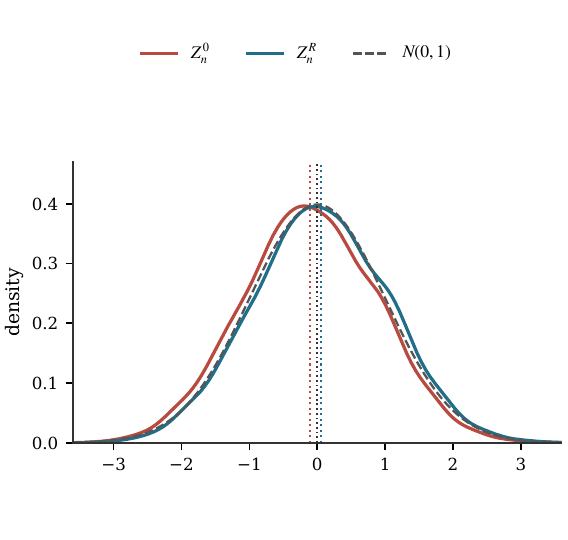}
\caption{Noisy}
\end{subfigure}
\caption{Empirical densities of $Z_n^0$ and $Z_n^R$ without and with observation noise.
Dotted lines mark zero (black) and empirical means (colored);
the dashed curve is the standard normal density.}
\label{fig:bias}
\end{figure}

Table~\ref{tab:bias} and
Figure~\ref{fig:bias} illustrate the finite-sample impact of
the block term \(R_n\) on the standardized statistic. The
standardized mean changes from $-0.497$ to $0.095$ without noise
and from $-0.102$ to $0.053$ with observation noise.
Under negative leverage, the finite-block term shifts the estimate further below zero and exaggerates leverage magnitude. The correction removes this sampling-induced component, aligning the estimated return--spot-volatility dependence with the continuous-time parameter.

%% file: sections/emp.tex
\section{Empirical Study}

We use high-frequency data for nine U.S. equities and ETFs over 2011--2025:
AAPL, AMZN, DIA, IWM, MSFT, NVDA, QQQ, SPY, and TSLA.
For each asset, we obtain $\widehat H_n$ using a variogram-based procedure following \citet{Gatheral2018}.
Jump days are identified using the pre-averaging test of \citet{AitSahaliaJacodLi2012},
which extends \citet{AitSahaliaJacod2009} to account for microstructure noise.
We apply the universal threshold of \citet{Bajgrowicz2016} to reduce spurious detections
and exclude the identified jump days.

Using filtered 1-second observed returns, we apply fixed five-minute pre-averaging to compute the noisy estimator $\widehat\theta_n^{\mathrm N}$ and its single-scale variance estimator $Q_n^{\mathrm N}$ over weekly and monthly windows. Time is measured in years, with 252 trading days per year. For a window with $n$ retained returns, $T$ is the number of retained trading days divided by 252, and we set
\[
\Delta_n=\frac{T}{n},\qquad m_n=300,\qquad
\widehat k_n=\left\lceil\left(\frac{n}{m_n}\right)^{\frac{2\widehat H_n}{2\widehat H_n+1}}\right\rceil.
\]
Both choices of $H$ use exact finite-window correction, with 95\% confidence intervals
\[
\left[\widehat\theta_n^{\mathrm N}-1.96\sqrt{Q_n^{\mathrm N}},\quad
\widehat\theta_n^{\mathrm N}+1.96\sqrt{Q_n^{\mathrm N}}\right].
\]
The fitted $\widehat H_n$ is held fixed when constructing intervals and comparing sampling frequencies; $H=1/2$ uses the same construction with its corresponding $\widehat k_n$ as a semimartingale-scale reference.

\subsection{Main Results}
The fitted roughness parameters range from $0.147$ to $0.206$, well below $1/2$. In the resulting samples, estimates using $\widehat H_n$ have negative means and medians for every asset at both horizons, and negative estimates account for 51.7\%--79.8\% of monthly windows and 52.2\%--71.7\% of weekly windows.
In these monthly samples, the proportion of negative estimates is higher under $H=1/2$ for every asset, whereas comparisons of significantly negative proportions vary across assets and horizons.

SPY and QQQ have the largest proportions of negative estimates at both horizons, whereas TSLA and NVDA have the most negative medians; the medians for AAPL, DIA, and SPY are closer to zero. For every asset, the proportion of 95\% confidence intervals entirely below zero is higher for monthly than weekly windows.

\begin{table}[!htbp]
\centering
\normalsize
\linespread{1.6}\selectfont
\setlength{\tabcolsep}{4pt}
\caption{Summary statistics of fixed-five-minute pre-averaging rough leverage estimates.}
\label{tab:empirical}
\begin{tabular}{@{}lrrrrrrrrr@{}}
\toprule
Statistic & AAPL & AMZN & DIA & IWM & MSFT & NVDA & QQQ & SPY & TSLA \\
\midrule
$\widehat H_n$ & 0.153 & 0.181 & 0.204 & 0.206 & 0.179 & 0.147 & 0.184 & 0.200 & 0.150 \\
\midrule
\multicolumn{10}{l}{\textbf{A: Monthly windows}} \\
Median & -0.09 & -1.15 & -0.29 & -0.92 & -0.53 & -2.05 & -0.83 & -0.36 & -3.14 \\
Mean & -0.70 & -5.63 & -0.76 & -1.56 & -1.75 & -4.70 & -2.14 & -0.65 & -27.77 \\
SD & 13.71 & 30.27 & 3.42 & 8.64 & 14.37 & 42.17 & 11.79 & 4.18 & 140.37 \\
Neg. & 51.7 & 53.9 & 71.9 & 69.1 & 55.6 & 57.9 & 76.7 & 79.8 & 55.6 \\
\midrule
\multicolumn{10}{l}{\textbf{B: Weekly windows}} \\
Median & -0.05 & -0.12 & -0.05 & -0.22 & -0.11 & -0.44 & -0.15 & -0.06 & -0.39 \\
Mean & -1.23 & -0.84 & -0.16 & -0.63 & -0.45 & -1.64 & -0.91 & -0.22 & -4.76 \\
SD & 15.56 & 12.36 & 3.41 & 4.83 & 7.92 & 24.18 & 12.91 & 2.92 & 60.91 \\
Neg. & 52.2 & 52.5 & 64.3 & 67.0 & 53.9 & 53.8 & 71.7 & 71.7 & 52.9 \\
\bottomrule
\end{tabular}

\begin{minipage}{\textwidth}
\normalsize
\linespread{1}\selectfont
\raggedright
Note: Median, mean, and standard deviation (SD) are in units of $10^{-4}$. Neg. is the percentage of negative estimates.
\end{minipage}
\end{table}

The AMZN leverage series cover 2011--2025.
The top and bottom rows report monthly and weekly estimates; the left and right columns use $\widehat H_n$ and $H=1/2$, respectively.

\begin{figure}[!htbp]
	\centering

	\begin{minipage}[t]{0.48\textwidth}
		\centering
		\includegraphics[width=\textwidth]{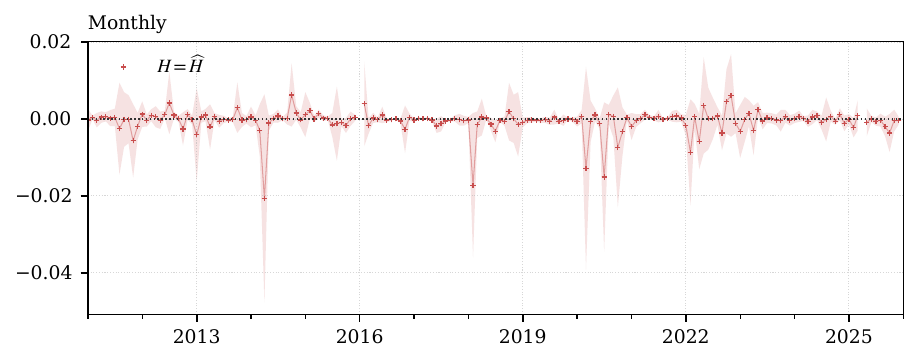}
	\end{minipage}
	\hfill
	\begin{minipage}[t]{0.48\textwidth}
		\centering
		\includegraphics[width=\textwidth]{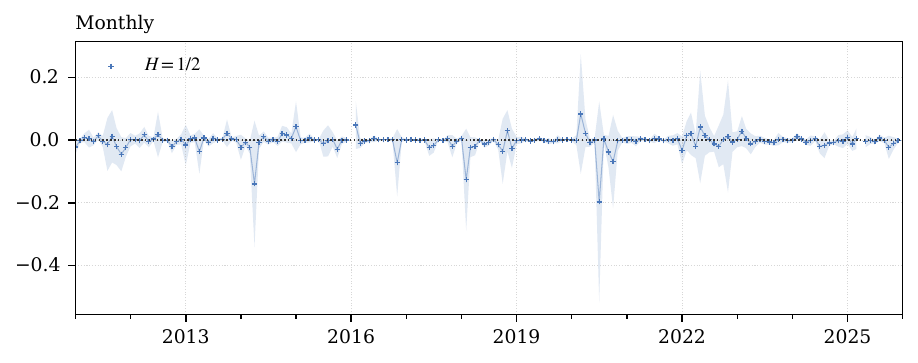}
	\end{minipage}

	\begin{minipage}[t]{0.48\textwidth}
		\centering
		\includegraphics[width=\textwidth]{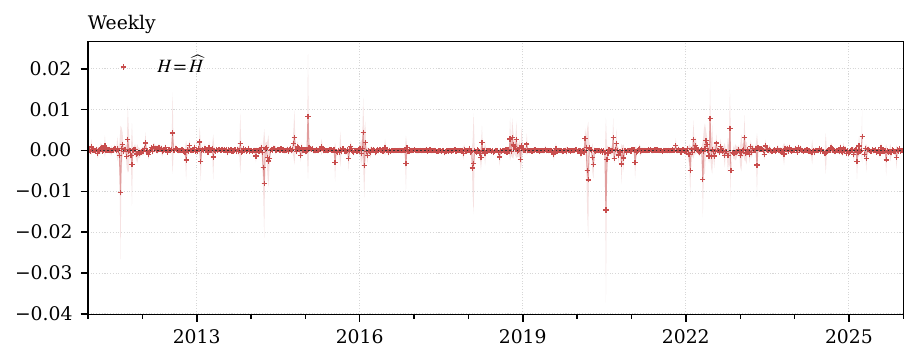}
	\end{minipage}
	\hfill
	\begin{minipage}[t]{0.48\textwidth}
		\centering
		\includegraphics[width=\textwidth]{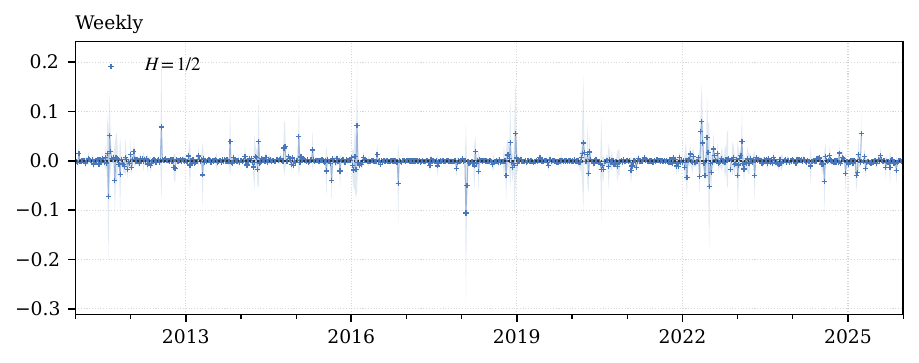}
	\end{minipage}

	\caption{Leverage estimates for AMZN over the 2011--2025 sample period.}

\end{figure}

\subsection{Signature Plots}
We examine the sensitivity of leverage estimates to the sampling interval for selected weekly and corresponding monthly windows of AAPL, AMZN, MSFT, and NVDA. For sampling intervals $s=1,\ldots,300$ seconds, we adjust $m_n=\lceil300/s\rceil$ and the corresponding $\widehat k_n$ at each frequency, using the estimator and confidence intervals described above.
The top and bottom rows report weekly and monthly windows; the left and right columns use $\widehat H_n$ and $H=1/2$, respectively.

In the selected weekly and monthly windows, the signature plots of the proposed statistic are more stable across sampling frequencies, with significantly negative leverage estimates at more frequencies. These results illustrate the practical value of the proposed statistic and its feasible inference procedure: for models whose leverage loadings vary with endogenous rough states, one can directly estimate leverage effects and conduct statistical inference from price observations subject to microstructure noise.

\begin{figure}[!htbp]
	\centering
	\begin{subfigure}[t]{0.48\textwidth}
		\centering
		\includegraphics[width=\linewidth]{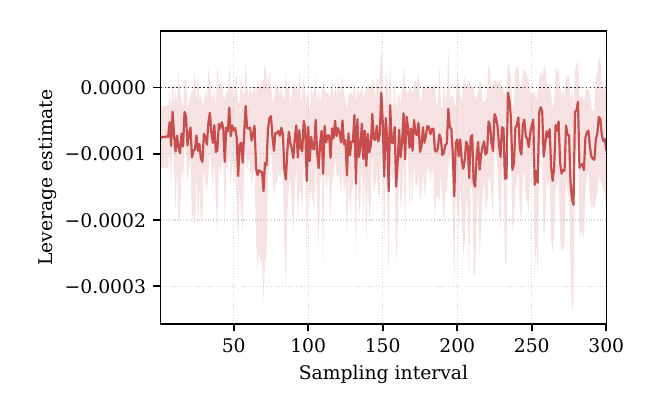}
		\caption{Weekly, $H=\widehat H_n$}
	\end{subfigure}
	\hfill
	\begin{subfigure}[t]{0.48\textwidth}
		\centering
		\includegraphics[width=\linewidth]{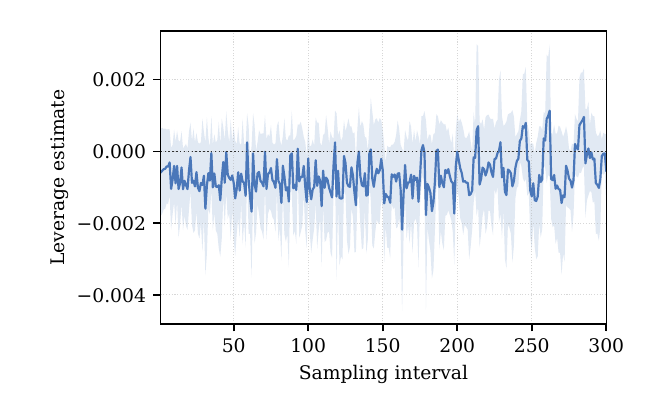}
		\caption{Weekly, $H=1/2$}
	\end{subfigure}

	\begin{subfigure}[t]{0.48\textwidth}
		\centering
		\includegraphics[width=\linewidth]{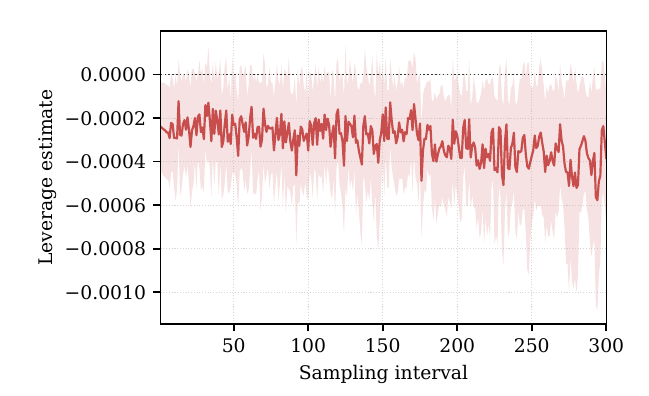}
		\caption{Monthly, $H=\widehat H_n$}
	\end{subfigure}
	\hfill
	\begin{subfigure}[t]{0.48\textwidth}
		\centering
		\includegraphics[width=\linewidth]{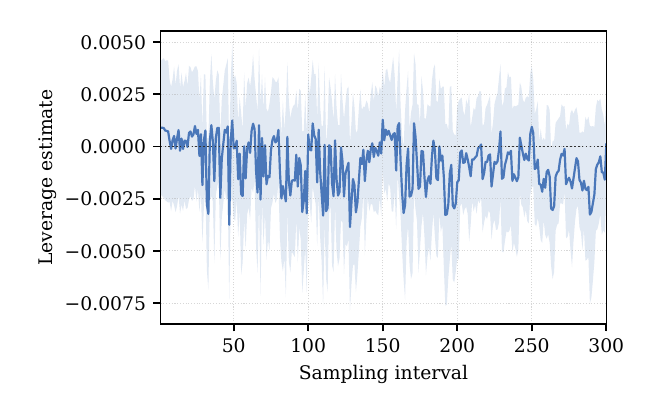}
		\caption{Monthly, $H=1/2$}
	\end{subfigure}
	\caption{AAPL: week 2017-08-14; month 2017-08.}
\end{figure}

\begin{figure}[!htbp]
	\centering
	\begin{subfigure}[t]{0.48\textwidth}
		\centering
		\includegraphics[width=\linewidth]{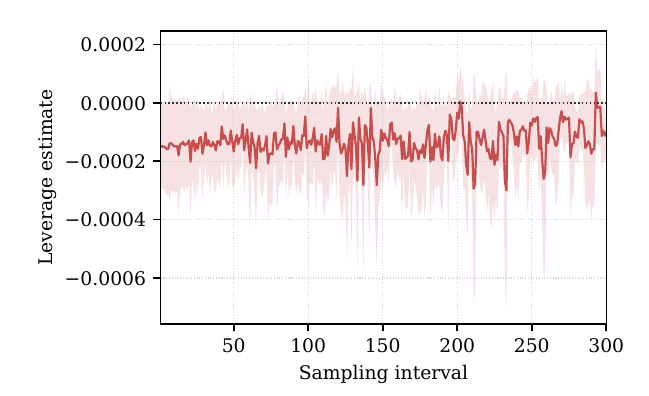}
		\caption{Weekly, $H=\widehat H_n$}
	\end{subfigure}
	\hfill
	\begin{subfigure}[t]{0.48\textwidth}
		\centering
		\includegraphics[width=\linewidth]{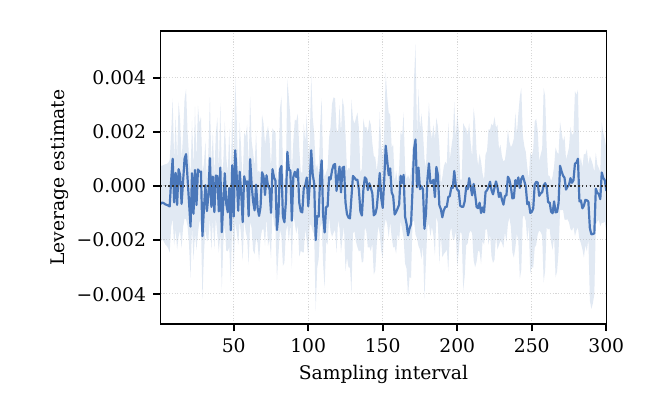}
		\caption{Weekly, $H=1/2$}
	\end{subfigure}

	\begin{subfigure}[t]{0.48\textwidth}
		\centering
		\includegraphics[width=\linewidth]{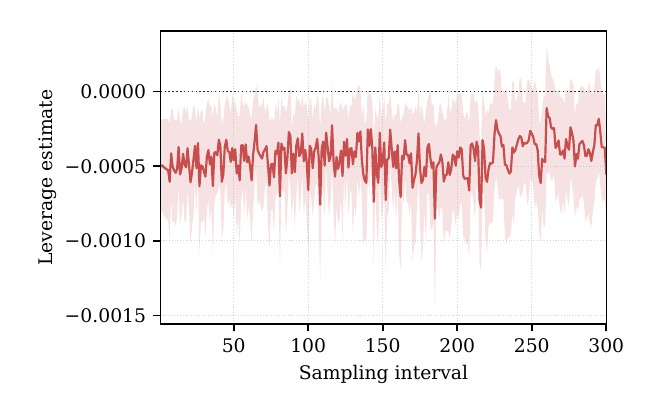}
		\caption{Monthly, $H=\widehat H_n$}
	\end{subfigure}
	\hfill
	\begin{subfigure}[t]{0.48\textwidth}
		\centering
		\includegraphics[width=\linewidth]{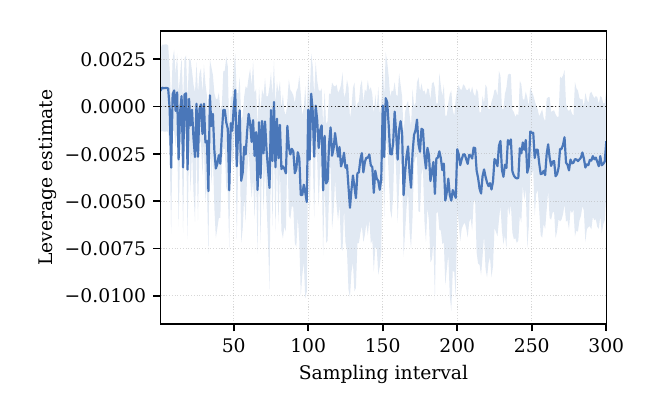}
		\caption{Monthly, $H=1/2$}
	\end{subfigure}
	\caption{AMZN: week 2017-08-07; month 2017-08.}
\end{figure}

\begin{figure}[!htbp]
	\centering
	\begin{subfigure}[t]{0.48\textwidth}
		\centering
		\includegraphics[width=\linewidth]{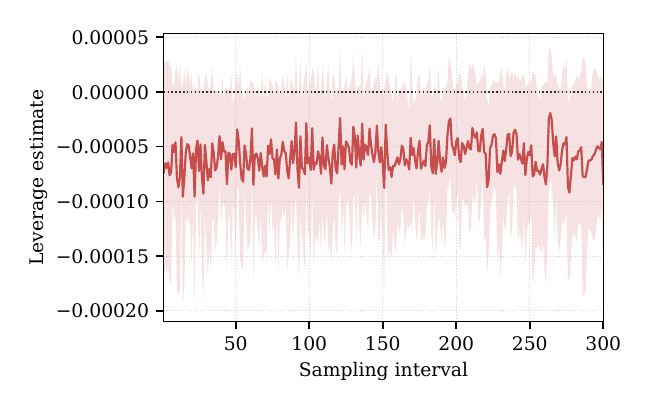}
		\caption{Weekly, $H=\widehat H_n$}
	\end{subfigure}
	\hfill
	\begin{subfigure}[t]{0.48\textwidth}
		\centering
		\includegraphics[width=\linewidth]{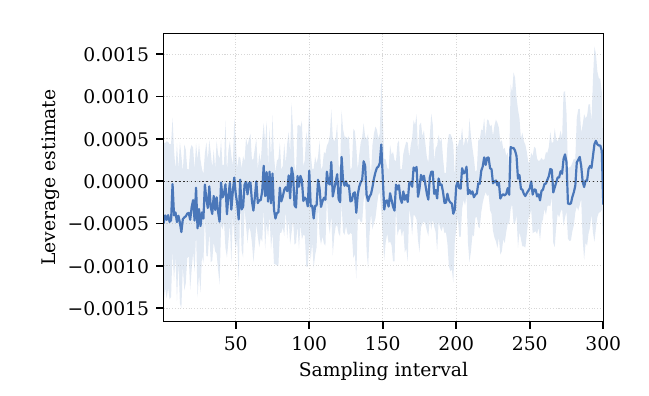}
		\caption{Weekly, $H=1/2$}
	\end{subfigure}

	\begin{subfigure}[t]{0.48\textwidth}
		\centering
		\includegraphics[width=\linewidth]{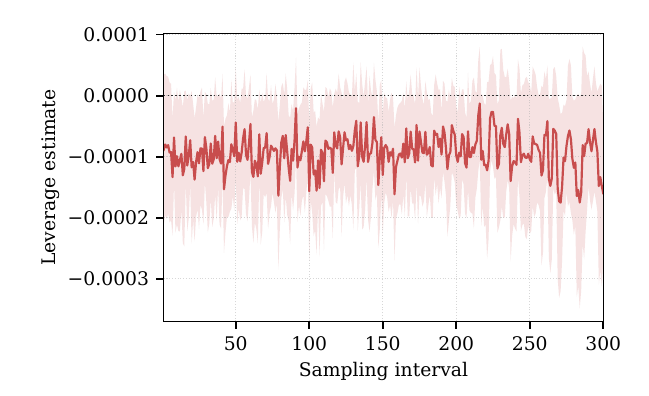}
		\caption{Monthly, $H=\widehat H_n$}
	\end{subfigure}
	\hfill
	\begin{subfigure}[t]{0.48\textwidth}
		\centering
		\includegraphics[width=\linewidth]{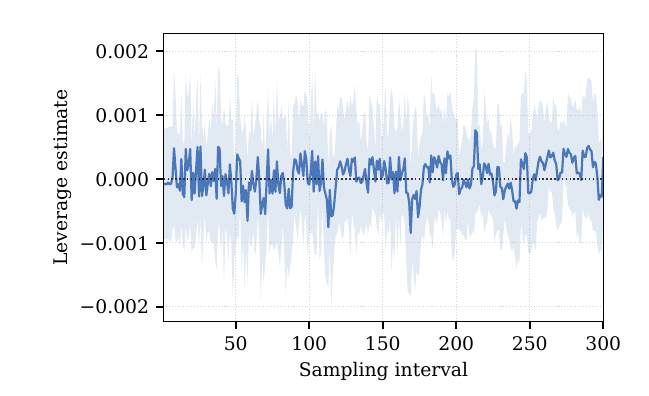}
		\caption{Monthly, $H=1/2$}
	\end{subfigure}
	\caption{MSFT: week 2017-04-24; month 2017-04.}
\end{figure}

\begin{figure}[!htbp]
	\centering
	\begin{subfigure}[t]{0.48\textwidth}
		\centering
		\includegraphics[width=\linewidth]{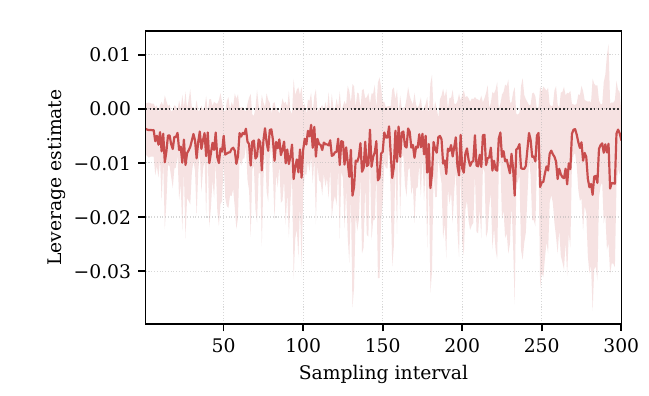}
		\caption{Weekly, $H=\widehat H_n$}
	\end{subfigure}
	\hfill
	\begin{subfigure}[t]{0.48\textwidth}
		\centering
		\includegraphics[width=\linewidth]{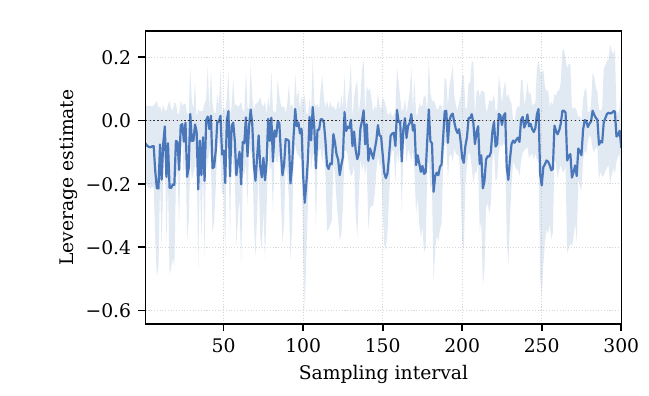}
		\caption{Weekly, $H=1/2$}
	\end{subfigure}

	\begin{subfigure}[t]{0.48\textwidth}
		\centering
		\includegraphics[width=\linewidth]{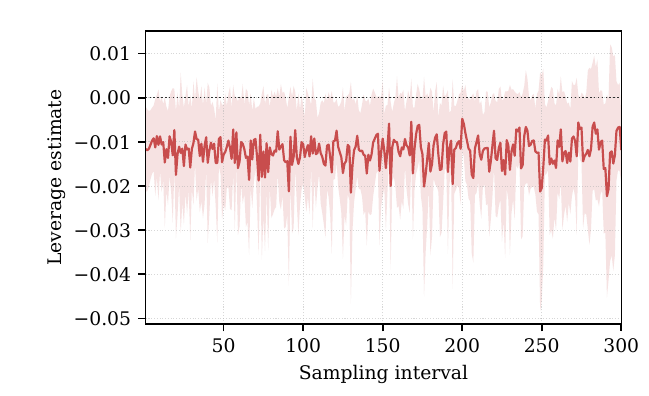}
		\caption{Monthly, $H=\widehat H_n$}
	\end{subfigure}
	\hfill
	\begin{subfigure}[t]{0.48\textwidth}
		\centering
		\includegraphics[width=\linewidth]{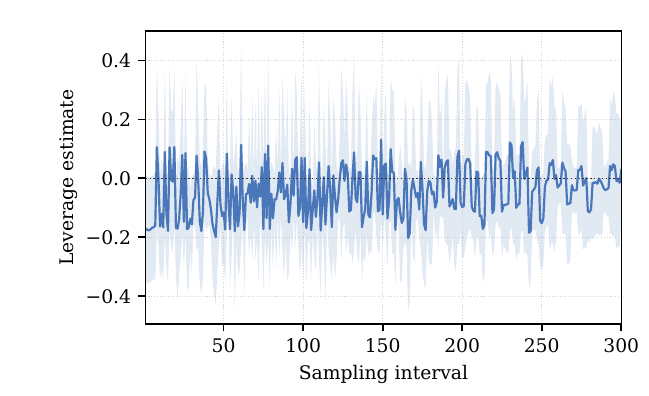}
		\caption{Monthly, $H=1/2$}
	\end{subfigure}
	\caption{NVDA: week 2024-07-22; month 2024-07.}
\end{figure}

%% file: appendices/A_loc.tex
\section{Localization and weighted sums}
Throughout the appendices, $K(t)=t^{H-1/2}$ for $t>0$, with $K(t)=0$ for $t\le0$.
The model is that of Assumption~\ref{ass:model}.

\subsection{Localization of the observed model}
\begin{pro}\label{prop:localization}
There are stopping times $\tau_N$ with $\P(\tau_N\le T)\to0$ and processes
$(X^{(N)},(\sigma^{(N)})^2)$, driven by the original Brownian motions, satisfying
\eqref{eq:model} and agreeing with $(X,\sigma^2)$ before $\tau_N$.
For each fixed $N$, the input coefficients and spot variance $(\sigma^{(N)})^2$
have deterministic bounds. For every finite $p\ge2$,
\begin{equation}\label{eq:variance-moments}
\begin{aligned}
&\|(\sigma_t^{(N)})^2-(\sigma_s^{(N)})^2\|_p\le C_{N,p}|t-s|^H,\\
&\|\sigma_t^{(N)}-\sigma_s^{(N)}\|_{2p}^2\le C_{N,p}|t-s|^H.
\end{aligned}
\end{equation}
For every $0<\eta<H$, the $C^\eta[0,T]$ norm of $(\sigma^{(N)})^2$ has all finite moments.
If LR holds, the stopping times may also be chosen so that
$L^{(N)}=\mathbf1_{[0,\tau_N)}\widetilde L^N$.
Probability and stable limits for the statistics, their pathwise corrections,
targets and variance functionals transfer from these processes to the original model.
\end{pro}
\begin{proof}
Choose coefficient localization times and intersect them with the upper exit
of $\sigma^2$ from $[0,N]$, and, under LR, with its representation time.
The initial event $\{\sigma_0^2>N\}$ is assigned $\tau_N=0$.
The resulting times still satisfy $\P(\tau_N\le T)\to0$.

Set
$k(t)=t^{H-1/2}/\Gamma(H+1/2)$ and
$\ell(t)=t^{-H-1/2}/\Gamma(1/2-H)$. The Beta identity gives $k*\ell=1$.
For the original model,
\[
\begin{aligned}
&\sigma^2=\sigma_0^2+k*dZ,\\
&dZ_t=\{\Gamma(H+1/2)g_t+(\ell*f)(t)\}\,dt
       +\Gamma(H+1/2)K_H^{-1}v_t\,dB_t.
\end{aligned}
\]
Define the stopped inputs and their convolution by
\begin{align*}
&f_t^{(N)}=0,\\
&v_t^{(N)}=\1_{\{t<\tau_N\}}v_t,\\
&\mu_t^{(N)}=\1_{\{t<\tau_N\}}\mu_t,\\
&\rho_t^{(N)}=\rho_t,\\
&g_t^{(N)}=\1_{\{t<\tau_N\}}\left\{g_t+
 \frac1{\Gamma(H+1/2)}\int_0^t\ell(t-s)\1_{\{s<\tau_N\}}f_sds\right\},\\
&(\sigma_t^{(N)})^2=(\sigma_0^{(N)})^2+\int_0^tK(t-s)g_s^{(N)}ds
 +K_H^{-1}\int_0^tK(t-s)v_s^{(N)}dB_s.
\end{align*}
Here $(\sigma_0^{(N)})^2=\sigma_0^2\,\mathbf{1}_{\{\sigma_0^2\le N\}}$.
All input coefficients have deterministic bounds at fixed $N$. The
stochastic integral uses the predictable integrand $\1_{\{0<t\le\tau_N\}}v_{t-}$;
the displayed c\`adl\`ag integrand gives the same integral.
Fubini and $k*\ell=1$ show equality with the original spot variance before
$\tau_N$.

Stopping the inputs also preserves nonnegativity. At any
deterministic $a<t$, the resolvent identity
$Z=\ell*(\sigma^2-\sigma_0^2)$ and integration by parts give
\begin{align*}
&\sigma_0^2+\int_0^a k(t-s)dZ_s
 =\left(1-\int_0^a w(t,a,r)dr\right)\sigma_0^2+\int_0^a w(t,a,r)\sigma_r^2dr,\\
&w(t,a,r)=\frac{\sin(\pi(H+1/2))}{\pi}
 \frac{(t-a)^{H+1/2}}{(a-r)^{H+1/2}(t-r)},
\qquad 0\le\int_0^a w(t,a,r)dr<1.
\end{align*}
Indeed, the coefficient of $\sigma_r^2-\sigma_0^2$ after the integration by parts is
$k(t-a)\ell(a-r)+\int_r^a k'(t-s)\ell(s-r)ds$;
direct integration gives the displayed $w$. With $x=(a-r)/(t-a)$ its
mass is bounded by
$(\sin(\pi(H+1/2))/\pi)\int_0^\infty x^{-H-1/2}(1+x)^{-1}dx=1$,
strictly for finite $a<t$. The weights are nonnegative.
The identity holds simultaneously for all $a<t$ by continuity, starting
with rational pairs; it therefore holds at $a=\tau_N$.
It proves $0\le (\sigma_t^{(N)})^2\le C_N$ also after exit, and continuity there
follows from the original convolution and the approximate-identity
limit of the weights. Take $\sigma^{(N)}\ge0$ and define
$X^{(N)}=X_0+\int\mu^{(N)}ds+\int\sigma^{(N)}dW$.
Under LR, $L_t^{(N)}=\sigma_t^{(N)}v_t^{(N)}\rho_t=\1_{\{t<\tau_N\}}\widetilde L_t^N$.

The kernel estimates
\begin{align*}
&\int_0^s|K(t-u)-K(s-u)|\,du+\int_s^tK(t-u)\,du\le C|t-s|^{H+1/2},\\
&\int_0^s|K(t-u)-K(s-u)|^2du+\int_s^tK(t-u)^2du\le C|t-s|^{2H}
\end{align*}
and BDG prove the first bound in \eqref{eq:variance-moments}.
The second follows from $|\sqrt{x}-\sqrt{y}|^2\le|x-y|$ for $x,y\ge0$.
For $0<\eta<H$ and any finite $p\ge2$, choose $q>\max\{p,(H-\eta)^{-1}\}$.
The Garsia--Rodemich--Rumsey inequality \citep[Lemma~1.1]{GarsiaRodemichRumsey1970} gives
\begin{align*}
\left\|\sup_{0\le s<t\le T}
\frac{|(\sigma_t^{(N)})^2-(\sigma_s^{(N)})^2|}{|t-s|^\eta}\right\|_p
&\le C_{q,\eta,T}\left\{\int_0^T\!\int_0^T
\frac{\E|(\sigma_t^{(N)})^2-(\sigma_s^{(N)})^2|^q}
{|t-s|^{2+q\eta}}\,ds\,dt\right\}^{1/q}\\
&\le C_{N,q,\eta,T}\left\{\int_0^T\!\int_0^T
|t-s|^{q(H-\eta)-2}\,ds\,dt\right\}^{1/q}<\infty.
\end{align*}
The last integral is finite since $q(H-\eta)>1$. The bounded initial
spot variance then gives the full $C^\eta[0,T]$ norm bound.
On $E_N=\{\tau_N>T\}$, the observed paths and all the displayed path
functionals agree. Construct the limiting stochastic integrals using the
same independent Brownian motion on one extension. Their integrands also
agree on $E_N$. Thus, for bounded $\F$-measurable $Z$ and bounded
continuous $f$, the original statistic $S_n$ and its proposed limit $S$
satisfy
\begin{align*}
&|\E[Zf(S_n)]-\overline\E[Zf(S)]|\\
&\le4\|Z\|_\infty\|f\|_\infty\P(E_N^c)
+|\E[Zf(S_n^{(N)})]-\overline\E[Zf(S^{(N)})]|.
\end{align*}
Let $n\to\infty$ and then $N\to\infty$. This proves stable transfer;
the same event argument transfers probability limits.
\end{proof}
\subsection{Zero-integral weights}
For $0<u<1$, define
\begin{equation*}
\begin{aligned}
&w_0(u)=K_H^{-1}\{(1-u)^{H-1/2}-(H+1/2)^{-1}\},\\
&w_1(u)=(H+3/2)(1-u)^{H+1/2}-1.
\end{aligned}
\end{equation*}
Both profiles have integral zero and finite $L^1(0,1)$ norm.
For a process $Z$ and a uniform mesh $h$, define
\begin{equation}\label{eq:weighted-cell}
\begin{aligned}
&I_{i,h}^{w}(Z)=\int_{(i-1)h}^{ih}w((s-(i-1)h)/h)(Z_s-Z_{(i-1)h})\,ds,\\
&p_{i,h}^{w}(Z)=\E[I_{i,h}^{w}(Z)\mid\F_{(i-1)h}].
\end{aligned}
\end{equation}
\begin{lemma}\label{lem:projection}
Let $Z$ be adapted and c\`adl\`ag, with $\E\sup_{t\le T}|Z_t|^2<\infty$,
and let $w\in\{w_0,w_1\}$. For any deterministic $|\vartheta_i|\le1$ and
partition into disjoint consecutive sets $J_\nu$,
\begin{equation*}
\sum_\nu\E\max_{M\in J_\nu}\left|
\sum_{i\in J_\nu,\,i\le M}\vartheta_i
\{I_{i,h}^{w}(Z)-p_{i,h}^{w}(Z)\}\right|^2=o(h).
\end{equation*}
Taking a single group gives the same bound for the full-grid maximal partial sum.
\end{lemma}
\begin{proof}
Set $\omega_{i,h}(Z)=\sup_{(i-1)h\le s\le ih}|Z_s-Z_{(i-1)h}|$.
The cell difference $I_{i,h}^{w}(Z)-p_{i,h}^{w}(Z)$ is a martingale difference.
Doob and conditional variance contraction imply
\begin{align*}
&\E\max_M\left|\sum_{i\le M}\vartheta_i
\{I_{i,h}^{w}(Z)-p_{i,h}^{w}(Z)\}\right|^2\\
&\le4\sum_i\E|I_{i,h}^{w}(Z)|^2
\le C h\sum_i h\E[\omega_{i,h}(Z)^2]=o(h).
\end{align*}
A c\`adl\`ag path can be uniformly approximated by a finite step function;
only finitely many cells cross its jump times. Hence
$\sum_i h\omega_{i,h}(Z)^2\to0$ pathwise. It is dominated by
$4T\sup|Z|^2$, so dominated convergence proves the last equality.
Doob within disjoint groups and orthogonality give the same $o(h)$ bound
for their squared maxima.
\end{proof}
\begin{lemma}\label{lem:lr-sums}
Fix $N$ in LR, and let $w\in\{w_0,w_1\}$.
For any deterministic $|\vartheta_i|\le1$ and disjoint consecutive index
sets $J_\nu$, the raw integrals of $\widetilde L^N$ satisfy
\begin{equation}\label{eq:lr-raw}
\begin{aligned}
&\left\|\max_{M\le T/h}\left|\sum_{i=1}^M\vartheta_i I_{i,h}^{w}(\widetilde L^N)\right|\right\|_2
\le C_N\{h+h^{\zeta_N}+h^{H_N+1/2}\log(2+T/h)\},\\
&\sum_\nu\left\|\max_{M\in J_\nu}\left|
\sum_{i\in J_\nu,\,i\le M}\vartheta_i I_{i,h}^{w}(\widetilde L^N)\right|\right\|_2^2
\le C_N\{h^2+h^{2\zeta_N}+h^{2H_N+1}\log^2(2+T/h)\}.
\end{aligned}
\end{equation}
The same bounds hold for $Z=\mathbf1_{[0,\tau)}\widetilde L^N$,
where $\tau$ is any stopping time. For both choices of $Z$,
the conditional integrals satisfy
\begin{equation}\label{eq:lr-projection}
\begin{aligned}
&\left\|\max_{M\le T/h}\left|\sum_{i=1}^M\vartheta_i p_{i,h}^{w}(Z)\right|\right\|_2=o(\sqrt h),\\
&\sum_\nu\left\|\max_{M\in J_\nu}\left|
\sum_{i\in J_\nu,\,i\le M}\vartheta_i p_{i,h}^{w}(Z)\right|\right\|_2^2=o(h).
\end{aligned}
\end{equation}
The bounds are uniform in the masks and the partitions.
\end{lemma}
\begin{proof}
Fix $N$ and suppress its index except in $H_N$. For $A^{\mathrm{BV}}$, the sum of
the absolute cell integrals is at most $Ch\operatorname{TV}(A^{\mathrm{BV}};[0,T])$.
For $A^{\mathrm{reg}}$, Minkowski gives a sum of $L^2$ norms at most
$C(T/h)h^{1+\zeta}$. These estimates also bound the sum of grouped squared
norms by the square of the respective total bound.
For the stochastic part, the LR amplitude bound gives
\begin{equation*}
\int_{(i-1)h}^{ih}|w((s-(i-1)h)/h)|
\left(\int_0^s\|\psi(s,u)\|_2^2du\right)^{1/2}ds\le C_T h.
\end{equation*}
This bound, and the same bound for the value at $(i-1)h$, justify stochastic
Fubini. The stochastic part has cell integral $\int_0^{ih}\Psi_i(u)\,d\mathbf B_u$, where
\begin{align*}
\Psi_i(u)=&\mathbf1_{\{u<(i-1)h\}}\int_{(i-1)h}^{ih}w((s-(i-1)h)/h)
\{\psi(s,u)-\psi((i-1)h,u)\}\,ds\\
&+\mathbf1_{\{(i-1)h\le u<ih\}}\int_u^{ih}w((s-(i-1)h)/h)\psi(s,u)\,ds.
\end{align*}
Each $\Psi_i$ is predictable in $u$. The kernel difference bound in
\eqref{eq:lr-bounds} yields, with
$E_{H_N}(x)=x^{H_N-1/2}\mathbf1_{\{0<x\le1\}}+x^{H_N-3/2}\mathbf1_{\{x>1\}}$,
\begin{equation}\label{eq:cell-kernel}
\begin{aligned}
\|\Psi_i(u)\|_2\le& C\bigl[h^{H_N+1/2}
\{E_{H_N}(((i-1)h-u)/h)\mathbf1_{\{u<(i-1)h\}}+\mathbf1_{\{(i-1)h\le u<ih\}}\}\\
&+h^2\mathbf1_{\{u<(i-1)h\}}+h\mathbf1_{\{(i-1)h\le u<ih\}}\bigr].
\end{aligned}
\end{equation}
For $u<(i-1)h$, integrate the kernel difference over $s$ and split at
$(i-1)h-u=h$; the power is respectively $((i-1)h-u)^{H_N-1/2}$ or
$h((i-1)h-u)^{H_N-3/2}$. For $(i-1)h\le u<ih$, the only unbounded weight is the
first term in $w_0$. Its integral is
\begin{align*}
&h^{1/2-H}\int_u^{ih}(ih-s)^{H-1/2}(s-u)^{H_N-1/2}ds\\
&=h^{1/2-H}\mathrm B(H+1/2,H_N+1/2)(ih-u)^{H+H_N}\le C h^{H_N+1/2}.
\end{align*}
This proves \eqref{eq:cell-kernel}. Since $E_{H_N}\in L^1\cap L^2$,
the overlap integral of two shifted copies is at most
$C(1+|i-j|)^{H_N-3/2}$: for separated cells use the tail bound for the more
distant factor and the $L^1$ bound for the other; for adjacent cells use
Cauchy--Schwarz. Thus It\^o isometry and Minkowski, for any consecutive
set $I$ of $m$ indices, give
\begin{equation}\label{eq:kernel-energy}
\begin{aligned}
\left\|\sum_{i\in I}\vartheta_i\int_0^{ih}\Psi_i(u)\,d\mathbf B_u\right\|_2^2
&\le\int_0^T\left(\sum_{i\in I}\|\Psi_i(u)\|_2\right)^2du\\
&\le C\{m h^{2H_N+2}+m h^3+m^2h^4\}\le C_T m h^{2H_N+2}.
\end{aligned}
\end{equation}
Here $mh\le T$ and $H_N<1/2$. Decompose a prefix into at most one
interval from each binary level of a deterministic partition. At each
level, the sum of the squared norms in \eqref{eq:kernel-energy} is at
most $C_T h^{2H_N+1}$. The $L^2$ norm of the maximum is bounded by the square
root of this sum. Minkowski over at most $C\log(2+T/h)$ levels gives the
first bound in \eqref{eq:lr-raw}. Applying the same decomposition
within each $J_\nu$ and summing its energies gives the second bound.
For the stopped process $Z=\mathbf1_{[0,\tau)}\widetilde L$, every prefix
consists of the original complete-cell prefix before $\tau$ and at most
one partial cell. Because $\int_0^1w=0$, that partial integral is bounded
by $Ch\sup_{t\le T}|\widetilde L_t|$. The within-group maxima handle the
stopped prefixes, and only one group contains the partial cell. This adds
at most $Ch^2\E\sup|\widetilde L|^2$ to the squared estimates.
Lemma~\ref{lem:projection} applies to both processes $Z$.
Together with \eqref{eq:lr-raw}, $H_N>0$ and $\zeta>1/2$, it proves
\eqref{eq:lr-projection}.
\end{proof}

%% file: appendices/B_lr.tex
\section{Verification of LR}
This section proves Proposition~\ref{prop:lr}, in the order (i)--(iii).
\subsection{Direct Volterra representations}
\begin{proof}
For Proposition~\ref{prop:lr}(\ref{cond:volterra}),
intersect coefficient localization times with the exit time of the running
$\zeta$-H\"older seminorm of $A^L$. Set $\tau_N=0$ when $|L_0|>N$.
One can arrange $\P(\tau_N\le T)\to0$ and
\begin{equation}\label{eq:direct-stopping}
\begin{aligned}
&|\mathbf1_{\{u<\tau_N\}}\gamma_u^L|\le C_N,\\
&|A_{t\wedge\tau_N}^L-A_{s\wedge\tau_N}^L|\le N|t-s|^\zeta.
\end{aligned}
\end{equation}
The seminorm is measurable using rational time pairs, and continuity
preserves the second bound at exit. Define
\begin{align*}
\widetilde L_t^N=&L_0\mathbf1_{\{|L_0|\le N\}}+A_{t\wedge\tau_N}^L\\
&+K_H^{-1}\int_0^t(t-u)^{H-1/2}
\gamma_u^L\mathbf1_{\{|\gamma_u^L|\le C_N\}}\,d\mathbf B_u.
\end{align*}
This equals $L_t$ for $t<\tau_N$. In~\eqref{eq:lr}, take the truncated
initial value as $A^{N,\mathrm{BV}}$, the stopped $A^L$ as $A^{N,\mathrm{reg}}$,
and
\begin{equation*}
\psi_N(t,u)=K_H^{-1}(t-u)^{H-1/2}\gamma_u^L\mathbf1_{\{|\gamma_u^L|\le C_N\}}.
\end{equation*}
The coefficient $\gamma_u^L\mathbf1_{\{|\gamma_u^L|\le C_N\}}$ is
predictable, bounded by $C_N$, and equals $\gamma_u^L$ before $\tau_N$.
Its bound and $K'(v-u)=(H-1/2)(v-u)^{H-3/2}$ give
\begin{align*}
&\|\psi_N(t,u)\|_2\le C_NK_H^{-1}(t-u)^{H-1/2},\\
&\|\psi_N(t,u)-\psi_N(s,u)\|_2
\le C_NK_H^{-1}(1/2-H)\int_s^t(v-u)^{H-3/2}\,dv,\qquad u<s<t.
\end{align*}
These are the two kernel bounds in~\eqref{eq:lr-bounds}, with $H_N=H$.
To check the supremum moment, first split the increment at $s$:
\begin{align*}
\widetilde L_t^N-\widetilde L_s^N
=&A_{t\wedge\tau_N}^L-A_{s\wedge\tau_N}^L\\
&+K_H^{-1}\int_0^s\{K(t-u)-K(s-u)\}
\gamma_u^L\mathbf1_{\{|\gamma_u^L|\le C_N\}}\,d\mathbf B_u\\
&+K_H^{-1}\int_s^tK(t-u)
\gamma_u^L\mathbf1_{\{|\gamma_u^L|\le C_N\}}\,d\mathbf B_u.
\end{align*}
Changing variables in the two deterministic square integrals gives
\begin{align*}
&\int_0^s|K(t-u)-K(s-u)|^2du+\int_s^tK(t-u)^2du\\
&=(t-s)^{2H}\left\{
\int_0^{s/(t-s)}[(x+1)^{H-1/2}-x^{H-1/2}]^2dx+\frac1{2H}\right\}\\
&\le C(t-s)^{2H}.
\end{align*}
The integral over $x\in(0,\infty)$ is finite: the integrand is
$O(x^{2H-1})$ near zero and $O(x^{2H-3})$ at infinity. BDG applied to
the two displayed It\^o integrals, together with $\zeta>1/2>H$, yields
\begin{align*}
\|\widetilde L_t^N-\widetilde L_s^N\|_p
&\le N|t-s|^\zeta+C_{N,p}|t-s|^H\\
&\le C_{N,p,T}|t-s|^H,\qquad p\ge2.
\end{align*}
Choose $p>\max\{2,1/H\}$ and $1/p<\alpha<H$. The increment bound
gives a continuous modification. Applying the Garsia--Rodemich--Rumsey
inequality \citep[Lemma~1.1]{GarsiaRodemichRumsey1970} to this modification,
and using $|\widetilde L_0^N|\le N$, gives
\begin{align*}
\left\|\sup_{t\le T}|\widetilde L_t^N|\right\|_2
&\le\left\|\sup_{t\le T}|\widetilde L_t^N|\right\|_p\\
&\le N+C_{p,\alpha,T}\left\{
\int_0^T\!\int_0^T
\frac{\E|\widetilde L_t^N-\widetilde L_s^N|^p}{|t-s|^{1+\alpha p}}\,ds\,dt\right\}^{1/p}\\
&\le N+C_{N,p,\alpha,T}\left\{
\int_0^T\!\int_0^T|t-s|^{p(H-\alpha)-1}\,ds\,dt\right\}^{1/p}<\infty.
\end{align*}
The last integral is finite because $p(H-\alpha)>0$. Together with
\eqref{eq:direct-stopping}, this verifies every condition
in LR with $H_N=H$ and $\zeta_N=\zeta$.
In particular, $L_t=a\sigma_t^2+b_0$ with deterministic $a,b_0$
satisfies (i) by~\eqref{eq:model}, with $\zeta=H+1/2$.
Rough Heston is a particular instance of the direct Volterra condition.
Here $\sigma$ is spot volatility and $\sigma^2$ is spot variance.
Consider the nonnegative continuous weak solution
\begin{align*}
\sigma_t^2=&\sigma_0^2+\frac{\kappa}{\Gamma(H+1/2)}
\int_0^t(t-s)^{H-1/2}(\vartheta-\sigma_s^2)\,ds\\
&+\frac{\nu}{K_H}\int_0^t(t-s)^{H-1/2}\sigma_s\,dB_s,
\end{align*}
where $\sigma_0^2>0$, $\kappa,\vartheta\ge0$, $\nu>0$, and
$\rho_t\equiv\rho\in[-1,1]$. The constants $\kappa,\vartheta,\nu$ and $\sigma_0^2$
are deterministic. Existence of the nonnegative continuous weak solution
follows from \citet[Theorem~6.1]{AbiJaberLarssonPulido2019}. In this model
\[
v_t=\nu\sigma_t,\qquad L_t=\nu\rho \sigma_t^2.
\]
In~\eqref{eq:direct-volterra}, take $m=1$, $\mathbf B=B$ and
\begin{align*}
&L_0=\nu\rho \sigma_0^2,\qquad \gamma_s^L=\nu^2\rho\sigma_s,\\
&A_t^L=\frac{\nu\rho\kappa}{\Gamma(H+1/2)}
\int_0^t(t-s)^{H-1/2}(\vartheta-\sigma_s^2)\,ds.
\end{align*}
Continuity of $\sigma^2$ gives the local bound on $\gamma^L$. The time integral satisfies
\begin{align*}
|A_t^L-A_s^L|
&\le C\left(1+\sup_{u\le T}\sigma_u^2\right)
\left\{\int_0^s|K(t-u)-K(s-u)|du+\int_s^tK(t-u)du\right\}\\
&\le C\left(1+\sup_{u\le T}\sigma_u^2\right)|t-s|^{H+1/2}.
\end{align*}
Thus~\eqref{eq:direct-regularity} holds with $\zeta=H+1/2$, and
Proposition~\ref{prop:lr}(\ref{cond:volterra}) applies. The variance integrands
in the preceding results are $\sigma^6$, $\sigma^2v^2=\nu^2\sigma^{4}$ and
$L^2=\nu^2\rho^2\sigma^{4}$. Continuity and $\sigma_0^2>0$ imply
$\int_0^T\sigma_s^4\,ds>0$ and $\int_0^T\sigma_s^6\,ds>0$ almost surely. Hence the
limiting variances are strictly positive, and
Proposition~\ref{prop:variance} applies.
\end{proof}
\subsection{\texorpdfstring{Smooth functions of closed rough states}{Smooth functions of closed rough states}}
Use the local $C_b^3$ extensions of \eqref{eq:closed-state} in
Proposition~\ref{prop:lr}(\ref{cond:closed-state}) and put
$\mathbf Y_t=(\mathbf V_t,\mathbf b_t)$. In this subsection
$\Gamma_{V,\cdot j}$ and $\gamma_{b,\cdot j}$ denote the $j$th Brownian
columns, $1\le j\le m$. The initial state is $\F_0$-measurable, and
$\mathbf B$ is Brownian with respect to $(\F_t)$; thus its increments are
independent of $\F_0$. Here $\mathcal H=L^2([0,T];\mathbb R^m)$ is the
Hilbert space for Brownian Malliavin derivatives. Write $\mathbb{D}$ for the
Malliavin derivative with respect to the original Brownian motion $\mathbf B$,
and $\mathbb{D}_r^{(j)}$ for its $j$th component at time $r$. It fixes the
initial state. Gradients of scalar functions are column vectors;
derivatives of vector-valued coefficients are Jacobian matrices. All estimates are first proved for
bounded $C^3$ coefficients and $F\in C_b^3$, with constants independent of
the initial state.

\begin{lemma}\label{lem:state-bounds}
For the extended coefficients, the state is a
strong solution determined by $(\mathbf Y_0,\mathbf B)$. For every fixed
$p\ge2$, $0\le s<t\le T$ and $0\le r<t$,
\begin{equation}\label{eq:state-bounds}
\begin{aligned}
&\|\mathbf Y_t-\mathbf Y_s\|_p\le C_p|t-s|^H,\\
&\|\mathbb{D}_r^{(j)}\mathbf V_t\|_p\le C_pK(t-r),\\
&\|\mathbb{D}_r^{(j)}\mathbf b_t\|_p\le C_p,\\
&\left\|\mathbb{D}_r^{(j)}\mathbf V_t-K_H^{-1}K(t-r)
       \Gamma_{V,\cdot j}(\mathbf V_r)\right\|_p
 \le C_p\{(t-r)^{2H-1/2}+(t-r)^{H+1/2}\},\\
&\left\|\mathbb{D}_r^{(j)}\mathbf b_t-\gamma_{b,\cdot j}(\mathbf Y_r)\right\|_p
 \le C_p\{(t-r)^H+(t-r)^{1/2}+(t-r)\}.
\end{aligned}
\end{equation}
For bounded initial states these are global Malliavin derivatives.
For an arbitrary finite initial state they are the derivatives conditional
on $\F_0$; the bounded observable $F(\mathbf Y_t)$ belongs to
$\mathbb D^{1,p}$ on this conditional Brownian space and its derivative has
a finite unconditional $L^p(\Omega;\mathcal H)$ norm.
\end{lemma}
\begin{proof}
For $a,\lambda>0$, iteration of
$f(t)\le A t^{\lambda-1}+C\int_0^t(t-s)^{a-1}f(s)ds$ gives
\begin{equation}\label{eq:resolvent}
f(t)\le A\Gamma(\lambda)t^{\lambda-1}
 \sum_{q=0}^{\infty}\frac{[C\Gamma(a)t^a]^q}{\Gamma(\lambda+qa)}.
\end{equation}
The Beta integral computes each convolution. The Gamma denominators make
the series finite on $[0,T]$ after division by $t^{\lambda-1}$; the iterated
remainder tends to zero, first for bounded approximations and then by the
same bounds. On $0<t\le T$,
$1+K(t)+K(t)^2\le C_{H,T}t^{2H-1}$.

Picard iteration for the two state equations, using BDG and weighted
Cauchy--Schwarz for the time integrals, bounds squared $L^p$ differences
by convolution with $C(1+K^2)$. Equation~\eqref{eq:resolvent} proves
convergence and uniqueness on each finite horizon. The iterations are
measurable functions of $(\mathbf Y_0,\mathbf B)$. With a random initial
state, condition on $\F_0$; bounded coefficient values make all increment
bounds uniform in that state. The deterministic bounds
\begin{align*}
&\int_0^s|K(t-u)-K(s-u)|du+\int_s^tK(t-u)du\le C(t-s)^{H+1/2},\\
&\int_0^s|K(t-u)-K(s-u)|^2du+\int_s^tK(t-u)^2du\le C(t-s)^{2H}
\end{align*}
then give the state increment estimate and a continuous version.

Differentiating the Picard equations gives, for $t>r$,
\begin{equation}\label{eq:state-derivatives}
\begin{aligned}
&\begin{aligned}[t]
\mathbb{D}_r^{(j)}\mathbf V_t=&K_H^{-1}K(t-r)\Gamma_{V,\cdot j}(\mathbf V_r)\\
&+\int_r^t\nabla\beta_0(\mathbf V_u)\mathbb{D}_r^{(j)}\mathbf V_u\,du\\
&+K_H^{-1}\int_r^tK(t-u)\nabla\beta_H(\mathbf V_u)\mathbb{D}_r^{(j)}\mathbf V_u\,du\\
&+K_H^{-1}\sum_{c=1}^m\int_r^tK(t-u)
 \nabla\Gamma_{V,\cdot c}(\mathbf V_u)\mathbb{D}_r^{(j)}\mathbf V_u\,dB_u^c,
\end{aligned}\\
&\begin{aligned}[t]
\mathbb{D}_r^{(j)}\mathbf b_t=&\gamma_{b,\cdot j}(\mathbf Y_r)
 +\int_r^t\nabla\mu_b(\mathbf Y_u)\mathbb{D}_r^{(j)}\mathbf Y_u\,du\\
&+\sum_{c=1}^m\int_r^t\nabla\gamma_{b,\cdot c}(\mathbf Y_u)
 \mathbb{D}_r^{(j)}\mathbf Y_u\,dB_u^c.
\end{aligned}
\end{aligned}
\end{equation}
For the bound on $\mathbb{D}_r^{(j)}\mathbf V_t$ in
\eqref{eq:state-bounds}, the first equation in
\eqref{eq:state-derivatives} and BDG give
\[
\|\mathbb{D}_r^{(j)}\mathbf V_t\|_p^2
\le C_p(t-r)^{2H-1}
 +C_p\int_r^t(t-u)^{2H-1}\|\mathbb{D}_r^{(j)}\mathbf V_u\|_p^2du.
\]
Apply \eqref{eq:resolvent} with $a=\lambda=2H$.
For $\mathbb{D}_r^{(j)}\mathbf b_t$, its ordinary stochastic integral instead gives
\begin{align*}
\|\mathbb{D}_r^{(j)}\mathbf b_t\|_p^2
&\le C_p\left(1+\int_r^tK(u-r)^2du\right)
 +C_p\int_r^t\|\mathbb{D}_r^{(j)}\mathbf b_u\|_p^2du\\
&\le C_{p,T}+C_p\int_r^t\|\mathbb{D}_r^{(j)}\mathbf b_u\|_p^2du.
\end{align*}
Ordinary Gronwall proves its uniform bound. For the remainder bound in
\eqref{eq:state-bounds}, estimate the three remaining integrals
in its first response equation separately:
\begin{align*}
&\int_r^t\|\mathbb{D}_r^{(j)}\mathbf V_u\|_pdu
 \le C_p(t-r)^{H+1/2},\\
&\int_r^tK(t-u)\|\mathbb{D}_r^{(j)}\mathbf V_u\|_pdu
 \le C_p\mathrm B(H+1/2,H+1/2)(t-r)^{2H},\\
&\left(\int_r^tK(t-u)^2\|\mathbb{D}_r^{(j)}\mathbf V_u\|_p^2du\right)^{1/2}
 \le C_p\sqrt{\mathrm B(2H,2H)}(t-r)^{2H-1/2}.
\end{align*}
On $0<t-r\le T$, $(t-r)^{2H}\le C_T(t-r)^{2H-1/2}$.
For the last line of \eqref{eq:state-bounds}, the two
integrals in the second response equation satisfy
\begin{align*}
&\int_r^t\|\mathbb{D}_r^{(j)}\mathbf Y_u\|_pdu
 \le C_p\{(t-r)^{H+1/2}+t-r\},\\
&\left(\int_r^t\|\mathbb{D}_r^{(j)}\mathbf Y_u\|_p^2du\right)^{1/2}
 \le C_p\{(t-r)^H+(t-r)^{1/2}\}.
\end{align*}
The coefficient bounds multiply these deterministic expressions.

The differentiated iterations have uniformly integrable squared kernels
$K(t-r)^2$. Closability of the Brownian Malliavin derivative and the chain
rule give $\mathbb{D}_r^{(j)}F(\mathbf Y_t)=\nabla F(\mathbf Y_t)^\top\mathbb{D}_r^{(j)}\mathbf Y_t$;
see \citet[Section~1.2]{Nualart2006}. One may first restrict $\mathbf Y_0$
to bounded $\F_0$-events and then pass to the limit: their Brownian derivatives
are zero, $F$ is bounded, and the derivative bound is uniform in $\mathbf Y_0$.
This proves the last assertion without a moment assumption on $\mathbf Y_0$.
\end{proof}

\begin{pro}\label{prop:conditional-kernel}
Define
\[
m_t=\E[F(\mathbf Y_t)\mid\F_0],\qquad
\psi^j(t,r)=\E[\mathbb{D}_r^{(j)}F(\mathbf Y_t)\mid\F_r],\quad r<t.
\]
There are versions, jointly measurable in their time arguments and
predictable in $r$ for fixed $t$, such that
\begin{equation}\label{eq:clark-ocone}
F(\mathbf Y_t)=m_t+\sum_{j=1}^m\int_0^t\psi^j(t,r)\,dB_r^j.
\end{equation}
For every fixed $p\ge2$,
\begin{equation}\label{eq:conditional-kernel}
\begin{aligned}
&\operatorname{TV}(m;[0,T])\le C_{H,T}\quad\hbox{a.s.},\\
&\|\psi^j(t,r)\|_p\le C_pK(t-r),\\
&\|\psi^j(t,r)-\psi^j(s,r)\|_p
 \le C_p\int_s^t(v-r)^{H-3/2}dv,\qquad r<s<t.
\end{aligned}
\end{equation}
The deterministic integral $\int_s^t(v-r)^{H-3/2}dv$ equals $\{K(s-r)-K(t-r)\}/(1/2-H)$.
Only coefficient and observable derivatives through order three are used
in this proposition. The constants are for fixed $H\in(0,1/2)$.
\end{pro}
\begin{proof}
Lemma~\ref{lem:state-bounds} proves the first-Sobolev input for the
conditional Clark--Ocone formula~\eqref{eq:clark-ocone}; see
\citet{Nualart2006}. The strong-solution construction and
\eqref{eq:state-derivatives} make $\mathbb{D}_r^{(j)}F(\mathbf Y_t)$ a measurable
function of $\mathbf Y_0$ and $(\mathbf B_u)_{u\le t}$.
Since $(\mathbf B_{r+u}-\mathbf B_r)_{u\ge0}$ is independent of $\F_r$,
\[
\E[\mathbb{D}_r^{(j)}F(\mathbf Y_t)\mid\F_r]
=\E[\mathbb{D}_r^{(j)}F(\mathbf Y_t)\mid\sigma(\mathbf Y_0,\mathbf B_u:0\le u\le r)].
\]
Thus the Clark--Ocone integrand obtained from the initial state and the
Brownian path is also $\psi^j(t,r)$ with respect to $(\F_r)$.
Conditional Jensen, the chain rule and \eqref{eq:state-bounds} give
\begin{align*}
\|\psi^j(t,r)\|_p&\le\|\mathbb{D}_r^{(j)}F(\mathbf Y_t)\|_p\\
&\le\|\nabla F\|_\infty
 \{\|\mathbb{D}_r^{(j)}\mathbf V_t\|_p+\|\mathbb{D}_r^{(j)}\mathbf b_t\|_p\}
 \le C_pK(t-r).
\end{align*}
Here $1\le C_{H,T}K(t-r)$ for $0<t-r\le T$ absorbs the It\^o-coordinate
bound. This proves the second line of \eqref{eq:conditional-kernel}.
It remains to bound the terminal-time increments of $\psi^j$ and the total
variation of $m$.

To prove the last line of \eqref{eq:conditional-kernel}, fix $r$ and
$0<\eta\le T-r$. On $\Omega\times\widetilde\Omega$, let
$\widetilde{\mathbf B}$ be a Brownian motion on $[0,1]$ independent of
$\F$. Write $\widetilde{\E}$ for expectation over
$\widetilde\Omega$, holding the original history on $[0,r]$ and all
$\F_r$-measurable quantities fixed. For each $\eta$, the future Brownian increment has the
conditional law
\[
 (\mathbf B_{r+\eta x}-\mathbf B_r)_{0\le x\le1}
 \ \stackrel{\mathcal L\mid\F_r}{=}\
 (\sqrt \eta\,\widetilde{\mathbf B}_x)_{0\le x\le1}.
\]
For each $0<\eta\le T-r$, define the auxiliary process
$\mathbf Y^{r,\eta}=(\mathbf V^{r,\eta},\mathbf b^{r,\eta})$ on
$\Omega\times\widetilde\Omega$ as the unique solution of the following
equations on $[0,1]$, adapted to
$(\F_r\vee\sigma(\widetilde{\mathbf B}_y:0\le y\le x))_{0\le x\le1}$,
with initial value $\mathbf Y_0^{r,\eta}=\mathbf Y_r$.
All values of $\eta$ use the same $\widetilde{\mathbf B}$ and the same
original history on $[0,r]$:
\begin{equation}\label{eq:scaled-state}
\begin{aligned}
&\begin{aligned}[t]
\mathbf V_x^{r,\eta}=&\mathbf V_r
 +K_H^{-1}\int_0^r[K(r+\eta x-u)-K(r-u)]\beta_H(\mathbf V_u)du\\
&+K_H^{-1}\sum_{c=1}^m\int_0^r[K(r+\eta x-u)-K(r-u)]
 \Gamma_{V,\cdot c}(\mathbf V_u)dB_u^c\\
&+\eta\int_0^x\beta_0(\mathbf V_y^{r,\eta})dy
 +\frac{\eta^{H+1/2}}{K_H}\int_0^xK(x-y)\beta_H(\mathbf V_y^{r,\eta})dy\\
&+\frac{\eta^H}{K_H}\sum_{c=1}^m\int_0^xK(x-y)
 \Gamma_{V,\cdot c}(\mathbf V_y^{r,\eta})d\widetilde B_y^c,
\end{aligned}\\
&\begin{aligned}[t]
\mathbf b_x^{r,\eta}=&\mathbf b_r+\eta\int_0^x\mu_b(\mathbf Y_y^{r,\eta})dy\\
&+\sqrt \eta\sum_{c=1}^m\int_0^x\gamma_{b,\cdot c}(\mathbf Y_y^{r,\eta})
 d\widetilde B_y^c.
\end{aligned}
\end{aligned}
\end{equation}
Brownian scaling and uniqueness for \eqref{eq:scaled-state} imply
\[
\mathcal L\bigl((\mathbf Y_x^{r,\eta})_{0\le x\le1}\mid\F_r\bigr)
=\mathcal L\bigl((\mathbf Y_{r+\eta x})_{0\le x\le1}\mid\F_r\bigr).
\]
In the auxiliary-process estimates, $L^p$ norms are taken over $\Omega\times\widetilde\Omega$,
including the original random history; after $\widetilde{\E}$ they are
taken over $\Omega$. In $Q$, $Z$ and $R$, suppress the parameters $r,\eta$
unless comparing different values of $\eta$; also suppress the Brownian
component $j$ in $Z$ and $R$.
Let $Q=(Q^V,Q^b)$ solve the following linear equations, obtained by
formal differentiation of \eqref{eq:scaled-state}. The parameter
difference quotients below will identify $Q=\eta\partial_\eta\mathbf Y^{r,\eta}$:
\begin{equation}\label{eq:q-equation}
\begin{aligned}
&\begin{aligned}[t]
Q_x^V=&\frac{(H-1/2)\eta x}{K_H}
 \int_0^r(r+\eta x-u)^{H-3/2}\beta_H(\mathbf V_u)du\\
&+\frac{(H-1/2)\eta x}{K_H}\sum_{c=1}^m
 \int_0^r(r+\eta x-u)^{H-3/2}\Gamma_{V,\cdot c}(\mathbf V_u)dB_u^c\\
&+\eta\int_0^x\{\beta_0(\mathbf V_y^{r,\eta})+
 \nabla\beta_0(\mathbf V_y^{r,\eta})Q_y^V\}dy\\
&+\frac{\eta^{H+1/2}}{K_H}\int_0^xK(x-y)
 \{(H+1/2)\beta_H(\mathbf V_y^{r,\eta})+
 \nabla\beta_H(\mathbf V_y^{r,\eta})Q_y^V\}dy\\
&+\frac{\eta^H}{K_H}\sum_{c=1}^m\int_0^xK(x-y)
 \{H\Gamma_{V,\cdot c}(\mathbf V_y^{r,\eta})+
 \nabla\Gamma_{V,\cdot c}(\mathbf V_y^{r,\eta})Q_y^V\}d\widetilde B_y^c,
\end{aligned}\\
&\begin{aligned}[t]
Q_x^b=&\eta\int_0^x\{\mu_b(\mathbf Y_y^{r,\eta})+
 \nabla\mu_b(\mathbf Y_y^{r,\eta})Q_y\}dy\\
&+\sqrt \eta\sum_{c=1}^m\int_0^x
 \{\tfrac12\gamma_{b,\cdot c}(\mathbf Y_y^{r,\eta})+
 \nabla\gamma_{b,\cdot c}(\mathbf Y_y^{r,\eta})Q_y\}d\widetilde B_y^c.
\end{aligned}
\end{aligned}
\end{equation}
The two historical terms are controlled by
\begin{align*}
&\eta x\int_0^r(r+\eta x-u)^{H-3/2}du\le C_{H,T}(\eta x)^{H+1/2},\\
&(\eta x)^2\int_0^r(r+\eta x-u)^{2H-3}du\le C_{H,T}(\eta x)^{2H}.
\end{align*}
The terms in \eqref{eq:q-equation} containing a coefficient value,
but no $Q$, satisfy
\begin{align*}
&\eta\int_0^x C_pdy=C_p\eta x,\qquad
 \eta^{H+1/2}\int_0^xK(x-y)C_pdy\le C_p(\eta x)^{H+1/2},\\
&\sqrt \eta\left(\int_0^x C_p^2dy\right)^{1/2}=C_p(\eta x)^{1/2},\qquad
 \eta^H\left(\int_0^xK(x-y)^2C_p^2dy\right)^{1/2}\le C_p(\eta x)^H.
\end{align*}
Together with the two historical estimates, these bounds give, after $t=\eta x$,
\[
\|Q_{t/\eta}\|_p^2\le C_pt^{2H}
 +C_p\int_0^t(t-u)^{2H-1}\|Q_{u/\eta}\|_p^2du.
\]
Equation~\eqref{eq:resolvent}, now with $a=2H$ and $\lambda=2H+1$, gives
\begin{equation}\label{eq:q-bound}
\|Q_x\|_p\le C_p(\eta x)^H.
\end{equation}
Fix a Brownian component $j$. On the same product space, define
$Z=(Z^V,Z^b)$ as the solution of the following linear equation, obtained
by Brownian time scaling in~\eqref{eq:state-derivatives}:
\begin{equation}\label{eq:z-equation}
\begin{aligned}
&\begin{aligned}[t]
Z_x^V=&\frac{\eta^{H-1/2}}{K_H}K(x)\Gamma_{V,\cdot j}(\mathbf V_r)\\
&+\eta\int_0^x\nabla\beta_0(\mathbf V_y^{r,\eta})Z_y^Vdy\\
&+\frac{\eta^{H+1/2}}{K_H}\int_0^xK(x-y)
 \nabla\beta_H(\mathbf V_y^{r,\eta})Z_y^Vdy\\
&+\frac{\eta^H}{K_H}\sum_{c=1}^m\int_0^xK(x-y)
 \nabla\Gamma_{V,\cdot c}(\mathbf V_y^{r,\eta})Z_y^Vd\widetilde B_y^c,
\end{aligned}\\
&\begin{aligned}[t]
Z_x^b=&\gamma_{b,\cdot j}(\mathbf Y_r)
 +\eta\int_0^x\nabla\mu_b(\mathbf Y_y^{r,\eta})Z_y dy\\
&+\sqrt \eta\sum_{c=1}^m\int_0^x
 \nabla\gamma_{b,\cdot c}(\mathbf Y_y^{r,\eta})Z_y d\widetilde B_y^c.
\end{aligned}
\end{aligned}
\end{equation}
By uniqueness, given $\F_r$, the processes $\mathbf Y^{r,\eta}$ and $Z$
have the joint law of the original state and its response $\mathbb{D}_r^{(j)}\mathbf Y$
at times $r+\eta x$, $0<x\le1$.
Write $\mathbb{D}_y^{(j),\mathrm{new}}$ for the Malliavin derivative with
respect to the auxiliary Brownian motion $\widetilde{\mathbf B}$ on the
product space. It fixes the original $\Omega$ coordinate, so every
$\F_r$-measurable history factor has derivative zero, and
\[
\mathbb{D}_y^{(j),\mathrm{new}}\widetilde B_x^c
=\1_{\{j=c\}}\1_{\{0<y\le x\}},\qquad 1\le j,c\le m.
\]
For $0<y<x$, Brownian scaling makes
$\mathbb{D}_y^{(j),\mathrm{new}}\mathbf Y_x^{r,\eta}$ equal to $\sqrt\eta$
times the response at terminal time $r+\eta x$ to the original Brownian
noise at $r+\eta y$, constructed with the same history and
$\widetilde{\mathbf B}$. At scaled time $x$, $Z_x$ represents the original
response $\mathbb{D}_r^{(j)}\mathbf Y_{r+\eta x}$; no finite value for its
rough component at $x=0$ is imposed.
By~\eqref{eq:state-bounds},
\begin{equation}\label{eq:z-bound}
\|Z_x\|_p\le C_p(\eta x)^{H-1/2},\qquad x>0.
\end{equation}
We use the bilinear Hessian convention
\begin{align*}
&\nabla^2F(\mathbf Y_x^{r,\eta})[Q_x,Z_x]
=Q_x^\top\nabla^2F(\mathbf Y_x^{r,\eta})Z_x,\\
&|\nabla^2F(\mathbf Y_x^{r,\eta})[Q_x,Z_x]|
\le C|Q_x|\,|Z_x|.
\end{align*}
Second derivatives of vector-valued coefficients use the same convention componentwise,
with $Q_x^V,Z_x^V$ in the rough-coordinate equations.
Let $R=(R^V,R^b)$ solve the following linear equation. The parameter
difference quotients below will identify $R=\eta\partial_\eta Z$:
\begin{equation}\label{eq:r-equation}
\begin{aligned}
&\begin{aligned}[t]
R_x^V=&\frac{(H-1/2)\eta^{H-1/2}}{K_H}K(x)
 \Gamma_{V,\cdot j}(\mathbf V_r)\\
&+\eta\int_0^x\left\{
 \nabla\beta_0(\mathbf V_y^{r,\eta})(Z_y^V+R_y^V)
 +\nabla^2\beta_0(\mathbf V_y^{r,\eta})[Q_y^V,Z_y^V]\right\}dy\\
&+\frac{\eta^{H+1/2}}{K_H}\int_0^xK(x-y)
 \left\{(H+1/2)\nabla\beta_H(\mathbf V_y^{r,\eta})Z_y^V
 +\nabla\beta_H(\mathbf V_y^{r,\eta})R_y^V\right.\\
&\left.+\nabla^2\beta_H(\mathbf V_y^{r,\eta})[Q_y^V,Z_y^V]
 \right\}dy\\
&+\frac{\eta^H}{K_H}\sum_{c=1}^m\int_0^xK(x-y)
 \left\{H\nabla\Gamma_{V,\cdot c}(\mathbf V_y^{r,\eta})Z_y^V
 +\nabla\Gamma_{V,\cdot c}(\mathbf V_y^{r,\eta})R_y^V\right.\\
&\left.+\nabla^2\Gamma_{V,\cdot c}(\mathbf V_y^{r,\eta})
 [Q_y^V,Z_y^V]\right\}d\widetilde B_y^c,
\end{aligned}\\
&\begin{aligned}[t]
R_x^b=&\eta\int_0^x\left\{
 \nabla\mu_b(\mathbf Y_y^{r,\eta})(Z_y+R_y)
 +\nabla^2\mu_b(\mathbf Y_y^{r,\eta})[Q_y,Z_y]\right\}dy\\
&+\sqrt \eta\sum_{c=1}^m\int_0^x\left\{
 \tfrac12\nabla\gamma_{b,\cdot c}(\mathbf Y_y^{r,\eta})Z_y
 +\nabla\gamma_{b,\cdot c}(\mathbf Y_y^{r,\eta})R_y\right.\\
&\left.+\nabla^2\gamma_{b,\cdot c}(\mathbf Y_y^{r,\eta})
 [Q_y,Z_y]\right\}d\widetilde B_y^c.
\end{aligned}
\end{aligned}
\end{equation}
In \eqref{eq:r-equation}, the initial factors remain fixed as $\eta$
varies. Bounded second derivatives, H\"older and
\eqref{eq:q-bound}--\eqref{eq:z-bound} give
\begin{align*}
\bigl\||Q_y|\,|Z_y|\bigr\|_p
&=\bigl(\E\widetilde{\E}[|Q_y|^p|Z_y|^p]\bigr)^{1/p}\\
&\le\|Q_y\|_{2p}\|Z_y\|_{2p}
\le C_p(\eta y)^{2H-1/2}.
\end{align*}
For the four integrals containing $Z$ and a first coefficient derivative,
the required estimates are
\begin{align*}
&\eta\int_0^x\|Z_y\|_pdy\le C_p(\eta x)^{H+1/2},\\
&\eta^{H+1/2}\int_0^xK(x-y)\|Z_y\|_pdy\le C_p(\eta x)^{2H},\\
&\sqrt \eta\left(\int_0^x\|Z_y\|_p^2dy\right)^{1/2}\le C_p(\eta x)^H,\\
&\eta^H\left(\int_0^xK(x-y)^2\|Z_y\|_p^2dy\right)^{1/2}
 \le C_p(\eta x)^{2H-1/2}.
\end{align*}
For the four integrals containing second coefficient derivatives applied
to $Q_y,Z_y$, replace $\|Z_y\|_p$ by
$\bigl\||Q_y|\,|Z_y|\bigr\|_p$ to obtain
\begin{align*}
&\eta\int_0^x\bigl\||Q_y|\,|Z_y|\bigr\|_pdy\le C_p(\eta x)^{2H+1/2},\\
&\eta^{H+1/2}\int_0^xK(x-y)\bigl\||Q_y|\,|Z_y|\bigr\|_pdy\le C_p(\eta x)^{3H},\\
&\sqrt \eta\left(\int_0^x\bigl\||Q_y|\,|Z_y|\bigr\|_p^2dy\right)^{1/2}\le C_p(\eta x)^{2H},\\
&\eta^H\left(\int_0^xK(x-y)^2\bigl\||Q_y|\,|Z_y|\bigr\|_p^2dy\right)^{1/2}
 \le C_p(\eta x)^{3H-1/2}.
\end{align*}
In the last line the deterministic calculation is
\begin{equation}\label{eq:derivative-products}
\eta^H\left(\int_0^x(x-y)^{2H-1}(\eta y)^{4H-1}dy\right)^{1/2}
=\sqrt{\mathrm B(2H,4H)}(\eta x)^{3H-1/2}.
\end{equation}
For the fourth integral containing $Z$, the same calculation uses
$\mathrm B(2H,2H)$. Both Beta integrals are finite for $H>0$.
Every displayed power is bounded by $C_{H,T}(\eta x)^{H-1/2}$.
The terms containing $R$ in \eqref{eq:r-equation} have bounded linear
coefficients, so, after $t=\eta x$,
\[
\|R_{t/\eta}\|_p^2\le C_pt^{2H-1}
 +C_p\int_0^t(t-u)^{2H-1}\|R_{u/\eta}\|_p^2du.
\]
Apply \eqref{eq:resolvent} with $a=\lambda=2H$ to get
\begin{equation}\label{eq:r-bound}
\|R_x\|_p\le C_p(\eta x)^{H-1/2}.
\end{equation}
We now justify the parameter derivatives, varying $\eta$ with $r$, the
original history and $\widetilde{\mathbf B}$ fixed. Fix
$0<\eta_-\le \eta\le \eta_+$ and let $\eta+\varepsilon$ stay in a slightly larger
compact interval in $(0,T-r]$. At $\eta=T-r$, parameter difference
quotients and derivatives are understood one-sided, from within this interval.
For each coefficient $c$ appearing in
\eqref{eq:scaled-state}, Taylor's formula gives
\begin{align*}
&c(y+z)-c(y)-\nabla c(y)z
 =\int_0^1\{\nabla c(y+uz)-\nabla c(y)\}z\,du,\\
&\nabla c(y+z)-\nabla c(y)-\nabla^2c(y)[z]
 =\int_0^1\{\nabla^2c(y+uz)-\nabla^2c(y)\}[z]du.
\end{align*}
The norms of the two right sides are bounded by $C|z|^2$ under the
second- and third-derivative bounds, respectively. The amplitudes
$\eta$, $\eta^{H+1/2}$, $\eta^{1/2}$ and $\eta^H$ have uniformly bounded first and
second derivatives on this compact $\eta$ interval.
For the integrals over $[0,r]$, use the $L^1$ and squared-kernel bounds
for the first two terms of \eqref{eq:q-equation}.
These justify differentiation under the integrals and provide integrable
bounds for their difference quotients.

Subtract the two state equations at $\eta+\varepsilon$ and $\eta$.
BDG, the coefficient Lipschitz bounds and \eqref{eq:resolvent} first give
\[
\sup_{0<x\le1}x^{-H}
 \|\mathbf Y_x^{r,\eta+\varepsilon}-\mathbf Y_x^{r,\eta}\|_p
 \le C_{p,\eta_-,\eta_+}|\varepsilon|.
\]
Divide this difference by $\varepsilon/\eta$ and subtract
\eqref{eq:q-equation}. The displayed coefficient remainders converge
to zero in the weighted norm. The terms linear in the difference are
controlled by the same resolvent. Thus
\begin{equation}\label{eq:q-difference}
\lim_{\varepsilon\to0}\sup_{0<x\le1}x^{-H}
 \left\|\frac \eta\varepsilon
 (\mathbf Y_x^{r,\eta+\varepsilon}-\mathbf Y_x^{r,\eta})-Q_x\right\|_p=0.
\end{equation}
Apply the same subtraction to \eqref{eq:z-equation}, using the weight
$x^{1/2-H}$. The terms in its coefficient difference are
\begin{align*}
&\{\nabla c(\mathbf Y_y^{r,\eta+\varepsilon})
 -\nabla c(\mathbf Y_y^{r,\eta})\}Z_y^{r,\eta},\\
&\nabla c(\mathbf Y_y^{r,\eta+\varepsilon})
 (Z_y^{r,\eta+\varepsilon}-Z_y^{r,\eta}).
\end{align*}
They give
$\sup_{0<x\le1}x^{1/2-H}\|Z_x^{r,\eta+\varepsilon}-Z_x^{r,\eta}\|_p
\le C|\varepsilon|$.
After division by $\varepsilon/\eta$, the first line converges to
$\nabla^2c(\mathbf Y_y^{r,\eta})[Q_y,Z_y]$ and the second gives the linear
$R$ term. The Taylor bounds above and
\eqref{eq:q-difference} bound their differences by integrable
multiples of $y^{H-1/2}$ and $y^{2H-1/2}$.
The corresponding stochastic square integrals use $y^{2H-1}$ and
$y^{4H-1}$, exactly as in \eqref{eq:derivative-products}.
Dominated convergence and the resolvent therefore prove
\begin{equation*}
\lim_{\varepsilon\to0}\sup_{0<x\le1}x^{1/2-H}
 \left\|\frac \eta\varepsilon
 (Z_x^{r,\eta+\varepsilon}-Z_x^{r,\eta})-R_x\right\|_p=0.
\end{equation*}
Thus $Q=\eta\partial_\eta\mathbf Y^{r,\eta}$ and
$R=\eta\partial_\eta Z$ in $L^p$ for $x>0$.

The conditional-law construction gives
\[
\psi^j(r+\eta,r)=\widetilde{\E}[\nabla F(\mathbf Y_1^{r,\eta})^\top Z_1].
\]
Take $x=1$ in \eqref{eq:q-bound}, \eqref{eq:z-bound} and
\eqref{eq:r-bound}, and use $Q=\eta\partial_\eta\mathbf Y^{r,\eta}$
and $R=\eta\partial_\eta Z$. This gives
\begin{align*}
&\|\partial_\eta\mathbf Y_1^{r,\eta}\|_{2p}
=\eta^{-1}\|Q_1\|_{2p}\le C_p\eta^{H-1},\\
&\|Z_1\|_{2p}\le C_p\eta^{H-1/2},\\
&\|\partial_\eta Z_1\|_p
=\eta^{-1}\|R_1\|_p\le C_p\eta^{H-3/2}.
\end{align*}
These bounds and convergence of the difference quotients justify
differentiation under $\widetilde{\E}$. The chain rule, H\"older and
conditional Jensen give
\begin{equation*}
\begin{aligned}
&\partial_\eta\psi^j(r+\eta,r)
 =\eta^{-1}\widetilde{\E}\left[
 \nabla^2F(\mathbf Y_1^{r,\eta})[Q_1,Z_1]
 +\nabla F(\mathbf Y_1^{r,\eta})^\top R_1\right],\\
&\begin{aligned}[t]
\|\partial_\eta\psi^j(r+\eta,r)\|_p
&\le C_p\{\|\partial_\eta\mathbf Y_1^{r,\eta}\|_{2p}\|Z_1\|_{2p}
+\|\partial_\eta Z_1\|_p\}\\
&\le C_p\{\eta^{H-1}\eta^{H-1/2}+\eta^{H-3/2}\}
=C_p(\eta^{2H-3/2}+\eta^{H-3/2})\\
&\le C_{H,T,p}\eta^{H-3/2}.
\end{aligned}
\end{aligned}
\end{equation*}
The last line uses $\eta^H\le T^H$ for $0<\eta\le T$.
Integrating in $\eta$ proves the last line of~\eqref{eq:conditional-kernel}.

To prove the total-variation bound for $m$, take $r=0$ and $\eta=t$ in
\eqref{eq:scaled-state}. Then $\mathbf Y_0^{0,t}=\mathbf Y_0$, both
historical integrals vanish, and equality of the conditional laws gives
\[
m_t=\widetilde{\E}[F(\mathbf Y_1^{0,t})].
\]
The initial state $\mathbf Y_0$ remains fixed under $\widetilde{\E}$.
Throughout this paragraph, evaluate $Q_x$ at $r=0$, $\eta=t$, so that
$Q_x=t\partial_t\mathbf Y_x^{0,t}$.
With the initial state fixed, apply the same BDG and time-integral estimates
as in~\eqref{eq:q-bound} to~\eqref{eq:q-equation}, obtaining
\[
\widetilde{\E}|Q_x|^2
\le C_{H,T}(tx)^{2H}
+C_{H,T}t^{2H}\int_0^x(x-y)^{2H-1}\widetilde{\E}|Q_y|^2dy,
\qquad 0\le x\le1.
\]
Uniform boundedness of the coefficients and their first derivatives makes
these constants independent of the fixed initial state.
By~\eqref{eq:resolvent},
$\widetilde{\E}|Q_1|\le(\widetilde{\E}|Q_1|^2)^{1/2}\le C_{H,T}t^H$.
The parameter difference-quotient argument above permits differentiation
under $\widetilde{\E}$. Thus the absolutely continuous version of $m$
on $(0,T]$ satisfies, for almost every $t$,
\begin{align*}
&m_t'=t^{-1}\widetilde{\E}[\nabla F(\mathbf Y_1^{0,t})^\top Q_1],\\
&|m_t'|\le t^{-1}\|\nabla F\|_\infty\widetilde{\E}|Q_1|
\le C_{H,T}t^{H-1}.
\end{align*}
Again with the initial state fixed, the increment estimate in
Lemma~\ref{lem:state-bounds} gives
\[
|m_t-F(\mathbf Y_0)|
\le\|\nabla F\|_\infty\widetilde{\E}|\mathbf Y_1^{0,t}-\mathbf Y_0|
\le C_{H,T}t^H.
\]
Hence $m$ is continuous at $0$, and $H>0$ makes its derivative integrable
near $0$. Consequently,
\[
\operatorname{TV}(m;[0,T])\le\int_0^T|m_t'|dt
\le C_{H,T}\int_0^Tt^{H-1}dt
=C_{H,T}\frac{T^H}{H}\quad\hbox{a.s.}
\]
Jointly measurable versions follow from the measurable Picard iterates
and their parameter difference quotients. Predictable projections in $r$
may first be taken for rational terminal times; the local $L^p$ continuity
and absolute continuity just proved extend the versions to $r<t$.
They agree with the Clark--Ocone integrands for almost every $r$ and
preserve~\eqref{eq:clark-ocone}.
\end{proof}

\begin{proof}
The preceding proposition proves Proposition~\ref{prop:lr}(\ref{cond:closed-state}).
Indeed, choose the stated $C_b^3$ extensions on exhausting compact subsets,
preserving the dependence of the rough-state coefficients on $\mathbf V$.
Local pathwise uniqueness, from the Picard estimate in
Lemma~\ref{lem:state-bounds}, makes the extended state agree with
the original state before exit. The exit probabilities tend to zero by
compact containment. For each extension define
$\widetilde L_t^N=F_N(\mathbf V_t^N,\mathbf b_t^N)$.
Apply Proposition~\ref{prop:conditional-kernel} to this state. In
\eqref{eq:lr}, take
\begin{align*}
&A_t^{N,\mathrm{BV}}=\E[F_N(\mathbf Y_t^N)\mid\F_0],\qquad
A_t^{N,\mathrm{reg}}=0,\\
&\psi_N^j(t,u)=\E[\mathbb{D}_u^{(j)}F_N(\mathbf Y_t^N)\mid\F_u],\qquad H_N=H.
\end{align*}
The variation and kernel bounds in that proposition verify LR; the supremum moment follows from boundedness of $F_N$.
\end{proof}
\subsection{\texorpdfstring{General adapted state coefficients for $H>1/6$}{General adapted state coefficients for H>1/6}}
\begin{proof}
Fix a representative in Proposition~\ref{prop:lr}(\ref{cond:general-state})
and omit its index $N$ in this proof. All constants may depend on $N$.
The normalization $\mathbf A_0^V=0$ ensures that the initial state in the
It\^o--Young formula below is $(\mathbf V_0,\mathbf b_0)$.
Write the It\^o equation as
\begin{equation*}
\mathbf b_t=\mathbf b_0+\int_0^t\mu_{b,u}\,du+\int_0^t\gamma_{b,u}\,d\mathbf B_u.
\end{equation*}
For each terminal time $t$, define, for $0\le u\le t$,
\begin{equation*}
\mathbf V_{t,u}=\mathbf V_0+\mathbf A_u^V+K_H^{-1}\int_0^u(t-v)^{H-1/2}\gamma_v^V\,d\mathbf B_v.
\end{equation*}
It is adapted in $u$ and satisfies $\mathbf V_{t,t}=\mathbf V_t$.
For $u<t$, differentiating the deterministic kernel with respect to the
terminal time $t$ gives, in $L^p$,
\begin{equation*}
\partial_t\mathbf V_{t,u}
=K_H^{-1}(H-1/2)\int_0^u(t-v)^{H-3/2}\gamma_v^V\,d\mathbf B_v.
\end{equation*}
BDG and the kernel increment estimates then give
\begin{equation}\label{eq:parameter-bounds}
\begin{aligned}
&\|\partial_t\mathbf V_{t,u}\|_p
\le C_p\left(\int_0^u(t-v)^{2H-3}\,dv\right)^{1/2}
\le C_p(t-u)^{H-1},\\
&\|\mathbf V_{t,u}-\mathbf V_{s,u}\|_p
\le C_p\min\{(t-s)^H,(t-s)(s-u)^{H-1}\},\quad u<s<t,\\
&\|\mathbf V_{t,u}-\mathbf V_u\|_p\le C_p(t-u)^H.
\end{aligned}
\end{equation}
The increment condition on $\mathbf A^V$, at arbitrarily large finite
moment orders, gives a version with path H\"older exponent greater than $1/2$.
For fixed $t$, apply the It\^o formula with Young integration for $\mathbf A^V$
to $F(\mathbf V_{t,u},\mathbf b_u)$ on $[0,t-\varepsilon]$. Taking
$\varepsilon\downarrow0$ yields
\begin{equation}
\begin{aligned}
\widetilde L_t=&F(\mathbf V_0,\mathbf b_0)+Q_t+I_{0,t}+I_{1,t}+I_{A,t}\\
&+\int_0^t\psi(t,u)\,d\mathbf B_u,
\end{aligned}
\label{eq:ito-young}
\end{equation}
where gradients are column vectors and Hessian blocks are taken in the
indicated rough or It\^o coordinates. Every derivative of $F$ in the
following formulas is evaluated at $(\mathbf V_{t,u},\mathbf b_u)$:
\begin{align*}
&\begin{aligned}[t]
\psi(t,u)=&K_H^{-1}(t-u)^{H-1/2}(\nabla_{\mathbf V}F)^\top\gamma_u^V\\
&+(\nabla_{\mathbf b}F)^\top\gamma_{b,u},
\end{aligned}\\
&Q_t=(2K_H^2)^{-1}\int_0^t(t-u)^{2H-1}
\operatorname{tr}(\nabla_{\mathbf V\mathbf V}^2F\,\Sigma_u)\,du,\\
&I_{0,t}=\int_0^t\left\{(\nabla_{\mathbf b}F)^\top\mu_{b,u}
+\frac12\operatorname{tr}(\nabla_{\mathbf b\mathbf b}^2F\,\gamma_{b,u}\gamma_{b,u}^\top)\right\}du,\\
&I_{1,t}=K_H^{-1}\int_0^t(t-u)^{H-1/2}
\operatorname{tr}((\gamma_u^V)^\top\nabla_{\mathbf V\mathbf b}^2F\,\gamma_{b,u})\,du,\\
&I_{A,t}=\int_0^t\nabla_{\mathbf V}F(\mathbf V_{t,u},\mathbf b_u)^\top\,d\mathbf A_u^V.
\end{align*}
All stochastic integrands here are predictable in $u$. The estimates
below also justify the limit at $u=t$ in~\eqref{eq:ito-young}.
First, bounded derivatives of $F$, the coefficient moments and
\eqref{eq:parameter-bounds} imply
\begin{equation}\label{eq:general-kernel}
\begin{aligned}
&\|\psi(t,u)\|_2\le C\{1+(t-u)^{H-1/2}\},\\
&\begin{aligned}[t]
\|\partial_t\psi(t,u)\|_2
&\le C\{(t-u)^{H-3/2}+(t-u)^{2H-3/2}+(t-u)^{H-1}\}\\
&\le C_T\{1+(t-u)^{H-3/2}\}.
\end{aligned}
\end{aligned}
\end{equation}
Integrating the last estimate from $s$ to $t$ proves both kernel
bounds in LR with $H_N=H$. It remains to separate a bounded-variation part
from the time integrals in~\eqref{eq:ito-young}.
For the term $Q$, define
\begin{align*}
&c(t,u)=(2K_H^2)^{-1}\operatorname{tr}
\bigl(\nabla_{\mathbf V\mathbf V}^2F(\mathbf V_{t,u},\mathbf b_u)\Sigma_u\bigr),\\
&C_u=c(u,u).
\end{align*}
Fix $0\le s<t\le T$ and write $\delta=t-s$.
The $H$-order increment bound on $\Sigma$,
\eqref{eq:parameter-bounds}, and the It\^o increment estimate give
\begin{equation}\label{eq:coefficient-bounds}
\begin{aligned}
&\|c(s,u)-C_s\|_p\le C_p(s-u)^H,\qquad u<s,\\
&\|c(t,u)-c(s,u)\|_p
\le C_p\min\{\delta^H,\delta(s-u)^{H-1}\},\qquad u<s,\\
&\|c(t,u)-C_s\|_p\le C_p\delta^H,\qquad s\le u\le t,\\
&\|C_t-C_s\|_p\le C_p\delta^H.
\end{aligned}
\end{equation}
Subtract the same value $C_s$ inside the old and new time integrals.
The resulting exact identity is
\begin{align*}
&Q_t-Q_s-C_s\frac{t^{2H}-s^{2H}}{2H}\\
&=\int_0^s\{(t-u)^{2H-1}-(s-u)^{2H-1}\}\{c(s,u)-C_s\}\,du\\
&+\int_0^s(t-u)^{2H-1}\{c(t,u)-c(s,u)\}\,du
+\int_s^t(t-u)^{2H-1}\{c(t,u)-C_s\}\,du.
\end{align*}
Split each old integral at $s-u=\delta$. On the part where
$s-u\ge\delta$, the kernel difference is bounded by
$C\delta(s-u)^{2H-2}$; on the remaining part use the sum of the two kernels.
Together with~\eqref{eq:coefficient-bounds}, this gives
\begin{align*}
&\left\|Q_t-Q_s-C_s\frac{t^{2H}-s^{2H}}{2H}\right\|_p
\le C_p\left\{\delta^{3H}+\delta\int_\delta^T x^{3H-2}\,dx\right\},\\
&\left\|\int_s^t C_u u^{2H-1}\,du-C_s\frac{t^{2H}-s^{2H}}{2H}\right\|_p
\le C_p\int_s^t(u-s)^H u^{2H-1}\,du\le C_p\delta^{3H}.
\end{align*}
The first integral is of order $\delta^{3H}$ when $3H<1$,
$\delta|\log\delta|$ when $3H=1$, and $\delta$ when $3H>1$.
In all cases,
\begin{equation*}
\left\|Q_t-Q_s-\int_s^tC_u u^{2H-1}\,du\right\|_p
\le C_p\delta^{\min\{3H,H+1/2\}}.
\end{equation*}
For the other Lebesgue integrals, use the same parameter increment
estimate. The part over $[s,t]$ and the change in the integrand over $[0,s]$
satisfy
\begin{align*}
&\|I_{0,t}-I_{0,s}\|_p
\le C_p\delta+C_p\int_0^s\min\{\delta^H,\delta(s-u)^{H-1}\}\,du
\le C_p\delta,\\
&\begin{aligned}[t]
\|I_{1,t}-I_{1,s}\|_p
&\le C_p\delta^{H+1/2}
+C_p\int_0^s(x+\delta)^{H-1/2}\min\{\delta^H,\delta x^{H-1}\}\,dx\\
&\le C_p\delta^{H+1/2}.
\end{aligned}
\end{align*}
We give the Young-integral estimate explicitly. If, on an interval
$[a,b]$, an $\mathbb R^d$-valued process $f$ satisfies
$\|f_v-f_u\|_{2p}\le M|v-u|^{1/2}$, then the increment
condition on $\mathbf A^V$ and H\"older's inequality imply
\begin{equation}\label{eq:young-bound}
\begin{aligned}
&\|(f_v-f_u)^\top(\mathbf A_w^V-\mathbf A_v^V)\|_p\le C_pM(w-u)^{H+1},\\
&\left\|\int_a^b f_u^\top\,d\mathbf A_u^V-f_a^\top(\mathbf A_b^V-\mathbf A_a^V)\right\|_p
\le C_pM(b-a)^{H+1}.
\end{aligned}
\end{equation}
The second inequality follows by summing the errors when successive
Riemann sums are refined; the first has exponent $H+1>1$.
Partition each interval into subintervals whose length is comparable to its
distance from the singular endpoint. Denote that length by $x$.
For $u,v$ in one such subinterval, BDG gives
\begin{align*}
&\|\mathbf V_{t,v}-\mathbf V_{t,u}\|_{2p}
\le C_p(1+x^{H-1/2})|v-u|^{1/2},\\
&\|(\mathbf V_{t,v}-\mathbf V_{s,v})-(\mathbf V_{t,u}-\mathbf V_{s,u})\|_{2p}
\le C_p\delta x^{H-3/2}|v-u|^{1/2},\quad x\ge\delta.
\end{align*}
The second estimate uses the difference of the two deterministic
kernels; the $\mathbf A^V$ terms cancel. Bounded derivatives of $F$, the It\^o increment
bound, and~\eqref{eq:parameter-bounds} therefore give the following
bounds for the vector integrand in $I_A$. In the last two rows,
$f_u=\nabla_{\mathbf V}F(\mathbf V_{t,u},\mathbf b_u)-\nabla_{\mathbf V}F(\mathbf V_{s,u},\mathbf b_u)$:
\begin{equation*}
\begin{array}{c|c|c}
 &\|f_u\|_{2p}&\|f_v-f_u\|_{2p}/|v-u|^{1/2}\\ \hline
u,v\in[s,t]&C_p&C_p(1+x^{H-1/2})\\
u,v<s,\ x\le\delta&C_p\delta^H&C_px^{H-1/2}\\
u,v<s,\ x\ge\delta&C_p\delta x^{H-1}&C_p\delta x^{H-3/2}
\end{array}
\end{equation*}
In the first row take $f_u=\nabla_{\mathbf V}F(\mathbf V_{t,u},\mathbf b_u)$,
and measure the distance from $t$; in the other rows measure it from $s$.
Applying~\eqref{eq:young-bound} on each subinterval bounds its
integral respectively by
\begin{equation*}
C_p(x^{H+1/2}+x^{2H+1/2}+x^{H+1}),\qquad
C_p\delta^H x^{H+1/2},\qquad C_p\delta x^{2H-1/2}.
\end{equation*}
Summing over geometrically decreasing lengths gives
\begin{align*}
\|I_{A,t}-I_{A,s}\|_p
&\le C_p\{\delta^{H+1/2}+\delta^{2H+1/2}
+\delta(1+|\log\delta|)\}\\
&\le C_{p,T}\delta^{H+1/2}.
\end{align*}
These estimates also show convergence of the Young integral at its
terminal endpoint. The other terms in~\eqref{eq:ito-young}
converge there by BDG and the integrability of the displayed powers.
Thus the identity holds on $[0,T]$. Now define
\begin{align*}
&A_t^{\mathrm{BV}}=F(\mathbf V_0,\mathbf b_0)+\int_0^t C_u u^{2H-1}\,du,\\
&A_t^{\mathrm{reg}}=Q_t-\int_0^t C_u u^{2H-1}\,du+I_{0,t}+I_{1,t}+I_{A,t}.
\end{align*}
They are adapted. The uniform moments of $C$ and the preceding bounds
imply
\begin{equation}\label{eq:remainder-regularity}
\begin{aligned}
&\|\operatorname{TV}(A^{\mathrm{BV}};[0,T])\|_2
\le\int_0^T\|C_u\|_2u^{2H-1}\,du\le C_T,\\
&\|A_t^{\mathrm{reg}}-A_s^{\mathrm{reg}}\|_2
\le C_T|t-s|^{\min\{3H,H+1/2\}}.
\end{aligned}
\end{equation}
Since $H>1/6$, the exponent in the second line exceeds $1/2$.
Also $F_N\in C_b^4$ gives the required square-integrable supremum of
$\widetilde L^N$. Restoring $N$, equations~\eqref{eq:ito-young},
\eqref{eq:general-kernel}, and
\eqref{eq:remainder-regularity} verify LR for each local
representative, with $H_N=H$ and $\zeta_N=\min\{3H,H+1/2\}>1/2$.
Hence $L$ satisfies LR.
\end{proof}

%% file: appendices/C_limit.tex
\section{Proofs of the limits for \texorpdfstring{$T_n$}{Tn}}
We prove Theorems~\ref{thm:corrected-clt} and~\ref{thm:rcv-clt}.
Fix the localization level in Proposition~\ref{prop:localization} and omit
$(N)$ from the processes in the following estimates. Constants may depend on $N$.
Write the price decomposition as
\begin{equation*}
\begin{aligned}
&X_t=X_0+M_t+A_t,\\
&M_t=\int_0^t\sigma_s\,dW_s,\\
&A_t=\int_0^t\mu_s\,ds.
\end{aligned}
\end{equation*}
Define
\begin{equation*}
P_n^L(t)=\Delta_n^{-1/2}\sum_{i:t_i\le t}\int_{t_{i-1}}^{t_i}
\Lambda_i(s)\E[L_s-L_{t_{i-1}}\mid\F_{t_{i-1}}]\,ds.
\end{equation*}
\begin{lemma}\label{lem:stable-clt}
The following form of the martingale limit theorem is due to \citet{Jacod1997}.
Let $\widetilde W$ be a Brownian motion with respect to $(\F_t)$.
In the vector case, $\vartheta_s^2$ in the variance condition denotes the
squared Euclidean norm of the Brownian loading.
Let \(Y_t^n=\sum_{i=1}^{\lfloor t/\Delta_n\rfloor}X_{in}\), \(t\ge0\),
where $X_{in}\in L^2(\F_{i\Delta_n})$.
Assume there exist absolutely continuous processes $F$, $G$, and a continuous
process $\Upsilon$ with finite variation such that
\begin{align}
&\sum_{i=1}^{\lfloor t/\Delta_n\rfloor}
\E\bigl(X_{in}\mid \F_{(i-1)\Delta_n}\bigr)
\convucp\Upsilon_t,
\label{eq:clt-mean}\\
&\sum_{i=1}^{\lfloor t/\Delta_n\rfloor}
\Bigl(\E\bigl(X_{in}^2\mid \F_{(i-1)\Delta_n}\bigr)
-\E\bigl(X_{in}\mid \F_{(i-1)\Delta_n}\bigr)^2\Bigr)
\convp F_t=\int_0^t(\vartheta_s^2+\varpi_s^2)\,ds,
\label{eq:clt-variance}\\
&\sum_{i=1}^{\lfloor t/\Delta_n\rfloor}
\E\bigl(X_{in}\Delta_i^n\widetilde W\mid \F_{(i-1)\Delta_n}\bigr)
\convp G_t=\int_0^t\vartheta_s\,ds,
\label{eq:clt-brownian}\\
&\sum_{i=1}^{\lfloor t/\Delta_n\rfloor}
\E\bigl(X_{in}^2\mathbf{1}_{\{|X_{in}|>\eps\}}
\mid \F_{(i-1)\Delta_n}\bigr)
\convp0,\qquad\forall\eps>0,
\label{eq:clt-lindeberg}\\
&\sum_{i=1}^{\lfloor t/\Delta_n\rfloor}
\E\bigl(X_{in}\Delta_i^nN\mid \F_{(i-1)\Delta_n}\bigr)
\convp0.
\label{eq:clt-martingale}
\end{align}
Here $(\vartheta_s)_{s\ge0}$ and $(\varpi_s)_{s\ge0}$ are predictable processes,
and the last condition holds for every bounded $\F_t$-martingale $N$ with
$N_0=0$ and $\langle\widetilde W,N\rangle=0$. Then
\[
Y_t^n\convst Y_t=\Upsilon_t+\int_0^t\vartheta_s\,d\widetilde W_s
+\int_0^t\varpi_s\,d\widetilde W_s^\perp.
\]
where $\widetilde W^\perp$ is a Brownian motion defined on an extension of
$(\Omega,\F,\P)$ and independent of $\F$.
When $\widetilde W$ is vector-valued, the third condition and
$\langle\widetilde W,N\rangle=0$ are understood componentwise, and
$\int_0^t\vartheta_s\,d\widetilde W_s$ denotes the corresponding vector stochastic integral.
\end{lemma}

\begin{lemma}\label{lem:compensator}
Let $D_t=\int_0^tK(t-s)g_s\,ds$. At the fixed localization level,
$\|\Delta_i^nD\|_p\le C_p\Delta_n^{H+1/2}$ and
\[
\Delta_n^{-H}\sup_{N\le n}\left|\sum_{i=1}^N
\Delta_i^nX\,\Delta_i^nD\right|\convp0.
\]
Moreover, the true conditional compensator satisfies
\begin{equation}\label{eq:compensator}
\sup_{t\le T}\left|\Delta_n^{-1/2}\sum_{i:t_i\le t}
\E[U_i\mid\F_{t_{i-1}}]-P_n^L(t)\right|\convp0.
\end{equation}
\end{lemma}
\begin{proof}
For a c\`adl\`ag coefficient $c$ with $\E\sup_{s\le T}|c_s|^p<\infty$ and fixed $p\ge2$,
\begin{equation}\label{eq:cadlag-sum}
\sum_i\Delta_n\left\|\sup_{s\in[t_{i-1},t_i]}|c_s-c_{t_{i-1}}|\right\|_p^2
\longrightarrow0.
\end{equation}
Indeed, the oscillations of a c\`adl\`ag path on shrinking grid cells tend
to zero at almost every time and are bounded by $2\sup_{s\le T}|c_s|$.
Dominated convergence and H\"older give
\begin{align*}
&\sum_i\Delta_n\left\|\sup_{s\in[t_{i-1},t_i]}|c_s-c_{t_{i-1}}|\right\|_p^2\\
&\le T^{1-2/p}\left(\sum_i\Delta_n\E\left[
\sup_{s\in[t_{i-1},t_i]}|c_s-c_{t_{i-1}}|^p\right]\right)^{2/p}
\longrightarrow0.
\end{align*}
In particular this applies to $\sigma$ and
$v\sqrt{1-\rho^2}$ at this localization level.

The exact decomposition is
\begin{align*}
\Delta_i^nD=&
\underbrace{\int_0^{t_{i-1}}[K(t_i-s)-K(t_{i-1}-s)]g_s\,ds
+\frac{\Delta_n^{H+1/2}}{H+1/2}g_{t_{i-1}}}_{\mathrm I_{1,i}}\\
&+\underbrace{\int_{t_{i-1}}^{t_i}K(t_i-s)(g_s-g_{t_{i-1}})\,ds}_{\mathrm I_{2,i}}.
\end{align*}
Here $\mathrm I_{1,i}$ is $\F_{t_{i-1}}$-measurable and
$|\mathrm I_{1,i}|\le C\Delta_n^{H+1/2}$. Doob's inequality gives
\[
\left\|\Delta_n^{-H}\max_N\left|\sum_{i\le N}\Delta_i^nM\,\mathrm I_{1,i}\right|\right\|_2
\le C\Delta_n^{1/2}.
\]
For $\mathrm I_{2,i}$, Cauchy--Schwarz and \eqref{eq:cadlag-sum} give
\[
\Delta_n^{-H}\sum_i\|\Delta_i^nM\,\mathrm I_{2,i}\|_1
\le C\sum_i\Delta_n
\left\|\sup_{s\in[t_{i-1},t_i]}|g_s-g_{t_{i-1}}|\right\|_2=o(1).
\]
The price drift contributes at most $C\Delta_n^{1/2}$ to this normalized sum.
This proves the first assertion and its conditional-mean version.

Conditional It\^o isometry, with $d\langle W,B\rangle_s=\rho_sds$, gives
\begin{align*}
&\E\left[\Delta_i^nM\,K_H^{-1}\int_{t_{i-1}}^{t_i}K(t_i-s)v_s\,dB_s
\Bigm|\F_{t_{i-1}}\right]\\
&=K_H^{-1}\E\left[\int_{t_{i-1}}^{t_i}K(t_i-s)L_s\,ds\Bigm|\F_{t_{i-1}}\right].
\end{align*}
The stochastic integral over $[0,t_{i-1}]$ is $\F_{t_{i-1}}$-measurable and its
product with $\Delta_i^nM$ has conditional mean zero. Thus the difference in
\eqref{eq:compensator} is exactly the sum of the conditional
expectations of
\[
\Delta_n^{-H}\left\{
\underbrace{\left(\int_{t_{i-1}}^{t_i}\mu_s\,ds\right)\Delta_i^n\sigma^2}_{\mathrm J_{1,i}}
+\underbrace{\Delta_i^nM\int_{t_{i-1}}^{t_i}f_s\,ds}_{\mathrm J_{2,i}}
+\underbrace{\Delta_i^nM\,\Delta_i^nD}_{\mathrm J_{3,i}}\right\}.
\]
The absolute $L^1$ sum of $\Delta_n^{-H}\mathrm J_{2,i}$ is
$O(\Delta_n^{1/2-H})$; $\mathrm J_{3,i}$ has just been treated.
For $\mathrm J_{1,i}$, first take a bounded elementary predictable drift on a
fixed deterministic partition. On each partition interval its value is a
bounded variable measurable at the interval's left endpoint. Away from the
finitely many partition boundaries,
\[
\Delta_n^{-H}\sum_i\mathrm J_{1,i}
=\Delta_n^{1-H}\mu^{\rm step}\sum_i\Delta_i^n\sigma^2.
\]
The last sum telescopes and is bounded in probability uniformly over its upper
endpoint. The difference between this sum and its predictable projection is a
martingale with $L^2$ maximal norm at most $C\Delta_n^{1/2}$, since
$\|\Delta_i^n\sigma^2\|_p\le C_p\Delta_n^H$. Finitely many boundary terms
contribute $O_{L^1}(\Delta_n)$. Finally choose elementary predictable drifts
converging to $\mu$ in $L^2(\Omega\times[0,T])$, uniformly bounded. For their
difference $\nu$,
\[
\Delta_n^{-H}\sum_i
\left\|\left(\int_{t_{i-1}}^{t_i}\nu_s\,ds\right)\Delta_i^n\sigma^2\right\|_1
\le C\|\nu\|_{L^2(\Omega\times[0,T])}.
\]
Conditional Jensen preserves this bound. Let $n\to\infty$ before letting the
approximation error tend to zero. The zero integral of $\Lambda_i$ then proves
\eqref{eq:compensator}.
\end{proof}

\begin{lemma}
Let
\[
\zeta_{i,n}=K_H^{-1}\int_0^{t_{i-1}}
[K(t_i-s)-K(t_{i-1}-s)]v_s\,dB_s.
\]
With $C_H=\int_0^\infty[(x+1)^{H-1/2}-x^{H-1/2}]^2dx=K_H^2-(2H)^{-1}$, for every bounded adapted
c\`adl\`ag process $c$,
\begin{align}
&\Delta_n^{1-H}\sup_{N\le n}\left|\sum_{i=1}^Nc_{t_{i-1}}\zeta_{i,n}\right|
\convp0,\label{eq:memory-linear}\\
&\Delta_n^{1-2H}\sum_{i:t_i\le t}c_{t_{i-1}}\zeta_{i,n}^2
\convucp\frac{C_H}{K_H^2}\int_0^tc_sv_s^2\,ds.
\label{eq:memory-square}
\end{align}
\end{lemma}
\begin{proof}
The historical kernel differences satisfy
\begin{align*}
&\int_0^{t_{i-1}}[K(t_i-s)-K(t_{i-1}-s)]^2\,ds\le C\Delta_n^{2H},\\
&\begin{aligned}[t]
&\int_0^{t_{i-1}\wedge t_{j-1}}
|K(t_i-s)-K(t_{i-1}-s)|\,|K(t_j-s)-K(t_{j-1}-s)|\,ds\\
&\le C\Delta_n^{2H}(1+|i-j|)^{H-3/2}.
\end{aligned}
\end{align*}
For the second inequality, scale by $\Delta_n$, put
$f(x)=(1+x)^{H-1/2}-x^{H-1/2}$, and split
$\int_0^\infty|f(x)f(x+r)|\,dx$ at $r/2$.
On the first part use $f\in L^1(0,\infty)$ and
$|f(x+r)|\le Cr^{H-3/2}$; on the second use
$|f(x)|\le Cx^{H-3/2}$. The resulting exponents are summable.
BDG also gives $\|\zeta_{i,n}\|_p\le C_p\Delta_n^H$.

For \eqref{eq:memory-linear}, approximate $c$ by its bounded predictable
step values on a deterministic coarse partition. On each such interval use
\[
\zeta_{i,n}=K_H^{-1}\Delta_i^n\!\left(\int_0^\cdot K(\cdot-s)v_s\,dB_s\right)
-K_H^{-1}\int_{t_{i-1}}^{t_i}K(t_i-s)v_s\,dB_s.
\]
The first sum telescopes, and the second is a martingale sum whose normalized
maximal $L^2$ norm is $O(\sqrt{\Delta_n})$. The coefficient-approximation error
is bounded in $L^1$ by $C\sum_i\Delta_n\|c_{t_{i-1}}-c^{\rm step}_{t_{i-1}}\|_2$.
This tends to zero upon refinement of the coarse partition.

For \eqref{eq:memory-square}, fix a deterministic partition of mesh
$\varepsilon$. On a partition interval beginning at $b$, first discard the
integral over $[0,b]$. If $j$ counts the grid distance from $b$, its squared
$L^p$ norm is bounded by
$C_p\Delta_n^{2H}(1+j)^{2H-2}$. Summing $j$ and then the finitely many
partition intervals bounds its normalized square sum by $C\Delta_n/\varepsilon$;
Cauchy--Schwarz controls cross-products. Replace $v_s$ on $[b,t_{i-1}]$ by $v_b$
and $c_{t_{i-1}}$ by $c_b$. The errors tend to zero as $\varepsilon\downarrow0$
by the coarse-partition version of \eqref{eq:cadlag-sum}.
For indices with $t_{i-1}\ge b$, set
\[
d_i(s)=[K(t_i-s)-K(t_{i-1}-s)]\1_{\{b<s<t_{i-1}\}}.
\]
After replacing the coefficients, the comparison variable and its conditional
second moment are
\begin{align*}
&K_H^{-1}v_b\int_b^{t_{i-1}}d_i(s)dB_s,\\
&\E\!\left[\left(K_H^{-1}v_b\int_b^{t_{i-1}}d_i(s)dB_s\right)^2\middle|\F_b\right]
=K_H^{-2}v_b^2\int_b^{t_{i-1}}d_i(s)^2ds.
\end{align*}
Conditionally on $\F_b$, the integrals in these expressions are jointly
Gaussian with deterministic kernels. Their centered squares satisfy
\begin{align*}
&\operatorname{Cov}\left[\left(\int_b^{t_{i-1}}d_i(s)\,dB_s\right)^2,
\left(\int_b^{t_{j-1}}d_j(s)\,dB_s\right)^2\Bigm|\F_b\right]\\
&=2\left(\int_b^{t_{i-1}\wedge t_{j-1}}d_i(s)d_j(s)\,ds\right)^2.
\end{align*}
The normalized variance of the centered sum is bounded explicitly by
\[
C\Delta_n^{2-4H}\sum_{i,j}
 \left(\int_b^{t_{i-1}\wedge t_{j-1}}|d_i(s)d_j(s)|ds\right)^2
\le C\Delta_n^2\sum_{i,j}(1+|i-j|)^{2H-3}
\le C_T\Delta_n.
\]
The sums here range over the grid cells within the fixed partition interval;
conditioning and the bounded frozen coefficients preserve the bound. The conditional mean differs from
$(C_H/K_H^2)\sum_i\Delta_nc_bv_b^2$ by at most $C\Delta_n/\varepsilon$,
because the omitted deterministic squared tail has exponent $2H-2$.
For $c\ge0$ the partial sums are increasing; convergence at finitely many time
points and continuity of the limit imply uniform convergence. General $c$ is
the difference of its positive and negative parts. Let $n\to\infty$ and then
$\varepsilon\downarrow0$.
\end{proof}

For the remaining Jacod conditions, represent the Brownian pair driving the
price and spot variance on an independent extension as
\[
dB_s=\rho_s\,dW_s+\sqrt{1-\rho_s^2}\,dW_s^\perp.
\]
One construction uses $B-\int\rho\,dW$ on $\{|\rho|<1\}$, divided by its
instantaneous standard deviation, and an auxiliary independent Brownian motion
on $\{|\rho|=1\}$. The resulting pair $(W,W^\perp)$ has identity bracket and
is a two-dimensional Brownian motion. No derivative of $\sqrt{1-\rho^2}$ is
used. Define, on the current grid cell,
\[
\widetilde Y_{i,n}=
\rho_{t_{i-1}}\int_{t_{i-1}}^{t_i}K(t_i-s)\,dW_s
+\sqrt{1-\rho_{t_{i-1}}^2}\int_{t_{i-1}}^{t_i}K(t_i-s)\,dW_s^\perp
\]
and the conditionally centered comparison array
\begin{equation}\label{eq:rcv-frozen}
\widetilde\xi_{i,n}=
\Delta_n^{-H}\sigma_{t_{i-1}}
\left[\Delta_i^nW\,\zeta_{i,n}
+\frac{v_{t_{i-1}}}{K_H}
\left\{\Delta_i^nW\,\widetilde Y_{i,n}
-\rho_{t_{i-1}}\frac{\Delta_n^{H+1/2}}{H+1/2}\right\}\right].
\end{equation}
Only $(\Delta_i^nW,\widetilde Y_{i,n},\Delta_i^nW^\perp)$ is asserted to be
conditionally Gaussian. The historical quantity $\zeta_{i,n}$ remains
$\F_{t_{i-1}}$-measurable.

\begin{lemma}
For $\xi_{i,n}=\Delta_n^{-1/2}U_i$,
\begin{equation}\label{eq:rcv-transfer}
\begin{aligned}
&\sum_{i=1}^n
\|\xi_{i,n}-\E[\xi_{i,n}\mid\F_{t_{i-1}}]-\widetilde\xi_{i,n}\|_2^2=o(1),\\
&\sum_i\E|\widetilde\xi_{i,n}|^4=O(\Delta_n).
\end{aligned}
\end{equation}
\end{lemma}
\begin{proof}
For the first assertion of \eqref{eq:rcv-transfer}, define the errors
\begin{align*}
&e_{i,n}^{X}=\int_{t_{i-1}}^{t_i}\mu_sds
 +\int_{t_{i-1}}^{t_i}(\sigma_s-\sigma_{t_{i-1}})dW_s,\\
&\begin{aligned}[t]
e_{i,n}^{V}=&\int_{t_{i-1}}^{t_i}f_sds+D_{t_i}-D_{t_{i-1}}\\
&+\frac1{K_H}\int_{t_{i-1}}^{t_i}K(t_i-s)
 (v_s\rho_s-v_{t_{i-1}}\rho_{t_{i-1}})dW_s\\
&+\frac1{K_H}\int_{t_{i-1}}^{t_i}K(t_i-s)
 \left(v_s\sqrt{1-\rho_s^2}-v_{t_{i-1}}\sqrt{1-\rho_{t_{i-1}}^2}\right)dW_s^\perp,
\end{aligned}\\
&e_{i,n}^{\theta}=\frac1{K_H(H+1/2)}
 \int_{t_{i-1}}^{t_i}(L_s-L_{t_{i-1}})ds.
\end{align*}
The price and spot variance increments satisfy
\[
\begin{aligned}
&\Delta_i^nX=\sigma_{t_{i-1}}\Delta_i^nW+e_{i,n}^X,\\
&\Delta_i^n\sigma^2=\zeta_{i,n}+\frac{v_{t_{i-1}}}{K_H}\widetilde Y_{i,n}+e_{i,n}^V.
\end{aligned}
\]
The covariance subtracted in~\eqref{eq:rcv-frozen} is
$\E[\Delta_i^nW\,\widetilde Y_{i,n}\mid\F_{t_{i-1}}]$.
Multiplying these identities, substituting in
$\xi_{i,n}=\Delta_n^{-1/2}U_i$ and conditionally centering give
\begin{align*}
\xi_{i,n}-\E[\xi_{i,n}\mid\F_{t_{i-1}}]-\widetilde\xi_{i,n}
=&(I-\E[\,\cdot\mid\F_{t_{i-1}}])\\
&\quad\Big[\Delta_n^{-H}
 \Big\{e_{i,n}^X\left(\zeta_{i,n}+\frac{v_{t_{i-1}}}{K_H}\widetilde Y_{i,n}\right)\\
&\qquad+\sigma_{t_{i-1}}\Delta_i^nW\,e_{i,n}^V+e_{i,n}^Xe_{i,n}^V\Big\}
 -\Delta_n^{-1/2}e_{i,n}^{\theta}\Big].
\end{align*}
BDG and the kernel energy give
\begin{align*}
&\|e_{i,n}^X\|_p
\le C_p\left\{\Delta_n+
 \sqrt{\Delta_n}\left\|\sup_{t_{i-1}\le s\le t_i}
 |\sigma_s-\sigma_{t_{i-1}}|\right\|_p\right\},\\
&\begin{aligned}[t]
\|e_{i,n}^V\|_p
\le& C_p\Delta_n^{H+1/2}
 +C_p\Delta_n^H\left\{\left\|\sup_{t_{i-1}\le s\le t_i}
 |v_s\rho_s-v_{t_{i-1}}\rho_{t_{i-1}}|\right\|_p\right.\\
&\left.+\left\|\sup_{t_{i-1}\le s\le t_i}
 \left|v_s\sqrt{1-\rho_s^2}
 -v_{t_{i-1}}\sqrt{1-\rho_{t_{i-1}}^2}\right|\right\|_p\right\},
\end{aligned}\\
&\|e_{i,n}^{\theta}\|_p
\le C_p\Delta_n\left\|\sup_{t_{i-1}\le s\le t_i}
 |L_s-L_{t_{i-1}}|\right\|_p.
\end{align*}
Also
$\|\zeta_{i,n}+v_{t_{i-1}}\widetilde Y_{i,n}/K_H\|_p
\le C_p\Delta_n^H$ and $\|\Delta_i^nW\|_p=C_p\sqrt{\Delta_n}$.
In the following sum, $c$ ranges over
$\sigma$, $v\rho$, $v\sqrt{1-\rho^2}$ and $L$.
Since $\|Z-\E[Z\mid\F_{t_{i-1}}]\|_2\le2\|Z\|_2$, conditional centering
only changes the constant. Applying H\"older to the three products in the
expansion gives
\begin{align*}
&\sum_i\|\xi_{i,n}-\E[\xi_{i,n}\mid\F_{t_{i-1}}]-\widetilde\xi_{i,n}\|_2^2
\le C\sum_i\Delta_n\sum_c
 \left\|\sup_{t_{i-1}\le s\le t_i}|c_s-c_{t_{i-1}}|\right\|_8^2
 +C\Delta_n.
\end{align*}
Equation~\eqref{eq:cadlag-sum} proves that the right side tends to zero.

For the second assertion of \eqref{eq:rcv-transfer}, conditional Gaussian
fourth moments in \eqref{eq:rcv-frozen} and
$\E|\zeta_{i,n}|^4\le C\Delta_n^{4H}$ give
\[
\E|\widetilde\xi_{i,n}|^4
\le C\Delta_n^{-4H}
 \left\{\Delta_n^2\E|\zeta_{i,n}|^4+\Delta_n^{4H+2}\right\}
\le C\Delta_n^2.
\]
Summing over the $n$ cells proves the stated $O(\Delta_n)$ bound.
\end{proof}

\begin{proof}
\textit{(a) Conditional means, variance and Lindeberg condition.}
Apply Lemma~\ref{lem:stable-clt} to the centered array
$\xi_{i,n}-\E[\xi_{i,n}\mid\F_{t_{i-1}}]$.
Its conditional mean is zero, so \eqref{eq:clt-mean} holds
with $\Upsilon=0$.
We verify Jacod conditions \eqref{eq:clt-variance}
and \eqref{eq:clt-lindeberg}.
By conditional Gaussian pairing,
\begin{align*}
\E[\widetilde\xi_{i,n}^2\mid\F_{t_{i-1}}]
=&\sigma_{t_{i-1}}^2\Delta_n^{1-2H}\zeta_{i,n}^2\\
&+\frac{\sigma_{t_{i-1}}^2v_{t_{i-1}}^2}{K_H^2}
\left\{\frac1{2H}+\frac{\rho_{t_{i-1}}^2}{(H+1/2)^2}\right\}\Delta_n.
\end{align*}
The mixed term is zero by an odd future Gaussian moment. Equation
\eqref{eq:memory-square}, Riemann summation and
$C_H+(2H)^{-1}=K_H^2$ therefore yield
\[
\sum_{i:t_i\le t}\E[\widetilde\xi_{i,n}^2\mid\F_{t_{i-1}}]
\convucp
\int_0^t\sigma_s^2v_s^2\,ds+
\frac1{K_H^2(H+1/2)^2}\int_0^tL_s^2\,ds.
\]
The conditional variance transfer follows from
\begin{align*}
&\E\sum_i\left|\E[(\xi_{i,n}-\E[\xi_{i,n}\mid\F_{t_{i-1}}])^2
 -\widetilde\xi_{i,n}^{\,2}\mid\F_{t_{i-1}}]\right|\\
&\le
 \left(\sum_i\|\xi_{i,n}-\E[\xi_{i,n}\mid\F_{t_{i-1}}]
 -\widetilde\xi_{i,n}\|_2^2\right)^{1/2}\\
&\times
 \left(\sum_i\|\xi_{i,n}-\E[\xi_{i,n}\mid\F_{t_{i-1}}]
 +\widetilde\xi_{i,n}\|_2^2\right)^{1/2}=o(1).
\end{align*}
The first factor tends to zero by \eqref{eq:rcv-transfer}; the second
is bounded by that equation and the conditional variance calculation.
Thus the centered array has the same predictable variance limit.
For $\xi_{i,n}$, BDG and \eqref{eq:variance-moments} give
$\E|\xi_{i,n}|^4\le C\Delta_n^2$. Conditional Jensen therefore gives
\begin{align*}
&\sum_i\E|\xi_{i,n}-\E[\xi_{i,n}\mid\F_{t_{i-1}}]|^4
\le16\sum_i\E|\xi_{i,n}|^4\le C\Delta_n,\\
&\E\sum_i\E\left[
|\xi_{i,n}-\E[\xi_{i,n}\mid\F_{t_{i-1}}]|^2
\mathbf1_{\{|\xi_{i,n}-\E[\xi_{i,n}\mid\F_{t_{i-1}}]|>\eps\}}
\mid\F_{t_{i-1}}\right]\le C_\eps\Delta_n\longrightarrow0.
\end{align*}
\textit{(b) Orthogonality.}
We verify Jacod condition \eqref{eq:clt-brownian}.
Conditional Gaussian moments in \eqref{eq:rcv-frozen} give
\[
\begin{aligned}
&\E[\widetilde\xi_{i,n}\Delta_i^nW\mid\F_{t_{i-1}}]
=\Delta_n^{1-H}\sigma_{t_{i-1}}\zeta_{i,n},\\
&\E[\widetilde\xi_{i,n}\Delta_i^nW^\perp\mid\F_{t_{i-1}}]=0.
\end{aligned}
\]
Their sums tend to zero by \eqref{eq:memory-linear}.
For bounded elementary predictable integrands the same assertion follows by
multiplication on each deterministic partition interval. General predictable
$L^2$ integrands follow by approximation and conditional Cauchy--Schwarz,
using $\sum_i\E\widetilde\xi_{i,n}^2=O(1)$.
The squared-error bound \eqref{eq:rcv-transfer} transfers these assertions
to the centered array. Its conditional means have zero product with
Brownian increments. In particular the condition also holds for $B$ through
its displayed stochastic-integral representation.
We verify Jacod condition \eqref{eq:clt-martingale}.
Let $N$ be a bounded martingale orthogonal to $W,W^\perp$. On $(t_{i-1},t_i]$ the first
term of \eqref{eq:rcv-frozen} is an It\^o integral with coefficient
$\Delta_n^{-H}\sigma_{t_{i-1}}\zeta_{i,n}$. For the second term of \eqref{eq:rcv-frozen}, It\^o's formula
provides the following two integrals on the current grid cell:
\begin{align*}
&\Delta_i^nW\,\widetilde Y_{i,n}
 -\frac{\rho_{t_{i-1}}\Delta_n^{H+1/2}}{H+1/2}\\
&=\int_{t_{i-1}}^{t_i}(W_s-W_{t_{i-1}})K(t_i-s)
 \{\rho_{t_{i-1}}dW_s+\sqrt{1-\rho_{t_{i-1}}^2}\,dW_s^\perp\}\\
&+\int_{t_{i-1}}^{t_i}\left[\int_{t_{i-1}}^{s}K(t_i-u)
 \{\rho_{t_{i-1}}dW_u+\sqrt{1-\rho_{t_{i-1}}^2}\,dW_u^\perp\}\right]dW_s.
\end{align*}
Both integrands are adapted and square integrable; the coefficients
$\sigma_{t_{i-1}}$, $v_{t_{i-1}}$ and $\rho_{t_{i-1}}$ are fixed by
$\F_{t_{i-1}}$. Hence
$\E[\widetilde\xi_{i,n}\Delta_i^nN\mid\F_{t_{i-1}}]=0$.
For the transfer to the centered array, conditional
Cauchy--Schwarz followed by ordinary Cauchy--Schwarz gives
\begin{align*}
&\E\sum_i\left|\E\!\left[
 (\xi_{i,n}-\E[\xi_{i,n}\mid\F_{t_{i-1}}]-\widetilde\xi_{i,n})
 \Delta_i^nN\mid\F_{t_{i-1}}\right]\right|\\
&\le
 \left(\sum_i\|\xi_{i,n}-\E[\xi_{i,n}\mid\F_{t_{i-1}}]
 -\widetilde\xi_{i,n}\|_2^2\right)^{1/2}
 \left(\sum_i\E(\Delta_i^nN)^2\right)^{1/2}=o(1).
\end{align*}
The second factor is bounded by the square-integrability of the bounded
martingale $N$.
The martingale decomposition into integrals against $(W,W^\perp)$ and an
orthogonal remainder, together with the previous verification, covers every
bounded martingale with respect to $(\F_t)$.
No Brownian representation property of $(\F_t)$ is assumed.
\textit{(c) The corrected and uncorrected limits.}
Lemma~\ref{lem:stable-clt} now yields
\begin{equation*}
\sum_{i:t_i\le\cdot}\{\xi_{i,n}-\E[\xi_{i,n}\mid\F_{t_{i-1}}]\}
\convst\int_0^\cdot\eta_s\,dZ_s.
\end{equation*}
All estimates are uniform over prefixes. The unmatched part of
$\theta_t$ has length at most $\Delta_n$ and is $O(\Delta_n)$.
Lemma~\ref{lem:compensator} identifies the removed conditional sum as
$P_n^L+o_{\mathrm{ucp}}(1)$. Lemma~\ref{lem:projection}, with $Z=L$, $h=\Delta_n$ and $w=w_0$, gives
\begin{equation*}
\left\|\sup_{t\le T}|\widetilde P_n^L(t)-P_n^L(t)|\right\|_2=o(1).
\end{equation*}
This proves the corrected limit at each fixed localization level.
The correction $\widetilde P_n^L$ is a path integral, so
Proposition~\ref{prop:localization} transfers the entire corrected statistic
and its limiting variance to the original locally bounded model. This proves
Theorem~\ref{thm:corrected-clt}.
The same argument also gives first-order consistency.
Indeed, at a fixed localization level,
\[
\sqrt{\Delta_n}\sup_{t\le T}|\widetilde P_n^L(t)|
\le C\sum_i\Delta_n\sup_{t_{i-1}\le s\le t_i}|L_s-L_{t_{i-1}}|
\longrightarrow0
\]
pathwise by c\`adl\`ag regularity. The corrected fluctuation is tight;
localization therefore yields $T_n\convucp\theta$.
If LR holds, apply \eqref{eq:lr-raw} to the killed representative
$L^{(N)}=\mathbf1_{[0,\tau_N)}\widetilde L^N$, with $h=\Delta_n$ and $w=w_0$.
It gives $\widetilde P_n^{L^{(N)}}\convucp0$ for each $N$.
Equality before exit and then $N\to\infty$ yield
$\widetilde P_n^L\convucp0$ in the original model. Stable Slutsky proves
Theorem~\ref{thm:rcv-clt}.
\end{proof}

%% file: appendices/D_est.tex
\section{The proposed estimator and variance estimation}
Fix the localization level of Proposition~\ref{prop:localization} and omit
its index. All constants in this section may depend on that level.
Under \eqref{eq:bandwidth},
\begin{equation*}
\Delta_n(k_n\Delta_n)^{-2H}+k_n\Delta_n\asymp n^{-\beta}.
\end{equation*}
The moment bounds are \eqref{eq:variance-moments}, and coefficient
replacement uses \eqref{eq:cadlag-sum}. LR is used for the conditional
mean and for the integrated target in the variance estimator.
\subsection{Conditional increment estimates}

\begin{lemma}\label{lem:increment-variance}
For $0\le\tau<t$, $0<\delta\le t-\tau$, and every fixed $p\ge2$,
\begin{align*}
&\left\|\delta^{-2H}\E[|\sigma_{t+\delta}^2-\sigma_t^2|^2\mid\F_\tau]-v_\tau^2\right\|_p\\
&\le C_p\left\{
\left\|\sup_{s\in[\tau,t+\delta]}|v_s-v_\tau|\right\|_{2p}
+(\delta/(t-\tau))^{2-2H}+\delta^{1/2}\right\}.
\end{align*}
\end{lemma}
\begin{lemma}\label{lem:increment-covariance}
For every fixed $p\ge2$ and $0\le\tau<u\le w<s\le t$ with
$\delta=t-u\le u-\tau$, the $L^p$ distance between
\begin{equation*}
\delta^{-2H}\E[(\sigma_t^2-\sigma_w^2)(\sigma_s^2-\sigma_u^2)\mid\F_\tau]
\end{equation*}
and
\begin{equation*}
\frac{v_\tau^2}{2}\left\{1+\left|\frac{s-w}{\delta}\right|^{2H}
-\left|\frac{t-s}{\delta}\right|^{2H}
-\left|\frac{w-u}{\delta}\right|^{2H}\right\}
\end{equation*}
is bounded by
$C_p\{\|\sup_{r\in[\tau,t]}|v_r-v_\tau|\|_{2p}
+(\delta/(u-\tau))^{2-2H}+\delta^{1/2}\}$.
\end{lemma}
\begin{proof}
We prove Lemmas~\ref{lem:increment-variance}--\ref{lem:increment-covariance} together.
The ordinary and Volterra time drifts have increment norms $C\delta$ and
$C\delta^{H+1/2}$; their normalized quadratic contributions and cross-products
are bounded by $C\delta^{1/2}$. The stochastic Volterra history before
$\tau$ is $\F_\tau$-measurable. BDG in the outer $L^{2p}$ norm gives
\begin{align*}
&\left\|\int_0^\tau[K(t+\delta-r)-K(t-r)]v_r\,dB_r\right\|_{2p}^2\\
&\le C_p\delta^2\int_{t-\tau}^\infty x^{2H-3}dx
=C_p\delta^{2H}(\delta/(t-\tau))^{2-2H}.
\end{align*}
Its conditional cross-product with the future It\^o integral is zero.
For the future square, conditional Jensen and H\"older give
\begin{equation*}
\|\E[v_r^2-v_\tau^2\mid\F_\tau]\|_p
\le\|v_r-v_\tau\|_{2p}\|v_r+v_\tau\|_{2p}
\le C_p\left\|\sup_{u\in[\tau,t+\delta]}|v_u-v_\tau|\right\|_{2p}.
\end{equation*}
Conditional It\^o isometry and this bound yield
\begin{equation*}
\left\|\int_\tau^{t+\delta}[K(t+\delta-r)-K(t-r)]^2
\E[v_r^2-v_\tau^2\mid\F_\tau]dr\right\|_p
\le C_p\delta^{2H}\left\|\sup_{r\in[\tau,t+\delta]}|v_r-v_\tau|\right\|_{2p}.
\end{equation*}
Completing the deterministic integral to $(-\infty,t+\delta]$ costs the same
squared-tail bound. Its full value is $K_H^2\delta^{2H}$.
For two different increments apply the same reasoning to the product of their
two kernels. Polarization gives the full integral
\begin{equation*}
\frac{K_H^2}{2}\{(t-u)^{2H}+(s-w)^{2H}-(t-s)^{2H}-(w-u)^{2H}\}.
\end{equation*}
Division by $K_H^2\delta^{2H}$ proves both statements. When summed over
blocks, the coefficient errors are controlled by \eqref{eq:cadlag-sum}.
\end{proof}

\begin{lemma}\label{lem:gaussian-moments}
For centered jointly Gaussian $X,Y,Z$, the moment formula of
\citet{Isserlis1918} gives
\begin{align*}
&\begin{aligned}[t]
\E[X^2Y^2Z^2]=&\E[X^2]\E[Y^2]\E[Z^2]+8\E[XY]\E[YZ]\E[ZX]\\
&+2\{\E[X^2]\E[YZ]^2+\E[Y^2]\E[ZX]^2+\E[Z^2]\E[XY]^2\},
\end{aligned}\\
&\E[X^2Y^4]=3\E[X^2]\E[Y^2]^2+12\E[Y^2]\E[XY]^2.
\end{align*}
\end{lemma}

\subsection{Price drift and the complete conditional mean}

\begin{lemma}\label{lem:price-drift}
For a price path $Z$, write $\widehat\sigma_{t_j}^2(Z)$ and
$\widehat\theta_n(Z;t)$ for the estimators evaluated on $Z$, with
coarse indices restricted to $t_{ik_n}\le t$. Define the unnormalized
coarse contribution by
\begin{equation*}
U_{i,n}(Z)=\frac{2H+3}{2}(k_n\Delta_n)^{1/2-H}
\Delta_{ik_n}^{n,k_n}Z\,
\Delta_{ik_n}^{n,k_n}\widehat\sigma^2(Z).
\end{equation*}
For the price martingale $M$,
\begin{equation}\label{eq:drift-statistic}
\sup_{t\le T}|\widehat\theta_n(X;t)-\widehat\theta_n(M;t)|
=o_p(n^{-\beta/2}).
\end{equation}
The assertion also holds after taking the conditional expectation of each
local difference given $\F_{t_{(i-1)k_n}}$, or subtracting that expectation.
Uniformly over deterministic families of disjoint consecutive coarse-index
groups, the squared $L^2$ norms of the grouped differences sum to
$o(n^{-\beta})$.
\end{lemma}
\begin{proof}
First establish the local bounds. It\^o's formula gives
\begin{align*}
\widehat\sigma_{t_{ik_n}}^2(M)
=&\frac1{k_n\Delta_n}\int_{t_{(i-1)k_n}}^{t_{ik_n}}\sigma_s^2\,ds\\
&+\frac2{k_n\Delta_n}\sum_{j=(i-1)k_n+1}^{ik_n}
\int_{t_{j-1}}^{t_j}(M_s-M_{t_{j-1}})\sigma_s\,dW_s.
\end{align*}
\begin{align*}
&\left\|\widehat\sigma_{t_{ik_n}}^2(M)
-\frac1{k_n\Delta_n}\int_{t_{(i-1)k_n}}^{t_{ik_n}}\sigma_s^2\,ds\right\|_p\\
&\le\frac{C_p}{k_n\Delta_n}
\left(\sum_{j=(i-1)k_n+1}^{ik_n}\int_{t_{j-1}}^{t_j}(s-t_{j-1})\,ds\right)^{1/2}
\le C_pk_n^{-1/2}.
\end{align*}
BDG gives the inequality, and \eqref{eq:variance-moments} controls
the difference of adjacent spot variance averages. Hence
\begin{align*}
&\|\widehat\sigma_{t_{ik_n}}^2(M)\|_p\le C_p,\qquad
\|\Delta_{ik_n}^{n,k_n}M\|_p\le C_p\sqrt{k_n\Delta_n},\\
&\|\Delta_{ik_n}^{n,k_n}\widehat\sigma^2(M)\|_p
\le C_p\{(k_n\Delta_n)^H+k_n^{-1/2}\},\\
&(k_n\Delta_n)^{1/2-H}\{(k_n\Delta_n)^H+k_n^{-1/2}\}
=\sqrt{k_n\Delta_n}+\sqrt{\Delta_n(k_n\Delta_n)^{-2H}}\le C_p n^{-\beta/2}.
\end{align*}
The price decomposition gives
$\Delta_j^nX=\Delta_j^nM+\Delta_j^nA$. Define the difference of the two
spot variance estimates by
\begin{align*}
&\begin{aligned}[t]
D_i^\mu
&:=\widehat\sigma_{t_{ik_n}}^2(X)-\widehat\sigma_{t_{ik_n}}^2(M)\\
&=\frac1{k_n\Delta_n}\sum_{j=(i-1)k_n+1}^{ik_n}
\{2\Delta_j^nM\,\Delta_j^nA+(\Delta_j^nA)^2\},
\end{aligned}\\
&\|D_i^\mu\|_p
\le\frac{C_pk_n}{k_n\Delta_n}\{\Delta_n^{3/2}+\Delta_n^2\}
\le C_p\sqrt{\Delta_n},\qquad
\|\Delta_{ik_n}^{n,k_n}A\|_p\le C_pk_n\Delta_n.
\end{align*}
Since $k_n\Delta_n\le1$ eventually, $\sqrt{\Delta_n}\le k_n^{-1/2}$, and hence
\begin{align*}
\|\Delta_{ik_n}^{n,k_n}\widehat\sigma^2(X)\|_p
&\le\|\Delta_{ik_n}^{n,k_n}\widehat\sigma^2(M)\|_p
+\|D_i^\mu\|_p+\|D_{i-1}^\mu\|_p\\
&\le C_p\{(k_n\Delta_n)^H+k_n^{-1/2}\}.
\end{align*}
Decompose the local difference as
\begin{align*}
&U_{i,n}(X)-U_{i,n}(M)=\sum_{a=1}^3\mathrm I_{a,i},\\
&\mathrm I_{1,i}=\frac{2H+3}{2}(k_n\Delta_n)^{1/2-H}
\Delta_{ik_n}^{n,k_n}A\,\Delta_{ik_n}^{n,k_n}\widehat\sigma^2(M),\\
&\mathrm I_{2,i}=\frac{2H+3}{2}(k_n\Delta_n)^{1/2-H}
\Delta_{ik_n}^{n,k_n}M(D_i^\mu-D_{i-1}^\mu),\\
&\mathrm I_{3,i}=\frac{2H+3}{2}(k_n\Delta_n)^{1/2-H}
\Delta_{ik_n}^{n,k_n}A(D_i^\mu-D_{i-1}^\mu).
\end{align*}
H\"older and the local bounds above give, uniformly in $i$,
\begin{align*}
&\|\mathrm I_{1,i}\|_p
\le C_p\{(k_n\Delta_n)^{3/2}+(k_n\Delta_n)^{3/2-H}k_n^{-1/2}\}\le C_pk_n\Delta_nn^{-\beta/2},\\
&\|\mathrm I_{2,i}\|_p
\le C_p(k_n\Delta_n)^{1-H}\sqrt{\Delta_n}
=C_p(k_n\Delta_n)^{3/2-H}k_n^{-1/2}\le C_pk_n\Delta_nn^{-\beta/2},\\
&\|\mathrm I_{3,i}\|_p
\le C_p(k_n\Delta_n)^{3/2-H}\sqrt{\Delta_n}\le C_p(k_n\Delta_n)^{3/2}n^{-\beta/2}.
\end{align*}
For a bounded elementary predictable drift equal to $c$ on both adjacent
coarse intervals, the squared-return difference simplifies to
\begin{equation*}
D_i^\mu=\frac{2c}{k_n}\Delta_{ik_n}^{n,k_n}M+c^2\Delta_n.
\end{equation*}
For consecutive admissible indices $j_1\le i\le j_2$ within this interval,
the first term telescopes:
\begin{equation*}
\sum_{i=j_1}^{j_2}\mathrm I_{1,i}
=\frac{2H+3}{2} c (k_n\Delta_n)^{3/2-H}
\{\widehat\sigma_{t_{j_2k_n}}^2(M)-\widehat\sigma_{t_{(j_1-1)k_n}}^2(M)\}.
\end{equation*}
Since $\|\max_i|\widehat\sigma_{t_{ik_n}}^2(M)|\|_p
\le C_p(k_n\Delta_n)^{-1/p}$, choosing $p>2$ bounds its maximal sum by
$C_p(k_n\Delta_n)^{3/2-H-1/p}=o(n^{-\beta/2})$. For the other two terms, the constant
$c^2\Delta_n$ cancels between adjacent estimates, so
\begin{align*}
&\sum_i\|\mathrm I_{2,i}\|_p
\le C_pk_n^{-1}(k_n\Delta_n)^{1/2-H}=o(n^{-\beta/2}),\\
&\sum_i\|\mathrm I_{3,i}\|_p
\le C_pk_n^{-1}(k_n\Delta_n)^{1-H}=o(n^{-\beta/2}).
\end{align*}
There are only finitely many coarse intervals meeting the fixed step boundaries.
Their total norm is $O(k_n\Delta_nn^{-\beta/2})=o(n^{-\beta/2})$ by the generic local bounds.
This proves the first assertion for a fixed elementary drift.
Choose uniformly bounded elementary predictable $\mu^{(m)}$ with
$\|\mu-\mu^{(m)}\|_{L^{8p}(\Omega\times[0,T])}\to0$.
Define $A_t^{(m)}=\int_0^t\mu_s^{(m)}ds$ and $X_t^{(m)}=X_0+M_t+A_t^{(m)}$.
Define $D_i^{\mu^{(m)}}$ by replacing $X$ with $X^{(m)}$ in the definition of $D_i^\mu$.
On each fine interval, Minkowski gives
\begin{equation*}
\left\|\int_{t_{j-1}}^{t_j}(\mu_s-\mu_s^{(m)})ds\right\|_{4p}
\le\int_{t_{j-1}}^{t_j}\|\mu_s-\mu_s^{(m)}\|_{4p}\,ds.
\end{equation*}
The exact squared-return difference and its bound are
\begin{align*}
&D_i^\mu-D_i^{\mu^{(m)}}
=\frac1{k_n\Delta_n}\sum_{j=(i-1)k_n+1}^{ik_n}
(2\Delta_j^nM+\Delta_j^nA+\Delta_j^nA^{(m)})
\int_{t_{j-1}}^{t_j}(\mu_s-\mu_s^{(m)})ds,\\
&\|D_i^\mu-D_i^{\mu^{(m)}}\|_{2p}
\le\frac{C_p\sqrt{\Delta_n}}{k_n\Delta_n}
\int_{t_{(i-1)k_n}}^{t_{ik_n}}\|\mu_s-\mu_s^{(m)}\|_{4p}\,ds.
\end{align*}
For the statistic, expand the product using the return on $X^{(m)}$ and the
spot variance contrast on $X$. H\"older gives the following two contributions:
\begin{align*}
&\sum_i\Bigl\|\frac{2H+3}{2}(k_n\Delta_n)^{1/2-H}
\{\Delta_{ik_n}^{n,k_n}X\,\Delta_{ik_n}^{n,k_n}\widehat\sigma^2(X)
-\Delta_{ik_n}^{n,k_n}X^{(m)}\,\Delta_{ik_n}^{n,k_n}\widehat\sigma^2(X^{(m)})\}\Bigr\|_p\\
&\le C_p(k_n\Delta_n)^{1/2-H}((k_n\Delta_n)^H+k_n^{-1/2})
\int_0^T\|\mu_s-\mu_s^{(m)}\|_{4p}\,ds\\
&+C_p(k_n\Delta_n)^{1/2-H}\sqrt{k_n\Delta_n}\frac{\sqrt{\Delta_n}}{k_n\Delta_n}
\int_0^T\|\mu_s-\mu_s^{(m)}\|_{4p}\,ds\\
&\le C_pn^{-\beta/2}\|\mu-\mu^{(m)}\|_{L^{8p}(\Omega\times[0,T])}.
\end{align*}
The first term contains the return difference
$\Delta_{ik_n}^{n,k_n}(A-A^{(m)})$; the second contains
$(D_i^\mu-D_i^{\mu^{(m)}})-(D_{i-1}^\mu-D_{i-1}^{\mu^{(m)}})$.
Each fine interval occurs at most twice in the latter sum.
The identity $(k_n\Delta_n)^{-H}\sqrt{\Delta_n}=(k_n\Delta_n)^{1/2-H}k_n^{-1/2}$
and the bandwidth bound $k_n\Delta_n+\Delta_n(k_n\Delta_n)^{-2H}=O(n^{-\beta})$
give the last line.
First let $n\to\infty$ at fixed $m$, then let $m\to\infty$.
This proves \eqref{eq:drift-statistic}.
To transfer conditional means, center $\mathrm I_{1,i}$ at the coarse left
endpoint. These are coarse martingale differences. Doob gives
\begin{align*}
&\left\|\max_{b\le T/(k_n\Delta_n)}\left|\sum_{i=2}^b
\{\mathrm I_{1,i}-\E[\mathrm I_{1,i}\mid\F_{t_{(i-1)k_n}}]\}\right|\right\|_2\\
&\le2\left(\sum_i\|\mathrm I_{1,i}\|_2^2\right)^{1/2}
\le C\{(k_n\Delta_n)^{-1}(k_n\Delta_nn^{-\beta/2})^2\}^{1/2}
=C\sqrt{k_n\Delta_n}n^{-\beta/2}=o(n^{-\beta/2}).
\end{align*}
Its predictable sum is the raw sum minus this martingale sum, so the
preceding telescope applies. For any consecutive groups, split additionally at
the finitely many step boundaries. The telescope bounds the sum of grouped
squared norms by $C_m(k_n\Delta_n)^{-1}(k_n\Delta_n)^{3-2H}=o(n^{-\beta})$; the centered groups
contribute at most $C_mk_n\Delta_nn^{-\beta}$. The other two terms and the approximation
error are controlled by their sums of $L^2$ norms, also after conditional
Jensen. Since $a_n\to1$, division by $a_n$ preserves these bounds and proves all claimed transfers.
\end{proof}

\begin{lemma}\label{lem:block-mean}
Under LR, the complete local conditional means satisfy
\begin{equation}\label{eq:block-mean}
\begin{aligned}
&\sup_{b\le T/(k_n\Delta_n)}\Biggl|\sum_{i=2}^b\E\Bigl[
U_{i,n}(M)-a_n\Delta_{ik_n}^{n,k_n}\theta
\Bigm|\F_{t_{(i-1)k_n}}\Bigr]\Biggr|=o_p(n^{-\beta/2}).
\end{aligned}
\end{equation}
Uniformly over deterministic families of disjoint consecutive coarse-index
groups, the sum of the squared $L^2$ norms of these grouped conditional
means is $o(n^{-\beta})$.
\end{lemma}
\begin{proof}
Apply Lemma~\ref{lem:lr-sums} to the localized loading
$L=\mathbf1_{[0,\tau_N)}\widetilde L^N$. With $\delta\in\{k_n\Delta_n,\Delta_n\}$,
\begin{equation}\label{eq:block-weight-sums}
\begin{aligned}
&\left\|\max_b\left|\sum_{j=1}^b p_{j,\delta}^{w_1}(L)\right|\right\|_2=o(\sqrt\delta),\\
&\sum_\nu\left\|\sum_{j\in J_\nu}p_{j,\delta}^{w_1}(L)\right\|_2^2=o(\delta).
\end{aligned}
\end{equation}
Here $J_\nu$ are disjoint consecutive index groups and the summands are defined in
\eqref{eq:weighted-cell}.
Fix $i\ge2$. For $(i-1)k_n<j\le ik_n$ and $t_{j-1}\le s\le t_j$, define
\begin{equation*}
Q_{i,s}=\sum_{v=(i-1)k_n+1}^{j-1}(\Delta_v^nM)^2
+(M_s-M_{t_{j-1}})^2.
\end{equation*}
The definitions agree at fine-grid endpoints, and
$Q_{i,t_{ik_n}}=k_n\Delta_n\widehat\sigma_{t_{ik_n}}^2(M)$.
On the fine interval, It\^o's formula gives
\begin{align*}
&dQ_{i,s}=2(M_s-M_{t_{j-1}})\,dM_s+\sigma_s^2\,ds,\\
&\begin{aligned}[t]
d\{(M_s-M_{t_{(i-1)k_n}})Q_{i,s}\}
=&\{Q_{i,s}+2(M_s-M_{t_{(i-1)k_n}})(M_s-M_{t_{j-1}})\}\,dM_s\\
&+\{M_s-M_{t_{(i-1)k_n}}+2(M_s-M_{t_{j-1}})\}\sigma_s^2\,ds.
\end{aligned}
\end{align*}
The stochastic integral has zero conditional mean given $\F_{t_{(i-1)k_n}}$.
The preceding spot variance estimate is measurable with respect to the same sigma-field,
whereas the current return has conditional mean zero. Consequently,
\begin{align*}
&\E[\Delta_{ik_n}^{n,k_n}M\,\Delta_{ik_n}^{n,k_n}\widehat\sigma^2(M)
\mid\F_{t_{(i-1)k_n}}]\\
&=\frac1{k_n\Delta_n}\sum_{j=(i-1)k_n+1}^{ik_n}\int_{t_{j-1}}^{t_j}
\E[\{M_s-M_{t_{(i-1)k_n}}+2(M_s-M_{t_{j-1}})\}\sigma_s^2
\mid\F_{t_{(i-1)k_n}}]\,ds.
\end{align*}
For $b=t_{(i-1)k_n}$ or $b=t_{j-1}$, the Volterra integral over $[0,b]$
is $\F_b$-measurable. Its product with $M_s-M_b$ has zero conditional mean
given $\F_b$, and hence given $\F_{t_{(i-1)k_n}}$. Conditional It\^o isometry
for the remaining interval gives
\begin{align*}
&\E\left[(M_s-M_b)\frac1{K_H}\int_b^sK(s-u)v_u\,dB_u
\mid\F_{t_{(i-1)k_n}}\right]\\
&=\frac1{K_H}\E\left[\int_b^sK(s-u)L_u\,du
\mid\F_{t_{(i-1)k_n}}\right].
\end{align*}
Integrating in $s$ is justified by bounded coefficient moments and
$\int_u^{t_j}K(s-u)ds=(t_j-u)^{H+1/2}/(H+1/2)$.
Substitution of the variance equation and ordinary Fubini yield
\begin{equation}\label{eq:block-weight}
\begin{aligned}
\E[U_{i,n}(M)\mid\F_{t_{(i-1)k_n}}]
=&\frac{\frac{2H+3}{2}(k_n\Delta_n)^{-H-1/2}}{K_H(H+1/2)}\\
&\times\E\Biggl[\int_{t_{(i-1)k_n}}^{t_{ik_n}}(t_{ik_n}-u)^{H+1/2}L_u\,du\\
&+2\sum_{j=(i-1)k_n+1}^{ik_n}\int_{t_{j-1}}^{t_j}(t_j-u)^{H+1/2}L_u\,du
\Bigm|\F_{t_{(i-1)k_n}}\Biggr]+e_i^f+e_i^g.
\end{aligned}
\end{equation}
The time-drift contributions are
\begin{align*}
&\begin{aligned}[t]
e_i^f=&\frac{2H+3}{2}(k_n\Delta_n)^{-H-1/2}\sum_{j=(i-1)k_n+1}^{ik_n}
\int_{t_{j-1}}^{t_j}\E\Bigl[
\{M_s-M_{t_{(i-1)k_n}}+2(M_s-M_{t_{j-1}})\}\\
&\times\int_{t_{(i-1)k_n}}^s f_u\,du
\Bigm|\F_{t_{(i-1)k_n}}\Bigr]ds,
\end{aligned}\\
&\begin{aligned}[t]
e_i^g=&\frac{2H+3}{2}(k_n\Delta_n)^{-H-1/2}\sum_{j=(i-1)k_n+1}^{ik_n}
\int_{t_{j-1}}^{t_j}\E\Bigl[
\{M_s-M_{t_{(i-1)k_n}}+2(M_s-M_{t_{j-1}})\}\\
&\times\int_{t_{(i-1)k_n}}^sK(s-u)(g_u-g_{t_{(i-1)k_n}})du
\Bigm|\F_{t_{(i-1)k_n}}\Bigr]ds.
\end{aligned}
\end{align*}
The histories of both time integrals, and the part with constant
$g_{t_{(i-1)k_n}}$, are $\F_{t_{(i-1)k_n}}$-measurable and have zero product mean.
H\"older and the return increment bound give
\begin{align*}
&\|e_i^f\|_p\le C_p(k_n\Delta_n)^{-H-1/2}\int_0^{k_n\Delta_n}x^{3/2}dx\le C_p(k_n\Delta_n)^{2-H},\\
&\|e_i^g\|_p\le C_p(k_n\Delta_n)^{-H-1/2}\int_0^{k_n\Delta_n}x^{H+1}dx
\left\|\sup_{t_{(i-1)k_n}\le u\le t_{ik_n}}|g_u-g_{t_{(i-1)k_n}}|\right\|_{2p}.
\end{align*}
Summing and using \eqref{eq:cadlag-sum} on the coarse grid gives
\begin{align*}
&\sum_i\|e_i^f\|_p\le C_p(k_n\Delta_n)^{1-H}=o(\sqrt{k_n\Delta_n}),\\
&\sum_i\|e_i^g\|_p\le C_p\sqrt{k_n\Delta_n}\sum_i k_n\Delta_n
\left\|\sup_{t_{(i-1)k_n}\le u\le t_{ik_n}}|g_u-g_{t_{(i-1)k_n}}|\right\|_{2p}
=o(\sqrt{k_n\Delta_n}).
\end{align*}
The grouped squared norms are bounded by the square of these total norms.
The two deterministic weight masses in \eqref{eq:block-weight} are
\begin{align*}
&\frac{2H+3}{2}(k_n\Delta_n)^{-H-1/2}\int_{t_{(i-1)k_n}}^{t_{ik_n}}(t_{ik_n}-u)^{H+1/2}du=k_n\Delta_n,\\
&2\frac{2H+3}{2}(k_n\Delta_n)^{-H-1/2}\sum_{j=(i-1)k_n+1}^{ik_n}
\int_{t_{j-1}}^{t_j}(t_j-u)^{H+1/2}du=2k_n\Delta_nk_n^{-H-1/2}.
\end{align*}
After subtracting the conditional expectation of
$a_n\Delta_{ik_n}^{n,k_n}\theta$, the remaining expression is exactly
\begin{align*}
&\frac{p_{i,k_n\Delta_n}^{w_1}(L)}{K_H(H+1/2)}
+e_i^f+e_i^g\\
&+\frac{2k_n^{-H-1/2}}{K_H(H+1/2)}
\E\left[\sum_{j=(i-1)k_n+1}^{ik_n}
I_{j,\Delta_n}^{w_1}(L)\Bigm|\F_{t_{(i-1)k_n}}\right].
\end{align*}
The zero integrals of $w_1$ allow subtraction of $L$ at the corresponding
coarse or fine left endpoints. The first term is controlled by
\eqref{eq:block-weight-sums} with $\delta=k_n\Delta_n$.
For the fine-grid term, the tower property replaces each
$I_{j,\Delta_n}^{w_1}(L)$ by $p_{j,\Delta_n}^{w_1}(L)$ inside the displayed
conditional expectation. The projection difference
\begin{equation*}
\sum_{j=(i-1)k_n+1}^{ik_n}p_{j,\Delta_n}^{w_1}(L)
-\E\left[\sum_{j=(i-1)k_n+1}^{ik_n}p_{j,\Delta_n}^{w_1}(L)
\Bigm|\F_{t_{(i-1)k_n}}\right]
\end{equation*}
is a coarse martingale difference. Since
$\|p_{j,\Delta_n}^{w_1}(L)\|_2\le C_N\Delta_n$, the sum of its squared norms,
after multiplication by $k_n^{-H-1/2}$, is bounded by
\begin{equation*}
C_N k_n^{-2H-1}\frac{T}{k_n\Delta_n}(k_n\Delta_n)^2
\le C_Nk_n^{-2H-1}k_n\Delta_n=o(k_n\Delta_n).
\end{equation*}
Doob and martingale orthogonality give the same maximal and grouped bounds.
The unprojected fine sums are controlled by \eqref{eq:block-weight-sums}
at $\delta=\Delta_n$. Since
\begin{equation*}
\sqrt{k_n\Delta_n}\le C n^{-\beta/2},\qquad
n^{\beta/2}k_n^{-H-1/2}o(\sqrt{\Delta_n})
\le Ck_n^{-H-1}o(1)=o(1),
\end{equation*}
these estimates and the bounds on $e_i^f,e_i^g$ prove
\eqref{eq:block-mean} and its grouped version.
Division by $a_n$ removes the finite-window bias
$(a_n-1)\theta_T=R_n$. Apart from the displayed rate comparison,
the argument uses only $k_n\to\infty$ and $k_n\Delta_n\to0$ under LR.
\end{proof}

\subsection{Proof of the feasible limit}
\begin{proof}
We prove Theorem~\ref{thm:feasible-clt}.
By Lemma~\ref{lem:price-drift}, it suffices to work with $M$.
The auxiliary sequences satisfy
$s_n\to\infty$, $s_n/l_n\to0$, and
$(s_n+l_n)k_n\Delta_n\to0$. Each complete auxiliary block contains
$s_n+l_n$ coarse intervals, with
$0\le\nu<\lfloor T/((s_n+l_n)k_n\Delta_n)\rfloor$.
The indices whose return right endpoint belongs to a complete auxiliary block are
$(s_n+l_n)\nu+1\le i\le (s_n+l_n)(\nu+1)$, intersected with $i\ge2$. Thus use
\begin{align*}
&I^s(\nu)=\{(s_n+l_n)\nu+1,\ldots,(s_n+l_n)\nu+s_n\},\\
&I^l(\nu)=\{(s_n+l_n)\nu+s_n+1,\ldots,(s_n+l_n)(\nu+1)\}.
\end{align*}
The full contribution on an auxiliary block and its centered decomposition are
\begin{equation}\label{eq:block-statistic}
\begin{aligned}
S_{\nu,n}^{\rm raw}
=&\sum_{\substack{i\in I^s(\nu)\cup I^l(\nu)\\i\ge2}}
\{U_{i,n}(M)-a_n\Delta_{ik_n}^{n,k_n}\theta\}\\
=&\sum_{\substack{i\in I^s(\nu)\cup I^l(\nu)\\i\ge2}}
(I-\E[\,\cdot\mid\F_{t_{(i-1)k_n}}])(U_{i,n}(M)-\Delta_{ik_n}^{n,k_n}\theta)\\
&+\sum_{\substack{i\in I^s(\nu)\cup I^l(\nu)\\i\ge2}}
\E[U_{i,n}(M)-a_n\Delta_{ik_n}^{n,k_n}\theta\mid\F_{t_{(i-1)k_n}}]
\\
&+2k_n^{-H-1/2}
\sum_{\substack{i\in I^s(\nu)\cup I^l(\nu)\\i\ge2}}
\{\E[\Delta_{ik_n}^{n,k_n}\theta\mid\F_{t_{(i-1)k_n}}]-\Delta_{ik_n}^{n,k_n}\theta\}.
\end{aligned}
\end{equation}
The sum of these raw blocks equals $a_n(\widehat\theta_n-\theta_T)$ up to endpoint contributions.
The last sum consists of martingale differences. Its total squared $L^2$ norm is
$O(k_n^{-2H-1}k_n\Delta_n)=o(n^{-\beta})$ at each fixed localization level.
Lemma~\ref{lem:block-mean} treats the sum of
$\E[U_{i,n}(M)-a_n\Delta_{ik_n}^{n,k_n}\theta\mid\F_{t_{(i-1)k_n}}]$. Only centered random terms are omitted when passing to
$I^l(\nu)$. Their small-block and terminal-block squared contribution is
$O(n^{-\beta}\{s_n/(s_n+l_n)+(s_n+l_n)k_n\Delta_n\})$; their target and finite-$k_n$ correction remain
in \eqref{eq:block-statistic}. The first missing coarse interval and final partial
interval contribute $O_p(k_n\Delta_n)=o_p(n^{-\beta/2})$ to the target.

\textit{(a) Conditional means.}
Lemma~\ref{lem:block-mean} controls the full coarse predictable sum.
Its grouped bound shows that subtracting each group's conditional mean
at the auxiliary block left endpoint gives martingale differences whose
squared $L^2$ norms sum to $o(n^{-\beta})$. For every index in
\eqref{eq:block-statistic}, $t_{(i-1)k_n}\ge t_{(s_n+l_n)\nu k_n}$.
The tower property therefore makes both centered sums in that display
conditionally mean zero at the auxiliary block left endpoint. Thus
\begin{equation}\label{eq:group-mean}
\begin{aligned}
&\max_N\left|\sum_{\nu<N}\E[S_{\nu,n}^{\rm raw}\mid\F_{t_{(s_n+l_n)\nu k_n}}]\right|=o_p(n^{-\beta/2}),\\
&\sum_\nu\|\E[S_{\nu,n}^{\rm raw}\mid\F_{t_{(s_n+l_n)\nu k_n}}]\|_2^2=o(n^{-\beta}).
\end{aligned}
\end{equation}
\textit{(b) Conditional variance.} We verify \eqref{eq:clt-variance}.
For each auxiliary block freeze the loadings on the independent Brownian drivers at $t_{(s_n+l_n)\nu k_n}$:
$dB_s^{(\nu)}=\rho_{t_{(s_n+l_n)\nu k_n}}dW_s+
\sqrt{1-\rho_{t_{(s_n+l_n)\nu k_n}}^2}\,dW_s^\perp$, $s>t_{(s_n+l_n)\nu k_n}$.
The superscript $(\nu)$ is used only on that interval. For $i\in I^l(\nu)$
define the following variables
\begin{align*}
&\psi_{ik}=\int_{t_{(s_n+l_n)\nu k_n}}^{t_{(i-1)k_n+k}}
\{K(t_{(i-1)k_n+k}-s)-K(t_{(i-2)k_n+k}-s)\}\,dB_s^{(\nu)},\\
&\xi_{ik}=\sigma_{t_{(s_n+l_n)\nu k_n}}^3\Delta_{ik_n}^{n,k_n}W
\{(\Delta_{(i-1)k_n+k}^nW)^2-(\Delta_{(i-2)k_n+k}^nW)^2\},\\
&\eta_{ik}=K_H^{-1}\sigma_{t_{(s_n+l_n)\nu k_n}}v_{t_{(s_n+l_n)\nu k_n}}
\{\Delta_{ik_n}^{n,k_n}W\psi_{ik}-\rho_{t_{(s_n+l_n)\nu k_n}}(k\Delta_n)^{H+1/2}/(H+1/2)\},\\
&\begin{aligned}[t]
F_{i,n}=&\frac{2H+3}{2}(k_n\Delta_n)^{-H-1/2}\sum_{k=1}^{k_n}\xi_{ik}\\
&+\frac{2H+3}{2}k_n^{-H-1/2}\Delta_n^{1/2-H}\sum_{k=1}^{k_n}\eta_{ik},
\end{aligned}\\
&L_{\nu,n}=\sum_{i\in I^l(\nu)}F_{i,n}.
\end{align*}
All these future Gaussian formulas are conditional on $\F_{t_{(s_n+l_n)\nu k_n}}$.
Their relation to the true array, including the omitted historical part,
is
\begin{equation}\label{eq:block-transfer}
\sum_\nu\|S_{\nu,n}^{\rm raw}-L_{\nu,n}\|_2^2=o(n^{-\beta}).
\end{equation}
For $i\in I^l(\nu)$ define the actual centered comparison error
\begin{align*}
e_{i,n}^{(\nu)}
=&(I-\E[\,\cdot\mid\F_{t_{(i-1)k_n}}])(U_{i,n}(M)-\Delta_{ik_n}^{n,k_n}\theta)-F_{i,n}.
\end{align*}
The frozen summands have zero conditional mean at $t_{(i-1)k_n}$: the measurement product has an odd current
Gaussian factor, and the covariance subtracted in $\eta_{ik}$ is
\begin{equation*}
\E[\Delta_{ik_n}^{n,k_n}W\psi_{ik}\mid\F_{t_{(i-1)k_n}}]
 =\frac{\rho_{t_{(s_n+l_n)\nu k_n}}(k\Delta_n)^{H+1/2}}{H+1/2}.
\end{equation*}
Consequently $\E[e_{i,n}^{(\nu)}\mid\F_{t_{(i-1)k_n}}]=0$.
Subtracting $L_{\nu,n}$ from \eqref{eq:block-statistic} now gives the exact
four-part identity
\begin{equation}\label{eq:group-error}
\begin{aligned}
S_{\nu,n}^{\rm raw}-L_{\nu,n}
=&\underbrace{\sum_{i\in I^l(\nu)}e_{i,n}^{(\nu)}}_{\mathrm I_{1,\nu}}\\
&+\underbrace{\sum_{i\in I^s(\nu)}
 (I-\E[\,\cdot\mid\F_{t_{(i-1)k_n}}])(U_{i,n}(M)-\Delta_{ik_n}^{n,k_n}\theta)}_{\mathrm I_{2,\nu}}\\
&+\underbrace{\sum_{i\in I^s(\nu)\cup I^l(\nu)}
 \E[U_{i,n}(M)-a_n\Delta_{ik_n}^{n,k_n}\theta\mid\F_{t_{(i-1)k_n}}]}_{\mathrm I_{3,\nu}}\\
&+\underbrace{2k_n^{-H-1/2}\sum_{i\in I^s(\nu)\cup I^l(\nu)}
 \{\E[\Delta_{ik_n}^{n,k_n}\theta\mid\F_{t_{(i-1)k_n}}]-\Delta_{ik_n}^{n,k_n}\theta\}}_{\mathrm I_{4,\nu}}.
\end{aligned}
\end{equation}
All sets in this display are intersected with the original admissible
indices $i\ge2$. Lemma~\ref{lem:block-mean} controls
$\sum_\nu\|\mathrm I_{3,\nu}\|_2^2$; the calculation following
\eqref{eq:block-statistic} controls $\sum_\nu\|\mathrm I_{4,\nu}\|_2^2$.
The number of coarse indices omitted in the small blocks and the local moment bound give
\begin{equation*}
\sum_\nu\|\mathrm I_{2,\nu}\|_2^2
\le C n^{-\beta}\frac{s_n}{(s_n+l_n)}.
\end{equation*}
The summands of $\mathrm I_{1,\nu}$ are coarse martingale differences,
so their squared norms add. Use the It\^o decomposition of the spot variance
estimate from Lemma~\ref{lem:price-drift}. For $i\in I^l(\nu)$, define
\begin{equation}\label{eq:group-remainders}
\begin{aligned}
&\varepsilon_{i,n}^{X,\nu}
=\int_{t_{(i-1)k_n}}^{t_{ik_n}}(\sigma_s-\sigma_{t_{(s_n+l_n)\nu k_n}})dW_s,\\
&\begin{aligned}[t]
\varepsilon_{i,n}^{RV,\nu}
=&\frac2{k_n\Delta_n}\sum_{j=(i-2)k_n+1}^{ik_n}
 \bigl(\1_{\{j>(i-1)k_n\}}-\1_{\{j\le(i-1)k_n\}}\bigr)\\
&\times\int_{t_{j-1}}^{t_j}
 \{(M_s-M_{t_{j-1}})\sigma_s
       -\sigma_{t_{(s_n+l_n)\nu k_n}}^2(W_s-W_{t_{j-1}})\}dW_s,
\end{aligned}\\
&\varepsilon_{i,n}^{V,\nu}
=\frac1{k_n\Delta_n}\int_{t_{(i-1)k_n}}^{t_{ik_n}}(\sigma_s^2-\sigma_{s-k_n\Delta_n}^2)ds
 -\frac{v_{t_{(s_n+l_n)\nu k_n}}}{K_Hk_n}\sum_{k=1}^{k_n}\psi_{ik}.
\end{aligned}
\end{equation}
Expanding one product, and then centering at $t_{(i-1)k_n}$, gives
\begin{equation}\label{eq:block-products}
\begin{aligned}
e_{i,n}^{(\nu)}
=&(I-\E[\,\cdot\mid\F_{t_{(i-1)k_n}}])\Big[
 \frac{2H+3}{2}(k_n\Delta_n)^{1/2-H}\big\{
 \varepsilon_{i,n}^{X,\nu}\Delta_{ik_n}^{n,k_n}\widehat\sigma^2(M)\\
&+\sigma_{t_{(s_n+l_n)\nu k_n}}\Delta_{ik_n}^{n,k_n}W
 (\varepsilon_{i,n}^{RV,\nu}+\varepsilon_{i,n}^{V,\nu})\big\}\\
&-(\Delta_{ik_n}^{n,k_n}\theta-\frac1{K_H(H+1/2)}k_n\Delta_nL_{t_{(i-1)k_n}})\Big].
\end{aligned}
\end{equation}
The covariance subtracted in $\eta_{ik}$ equals the conditional mean of
the frozen product, as computed above. Thus this equality includes the
target-centering error as well as the price and RV errors.

For $\varepsilon_{i,n}^{X,\nu}$ and $\varepsilon_{i,n}^{RV,\nu}$,
BDG and the identity
\begin{align*}
&(M_s-M_{t_{j-1}})\sigma_s-\sigma_{t_{(s_n+l_n)\nu k_n}}^2(W_s-W_{t_{j-1}})\\
&=(\sigma_s-\sigma_{t_{(s_n+l_n)\nu k_n}})(M_s-M_{t_{j-1}})\\
&+\sigma_{t_{(s_n+l_n)\nu k_n}}\int_{t_{j-1}}^s(\sigma_u-\sigma_{t_{(s_n+l_n)\nu k_n}})dW_u
\end{align*}
on each fine grid cell give
\begin{align*}
&\|\varepsilon_{i,n}^{X,\nu}\|_p
\le C_p\sqrt{k_n\Delta_n}
 \left\|\sup_{t_{(s_n+l_n)\nu k_n}\le s\le t_{(s_n+l_n)(\nu+1)k_n}}|\sigma_s-\sigma_{t_{(s_n+l_n)\nu k_n}}|\right\|_{2p},\\
&\|\varepsilon_{i,n}^{RV,\nu}\|_p
\le C_pk_n^{-1/2}
 \left\|\sup_{t_{(s_n+l_n)\nu k_n}\le s\le t_{(s_n+l_n)(\nu+1)k_n}}|\sigma_s-\sigma_{t_{(s_n+l_n)\nu k_n}}|\right\|_{4p}.
\end{align*}
We also use the local bounds in Lemma~\ref{lem:price-drift} and
$\|\Delta_{ik_n}^{n,k_n}W\|_p=C_p\sqrt{k_n\Delta_n}$.
By \eqref{eq:variance-moments}, the spot volatility oscillation above is bounded by
$C_{p,\eta}((s_n+l_n)k_n\Delta_n)^{\eta/2}$ for $0<\eta<H$; the other coefficient
oscillations converge after summation by \eqref{eq:cadlag-sum}.

Substituting the actual variance equation into the last line of
\eqref{eq:group-remainders} gives
\begin{equation*}
\varepsilon_{i,n}^{V,\nu}
=\mathrm J_{1,i}^{\nu}+\mathrm J_{2,i}^{\nu}
 +\mathrm J_{3,i}^{\nu}+\mathrm J_{4,i}^{\nu},
\end{equation*}
where
\begin{align*}
&\mathrm J_{1,i}^{\nu}
=\frac1{K_Hk_n\Delta_n}\int_{t_{(i-1)k_n}}^{t_{ik_n}}
 \int_0^{t_{(s_n+l_n)\nu k_n}}[K(s-u)-K(s-k_n\Delta_n-u)]v_udB_u\,ds,\\
&\begin{aligned}[t]
\mathrm J_{2,i}^{\nu}
=&\frac1{K_Hk_n\Delta_n}\int_{t_{(i-1)k_n}}^{t_{ik_n}}
 \Big\{\int_{t_{(s_n+l_n)\nu k_n}}^{s}[K(s-u)-K(s-k_n\Delta_n-u)]\\
&\times(v_u\rho_u-v_{t_{(s_n+l_n)\nu k_n}}\rho_{t_{(s_n+l_n)\nu k_n}})dW_u\\
&+\int_{t_{(s_n+l_n)\nu k_n}}^{s}[K(s-u)-K(s-k_n\Delta_n-u)]\\
&\times
 (v_u\sqrt{1-\rho_u^2}-v_{t_{(s_n+l_n)\nu k_n}}\sqrt{1-\rho_{t_{(s_n+l_n)\nu k_n}}^2})dW_u^\perp\Big\}ds,
\end{aligned}\\
&\begin{aligned}[t]
\mathrm J_{3,i}^{\nu}
=&\frac{v_{t_{(s_n+l_n)\nu k_n}}}{K_H}
 \Big\{\frac1{k_n\Delta_n}\int_{t_{(i-1)k_n}}^{t_{ik_n}}
 \int_{t_{(s_n+l_n)\nu k_n}}^{s}[K(s-u)-K(s-k_n\Delta_n-u)]dB_u^{(\nu)}\,ds\\
&-\frac1{k_n}\sum_{k=1}^{k_n}\psi_{ik}\Big\},
\end{aligned}\\
&\mathrm J_{4,i}^{\nu}
=\frac1{k_n\Delta_n}\int_{t_{(i-1)k_n}}^{t_{ik_n}}
 \Big\{\int_{s-k_n\Delta_n}^{s}f_u du
 +\int_0^{s}[K(s-u)-K(s-k_n\Delta_n-u)]g_u du\Big\}ds.
\end{align*}
The convention $K(x)=0$ for $x\le0$ is used in these integrals. For all
sufficiently large $n$, $s_n\ge1$; thus both adjacent coarse intervals of
$i\in I^l(\nu)$ lie after $t_{(s_n+l_n)\nu k_n}$.

For $\mathrm J_{1,i}^{\nu}$ put $j=i-1-(s_n+l_n)\nu\ge s_n$.
The square-kernel tail bound and Minkowski give
\begin{equation*}
\|\mathrm J_{1,i}^{\nu}\|_p^2
\le C_p(k_n\Delta_n)^{2H}(1+j)^{2H-2}.
\end{equation*}
Since $\mathrm J_{1,i}^{\nu}\in L^p(\F_{t_{(s_n+l_n)\nu k_n}})$, conditional It\^o isometry gives
\begin{equation*}
\E\left|\frac{2H+3}{2}(k_n\Delta_n)^{1/2-H}\sigma_{t_{(s_n+l_n)\nu k_n}}\Delta_{ik_n}^{n,k_n}W
          \mathrm J_{1,i}^{\nu}\right|^2
\le C(k_n\Delta_n)^2(1+j)^{2H-2}.
\end{equation*}
Summing over $i$ and $\nu$ gives $Ck_n\Delta_n/(s_n+l_n)$, since the exponent
$2H-2<-1$ is summable.

For $\mathrm J_{2,i}^{\nu}$ use the two independent Brownian directions
separately. For $\mathrm J_{3,i}^{\nu}$ compare the continuous integral in
$s$ with its right-endpoint Riemann sum, using the same frozen Brownian
motion. BDG, Minkowski and the increment energy of $K$ give
\begin{align*}
&\|\mathrm J_{2,i}^{\nu}\|_p
\le C_p(k_n\Delta_n)^H
 \sum_{c\in\{v\rho,\,v\sqrt{1-\rho^2}\}}
 \left\|\sup_{t_{(s_n+l_n)\nu k_n}\le s\le t_{(s_n+l_n)(\nu+1)k_n}}|c_s-c_{t_{(s_n+l_n)\nu k_n}}|\right\|_{2p},\\
&\|\mathrm J_{3,i}^{\nu}\|_p
\le \frac{C_p}{k_n\Delta_n}\sum_{k=1}^{k_n}
 \int_{t_{(i-1)k_n+k-1}}^{t_{(i-1)k_n+k}}
 |t_{(i-1)k_n+k}-s|^Hds
 \le C_p(k_n\Delta_n)^Hk_n^{-H},\\
&\|\mathrm J_{4,i}^{\nu}\|_p
\le C_p(k_n\Delta_n+(k_n\Delta_n)^{H+1/2}).
\end{align*}
The target term in \eqref{eq:block-products} satisfies
\begin{align*}
&\left\|\Delta_{ik_n}^{n,k_n}\theta
-\frac{k_n\Delta_nL_{t_{(i-1)k_n}}}{K_H(H+1/2)}\right\|_p\\
&\le C_pk_n\Delta_n
\left\|\sup_{t_{(s_n+l_n)\nu k_n}\le s\le t_{(s_n+l_n)(\nu+1)k_n}}
|L_s-L_{t_{(s_n+l_n)\nu k_n}}|\right\|_p.
\end{align*}
Conditional centering increases each $L^2$ norm by at most two. In
\eqref{eq:block-products}, the price and RV-martingale products
therefore have squared norms bounded by $Ck_n\Delta_nn^{-\beta}$ times the squared
coefficient oscillation. The terms with $\mathrm J_{3,i}^{\nu}$ and
$\mathrm J_{4,i}^{\nu}$ have squared norms at most
$C(k_n\Delta_n)^2k_n^{-2H}$ and $C(k_n\Delta_n)^3$. The term with $\mathrm J_{2,i}^{\nu}$
and the target term are bounded by $C(k_n\Delta_n)^2$ times their squared
coefficient oscillations. Collect these errors in
\begin{align*}
\omega_n=&\sum_\nu(s_n+l_n)k_n\Delta_n
\sum_{c\in\{\sigma,\,v\rho,\,v\sqrt{1-\rho^2},\,L\}}\\
&\times\left\|\sup_{t_{(s_n+l_n)\nu k_n}\le s\le t_{(s_n+l_n)(\nu+1)k_n}}
|c_s-c_{t_{(s_n+l_n)\nu k_n}}|\right\|_{16}^{\,2}.
\end{align*}
It tends to zero by \eqref{eq:cadlag-sum} on the deterministic auxiliary block
mesh $(s_n+l_n)k_n\Delta_n\to0$.
Adding the displayed squared norms over the coarse martingale differences
and using $k_n\Delta_n=O(n^{-\beta})$ gives
\begin{equation*}
\sum_\nu\|\mathrm I_{1,\nu}\|_2^2
\le Cn^{-\beta}\{\omega_n+k_n^{-2H}+k_n\Delta_n+(s_n+l_n)^{-1}\}.
\end{equation*}
The bounds on $\mathrm I_{2,\nu}$, $\mathrm I_{3,\nu}$ and
$\mathrm I_{4,\nu}$ in \eqref{eq:group-error}, together with
the incomplete auxiliary block, now yield
\begin{align*}
&\sum_\nu\|S_{\nu,n}^{\rm raw}-L_{\nu,n}\|_2^2\\
&\le Cn^{-\beta}\Bigl\{\omega_n+k_n^{-2H}+k_n^{-1}+k_n\Delta_n\\
&+(s_n+l_n)^{-1}+s_n/(s_n+l_n)+(s_n+l_n)k_n\Delta_n\Bigr\}+o(n^{-\beta}).
\end{align*}
Every term on the right tends to zero relative to $n^{-\beta}$ under the stated
conditions. This proves \eqref{eq:block-transfer}.

For the Gaussian measurement terms, disjoint fine increments give zero unless
$i=j$ and $k=l$. For $i=j,k=l$, write the current standardized increments
as $Z_1,\ldots,Z_{k_n}$ and the corresponding previous increment as $Z_k'$.
Then
\begin{align*}
&\E\left[\left(\sum_{v=1}^{k_n}Z_v\right)^2
\{Z_k^2-(Z_k')^2\}^2\right]\\
&=\E[Z_k^2\{Z_k^2-(Z_k')^2\}^2]
+(k_n-1)\E[\{Z_k^2-(Z_k')^2\}^2]\\
&=(15-6+3)+4(k_n-1)=4(k_n+2).
\end{align*}
Thus the measurement calculation in Lemma~\ref{lem:gaussian-moments} gives exactly
\begin{equation*}
\E[\xi_{ik}\xi_{jl}\mid\F_{t_{(s_n+l_n)\nu k_n}}]
=\1_{\{i=j,k=l\}}4\sigma_{t_{(s_n+l_n)\nu k_n}}^6(k_n+2)\Delta_n^3.
\end{equation*}
For the rough terms, the two pairings are
\begin{align*}
&\operatorname{Cov}(\Delta_{ik_n}^{n,k_n}W\psi_{ik},\Delta_{jk_n}^{n,k_n}W\psi_{jl}\mid\F_{t_{(s_n+l_n)\nu k_n}})\\
&=\E[\Delta_{ik_n}^{n,k_n}W\Delta_{jk_n}^{n,k_n}W\mid\F_{t_{(s_n+l_n)\nu k_n}}]
\E[\psi_{ik}\psi_{jl}\mid\F_{t_{(s_n+l_n)\nu k_n}}]\\
&+\E[\Delta_{ik_n}^{n,k_n}W\psi_{jl}\mid\F_{t_{(s_n+l_n)\nu k_n}}]
\E[\Delta_{jk_n}^{n,k_n}W\psi_{ik}\mid\F_{t_{(s_n+l_n)\nu k_n}}].
\end{align*}
For $i\ne j$, one factor in each pairing is zero by the time supports.
For $i=j$, define the deterministic kernel on the real line by
\begin{equation*}
d_{ik}(s)=K(t_{(i-1)k_n+k}-s)-K(t_{(i-2)k_n+k}-s).
\end{equation*}
The finite historical covariance is the full-line covariance minus the
ordinary integral $\int_{-\infty}^{t_{(s_n+l_n)\nu k_n}}d_{ik}(s)d_{il}(s)ds$.
Consequently,
\begin{align*}
&\begin{aligned}[t]
&\E[\eta_{ik}\eta_{il}\mid\F_{t_{(s_n+l_n)\nu k_n}}]\\
&=\sigma_{t_{(s_n+l_n)\nu k_n}}^2v_{t_{(s_n+l_n)\nu k_n}}^2(k_n\Delta_n)^{2H+1}
\left\{\gamma_n(k,l)
+\frac{\rho_{t_{(s_n+l_n)\nu k_n}}^2}{K_H^2(H+1/2)^2}\kappa_n(k,l)\right\}\\
&-\frac{\sigma_{t_{(s_n+l_n)\nu k_n}}^2v_{t_{(s_n+l_n)\nu k_n}}^2k_n\Delta_n}{K_H^2}
 \int_{-\infty}^{t_{(s_n+l_n)\nu k_n}}d_{ik}(s)d_{il}(s)ds,
\end{aligned}\\
&\gamma_n(k,l)=\tfrac12\{(1+|k-l|/k_n)^{2H}+(1-|k-l|/k_n)^{2H}
-2(|k-l|/k_n)^{2H}\},\\
&\kappa_n(k,l)=(kl/k_n^2)^{H+1/2}.
\end{align*}
The tails, after the displayed normalizations and summation, are $o(n^{-\beta})$
by the same squared-tail estimate used in \eqref{eq:block-transfer}.
The mixed measurement/rough expectation is zero because it has odd total
Gaussian degree. The limiting deterministic functions are
\begin{equation*}
\gamma(x,y)=\tfrac12\{(1+|x-y|)^{2H}+(1-|x-y|)^{2H}-2|x-y|^{2H}\},
\qquad \kappa(x,y)=(xy)^{H+1/2}.
\end{equation*}
Their integrals are
\begin{align*}
&\begin{aligned}[t]
\int_0^1\!\int_0^1\gamma(x,y)dxdy
&=\int_0^1(1-z)\{(1+z)^{2H}+(1-z)^{2H}-2z^{2H}\}dz\\
&=\frac{4(2^{2H}-1)}{(2H+1)(2H+2)},
\end{aligned}\\
&\int_0^1\!\int_0^1\kappa(x,y)dxdy=\left(\int_0^1x^{H+1/2}dx\right)^2
=\frac{4}{(2H+3)^2}.
\end{align*}
It follows that
\begin{equation}\label{eq:group-bracket}
\begin{aligned}
\sum_\nu\E[L_{\nu,n}^2\mid\F_{t_{(s_n+l_n)\nu k_n}}]
=&(2H+3)^2\Delta_n(k_n\Delta_n)^{-2H}\int_0^T\sigma_s^6ds\\
&+\frac{(2H+3)^2(2^{2H}-1)k_n\Delta_n}{(2H+1)(2H+2)}\int_0^T\sigma_s^2v_s^2ds\\
&+\frac{k_n\Delta_n}{K_H^2(H+1/2)^2}\int_0^TL_s^2ds+o_p(n^{-\beta}).
\end{aligned}
\end{equation}
Conditional Jensen and \eqref{eq:block-transfer} show that the squared
conditional means at auxiliary block left endpoints sum to $o_p(n^{-\beta})$. Conditional Cauchy--Schwarz transfers
\eqref{eq:group-bracket} to the conditional variances of
$S_{\nu,n}^{\rm raw}$. Multiplication by $n^\beta$ gives the stated variance
in Theorem~\ref{thm:feasible-clt}, including both components at $b=2H/(2H+1)$.

\textit{(c) Orthogonality and Lindeberg condition.}
The frozen rough contribution is a centered product of two first
Gaussian integrals; its orthogonal projection onto the linear Gaussian
coordinates is zero. For the measurement contribution, set
\begin{equation*}
P=k_n^{-1/2}\sum_{j=1}^{k_n}Z_j,\qquad
E=k_n^{-1/2}\sum_{j=1}^{k_n}(Z_j^2-{Z_j'}^2),
\end{equation*}
where all displayed coordinates are independent standard Gaussian
variables. For each $1\le l\le k_n$, direct fourth moments give
\begin{equation*}
\E[PEZ_l]=\frac2{k_n},\qquad \E[PEZ_l']=0,
\qquad
\operatorname{proj}_1(PE)=\frac2{k_n}\sum_{l=1}^{k_n}Z_l
=\frac{2P}{\sqrt{k_n}}.
\end{equation*}
After the prefactor in $L_{\nu,n}$, the sum over all auxiliary blocks of the
squared norms of these linear projections is at most $C\Delta_n(k_n\Delta_n)^{-2H}/k_n$.
Since $\Delta_n(k_n\Delta_n)^{-2H}=O(n^{-\beta})$, conditional Cauchy--Schwarz gives an
$O(k_n^{-1/2})$ bound for the normalized cumulative covariation with
each Brownian driver. For a general predictable square-integrable integrand, first take bounded
step integrands and then use their $L^2$ approximation, with the variance
bound \eqref{eq:group-bracket}. For a bounded martingale orthogonal to the
drivers, the centered Gaussian polynomial has an adapted Brownian integral
representation on its auxiliary block and hence conditional product zero.
The squared-error estimate \eqref{eq:block-transfer} transfers all these
assertions to the actual array centered at auxiliary block left endpoints.

The frozen polynomial has degree at most three. Conditional Gaussian
hypercontractivity gives
$\|L_{\nu,n}\|_{4\mid\F_{t_{(s_n+l_n)\nu k_n}}}
\le3^{3/2}\|L_{\nu,n}\|_{2\mid\F_{t_{(s_n+l_n)\nu k_n}}}$,
where these norms integrate only the future Gaussian coordinates.
Consequently
\begin{align*}
&\E[L_{\nu,n}^4\mid\F_{t_{(s_n+l_n)\nu k_n}}]\\
&\le C\{\E[L_{\nu,n}^2\mid\F_{t_{(s_n+l_n)\nu k_n}}]\}^2\\
&\le C((s_n+l_n)k_n\Delta_n)^2
\Bigl\{\Delta_n^2(k_n\Delta_n)^{-4H}\sigma_{t_{(s_n+l_n)\nu k_n}}^{12}\\
&+(k_n\Delta_n)^2\sigma_{t_{(s_n+l_n)\nu k_n}}^4v_{t_{(s_n+l_n)\nu k_n}}^4\Bigr\}.
\end{align*}
At the fixed localization level, Proposition~\ref{prop:localization} gives
\begin{equation*}
\E\sigma_t^{12}\le C,\qquad
\E[\sigma_t^4v_t^4]\le(\E \sigma_t^8)^{1/2}(\E v_t^8)^{1/2}\le C.
\end{equation*}
Taking the outer expectation, using $\Delta_n(k_n\Delta_n)^{-2H}+k_n\Delta_n=O(n^{-\beta})$ and summing gives
\begin{equation*}
n^{2\beta}\sum_\nu\E|L_{\nu,n}|^4
\le C (s_n+l_n)k_n\Delta_n\sup_{t\le T}
 \E[\sigma_t^{12}+\sigma_t^4v_t^4]
\le C (s_n+l_n)k_n\Delta_n\longrightarrow0.
\end{equation*}
For the actual array centered at auxiliary block left endpoints, apply the deterministic inequality
\begin{equation*}
x^2\1_{\{|x|>\varepsilon\}}
\le4(x-y)^2+2y^2\1_{\{|y|>\varepsilon/2\}}
\end{equation*}
with
\begin{equation*}
x=n^{\beta/2}\{S_{\nu,n}^{\rm raw}-\E[S_{\nu,n}^{\rm raw}\mid\F_{t_{(s_n+l_n)\nu k_n}}]\},
\qquad y=n^{\beta/2}L_{\nu,n}.
\end{equation*}
Conditional Jensen and \eqref{eq:block-transfer} make the sum of the first
terms converge to zero in $L^1$; the fourth-moment bound just proved treats
the second terms. This proves conditional Lindeberg.
Equation~\eqref{eq:group-mean} restores the conditional means. The omitted centered small auxiliary blocks and
last incomplete auxiliary block have maximal $L^2$ norm
$O(n^{-\beta/2}\sqrt{s_n/(s_n+l_n)+(s_n+l_n)k_n\Delta_n})$ by Doob; the predictable part is retained on
the full index set and has the established maximal bound. Lemma~\ref{lem:stable-clt}
gives the stable limit for the unnormalized centered sum. Since
\begin{equation*}
a_n(\widehat\theta_n-\theta_T)
=\sum_\nu S_{\nu,n}^{\rm raw}+o_p(n^{-\beta/2}),
\qquad a_n\longrightarrow1,
\end{equation*}
stable Slutsky proves Theorem~\ref{thm:feasible-clt}. Lemma~\ref{lem:price-drift} and Proposition~\ref{prop:localization}
restore the stated price process.
\end{proof}
\subsection{Variance estimation and studentization}

We establish quadratic-form estimates. For integers $1\le B\le N/2$ and a vector
$x=(x_0,\ldots,x_{N-1})$, define
\begin{equation}\label{eq:q-form}
\mathcal Q_{N,B}(x)=\frac1{2B}\sum_{s=0}^{N-1}
\left(\sum_{j=s+B}^{s+2B-1}x_{j\bmod N}
     -\sum_{j=s}^{s+B-1}x_{j\bmod N}\right)^2.
\end{equation}
When a coarse array $z_i$ is placed at distinct increasing deterministic
completion cells $j_{i,n}$, write $\mathcal Q_{N_n,B_n}(z)$ for this form
with $x_{j_{i,n}}=z_i$ and all other cells zero.
For the statistic on the observation grid, $j_{i,n}=ik_n-1$.
Write $U=(U_{i,n}(X))_i$. Then $\sum_iU_{i,n}(X)=a_n\widehat\theta_n$ and
$\mathcal Q_{n,B_n}(U)=a_n^2Q_n$.

\begin{lemma}\label{lem:q-energy}
The quadratic form satisfies
\begin{align}
&|\sqrt{\mathcal Q_{N,B}(x)}-\sqrt{\mathcal Q_{N,B}(y)}|
 \le\sqrt{\mathcal Q_{N,B}(x-y)},\notag\\
&|\mathcal Q_{N,B}(x+y)-\mathcal Q_{N,B}(x)-\mathcal Q_{N,B}(y)|
 \le2\sqrt{\mathcal Q_{N,B}(x)\mathcal Q_{N,B}(y)},\notag\\
&\sqrt{\mathcal Q_{N,B}(x)}\le\sum_j|x_j|.
\label{eq:q-triangle}
\end{align}
A vector supported at one cell has quadratic form equal to the square
of its nonzero entry. If $z_i\in L^2(\mathcal H_i)$ and
$\E[z_i\mid\mathcal H_{i-q}]=0$ for a filtration $(\mathcal H_i)$
and a fixed integer $q\ge1$, then
\begin{equation}\label{eq:q-color}
\E\mathcal Q_{N_n,B_n}(z)\le q\sum_i\|z_i\|_2^2.
\end{equation}
For any square-integrable coarse array,
\begin{equation}\label{eq:q-groups}
\E\mathcal Q_{N_n,B_n}(z)
\le C\sup_{(J_\nu)}\sum_\nu
       \left\|\sum_{i\in J_\nu}z_i\right\|_2^2,
\end{equation}
where the supremum is over deterministic families of disjoint consecutive
coarse-index groups.
\end{lemma}
\begin{proof}
The square root of \eqref{eq:q-form} is the norm of a linear map.
The triangle and Cauchy--Schwarz inequalities give the first two bounds.
Each cell occurs in exactly $2B$ contrasts with coefficient $1$ or $-1$;
its squared norm is one. Summing the norms of the individual cells proves
the remaining assertion in \eqref{eq:q-triangle}.

For deterministic coefficients $v_i$, divide the indices into their $q$
residue classes. Within each class the $z_i$ are martingale differences.
Orthogonality within a class and Cauchy--Schwarz between the classes give
$\E|\sum_i v_i z_i|^2\le q\sum_i v_i^2\E z_i^2$.
Apply this to every contrast, and count the $2B_n$ appearances of each
completion cell, to obtain \eqref{eq:q-color}.

For \eqref{eq:q-groups}, the squared difference of two window sums is
at most twice the sum of their squares. Sort the starting cells by their
residue modulo $B_n$. Within one residue class the length-$B_n$ windows
are disjoint until the final window wraps around $N_n$. Split that window
at the endpoint and use $|a+b|^2\le2a^2+2b^2$. The resulting ordinary
intervals form at most two families of disjoint intervals, each selecting
consecutive coarse indices. Sum first within a residue class and then
over the $B_n$ classes. Division by $2B_n$ gives a constant independent
of both $N_n$ and $B_n$.
\end{proof}

\begin{lemma}\label{lem:q-martingale}
Suppose $j_{i+1,n}-j_{i,n}=k_n$, $k_n/B_n\to0$, and $B_n/N_n\to0$.
Let $F_{i,n}$ be martingale differences at successive coarse times, with
\begin{equation}\label{eq:q-md-moments}
\E F_{i,n}^4\le C(k_n/N_n)^2.
\end{equation}
Zero terms at discarded coarse indices are allowed. Then
\begin{equation}\label{eq:q-md-diagonal}
\mathcal Q_{N_n,B_n}(F)=\sum_iF_{i,n}^2+o_p(1).
\end{equation}
Suppose further that the retained indices form deterministic consecutive
groups $I^l(\nu)$, each contained between a conditioning time and a later
group endpoint, and containing at most $l_n+s_n$ coarse indices. If
$(l_n+s_n)k_n/N_n\to0$, then
\begin{equation}\label{eq:q-md-bracket}
\mathcal Q_{N_n,B_n}(F)
=\sum_\nu\E\left[\left(\sum_{i\in I^l(\nu)}F_{i,n}\right)^2
                  \Bigm|\mathcal H_{\nu,n}^{\rm left}\right]+o_p(1),
\end{equation}
where $\mathcal H_{\nu,n}^{\rm left}$ denotes the sigma-field at that
conditioning time. The groups and their endpoints are ordered without
overlap.
\end{lemma}
\begin{proof}
For $4B<N$, the coefficient of two cells separated by $d$ is
\begin{equation*}
K_B(d)=
\begin{cases}
1-3d/(2B),&0\le d\le B,\\
-1+d/(2B),&B\le d\le2B,\\
0,&d\ge2B.
\end{cases}
\end{equation*}
Indeed, at separation $d\le B$ the two cells have the same sign in
$2(B-d)$ contrasts and opposite signs in $d$ contrasts. At separation
$B\le d\le2B$ they have opposite signs in $2B-d$ contrasts.
Consequently,
\begin{equation}\label{eq:q-lags}
\mathcal Q_{N_n,B_n}(z)=\sum_i z_i^2
 +2\sum_{i<l}\{K_{B_n}(j_{l,n}-j_{i,n})
             +K_{B_n}(N_n-j_{l,n}+j_{i,n})\}z_i z_l.
\end{equation}
Martingale orthogonality and \eqref{eq:q-md-moments} imply
$\E\mathcal Q_{N_n,B_n}(F)=\sum_i\E F_{i,n}^2=O(1)$.
The coarse indices within $2B_n$ cells of either endpoint have total
quadratic energy at most $C(B_n+k_n)/N_n=o(1)$ in expectation.
By \eqref{eq:q-triangle}, deleting them changes the form by $o_p(1)$.
Their sum of squares is also $o_p(1)$, and the wraparound coefficient
in \eqref{eq:q-lags} then vanishes.

For the remaining indices put
$S_{i,n}=\sum_{l<i}K_{B_n}(j_{i,n}-j_{l,n})F_{l,n}$.
There are at most $2B_n/k_n+2$ nonzero terms. BDG and Minkowski give
\begin{equation*}
\|S_{i,n}\|_4^2
\le C\sum_{0<j_{i,n}-j_{l,n}<2B_n}\|F_{l,n}\|_4^2
\le C\frac{B_n+k_n}{N_n}.
\end{equation*}
The products $F_{i,n}S_{i,n}$ are martingale differences. Hence
\begin{equation*}
\E\left|2\sum_iF_{i,n}S_{i,n}\right|^2
\le4\sum_i\|F_{i,n}\|_4^2\|S_{i,n}\|_4^2
\le C\frac{B_n+k_n}{N_n}\longrightarrow0.
\end{equation*}
This proves \eqref{eq:q-md-diagonal}.

For the second assertion set $A_{\nu,n}=\sum_{i\in I^l(\nu)}F_{i,n}^2$.
Conditional orthogonality identifies its conditional expectation with
that in \eqref{eq:q-md-bracket}. Also,
\begin{equation*}
\E A_{\nu,n}^2
\le\left(\sum_{i\in I^l(\nu)}\|F_{i,n}\|_4^2\right)^2.
\end{equation*}
After summing over groups, the right side is at most
$C(l_n+s_n)k_n/N_n=o(1)$: use the maximal group size for one factor,
and the total number $O(N_n/k_n)$ of indices for the other.
The centered $A_{\nu,n}$ are martingale differences at the group endpoints.
Their total variance tends to zero, proving \eqref{eq:q-md-bracket}.
\end{proof}

\begin{lemma}\label{lem:q-target}
Fix a localization level $J$ in LR. Let a grid of spacing $\Delta_n$
have $N_n$ cells, where $T\le N_n\Delta_n<T+\Delta_n$.
Place $\theta_{t_{ik_n}}-\theta_{t_{(i-1)k_n}}$ at cells $j_{i,n}$
with $j_{i+1,n}-j_{i,n}=k_n$ and
$|(j_{i,n}+1)\Delta_n-t_{ik_n}|\le\Delta_n$.
The covered intervals may omit an initial and terminal interval of total
length $O(k_n\Delta_n)$. Then, for $2B_n\le N_n$,
\begin{equation*}
\begin{aligned}
&\E\mathcal Q_{N_n,B_n}
       ((\theta_{t_{ik_n}}-\theta_{t_{(i-1)k_n}})_i)\\
&\qquad\le C_J\left\{(B_n\Delta_n)^{1+2H_J}
       +\frac{(k_n\Delta_n)^2}{B_n\Delta_n}
       +(B_n\Delta_n)^2\right\}.
\end{aligned}
\end{equation*}
The assertion applies to the killed loading
$L=\mathbf1_{[0,\tau_J)}\widetilde L^J$ from
Proposition~\ref{prop:localization}.
\end{lemma}
\begin{proof}
First use the unstopped LR representative. Its stochastic component has
$L^2$ increments bounded by $C_J|t-s|^{H_J}$. In fact, It\^o isometry
and \eqref{eq:lr-bounds}, split at $s-u=t-s$, give
\begin{align*}
&\left\|\int_0^t\psi_J(t,u)\,d\mathbf B_u^J
       -\int_0^s\psi_J(s,u)\,d\mathbf B_u^J\right\|_2^2\\
&\quad\le C_J\left\{(t-s)+(t-s)^{2H_J}
 +(t-s)^2\int_{t-s}^T(1+x^{H_J-3/2})^2dx\right\}
\le C_J|t-s|^{2H_J}.
\end{align*}
For a contrast contained in $[0,T]$, the difference of the two integrated
loading windows is
\begin{equation*}
\frac1{K_H(H+1/2)}\int_0^{B_n\Delta_n}
\{\widetilde L^J_{(s+B_n)\Delta_n+u}
 -\widetilde L^J_{s\Delta_n+u}\}\,du.
\end{equation*}
Minkowski bounds its stochastic component in $L^2$ by
$C_J(B_n\Delta_n)^{1+H_J}$. The regular component is bounded by
$C_J(B_n\Delta_n)^{1+\zeta_J}$. Squaring, summing over starting cells,
and dividing by $2B_n$ gives
$C_J\{(B_n\Delta_n)^{1+2H_J}+(B_n\Delta_n)^2\}$.

For the bounded-variation component, let $\mu_A$ be its total variation
measure and write its total mass as $K$ in this calculation. The absolute
contrast is at most
$C_JB_n\Delta_n\mu_A((s\Delta_n,(s+2B_n)\Delta_n])$.
Its square is bounded by $C_J(B_n\Delta_n)^2K$ times the same interval
mass. Each time point belongs to at most $2B_n+1$ windows, so summing
pathwise before taking expectations bounds the resulting energy by
$C_J(B_n\Delta_n)^2\E K^2$.

Killing the loading requires no increment condition on the stopping time.
For any $a>0$,
\begin{equation*}
\mathbf1_{\{t+a<\tau_J\}}\widetilde L^J_{t+a}
-\mathbf1_{\{t<\tau_J\}}\widetilde L^J_t
=\mathbf1_{\{t+a<\tau_J\}}
 (\widetilde L^J_{t+a}-\widetilde L^J_t)
 -\mathbf1_{\{t<\tau_J\le t+a\}}\widetilde L^J_t.
\end{equation*}
The first term obeys the componentwise estimates just used. The second
adds a contrast bounded by
$C_JB_n\Delta_n\sup_{t\le T}|\widetilde L_t^J|$
on windows meeting $\tau_J$. There are at most $2B_n+1$ such windows
on each path, so its expected energy is $O((B_n\Delta_n)^2)$.
Likewise, the cyclic boundary contrasts and the possible extension from
$T$ to $N_n\Delta_n$, with $\theta$ held constant there, have energy
$O((B_n\Delta_n)^2)$.

Finally compare the exact integrated target at a grid time with the
cumulative sum of completed target increments. The discrepancy is bounded
by $C_Jk_n\Delta_n\sup_{t\le T}|\widetilde L_t^J|$: it consists of
initial, unfinished and terminal pieces of total length $O(k_n\Delta_n)$.
After additive cyclic extension the same bound holds for the difference
of the two contrasts. Its squared energy is at most
$C_JN_n(k_n\Delta_n)^2/B_n
\le C_J(k_n\Delta_n)^2/(B_n\Delta_n)$ in expectation.
The quadratic triangle inequality completes the proof.
\end{proof}

\begin{proof}[Proof of Proposition~\ref{prop:variance}]
Use the frozen coarse contribution $F_{i,n}$ from the proof of
\eqref{eq:block-transfer}, extended by zero at discarded coarse indices.
It is measurable at $t_{ik_n}$ and has zero conditional mean given
$\F_{t_{(i-1)k_n}}$. The Gaussian covariance calculation and the
fourth-moment bound there give
\begin{equation}\label{eq:q-frozen-moments}
\E F_{i,n}^2\le Ck_n\Delta_n n^{-\beta},\qquad
\E F_{i,n}^4\le C(k_n\Delta_n)^2n^{-2\beta}.
\end{equation}
Indeed, the same-index measurement variance is bounded by
$C\Delta_n(k_n\Delta_n)^{1-2H}$, the rough variance by
$C(k_n\Delta_n)^2$, and the centered Gaussian polynomial has degree
at most three. The Gaussian moment bound for $L_{\nu,n}$ thus applies to each $F_{i,n}$.

We next transfer the observed contributions, including their conditional
means. The local identity underlying \eqref{eq:group-error} is
\begin{align*}
U_{i,n}(M)-a_n\Delta_{ik_n}^{n,k_n}\theta-F_{i,n}
=&e_{i,n}+\E[U_{i,n}(M)-a_n\Delta_{ik_n}^{n,k_n}\theta\mid\F_{t_{(i-1)k_n}}]\\
&+2k_n^{-H-1/2}
 \{\E[\Delta_{ik_n}^{n,k_n}\theta\mid\F_{t_{(i-1)k_n}}]
                             -\Delta_{ik_n}^{n,k_n}\theta\}.
\end{align*}
Here $e_{i,n}=e_{i,n}^{(\nu)}$ on retained indices; on the discarded
indices it is the centered actual term from \eqref{eq:group-error}.
The individual product bounds in \eqref{eq:block-products}, before
summing into auxiliary blocks, imply
\begin{equation*}
\sum_i\|e_{i,n}\|_2^2=o(n^{-\beta}).
\end{equation*}
The retained sum is bounded by
$Cn^{-\beta}\{\omega_n+k_n^{-2H}+k_n\Delta_n+(s_n+l_n)^{-1}\}$;
the discarded sum costs
$Cn^{-\beta}\{s_n/(s_n+l_n)+(s_n+l_n)k_n\Delta_n\}$.
Each $e_{i,n}$ is a coarse martingale difference. The final target term
is also a coarse martingale difference, with total squared norm at most
$Ck_n^{-2H-1}k_n\Delta_n=o(n^{-\beta})$.
Lemma~\ref{lem:q-energy} controls their quadratic energies. The grouped
bound in Lemma~\ref{lem:block-mean} controls the conditional-mean term
by \eqref{eq:q-groups}. The grouped bound in
Lemma~\ref{lem:price-drift} treats $U_{i,n}(X)-U_{i,n}(M)$ in the
same way. Consequently,
\begin{equation}\label{eq:q-centered-transfer}
\mathcal Q_{n,B_n}
 (U-a_n(\Delta_{ik_n}^{n,k_n}\theta)_i-F)=o_p(n^{-\beta}).
\end{equation}
Lemma~\ref{lem:q-target} and \eqref{eq:q-bandwidth} give
$\mathcal Q_{n,B_n}((\Delta_{ik_n}^{n,k_n}\theta)_i)=o_p(n^{-\beta})$.
In particular, the placement term is negligible because
\begin{equation*}
\frac{(k_n\Delta_n)^2}{B_n\Delta_n}
=\frac{k_n}{B_n}\,k_n\Delta_n=o(n^{-\beta}),
\end{equation*}
and $(B_n\Delta_n)^2=o((B_n\Delta_n)^{1+2H_J})$ for every fixed
$H_J<1/2$. Thus \eqref{eq:q-centered-transfer} also holds without
the target subtraction.

Apply Lemma~\ref{lem:q-martingale} to $n^{\beta/2}F_{i,n}$ on the
$n$-cell observation grid with $j_{i,n}=ik_n-1$. Equation~\eqref{eq:q-frozen-moments}
verifies its local fourth-moment condition, and the auxiliary
block choice verifies its group condition. It follows that
\begin{equation*}
\mathcal Q_{n,B_n}(F)
=\sum_\nu\E[L_{\nu,n}^2\mid\F_{t_{(s_n+l_n)\nu k_n}}]
 +o_p(n^{-\beta}).
\end{equation*}
The bracket is \eqref{eq:group-bracket} and is $O_p(n^{-\beta})$.
The triangle inequality now gives
\begin{align*}
a_n^2Q_n=&(2H+3)^2\Delta_n(k_n\Delta_n)^{-2H}\int_0^T\sigma_s^6\,ds\\
&+\frac{(2H+3)^2(2^{2H}-1)k_n\Delta_n}{(2H+1)(2H+2)}
          \int_0^T\sigma_s^2v_s^2\,ds\\
&+\frac{k_n\Delta_n}{K_H^2(H+1/2)^2}\int_0^TL_s^2\,ds
 +o_p(n^{-\beta}).
\end{align*}
Multiplying by $n^\beta$ and using $a_n\to1$ proves the stated variance limit at the fixed
localization level. Proposition~\ref{prop:localization} then transfers
both the variance estimator and its limit to the original model.
\end{proof}

\begin{proof}[Proof of studentization in Proposition~\ref{prop:variance}]
Combine the preceding consistency result with
Theorem~\ref{thm:feasible-clt} and stable Slutsky. Positivity of
$\mathcal V_T$ also implies $\P(Q_n=0)\to0$: for any $a>0$,
\begin{equation*}
\P(Q_n=0)\le\P(\mathcal V_T\le a)
 +\P(|n^\beta Q_n-\mathcal V_T|>a/2),
\end{equation*}
and then take $n\to\infty$ followed by $a\downarrow0$.
The confidence interval follows from this studentized limit.
\end{proof}

%% file: appendices/E_noise.tex
\section{Microstructure noise}
We prove Theorem~\ref{noisy:clt} and
Proposition~\ref{noisy:variance}. Fix the localization level in
Proposition~\ref{prop:localization} and omit its index, except in $H_N$.
All constants may depend on this level. The bandwidth condition gives
\begin{equation*}
m_n\Delta_n(k_nm_n\Delta_n)^{-2H}+k_nm_n\Delta_n
\asymp n^{-\beta/2}.
\end{equation*}
We use Gaussian observation noise as a reference law and transfer the results
to Assumption~\ref{noisy:noise} by coupling. Only the reference case uses
moments of arbitrarily high order.
All arguments use the original model and LR.

\subsection{Conditional increment estimates}

Use the kernels and correction in \eqref{noisy:correction-kernels}
and \eqref{noisy:correction-factor}. The deterministic kernels are
staircase functions, so the correction integral reduces to finite sums.

For a price path $Z$, write $R_j^n(Z)$ for the pre-averaged return in
\eqref{noisy:pa} evaluated on $Z$. Denote the coarse return, spot-variance difference
and block contribution by
\begin{align*}
&P_i=\sum_{j=(i-1)k_n}^{ik_n-1}R_j^n,\\
&D_i=\widehat\sigma_{t_{ik_nm_n}}^{2,\mathrm N}
       -\widehat\sigma_{t_{(i-1)k_nm_n}}^{2,\mathrm N},\\
&U_{i,n}=\frac{2H+3}{2}(k_nm_n\Delta_n)^{1/2-H}P_iD_i,\\
&T_{i,n}=U_{i,n}-a_n^{\mathrm N}\Delta_{ik_nm_n}^{n,k_nm_n}\theta.
\end{align*}
These contributions satisfy
\[
a_n^{\mathrm N}\widehat\theta_n^{\mathrm N}
=\sum_{i\in\mathcal I_n^{\mathrm N}}U_{i,n}.
\]
We initially use indices with $t_{ik_nm_n}+(m_n\Delta_n)\le T$. The exact set in
\eqref{noisy:indices} has at most one additional index, which the
observed statistic retains.

For the kernel $g_{m_n}$ in \eqref{noisy:correction-kernels}, write
$a_j(t)=g_{m_n}(t/(m_n\Delta_n)-j)$ and $N_j=R_j^n-R_j^n(X)$.

On an extension, let $W^{\epsilon}$ be a Brownian motion on $[0,T+1]$, independent of
$\F$, and realize the Gaussian noise by
\begin{equation}
\label{noisy:noise-embedding}
\epsilon_j^n=\sqrt{\nu/\Delta_n}
                  (W^{\epsilon}_{(j+1)\Delta_n}-W^{\epsilon}_{j\Delta_n}).
\end{equation}
Each row has exactly the prescribed conditional distribution given
$\F$. Thus this replacement preserves the joint law of the statistic
and every bounded $\F$-measurable variable. Set
\[
\mathcal G_t=\F_{t\wedge T}\vee\sigma(W^{\epsilon}_s:0\le s\le t),
\qquad 0\le t\le T+1,
\]
with the usual augmentation understood. Then
\begin{equation*}
\begin{aligned}
N_j=&\sqrt{\frac{c_{m_n}^2\nu}{m_n^2\Delta_n}}\bigl\{W^{\epsilon}_{t_{(j+2)m_n}}-W^{\epsilon}_{t_{(j+1)m_n}}\\
&\qquad-W^{\epsilon}_{t_{(j+1)m_n}}+W^{\epsilon}_{t_{jm_n}}\bigr\}.
\end{aligned}
\end{equation*}
As in Section~D, enlarge the space if necessary to write
$dB_t=\rho_t\,dW_t+\sqrt{1-\rho_t^2}\,dW_t^\perp$.
The three Brownian motions $W,W^\perp,W^{\epsilon}$ form a Brownian vector for
$(\mathcal G_t)$ on $[0,T]$.

\begin{lemma}\label{noisy:local-lemma}
\begin{align}
&0\le g_{m_n}\le c_{m_n}\le\sqrt{3/2},\notag\\
&\int_{\mathbb R}g_{m_n}^2=1,\notag\\
&\int_{\mathbb R}g_{m_n}=c_{m_n},\notag\\
&\sum_{j\in\mathbb Z}g_{m_n}(x-j)=c_{m_n},\notag\\
&\int_{\mathbb R}g_{m_n}(x)g_{m_n}(x-1)\,dx
   =\rho_{m_n}:=\frac{m_n^2-1}{4m_n^2+2},\notag\\
&1+2\rho_{m_n}=c_{m_n}^2.\label{noisy:kernel-identities}
\end{align}
Each $a_j$ is supported on $[t_{jm_n},t_{(j+2)m_n}]$, and at most two
of these kernels are nonzero at a time, apart from endpoints.
Here $\rho_{m_n}$ denotes the adjacent-kernel correlation, distinct
from the leverage process $\rho_t$.
The noise covariance satisfies
\begin{equation}
\label{noisy:noise-covariance}
\E[N_jN_l]
=\frac{c_{m_n}^2\nu}{m_n}\{2\mathbf1_{\{j=l\}}
                         -\mathbf1_{\{|j-l|=1\}}\}.
\end{equation}

First take the observations to be $M+\epsilon$. Define
\begin{align*}
&A_i=\frac1{k_nm_n\Delta_n}\int\sum_{j=(i-1)k_n}^{ik_n-1}a_j(t)^2\sigma_t^2\,dt,\\
&E_i=\widehat\sigma_{t_{ik_nm_n}}^{2,\mathrm N}-A_i-\frac{2c_{m_n}^2\nu}{m_n^2\Delta_n}.
\end{align*}
Then
\begin{align}
E_i=&\frac2{k_nm_n\Delta_n}\sum_{j=(i-1)k_n}^{ik_n-1}
\int a_j(t)\sigma_t
      \left(\int_{t_{jm_n}}^t a_j(u)\sigma_u\,dW_u\right)dW_t\notag\\
&+\frac2{k_nm_n\Delta_n}\sum_{j=(i-1)k_n}^{ik_n-1}R_j^n(M)N_j\notag\\
&+\frac1{k_nm_n\Delta_n}\sum_{j=(i-1)k_n}^{ik_n-1}(N_j^2-\frac{2c_{m_n}^2\nu}{m_n}).
\label{noisy:local-decomposition}
\end{align}
For every fixed $p\ge2$,
\begin{align}
&\|E_i\|_p\le C_pk_n^{-1/2},\notag\\
&\|A_i-A_{i-1}\|_p\le C_p(k_nm_n\Delta_n)^H,\notag\\
&\|P_i\|_p\le C_p\sqrt{k_nm_n\Delta_n},\notag\\
&\|\widehat\sigma_{t_{ik_nm_n}}^{2,\mathrm N}\|_p\le C_p,\notag\\
&\|D_i\|_p\le C_p((k_nm_n\Delta_n)^H+k_n^{-1/2}),\notag\\
&\|U_{i,n}\|_p\le C_p\sqrt{k_nm_n\Delta_n}\,n^{-\beta/4}.
\label{noisy:local-moments}
\end{align}
Moreover,
\begin{align}
&\left\|A_i-\frac1{k_nm_n\Delta_n}\int_{t_{(i-1)k_nm_n}}^{t_{ik_nm_n}}\sigma_t^2\,dt\right\|_p
       \le C_p(m_n\Delta_n)^H,\notag\\
&\|P_i-c_{m_n}\Delta_{ik_nm_n}^{n,k_nm_n}M\|_p\le C_p\sqrt{m_n\Delta_n}.
\label{noisy:box-comparison}
\end{align}
\end{lemma}
\begin{proof}
The first identities follow by integrating on the staircase intervals
of length $1/m_n$, using
$\sum_{a=1}^{m_n-1}a^2=(m_n-1)m_n(2m_n-1)/6$.
The translates of each unit-length indicator in $g_{m_n}$ sum to one.
For the noise covariance, set
$\bar\epsilon_j=(c_{m_n}/m_n)\sum_{a=0}^{m_n-1}\epsilon_{jm_n+a}$.
These variables are independent, with variance
$c_{m_n}^2\nu/m_n$, and
$N_j=\bar\epsilon_{j+1}-\bar\epsilon_j$.

It\^o's formula for $R_j^n(M)^2$ gives
\eqref{noisy:local-decomposition}. For its first term, BDG and the bounded
overlap of the kernels give squared $L^p$ norm at most
$C_p(k_nm_n\Delta_n)^{-2}(k_nm_n\Delta_n)(m_n\Delta_n)=C_p/k_n$.
Conditionally on $\F$, the second term is Gaussian, and its
conditional variance is bounded by
$C(m_n\Delta_n) (k_nm_n\Delta_n)^{-2}\sum_jR_j^n(M)^2$ by
\eqref{noisy:noise-covariance}; the $L^{p/2}$ norm of this bound is
$O(k_n^{-1})$.
In the third term, the even and odd subsequences separately consist of
independent centered Gaussian squares. Expanding their moments gives the
same bound. The difference of the two $A_i$ follows from translating their
positive weights by $(k_nm_n\Delta_n)$ and applying \eqref{eq:variance-moments}.
The noise in $P_i$ telescopes to
$\bar\epsilon_{ik_n}-\bar\epsilon_{(i-1)k_n}$, whose norm is $O(\sqrt{m_n\Delta_n})$.
BDG proves the price bound; the remaining bounds follow by H\"older and
\begin{align*}
(k_nm_n\Delta_n)^{1/2-H}((k_nm_n\Delta_n)^H+k_n^{-1/2})=&\sqrt{k_nm_n\Delta_n}\\
&+\sqrt{(m_n\Delta_n)(k_nm_n\Delta_n)^{-2H}}\\
\le& Cn^{-\beta/4}.
\end{align*}
The measures $a_j(t)^2dt$ and Lebesgue measure on
$[t_{jm_n},t_{(j+1)m_n}]$ both have mass $(m_n\Delta_n)$ and can be coupled within
distance $2(m_n\Delta_n)$. Integrating \eqref{eq:variance-moments} against such a
coupling and summing over the $k_n$ cells gives the first bound in
\eqref{noisy:box-comparison}. The translate identity in
\eqref{noisy:kernel-identities} shows that the price kernel differs from
$c_{m_n}\mathbf1_{(t_{(i-1)k_nm_n},t_{ik_nm_n}]}$ only on two strips of length $(m_n\Delta_n)$.
BDG and the telescoping noise prove the second bound.
\end{proof}

\begin{lemma}\label{noisy:color}\label{noisy:lr-uniform}
Suppose that $Z_i\in L^2(\mathcal G_{t_{(i+1)k_nm_n}})$ and
$\E[Z_i\mid\mathcal G_{t_{(i-2)k_nm_n}}]=0$.
For deterministic $|\vartheta_i|\le1$ and disjoint consecutive index sets
$J_r$,
\begin{equation*}
\sum_r\E\max_{q\in J_r}
 \left|\sum_{i\in J_r,\,i\le q}\vartheta_i Z_i\right|^2
\le C\sum_i\E|Z_i|^2.
\end{equation*}
In particular this holds for the maximal partial sum over all indices.

The conclusions of Lemmas~\ref{lem:projection} and~\ref{lem:lr-sums} hold
uniformly when their weight is replaced by a deterministic sequence $w_n$
with $\int_0^1w_n=0$ and $\sup_n\|w_n\|_\infty<\infty$.
They also hold for the killed LR representative
$L=\mathbf1_{[0,\tau_N)}\widetilde L^N$ supplied by
Proposition~\ref{prop:localization}.
\end{lemma}
\begin{proof}
Split the indices into their three residue classes. Within each class,
the preceding variable is measurable at the conditioning time of the next
one, so the sequence is a martingale difference sequence. Apply Doob's
inequality within each class and each group, then
$|x_1+x_2+x_3|^2\le3\sum_{a=1}^3|x_a|^2$.

For the LR assertions, the estimates \eqref{eq:cell-kernel} and \eqref{eq:kernel-energy} in
Section~A use the weight only through its zero integral and a bound on
its absolute integral; for the present bounded weights their constants
are uniform. The same stochastic Fubini calculation therefore gives,
for a consecutive set of $q$ cells and deterministic weights of absolute
value at most one, squared stochastic norm at most
$C\{q(k_nm_n\Delta_n)^{2H_N+2}+q(k_nm_n\Delta_n)^3+q^2(k_nm_n\Delta_n)^4\}\le C_Tq(k_nm_n\Delta_n)^{2H_N+2}$.
The binary decomposition in Lemma~\ref{lem:lr-sums}, together with the
bounded-variation and regular components, yields global and grouped bounds
\begin{align*}
&\begin{aligned}[t]
&C\bigl\{(k_nm_n\Delta_n)+(k_nm_n\Delta_n)^{\zeta_N}\\
&+(k_nm_n\Delta_n)^{H_N+1/2}
 \log(2+T/(k_nm_n\Delta_n))\bigr\}=o(\sqrt{k_nm_n\Delta_n}),
\end{aligned}\\
&\begin{aligned}[t]
&C\bigl\{(k_nm_n\Delta_n)^2+(k_nm_n\Delta_n)^{2\zeta_N}\\
&+(k_nm_n\Delta_n)^{2H_N+1}
 \log^2(2+T/(k_nm_n\Delta_n))\bigr\}=o(k_nm_n\Delta_n).
\end{aligned}
\end{align*}
Stopping changes a prefix into a stopped prefix plus one partial cell,
whose norm is at most $C(k_nm_n\Delta_n)\|\sup|\widetilde L^N|\|_2$.
Only one member of a disjoint group family contains that cell.
The proof of Lemma~\ref{lem:projection} likewise depends only on the
uniform weight bound and \eqref{eq:cadlag-sum}. This proves both
extensions, including the killed representative.
\end{proof}

\subsection{Price drift and the complete conditional mean}

\begin{lemma}\label{noisy:price-drift}
Evaluate the statistics at $X+\epsilon$ and $M+\epsilon$ with the same
noise. If $d_{i,n}=U_{i,n}(X+\epsilon)-U_{i,n}(M+\epsilon)$, then
\begin{align}
&\left\|\max_q\left|\sum_{i\le q}d_{i,n}\right|\right\|_2=o(n^{-\beta/4}),\notag\\
&\sup_{(J_r)}\sum_r
\left\|\max_{q\in J_r}
 \left|\sum_{i\in J_r,\,i\le q}d_{i,n}\right|\right\|_2^2
=o(n^{-\beta/2}).
\label{noisy:drift-groups}
\end{align}
The supremum is over deterministic disjoint consecutive
groups. Both bounds hold with $d_{i,n}$ replaced by its conditional
expectation given $\mathcal G_{t_{(i-2)k_nm_n}}$, or by its centered version.
\end{lemma}
\begin{proof}
Set $Z=M+\epsilon$. The exact identity
\begin{equation*}
\widehat\sigma_{t_{ik_nm_n}}^{2,\mathrm N}(Z+A)
 -\widehat\sigma_{t_{ik_nm_n}}^{2,\mathrm N}(Z)
=\frac1{k_nm_n\Delta_n}\sum_{j=(i-1)k_n}^{ik_n-1}
 \{2R_j^n(Z)R_j^n(A)+R_j^n(A)^2\}
\end{equation*}
gives local norm $C_p\sqrt{m_n\Delta_n}$, while $\|P_i(A)\|_p\le C(k_nm_n\Delta_n)$.
The three product differences in the proof of Lemma~\ref{lem:price-drift}
each have norm at most $C_pk_nm_n\Delta_n n^{-\beta/4}$ here.
For a bounded elementary predictable drift equal to a starting-time
measurable constant $z$ on both extended windows, the corresponding
identities become
\begin{align*}
&R_j^n(A)=zc_{m_n}(m_n\Delta_n),\\
&P_i(A)=zc_{m_n}(k_nm_n\Delta_n),\\
&\widehat\sigma_{t_{ik_nm_n}}^{2,\mathrm N}(Z+A)
 -\widehat\sigma_{t_{ik_nm_n}}^{2,\mathrm N}(Z)
=\frac{2zc_{m_n}}{k_n}P_i(Z)+z^2c_{m_n}^2(m_n\Delta_n).
\end{align*}
Thus the first product again telescopes and the constant squared-drift
terms cancel in the other two products. Their total norms are bounded by
$Ck_n^{-1}(k_nm_n\Delta_n)^{1/2-H}$ and $Ck_n^{-1}(k_nm_n\Delta_n)^{1-H}$; the summed squared endpoint
cost is $C(k_nm_n\Delta_n)^{2-2H}=o((k_nm_n\Delta_n))$. The finitely many windows crossing a breakpoint
have total norm $O((k_nm_n\Delta_n)n^{-\beta/4})$. These are precisely the bounds used in the
proof of Lemma~\ref{lem:price-drift}, with the bounded factor $c_{m_n}$ retained.

For uniformly bounded elementary approximations $\mu^{(q)}\to\mu$ in
$L^{8p}(\Omega\times[0,T])$, bounded overlap of the actual kernels gives
\begin{equation*}
\sum_i\|U_{i,n}(M+A+\epsilon)-U_{i,n}(M+A^{(q)}+\epsilon)\|_p
\le C_pn^{-\beta/4}\|\mu-\mu^{(q)}\|_{L^{8p}(\Omega\times[0,T])}.
\end{equation*}
This follows by inserting $R_j^n(A-A^{(q)})=\int a_j(t)
(\mu_t-\mu_t^{(q)})dt$ in the displayed identity and applying H\"older.
The approximation argument in Lemma~\ref{lem:price-drift} proves the
maximal and grouped bounds.
Finally $\sum_i\|d_{i,n}\|_2^2=O((k_nm_n\Delta_n)n^{-\beta/2})$, so the three-residue argument
of Lemma~\ref{noisy:color} also gives the projected and centered assertions.
\end{proof}

For the complete conditional mean, define the deterministic kernels
\begin{align}
&p_i(t)=\sum_{j=(i-1)k_n}^{ik_n-1}a_j(t),\notag\\
&e_i(t)=\frac1{k_nm_n\Delta_n}\left\{\sum_{j=(i-1)k_n}^{ik_n-1}a_j(t)^2
                  -\sum_{j=(i-2)k_n}^{(i-1)k_n-1}a_j(t)^2\right\},\notag\\
&\mathcal A_i(u,t)=\frac1{k_nm_n\Delta_n}\left\{
 \sum_{j=(i-1)k_n}^{ik_n-1}a_j(u)a_j(t)
 -\sum_{j=(i-2)k_n}^{(i-1)k_n-1}a_j(u)a_j(t)\right\},\notag\\
&\begin{aligned}[t]
\beta_i(u)=&(H+1/2)\frac{2H+3}{2}(k_nm_n\Delta_n)^{1/2-H}\\
&\times\left\{\begin{aligned}[t]
&p_i(u)\int_u^{t_{ik_nm_n}+(m_n\Delta_n)}e_i(t)K(t-u)\,dt\\
&+2\int_u^{t_{ik_nm_n}+(m_n\Delta_n)}p_i(t)\mathcal A_i(u,t)K(t-u)\,dt
\end{aligned}\right\}.
\end{aligned}
\label{noisy:mean-kernel}
\end{align}
\begin{lemma}\label{noisy:complete-mean}
The correction in \eqref{noisy:correction-factor} satisfies
\begin{equation}
\label{noisy:kernel-mass}
\begin{aligned}
&\int\beta_i(u)\,du=k_nm_n\Delta_n a_n^{\mathrm N},\\
&\sup_{i,k_n,m_n}\|\beta_i\|_\infty\le C,\\
&\operatorname{supp}\beta_i\subset
[t_{(i-1)k_nm_n}-m_n\Delta_n,t_{ik_nm_n}+m_n\Delta_n].
\end{aligned}
\end{equation}
Moreover, \eqref{noisy:calibration-properties} holds.

For the driftless Gaussian observations,
\begin{align}
\E[U_{i,n}\mid\mathcal G_{t_{(i-2)k_nm_n}}]=&\frac1{K_H(H+1/2)}
\E\left[\int\beta_i(u)L_u\,du
                      \,\middle|\,\mathcal G_{t_{(i-2)k_nm_n}}\right]\notag\\
&+e_i^f+e_i^g.\label{noisy:mean-identity}
\end{align}
The time-drift errors satisfy
\begin{align}
&\|e_i^f\|_p\le C_p(k_nm_n\Delta_n)^{2-H},\notag\\
&\|e_i^g\|_p\le C_p(k_nm_n\Delta_n)^{3/2}
 \left\|\sup_{t_{(i-2)k_nm_n}\le u\le t_{ik_nm_n}+m_n\Delta_n}
                    |g_u-g_{t_{(i-2)k_nm_n}}|\right\|_{2p}.
\label{noisy:mean-time-drifts}
\end{align}
Under LR, if $b_{i,n}=\E[T_{i,n}\mid\mathcal G_{t_{(i-2)k_nm_n}}]$, then
\begin{align}
&\left\|\max_q\left|\sum_{i\le q}b_{i,n}\right|\right\|_2=o(\sqrt{k_nm_n\Delta_n}),\notag\\
&\sup_{(J_r)}\sum_r
 \left\|\max_{q\in J_r}
  \left|\sum_{i\in J_r,\,i\le q}b_{i,n}\right|\right\|_2^2=o((k_nm_n\Delta_n)).
\label{noisy:mean-groups}
\end{align}
\end{lemma}
\begin{proof}
Translate the integrals by $t_{(i-1)k_nm_n}$ and rescale by $(m_n\Delta_n)$.
The first integral in \eqref{noisy:kernel-mass} becomes exactly
\eqref{noisy:correction-factor}. The first term of $\beta_i$ requires
$u$ to lie in the current price support. In the second, a previous-window
kernel can overlap $p_i$ only when its index is $(i-1)k_n-1$; thus
$u\ge t_{(i-1)k_nm_n}-(m_n\Delta_n)$. This proves the support assertion.
The bounds
$|e_i|\le C/(k_nm_n\Delta_n)$, $|p_i|\le C$, and
$|\mathcal A_i(u,t)|\le(C/(k_nm_n\Delta_n))\mathbf1_{\{|u-t|\le2m_n\Delta_n\}}$
bound the two integrals by $C(k_nm_n\Delta_n)^{H-1/2}$ and
$C(m_n\Delta_n)^{H+1/2}/(k_nm_n\Delta_n)$, respectively. Their normalization proves the
uniform bound.

To calculate the mass, replace the current price kernel by its box kernel
and transport each measure $a_j^2dt$ to the corresponding microcell of
equal mass. After the inner integral is taken, the power kernel has
H\"older exponent $H+1/2$, so this transport changes the normalized
mass by at most $Ck_n^{-H-1/2}$. The two price endpoint strips contribute
$O(k_n^{-1})$. The second contraction is supported on $|t-u|\le2(m_n\Delta_n)$;
its normalized total is bounded by
$C(k_nm_n\Delta_n)^{-H-3/2}k_n(m_n\Delta_n)^{H+3/2}=Ck_n^{-H-1/2}$.
The remaining previous-window overlap has the same bound.
On the scale $x=(u-t_{(i-1)k_nm_n})/(k_nm_n\Delta_n)$, the box contribution is
$c_{m_n}(H+3/2)(1-x)^{H+1/2}\mathbf1_{\{0<x<1\}}$, of integral $c_{m_n}$.
This proves $a_n^{\mathrm N}=c_{m_n}+O_H(k_n^{-H-1/2})$ uniformly in $m_n$.
When $m_n=1$, the kernels are disjoint ordinary increment indicators.
The box mass is exactly one, there is no previous-window overlap, and
the local contraction on each microcell has double integral
$(m_n\Delta_n)^{H+3/2}/\{(H+1/2)(H+3/2)\}$.
Its normalized sum is $2k_n^{-H-1/2}$, proving the exact identity.

For the complete conditional mean, take deterministic functions $f_1,f_2,f_3$ supported on $[a,b]$ and put
\begin{equation*}
M_j(t)=\int_a^tf_j(u)\sigma_u\,dW_u.
\end{equation*}
The product formula gives
\begin{equation*}
\E[M_1(b)M_2(b)M_3(b)\mid\F_a]
=\sum_{\mathrm{cyc}}\int_a^b f_j(t)f_l(t)
      \E[M_q(t)\sigma_t^2\mid\F_a]\,dt.
\end{equation*}
Insert the volatility equation. Its history before $a$, its initial value,
and its time-convolution term with $g_u$ replaced by $g_a$ are
$\F_a$-measurable and have zero product expectation with $M_q(t)$.
Conditional It\^o isometry gives the stochastic-convolution contribution
exactly as
\begin{equation*}
K_H^{-1}\E\left[\int_a^tf_q(u)K(t-u)L_u\,du
                                             \,\middle|\,\F_a\right].
\end{equation*}
Taking one kernel $p_i$ and two kernels $a_j$, and summing the signed
squares over the two windows, gives \eqref{noisy:mean-kernel} and the
latent part of \eqref{noisy:mean-identity} by Fubini.
For the two time-drift products, the price factor has $L^{2p}$ norm
$C\sqrt{k_nm_n\Delta_n}$ and each micro-return has norm $C\sqrt{m_n\Delta_n}$.
The summed squared-kernel weights are bounded by $C/(k_nm_n\Delta_n)$ on a set of
length $O((k_nm_n\Delta_n))$. The time drifts on that set have norms $C(k_nm_n\Delta_n)$ and
$C(k_nm_n\Delta_n)^{H+1/2}\|\sup|g-g_a|\|_{2p}$. Multiplying by $(k_nm_n\Delta_n)^{1/2-H}$ gives
\eqref{noisy:mean-time-drifts}. The grid oscillation estimate
\eqref{eq:cadlag-sum}, also valid for windows of bounded overlap, implies
\begin{align*}
&\sum_i\|e_i^f\|_2=O((k_nm_n\Delta_n)^{1-H})=o(\sqrt{k_nm_n\Delta_n}),\\
&\sum_i\|e_i^g\|_2=o(\sqrt{k_nm_n\Delta_n}).
\end{align*}
It remains to include the observation noise in the complete mean.
In the original level coordinates write $P=p^\top Y$ and $D=Y^\top A Y$.
The equal translated windows imply $\operatorname{tr}A=0$, and the
Gaussian third-moment identity gives
\begin{equation*}
\E_\epsilon[P(M+\epsilon)D(M+\epsilon)\mid M]
=P(M)D(M)+2\nu p^\top A M.
\end{equation*}
The last term is a deterministic linear combination of martingale
increments after $t_{(i-2)k_nm_n}$: its coefficients sum to zero because each
return kernel annihilates constants. Its conditional expectation given
$\F_{t_{(i-2)k_nm_n}}$ is therefore zero.
In \eqref{noisy:noise-embedding} all noise innovations in the two windows
also occur after $t_{(i-2)k_nm_n}$. Thus the noisy conditional mean given
$\mathcal G_{t_{(i-2)k_nm_n}}$ is exactly the latent one just computed.

For the LR step, set
\begin{equation*}
Z_i=\int\beta_i(u)L_u\,du
                 -a_n^{\mathrm N}\int_{t_{(i-1)k_nm_n}}^{t_{ik_nm_n}}L_u\,du.
\end{equation*}
Its deterministic kernel has zero integral by
\eqref{noisy:kernel-mass}. The translated kernel
$\beta_0(u):=\beta_i(t_{(i-1)k_nm_n}+u)$ is the same for every $i$,
with current cell $[0,k_nm_n\Delta_n]$. Periodizing it over cells of length
$k_nm_n\Delta_n$ gives, for $0<x<1$,
\begin{equation*}
w_n(x)=\frac1{a_n^{\mathrm N}}
          \sum_{q\in\mathbb Z}\beta_0((k_nm_n\Delta_n)(x-q))-1.
\end{equation*}
At most three summands are nonzero; $w_n$ is uniformly bounded and has
integral zero. A consecutive sum of the $Z_i$ differs from the corresponding
weighted-cell sum only on the two endpoint strips of width $(m_n\Delta_n)$,
where the total absolute kernel mass is $O((m_n\Delta_n))$.
The maximal $L^2$ cost of these strips is $O((m_n\Delta_n))$, and the summed
squared cost over disjoint groups is at most
$C(m_n\Delta_n)^2/(k_nm_n\Delta_n)=C(k_nm_n\Delta_n)/k_n^2=o((k_nm_n\Delta_n))$.
Lemma~\ref{noisy:lr-uniform} therefore gives the maximal $o(\sqrt{k_nm_n\Delta_n})$ and
grouped $o((k_nm_n\Delta_n))$ bounds for the raw $Z_i$.
Furthermore, the zero integral and \eqref{noisy:kernel-mass} give
\begin{align*}
&\|Z_i\|_2\le Ck_nm_n\Delta_n
 \left\|\sup_{t_{(i-2)k_nm_n}\le u\le t_{ik_nm_n}+m_n\Delta_n}
                   |L_u-L_{t_{(i-2)k_nm_n}}|\right\|_2,\\
&\sum_i\|Z_i\|_2^2=o(k_nm_n\Delta_n).
\end{align*}
Apply Lemma~\ref{noisy:color} to
$Z_i-\E[Z_i\mid\mathcal G_{t_{(i-2)k_nm_n}}]$.
The same bounds hold for the conditional projections. Combining them with
the two time-drift sums proves \eqref{noisy:mean-groups}.
\end{proof}

The exact endpoint omitted at the start of the proof has squared local
norm at most $C(k_nm_n\Delta_n)n^{-\beta/2}$ by \eqref{noisy:local-moments}.
The same estimates apply using its actual kernel support, which stops at
$t_{ik_nm_n}+(m_n\Delta_n)-\Delta_n\le T$; the auxiliary noise Brownian motion is defined
past $T$. This one index is negligible both for the centered local energy
and for the grouped conditional mean. The target intervals not covered by
the admissible indices have total length $O((k_nm_n\Delta_n))$ and contribute
$O((k_nm_n\Delta_n))=o(n^{-\beta/4})$, since $L$ is bounded at the fixed localization level.

\subsection{Proof of the feasible limit}

The price drift has been removed by Lemma~\ref{noisy:price-drift}.
Choose integers $s_n,l_n$ such that
\begin{align*}
&s_n\longrightarrow\infty,\\
&s_n/l_n\longrightarrow0,\\
&(l_n+s_n)(k_nm_n\Delta_n)\longrightarrow0.
\end{align*}
For example, $l_n=\lceil(k_nm_n\Delta_n)^{-1/2}\rceil$ and
$s_n=\lceil(k_nm_n\Delta_n)^{-1/4}\rceil$ suffice. In each complete
interval $[t_{(l_n+s_n)rk_nm_n},t_{(l_n+s_n)(r+1)k_nm_n}]$, retain
\begin{equation*}
I^l(r)=\{i:(l_n+s_n)r+s_n\le i-1,\quad
                      i+1\le(l_n+s_n)(r+1)\}
                      \cap\mathcal I_n^{\mathrm N}.
\end{equation*}
The other indices form the guard intervals and the final incomplete block.
Only their centered terms are discarded. Lemma~\ref{noisy:color} and
\eqref{noisy:local-moments} give squared total cost
\begin{equation}
\label{noisy:guard-cost}
Cn^{-\beta/2}\left\{\frac{s_n+2}{l_n+s_n}+(l_n+s_n)(k_nm_n\Delta_n)\right\}=o(n^{-\beta/2}).
\end{equation}
The complete conditional means continue to be summed over all admissible
indices, as in \eqref{noisy:mean-groups}.

Freeze the coefficients at $t_{(l_n+s_n)rk_nm_n}$, and write
\begin{equation*}
dB_u^r=\rho_{t_{(l_n+s_n)rk_nm_n}}\,dW_u
 +\sqrt{1-\rho_{t_{(l_n+s_n)rk_nm_n}}^2}\,dW_u^\perp.
\end{equation*}
For the frozen comparison only, use
\begin{align*}
&Z_j^r=\sigma_{t_{(l_n+s_n)rk_nm_n}}\int a_j(t)\,dW_t+N_j,\\
&E_i^r=\frac1{k_nm_n\Delta_n}\sum_{j=(i-1)k_n}^{ik_n-2}
                  \{(Z_j^r)^2-(m_n\Delta_n)(\sigma_{t_{(l_n+s_n)rk_nm_n}}^2+\frac{2c_{m_n}^2\nu}{m_n^2\Delta_n})\},\\
&d_i(u)=\frac1{k_nm_n\Delta_n}\int_{t_{(i-1)k_nm_n}}^{t_{ik_nm_n}}
                        \{K(t-u)-K(t-(k_nm_n\Delta_n)-u)\}\,dt.
\end{align*}
Deleting the last pre-averaged return in each $E_i^r$ ensures that every
remaining return is supported inside its own coarse interval. Its centered
square divided by $(k_nm_n\Delta_n)$ has norm $O(k_n^{-1})$. Its price product consequently
has total centered squared cost at most
$C(m_n\Delta_n) (k_nm_n\Delta_n)^{-2H}/k_n=o(n^{-\beta/2})$. No term is deleted from the observed statistic
or its exact finite-window correction.

For $i\in I^l(r)$ define
\begin{equation}
\label{noisy:frozen}
\begin{aligned}
F_{i,n}=&\frac{2H+3}{2}c_{m_n}(k_nm_n\Delta_n)^{1/2-H}
 \sigma_{t_{(l_n+s_n)rk_nm_n}}\Delta_{ik_nm_n}^{n,k_nm_n}W\\
&\times\left\{E_i^r-E_{i-1}^r
 +\frac{v_{t_{(l_n+s_n)rk_nm_n}}}{K_H}
   \int_{t_{(l_n+s_n)rk_nm_n}}^{t_{ik_nm_n}}d_i(u)\,dB_u^r\right\}\\
&-\frac{c_{m_n}k_nm_n\Delta_n}{K_H(H+1/2)}
 L_{t_{(l_n+s_n)rk_nm_n}}.
\end{aligned}
\end{equation}
Set $F_{i,n}=0$ at all other indices. Each term is
$\mathcal G_{t_{ik_nm_n}}$-measurable and centered given
$\mathcal G_{t_{(i-1)k_nm_n}}$. The sampling product is a Gaussian
cubic with zero mean. The remaining product has exactly the covariance
subtracted in the last line.

\begin{lemma}
For the exact admissible indices, there is the decomposition
\begin{equation}
\label{noisy:transfer-local}
\begin{aligned}
&U_{i,n}(X+\epsilon)-a_n^{\mathrm N}
             \Delta_{ik_nm_n}^{n,k_nm_n}\theta-F_{i,n}
          =e_{i,n}+b_{i,n}+d_{i,n},\\
&e_{i,n}=T_{i,n}(M+\epsilon)-b_{i,n}-F_{i,n},\\
&b_{i,n}=\E[T_{i,n}(M+\epsilon)\mid\mathcal G_{t_{(i-2)k_nm_n}}],\\
&d_{i,n}=U_{i,n}(X+\epsilon)-U_{i,n}(M+\epsilon),\\
&\sum_i\|e_{i,n}\|_2^2=o(n^{-\beta/2}).
\end{aligned}
\end{equation}
Each $e_{i,n}$ satisfies the measurability and conditional centering
conditions of Lemma~\ref{noisy:color}; $b_{i,n}$ and $d_{i,n}$ satisfy
\eqref{noisy:mean-groups} and \eqref{noisy:drift-groups}.
\end{lemma}
\begin{proof}
We give the local error estimates for the retained indices. Put
\begin{align*}
&\omega_{r,p}(c)=
 \left\|\sup_{t_{(l_n+s_n)rk_nm_n}\le t\le t_{(l_n+s_n)(r+1)k_nm_n}}
       |c_t-c_{t_{(l_n+s_n)rk_nm_n}}|\right\|_p,\\
&\omega_n=\sum_r(l_n+s_n)k_nm_n\Delta_n
 \sum_{c\in\{\sigma,v\rho,v\sqrt{1-\rho^2},L\}}
                      \omega_{r,16}(c)^2.
\end{align*}
By \eqref{eq:cadlag-sum}, $\omega_n\to0$.
Conditional centering increases an $L^p$ norm by at most two.
Replacing the price by its box return and $A_i$ by its box average in
\eqref{noisy:box-comparison} gives squared local errors bounded by
\begin{equation}
\label{noisy:pa-error}
\begin{aligned}
&C(k_nm_n\Delta_n)n^{-\beta/2}/k_n,\\
&C(k_nm_n\Delta_n)^2k_n^{-2H},
\end{aligned}
\end{equation}
respectively. For the price replacement use
$\|P_i-c_{m_n}\Delta_{ik_nm_n}^{n,k_nm_n}M\|_4\le C\sqrt{m_n\Delta_n}$ and
$\|D_i\|_4\le C((k_nm_n\Delta_n)^H+k_n^{-1/2})$.
For the variance replacement use its $C(m_n\Delta_n)^H$ norm and the
$C\sqrt{k_nm_n\Delta_n}$ price norm.

The remaining price and sampling errors satisfy
\begin{align}
&\|\Delta_{ik_nm_n}^{n,k_nm_n}M
        -\sigma_{t_{(l_n+s_n)rk_nm_n}}\Delta_{ik_nm_n}^{n,k_nm_n}W\|_p
\le C_p\sqrt{k_nm_n\Delta_n}\,\omega_{r,2p}(\sigma),\notag\\
&\|E_i-E_i^r\|_p
\le C_p\{k_n^{-1/2}\omega_{r,4p}(\sigma)+k_n^{-1}\}.
\label{noisy:sampling-error}
\end{align}
The first is BDG. For the second, expand the stochastic term of
\eqref{noisy:local-decomposition} into the two differences
\begin{align*}
&(\sigma_t-\sigma_{t_{(l_n+s_n)rk_nm_n}})
        \int_{t_{jm_n}}^ta_j(u)\sigma_u\,dW_u,\\
&\sigma_{t_{(l_n+s_n)rk_nm_n}}\int_{t_{jm_n}}^ta_j(u)
                         (\sigma_u-\sigma_{t_{(l_n+s_n)rk_nm_n}})\,dW_u.
\end{align*}
BDG and bounded overlap give squared norm
$C_pk_n^{-1}\omega_{r,4p}(\sigma)^2$.
The signal--noise difference, conditionally on $\F$, has
variance at most
\begin{equation*}
C(m_n\Delta_n) (k_nm_n\Delta_n)^{-2}\sum_j
 \left|R_j^n(M)-\sigma_{t_{(l_n+s_n)rk_nm_n}}\int a_j(u)\,dW_u\right|^2,
\end{equation*}
and hence the same bound. Pure-noise squares cancel at every retained
micro-index. The removed final centered square costs $C_p/k_n$.
This proves \eqref{noisy:sampling-error}.

For the box variance difference, the original model gives exactly
\begin{align*}
J_{1,i}^r+J_{2,i}^r+J_{3,i}^r=&\frac1{k_nm_n\Delta_n}\int_{t_{(i-1)k_nm_n}}^{t_{ik_nm_n}}
     (\sigma_t^2-\sigma_{t-k_nm_n\Delta_n}^2)\,dt\\
&-\frac{v_{t_{(l_n+s_n)rk_nm_n}}}{K_H}
       \int_{t_{(l_n+s_n)rk_nm_n}}^{t_{ik_nm_n}}d_i(u)\,dB_u^r,
\end{align*}
where
\begin{align*}
&J_{1,i}^r=K_H^{-1}\int_0^{t_{(l_n+s_n)rk_nm_n}}d_i(u)v_u\,dB_u,\\
&\begin{aligned}[t]
J_{2,i}^r=&K_H^{-1}\int_{t_{(l_n+s_n)rk_nm_n}}^{t_{ik_nm_n}}d_i(u)
       (v_u\rho_u-v_{t_{(l_n+s_n)rk_nm_n}}\rho_{t_{(l_n+s_n)rk_nm_n}})\,dW_u\\
&+K_H^{-1}\int_{t_{(l_n+s_n)rk_nm_n}}^{t_{ik_nm_n}}d_i(u)
       \bigl(v_u\sqrt{1-\rho_u^2}
        -v_{t_{(l_n+s_n)rk_nm_n}}\sqrt{1-\rho_{t_{(l_n+s_n)rk_nm_n}}^2}\bigr)\,dW_u^\perp,
\end{aligned}\\
&\begin{aligned}[t]
J_{3,i}^r=&\frac1{k_nm_n\Delta_n}\int_{t_{(i-1)k_nm_n}}^{t_{ik_nm_n}}
          \int_{t-k_nm_n\Delta_n}^tf_u\,du\,dt\\
&+\frac1{k_nm_n\Delta_n}\int_{t_{(i-1)k_nm_n}}^{t_{ik_nm_n}}
          \int_0^t[K(t-u)-K(t-k_nm_n\Delta_n-u)]g_u\,du\,dt.
\end{aligned}
\end{align*}
If $j=i-1-(l_n+s_n)r\ge s_n$, the kernel estimates in the proof of
Lemmas~\ref{lem:increment-variance}--\ref{lem:increment-covariance} give
\begin{align}
&\|J_{1,i}^r\|_p\le C_p(k_nm_n\Delta_n)^H(1+j)^{H-1},\notag\\
&\|J_{2,i}^r\|_p\le C_p(k_nm_n\Delta_n)^H
 \{\omega_{r,2p}(v\rho)
           +\omega_{r,2p}(v\sqrt{1-\rho^2})\},\notag\\
&\|J_{3,i}^r\|_p\le C_p((k_nm_n\Delta_n)+(k_nm_n\Delta_n)^{H+1/2}).
\label{noisy:rough-error}
\end{align}
Indeed, for $u<t_{(l_n+s_n)rk_nm_n}$,
$|d_i(u)|\le C(k_nm_n\Delta_n)(t_{(i-2)k_nm_n}-u)^{H-3/2}$; square integration proves the first
bound. The full bound $\int d_i(u)^2du\le C(k_nm_n\Delta_n)^{2H}$ proves the second,
and the $L^1$ kernel-increment bound proves the third.
After multiplication by the normalized price, the historical squared
error per index is $C(k_nm_n\Delta_n)^2(1+j)^{2H-2}$.
Since $2H-2<-1$, its sum over all retained indices is at most
$C(k_nm_n\Delta_n)/(l_n+s_n)$.

For the centered target, subtract
\begin{equation*}
\frac{a_n^{\mathrm N}k_nm_n\Delta_n}{K_H(H+1/2)}
 L_{t_{(l_n+s_n)rk_nm_n}}.
\end{equation*}
This quantity is measurable at the start of the auxiliary block and
therefore disappears under conditional centering.
Its squared local error is bounded by
$C(k_nm_n\Delta_n)^2\omega_{r,16}(L)^2$.
Expand each remaining product by replacing one factor at a time.
The unreplaced factors have norms
$C\sqrt{k_nm_n\Delta_n}$, $C((k_nm_n\Delta_n)^H+k_n^{-1/2})$ and $C(k_nm_n\Delta_n)^H$.
Thus price and sampling coefficient errors cost
$C(k_nm_n\Delta_n)n^{-\beta/2}\omega_{r,16}(\sigma)^2$ per index;
the two volatility-loading errors cost $C(k_nm_n\Delta_n)^2$ times their oscillations
squared; and the time-drift error costs $C(k_nm_n\Delta_n)^3$.
Together with \eqref{noisy:pa-error} and the historical estimate, this gives
\begin{equation}
\label{noisy:retained-error}
\sum_r\sum_{i\in I^l(r)}
  \|e_{i,n}\|_2^2
\le Cn^{-\beta/2}\{\omega_n+k_n^{-1}+k_n^{-2H}+(k_nm_n\Delta_n)+(l_n+s_n)^{-1}\}
=o(n^{-\beta/2}).
\end{equation}
Every difference is centered at $t_{(i-2)k_nm_n}$ and measurable by $t_{(i+1)k_nm_n}$.
The discarded indices cost \eqref{noisy:guard-cost}; the at-most-one exact
endpoint costs $O((k_nm_n\Delta_n)n^{-\beta/2}+(k_nm_n\Delta_n)^2)=o(n^{-\beta/2})$.
This proves the last assertion of \eqref{noisy:transfer-local}.
The other assertions follow from the exact decomposition and
Lemmas~\ref{noisy:price-drift} and~\ref{noisy:complete-mean}.
\end{proof}

Set $F_{r,n}=\sum_{i\in I^l(r)}F_{i,n}$.
The preceding lemma, Lemma~\ref{noisy:color}, and the omitted target
intervals imply
\begin{equation}
\label{noisy:global-transfer}
a_n^{\mathrm N}(\widehat\theta_n^{\mathrm N}-\theta_T)
                  =\sum_r F_{r,n}+o_p(n^{-\beta/4}).
\end{equation}

\begin{proof}[Proof of Theorem~\ref{noisy:clt} in the Gaussian case]
We apply Lemma~\ref{lem:stable-clt} to the array $F_{r,n}/a_n^{\mathrm N}$ on the grid
$t_{(l_n+s_n)rk_nm_n}$, with Brownian vector $(W,W^\perp,W^{\epsilon})$.

\textit{(a) Conditional means.}
The coarse terms in \eqref{noisy:frozen} are martingale differences, so
$\E[F_{r,n}\mid\mathcal G_{t_{(l_n+s_n)rk_nm_n}}]=0$ exactly.
All conditional means of the original array are already controlled by
\eqref{noisy:mean-groups}.

\textit{(b) Conditional variance.}
Condition on $\mathcal G_{t_{(l_n+s_n)rk_nm_n}}$ throughout this
calculation. The exact frozen return covariance is
\begin{equation*}
\begin{aligned}
\E[Z_j^r Z_l^r\mid\mathcal G_{t_{(l_n+s_n)rk_nm_n}}]=&m_n\Delta_n\sigma_{t_{(l_n+s_n)rk_nm_n}}^2\\
&\times\{\mathbf1_{\{j=l\}}+\rho_{m_n}\mathbf1_{\{|j-l|=1\}}\}\\
&+\frac{c_{m_n}^2\nu}{m_n}
 \{2\mathbf1_{\{j=l\}}-\mathbf1_{\{|j-l|=1\}}\}.
\end{aligned}
\end{equation*}
The two trimmed windows have disjoint supports. Each contains $k_n-1$
diagonal pairs and $k_n-2$ adjacent pairs. The Gaussian pairing identity
of Lemma~\ref{lem:gaussian-moments} therefore gives
\begin{equation*}
\begin{aligned}
\E[(E_i^r-E_{i-1}^r)^2
                  \mid\mathcal G_{t_{(l_n+s_n)rk_nm_n}}]=&\frac4{k_n}\left\{\begin{aligned}[t]
&\left(\sigma_{t_{(l_n+s_n)rk_nm_n}}^2
               +\frac{2c_{m_n}^2\nu}{m_n^2\Delta_n}\right)^2\\
&+2\left(\rho_{m_n}\sigma_{t_{(l_n+s_n)rk_nm_n}}^2
               -\frac{c_{m_n}^2\nu}{m_n^2\Delta_n}\right)^2
\end{aligned}\right\}\\
&+O(k_n^{-2}).
\end{aligned}
\end{equation*}
In the product with the squared coarse price increment, the pairing
that contracts the two price factors multiplies this expression by
$k_nm_n\Delta_n$. Every other pairing contains two price--return
covariances. Each is bounded by $Cm_n\Delta_n$, and only diagonal
or adjacent return pairs contribute. Thus the sum of the remaining
pairings is bounded by
\begin{equation*}
C(k_nm_n\Delta_n)^{-2}k_n(m_n\Delta_n)^3
=Ck_nm_n\Delta_n k_n^{-2}.
\end{equation*}
After multiplication by the squared normalization in
\eqref{noisy:frozen} and summation, the total error is
$O(n^{-\beta/2}/k_n)$.

The rough part has the same box-average kernels as in Section~D.
The calculation following \eqref{eq:block-transfer}, with the coarse
interval length $k_nm_n\Delta_n$, gives
\begin{align*}
&\frac{(k_nm_n\Delta_n)^{-2H}}{K_H^2}
 \int_{-\infty}^{t_{ik_nm_n}}d_i(u)^2\,du
=\frac{4(2^{2H}-1)}{(2H+1)(2H+2)},\\
&\int_{t_{(i-1)k_nm_n}}^{t_{ik_nm_n}}d_i(u)\,du
=\frac{2(k_nm_n\Delta_n)^{H+1/2}}{(H+1/2)(2H+3)}.
\end{align*}
The second identity proves the exact centering in \eqref{noisy:frozen}.
The first identity and
$\operatorname{Var}(AB)=\E[A^2]\E[B^2]
 +(\E[AB])^2$ for centered Gaussian $A,B$ give the rough
variance and the squared leverage-covariance term. Completing the history
from $t_{(l_n+s_n)rk_nm_n}$ to $-\infty$ costs
$O(k_nm_n\Delta_n/(l_n+s_n))$ after summation, by
\eqref{noisy:rough-error}. The sampling/rough covariance is zero because
their Gaussian degrees have opposite parity.

Now $\rho_{m_n}\to1/4$ and
$c_{m_n}^2\nu/(m_n^2\Delta_n)\to3\nu/(2\lambda^2T)$.
Consequently the sampling polynomial in the bracket converges, uniformly
on the bounded coefficient range, to
\begin{equation*}
\frac98\sigma_t^6+\frac{9\nu}{2\lambda^2T}\sigma_t^4
                  +\frac{27\nu^2}{2\lambda^4T^2}\sigma_t^2.
\end{equation*}
Since $c_{m_n}-a_n^{\mathrm N}\to0$, the same Riemann-sum argument as
in \eqref{eq:group-bracket} yields
\begin{equation}
\label{noisy:bracket}
\begin{aligned}
\sum_r\E[F_{r,n}^2
                   \mid\mathcal G_{t_{(l_n+s_n)rk_nm_n}}]=&(a_n^{\mathrm N})^2m_n\Delta_n(k_nm_n\Delta_n)^{-2H}(2H+3)^2\\
&\times\int_0^T\left\{\begin{aligned}[t]
&\frac98\sigma_t^6+\frac{9\nu}{2\lambda^2T}\sigma_t^4\\
&+\frac{27\nu^2}{2\lambda^4T^2}\sigma_t^2
\end{aligned}\right\}dt\\
&+(a_n^{\mathrm N})^2k_nm_n\Delta_n\int_0^T
 \left\{\begin{aligned}[t]
&\frac{(2H+3)^2(2^{2H}-1)}{(2H+1)(2H+2)}\sigma_t^2v_t^2\\
&+\frac{L_t^2}{K_H^2(H+1/2)^2}
\end{aligned}\right\}dt\\
&+o_p(n^{-\beta/2}).
\end{aligned}
\end{equation}
The same calculation up to each fixed terminal time gives the cumulative
conditional-variance convergence required by Lemma~\ref{lem:stable-clt}.
The limiting integrals are continuous and the partial conditional
variances are nondecreasing.

\textit{(c) Orthogonality and Lindeberg condition.}
The frozen terms are Gaussian polynomials of degree at most three.
The Gaussian moment argument in part (c) of the proof in Section~D gives
\begin{equation}
\label{noisy:frozen-moments}
\begin{aligned}
&\E F_{i,n}^2\le Ck_nm_n\Delta_n n^{-\beta/2},\\
&\E F_{i,n}^4\le C(k_nm_n\Delta_n)^2n^{-\beta},\\
&\E F_{r,n}^4
 \le C\{(l_n+s_n)k_nm_n\Delta_n\}^2n^{-\beta}.
\end{aligned}
\end{equation}
The rough centered quadratic has no linear Gaussian projection.
In the sampling product, its linear contraction is twice the sum of
price--return covariances times their return variables, divided by
$k_nm_n\Delta_n$. The covariance bound used in part (b), together
with bounded overlap of the kernels, therefore gives
\begin{equation*}
\sum_r\E|\operatorname{proj}_1F_{r,n}|^2
\le C\frac{m_n\Delta_n(k_nm_n\Delta_n)^{-2H}}{k_n}
\le Cn^{-\beta/2}/k_n.
\end{equation*}
The Brownian covariations thus vanish after normalization, first for
bounded step integrands and then for square-integrable predictable
integrands by approximation, exactly as in Section~D. The frozen
polynomials have adapted iterated It\^o representations; their conditional
cross expectations with martingales strongly orthogonal to
$(W,W^\perp,W^{\epsilon})$ are zero. These verify the orthogonality conditions.
Finally,
\begin{equation*}
n^{\beta}\sum_r\E|F_{r,n}|^4
\le C(l_n+s_n)k_nm_n\Delta_n\longrightarrow0,
\end{equation*}
which proves Lindeberg.
The bandwidth condition gives
\begin{align*}
&n^{\beta/2}(m_n\Delta_n)(k_nm_n\Delta_n)^{-2H}
\longrightarrow c^{-2H}\lambda^{1-2H+2Hb}T^{1-2H}
                  \mathbf1_{\{b\le2H/(2H+1)\}},\\
&n^{\beta/2}(k_nm_n\Delta_n)
\longrightarrow c\lambda^{1-b}T
                  \mathbf1_{\{b\ge2H/(2H+1)\}},\\
&(a_n^{\mathrm N})^2\longrightarrow3/2.
\end{align*}
Apply Lemma~\ref{lem:stable-clt} with normalization $n^{\beta/4}$.
Dividing the bracket in~\eqref{noisy:bracket} by $(a_n^{\mathrm N})^2$ gives the variance in
\eqref{noisy:limit-variance}.
Equation~\eqref{noisy:global-transfer} transfers the result to the actual
statistic, including the price drift.
The conditional-law identity \eqref{noisy:noise-embedding} transfers the
stable limit, relative to $\F$, to the original Gaussian level
noise. Proposition~\ref{prop:localization}, with the same independent
observation noise coupled to the localized and original paths, removes
the localization. This proves the Gaussian case of the theorem.
\end{proof}

\subsection{Observable variance and studentization}

For $0\le j\le N_n=\lceil n/m_n\rceil$, the cumulative estimator in
\eqref{noisy:Q} is, precisely,
\begin{equation*}
\widehat\theta_{n,j}^{\mathrm N}
=\frac1{a_n^{\mathrm N}}\sum_{\substack{i\in\mathcal I_n^{\mathrm N}\\
                   (ik_n+1)m_n-1\le jm_n}}U_{i,n}.
\end{equation*}
It includes the whole statistic at $j=N_n$. Its increments place the
contribution $U_{i,n}/a_n^{\mathrm N}$ at cell
\begin{align*}
j_{i,n}=
\begin{cases}ik_n,&m_n>1,\\ik_n-1,&m_n=1.\end{cases}
\end{align*}
Consecutive completion cells are $k_n$ cells apart. With the sparse-array
convention of \eqref{eq:q-form},
$(a_n^{\mathrm N})^2Q_n^{\mathrm N}=\mathcal Q_{N_n,B_n}(U)$.

\begin{proof}[Proof of Proposition~\ref{noisy:variance} for Gaussian noise]
Apply Lemma~\ref{lem:q-energy} to the local decomposition
\eqref{noisy:transfer-local}. Its centered errors are measurable at
$t_{(i+1)k_nm_n}$ and centered given $\mathcal G_{t_{(i-2)k_nm_n}}$,
so \eqref{eq:q-color} applies with three residue classes. The mean and
drift terms have the grouped bounds \eqref{noisy:mean-groups} and
\eqref{noisy:drift-groups}, to which \eqref{eq:q-groups} applies. Hence
\begin{equation*}
\mathcal Q_{N_n,B_n}
 \bigl(U-a_n^{\mathrm N}
   (\Delta_{ik_nm_n}^{n,k_nm_n}\theta)_i-F\bigr)
=o_p(n^{-\beta/2}).
\end{equation*}
Lemma~\ref{lem:q-target}, with micro-observation spacing $m_n\Delta_n$
and coarse width $k_nm_n\Delta_n$, gives at each localization level $J$
\begin{equation*}
\begin{aligned}
\E\mathcal Q_{N_n,B_n}
 ((\Delta_{ik_nm_n}^{n,k_nm_n}\theta)_i)&\le C_J\left\{\begin{aligned}[t]
&(B_nm_n\Delta_n)^{1+2H_J}\\
&+\frac{(k_nm_n\Delta_n)^2}{B_nm_n\Delta_n}\\
&+(B_nm_n\Delta_n)^2
\end{aligned}\right\}\\
&=o(n^{-\beta/2}).
\end{aligned}
\end{equation*}
The completion delay relative to $t_{ik_nm_n}$ is at most one
pre-averaging cell, as required by that lemma. Its proof already includes
the killed LR loading and cyclic endpoints. For the middle term use
$k_n/B_n\to0$ and $k_nm_n\Delta_n=O(n^{-\beta/2})$; the other two
terms are controlled by \eqref{noisy:Q-bandwidth} and $H_J<1/2$.
A sufficient choice of the window length is
$B_n=k_n\lceil\log(1+T/(k_nm_n\Delta_n))\rceil$.

By \eqref{noisy:frozen-moments}, the array $n^{\beta/4}F_{i,n}$ satisfies
\eqref{eq:q-md-moments}, since $k_n/N_n\asymp k_nm_n\Delta_n$.
The auxiliary blocks satisfy the group condition of
Lemma~\ref{lem:q-martingale}. Its conclusion and the quadratic triangle
inequality therefore give
\begin{equation*}
(a_n^{\mathrm N})^2Q_n^{\mathrm N}
=\sum_r\E[F_{r,n}^2\mid\mathcal G_{t_{(l_n+s_n)rk_nm_n}}]
 +o_p(n^{-\beta/2}),
\end{equation*}
where the bracket is \eqref{noisy:bracket}.
Multiplication by $n^{\beta/2}/(a_n^{\mathrm N})^2$ proves the stated limit at the fixed
localization level. Proposition~\ref{prop:localization} removes
localization. The last admissible contribution, if omitted in the
Gaussian construction, has squared norm
$O(k_nm_n\Delta_n n^{-\beta/2})$ by \eqref{noisy:local-moments};
its single-cell quadratic norm is its square. Restoring it therefore
preserves both the point and variance limits.
\end{proof}

\subsection{Non-Gaussian observation noise}

We transfer the Gaussian limits to Assumption~\ref{noisy:noise}
by coupling the pre-averaging batch means.
If $\nu=0$, the centered observation noise is zero almost surely, so
the Gaussian proof already applies. Assume henceforth that $\nu>0$.

Let $S_{m_n}=m_n^{-1/2}\sum_{a=1}^{m_n}\epsilon_a$, with distribution function
$F_{m_n}$. Write $F_{m_n}^{-1}(u)=\inf\{x:F_{m_n}(x)\ge u\}$ and let
$\Phi$ be the standard normal distribution function. Define
\[
\eta_{m_n}=\left(\int_0^1
 \bigl|F_{m_n}^{-1}(u)-\sqrt\nu\,\Phi^{-1}(u)\bigr|^8\,du\right)^{1/8}.
\]
By \citet[Lemma~2.3 and Theorem~3.2, with $r=8$]{JohnsonSamworth2005},
$\eta_{m_n}\to0$. Minkowski's inequality also gives
$\sup_n\|S_{m_n}\|_8<\infty$.

On an extension of $(\Omega,\F,\P)$, take a Brownian motion $W^{\epsilon}$
independent of $\F$ and, for each $n$, set
\begin{align*}
&Z_{j,n}=\sqrt{\nu/(m_n\Delta_n)}
          (W^{\epsilon}_{t_{(j+1)m_n}}-W^{\epsilon}_{t_{jm_n}}),\\
&S_{j,n}=F_{m_n}^{-1}\{\Phi(Z_{j,n}/\sqrt\nu)\},\\
&e_{j,n}=S_{j,n}-Z_{j,n}.
\end{align*}
The pairs $(S_{j,n},Z_{j,n})$ are independent within each row and
independent of $\F$. Each pair is
$\mathcal G_{t_{(j+1)m_n}}$-measurable.
Moreover,
\begin{align*}
&\E e_{j,n}=0,\\
&\|e_{j,n}\|_8=\eta_{m_n}\to0,\\
&\sup_{n,j}(\|S_{j,n}\|_8+\|Z_{j,n}\|_8)<\infty.
\end{align*}
There is no requirement that different rows be independent.
Define the coupled returns by
\begin{align}
&N_j=\frac{c_{m_n}}{\sqrt{m_n}}(S_{j+1,n}-S_{j,n}),\notag\\
&N_j^G=\frac{c_{m_n}}{\sqrt{m_n}}(Z_{j+1,n}-Z_{j,n}),\notag\\
&R_j^n=R_j^n(X)+N_j,\notag\\
&R_j^{n,G}=R_j^n(X)+N_j^G.
\label{noisy:coupled-returns}
\end{align}
The superscript $G$ on an estimator means that $R^{n,G}$ is inserted
into its formula, with the same indices and completion cells.
The observation errors enter \eqref{noisy:pa} only through the
two adjacent batch means. Thus the arrays $(R_j^n)_j$ and
$(R_j^{n,G})_j$ in \eqref{noisy:coupled-returns} have, conditionally on
$\F$, the joint laws of the observed pre-averaged returns and
their Gaussian reference counterparts, respectively.
This equality of conditional laws also preserves joint expectations
with every bounded $\F$-measurable variable.

\begin{lemma}\label{noisy:coupled-local}\label{noisy:coupled-mean}
At each fixed localization level, the coupling above satisfies
\begin{align}
&\|P_i-P_i^G\|_4\le C\eta_{m_n}\sqrt{m_n\Delta_n},\notag\\
&\|D_i-D_i^G\|_4\le C\eta_{m_n}k_n^{-1/2},\notag\\
&\|P_i\|_4+\|P_i^G\|_4\le C\sqrt{k_nm_n\Delta_n},\notag\\
&\|D_i\|_4+\|D_i^G\|_4\le C((k_nm_n\Delta_n)^H+k_n^{-1/2}),\notag\\
&\|U_{i,n}-U_{i,n}^G\|_2\le C\eta_{m_n}\sqrt{k_nm_n\Delta_n}\,n^{-\beta/4}.
\label{noisy:coupled-local-bounds}
\end{align}
These bounds hold for the complete price process, including its drift.

Let $\mu_3=\E\epsilon_0^3$. In the same coupling,
\begin{equation}
\label{noisy:coupled-skew}
\begin{aligned}
&\E[U_{i,n}-U_{i,n}^G\mid\mathcal G_{t_{(i-2)k_nm_n}}]=b_n^\epsilon,\\
&b_n^\epsilon=\frac{(2H+3)c_{m_n}^3\mu_3}{2m_n^2}(k_nm_n\Delta_n)^{-H-1/2}.
\end{aligned}
\end{equation}
The total conditional mean satisfies
\begin{equation}
\label{noisy:coupled-skew-sum}
|\mathcal I_n^{\mathrm N}|\,|b_n^\epsilon|
\le Ck_n^{-3/2}n^{-\beta/4}=o(n^{-\beta/4}).
\end{equation}

Under Assumption~\ref{noisy:noise}, the coupling gives
\begin{align}
&\widehat\theta_n^{\mathrm N}-\widehat\theta_n^G=o_p(n^{-\beta/4}),
\label{noisy:coupled-point}\\
&\mathcal Q_{N_n,B_n}(U-U^G)=o_p(n^{-\beta/2}),\notag\\
&Q_n^{\mathrm N}-Q_n^G=o_p(n^{-\beta/2}).
\label{noisy:coupled-quadratic}
\end{align}
There is no additional rate condition on $B_n\eta_{m_n}$.
\end{lemma}
\begin{proof}
The latent pre-averaged return satisfies
$\|R_j^n(X)\|_p\le C_p\sqrt{m_n\Delta_n}$ for each fixed $p$, by bounded
coefficients and BDG after localization. Also,
\begin{align*}
&m_n(m_n\Delta_n)=m_n^2\Delta_n\longrightarrow\lambda^2T,\\
&c_{m_n}/\sqrt{m_n}\asymp\sqrt{m_n\Delta_n}.
\end{align*}
The return difference telescopes over a coarse block:
\begin{equation*}
P_i-P_i^G
=\frac{c_{m_n}}{\sqrt{m_n}}
       (e_{ik_n,n}-e_{(i-1)k_n,n}).
\end{equation*}
This proves its bound in~\eqref{noisy:coupled-local-bounds}.

For a consecutive set $J$ of $k_n$ return indices, the difference of
the corresponding spot-variance estimates is
\begin{align*}
\frac1{k_nm_n\Delta_n}\sum_{j\in J}\{(R_j^n)^2-(R_j^{n,G})^2\}=&\frac2{k_nm_n\Delta_n}\sum_{j\in J}R_j^n(X)(N_j-N_j^G)\\
&+\frac1{k_nm_n\Delta_n}\sum_{j\in J}\{N_j^2-(N_j^G)^2\}.
\end{align*}
Conditionally on $\F$, collecting the independent
centered $e_{a,n}$ rewrites the first sum, before division by $(k_nm_n\Delta_n)$, as
\begin{equation*}
\frac{c_{m_n}}{\sqrt{m_n}}\sum_a
\{\mathbf1_{\{a-1\in J\}}R_{a-1}^n(X)
 -\mathbf1_{\{a\in J\}}R_a^n(X)\}e_{a,n}.
\end{equation*}
Expansion of the fourth moment of a sum of independent centered
variables bounds its conditional fourth moment by
$C\eta_{m_n}^4m_n^{-2}$ times the square of the sum of squared coefficients
in braces. That sum is at most $4\sum_{j\in J}R_j^n(X)^2$, and
\begin{equation*}
\left\|\sum_{j\in J}R_j^n(X)^2\right\|_2
\le\sum_{j\in J}\|R_j^n(X)\|_4^2\le Ck_n(m_n\Delta_n).
\end{equation*}
Thus the normalized mixed sum has $L^4$ norm at most
\[
\frac{C\eta_{m_n}\sqrt{k_n(m_n\Delta_n)}}
 {\sqrt{m_n}(k_nm_n\Delta_n)}
 \le C\eta_{m_n}k_n^{-1/2}.
\]
Put $B_j=N_j^2-(N_j^G)^2$ for this calculation. The two noises have
the same variance, so $\E B_j=0$, and H\"older gives
\begin{equation*}
\|B_j\|_4
\le\|N_j-N_j^G\|_8\|N_j+N_j^G\|_8
\le C\eta_{m_n}(m_n\Delta_n).
\end{equation*}
Each $B_j$ uses only the batch pairs numbered $j,j+1$.
The even and odd subcollections are separately independent. Expanding
the fourth moment within each subcollection therefore gives
\begin{equation*}
\left\|\frac1{k_nm_n\Delta_n}\sum_{j\in J}B_j\right\|_4
\le C\eta_{m_n}\frac{\sqrt{k_n}(m_n\Delta_n)}{(k_nm_n\Delta_n)}
=C\eta_{m_n} k_n^{-1/2}.
\end{equation*}
Apply these estimates to the two adjacent spot-variance estimates
to obtain the bound for $D_i-D_i^G$.
The uniform $P_i,D_i$ bounds then follow from the Gaussian bounds in
Lemma~\ref{noisy:local-lemma}, the triangle inequality, and the
boundedness of $\eta_{m_n}$. The price drift $A$ has $R_j^n(A)=O((m_n\Delta_n))$ and $P_i(A)=O((k_nm_n\Delta_n))$; its contribution to the spot-variance estimate
 has $L^4$ norm $O(\sqrt{m_n\Delta_n})\le O((k_nm_n\Delta_n)^H)$, as in
Lemma~\ref{noisy:price-drift}. Thus the same bounds hold for the full price.

Finally, expand $P_iD_i-P_i^GD_i^G$ into
$(P_i-P_i^G)D_i+P_i^G(D_i-D_i^G)$ and apply H\"older:
\begin{align*}
\|U_{i,n}-U_{i,n}^G\|_2
&\le C\eta_{m_n}(k_nm_n\Delta_n)^{1/2-H}
\left\{\begin{aligned}[t]
&\sqrt{m_n\Delta_n}((k_nm_n\Delta_n)^H+k_n^{-1/2})\\
&+\sqrt{k_nm_n\Delta_n}\,k_n^{-1/2}
\end{aligned}\right\}\\
&\le C\eta_{m_n}\sqrt{k_nm_n\Delta_n}\,n^{-\beta/4}.
\end{align*}
Every noise estimate here uses moments of order at most eight.

For the conditional mean, first prove
\[
\E\!\left[U_{i,n}-U_{i,n}^G\,\middle|\,
\F\vee\sigma(W^{\epsilon}_s:0\le s\le t_{(i-2)k_nm_n})\right]
=b_n^\epsilon.
\]
The pairs $(S_{j,n},Z_{j,n})$, $(i-2)k_n\le j\le ik_n$,
are independent of the conditioning $\sigma$-algebra. In the expansion of $P_iD_i$, terms
containing at most two observation noises have the same expectations
in both models, because the first two moments of all batches agree
and the two arrays share $R_j^n(X)$. Only the term containing three
noises can differ.

Translate the current return indices to $0,\ldots,k_n-1$ and put
$\zeta_j=c_{m_n} S_{j,n}/\sqrt{m_n}$. Then
\begin{align*}
&\E\zeta_j^3=\frac{c_{m_n}^3\mu_3}{m_n^2},\\
&P_i^{\epsilon}=\zeta_{k_n}-\zeta_0,
\end{align*}
and the noise-only spot-variance contrast is
\begin{equation*}
\begin{aligned}
(k_nm_n\Delta_n)D_i^{\epsilon}=&\sum_{j=0}^{k_n-1}(\zeta_{j+1}-\zeta_j)^2\\
&-\sum_{j=-k_n}^{-1}(\zeta_{j+1}-\zeta_j)^2.
\end{aligned}
\end{equation*}
Independence and centering imply
\begin{equation*}
\E[\zeta_a(\zeta_{j+1}-\zeta_j)^2]
=\E\zeta_0^3\{\mathbf1_{\{a=j\}}+
                         \mathbf1_{\{a=j+1\}}\}.
\end{equation*}
The two endpoints of the current sum cancel after multiplication by
$\zeta_{k_n}-\zeta_0$. The preceding sum gives one remaining positive
copy of $\E\zeta_0^3$. Consequently,
\begin{equation*}
\E[P_i^{\epsilon}D_i^{\epsilon}]
=\frac{c_{m_n}^3\mu_3}{m_n^2(k_nm_n\Delta_n)}.
\end{equation*}
Its Gaussian counterpart is zero. Multiplication by
$(2H+3)(k_nm_n\Delta_n)^{1/2-H}/2$ and conditional expectation with respect to
$\mathcal G_{t_{(i-2)k_nm_n}}$ give \eqref{noisy:coupled-skew}.

There are at most $C/(k_nm_n\Delta_n)$ coarse indices. Using
$\sqrt{m_n\Delta_n}(k_nm_n\Delta_n)^{-H}\le Cn^{-\beta/4}$ and
$m_n^2\Delta_n$ bounded away from zero,
\begin{equation*}
\begin{aligned}
\frac{|\mathcal I_n^{\mathrm N}|\,|b_n^\epsilon|}{n^{-\beta/4}}
\le&\frac{C}{m_n^2\sqrt{m_n\Delta_n}(k_nm_n\Delta_n)^{3/2}}\\
=&\frac{C}{m_n^2(m_n\Delta_n)^2k_n^{3/2}}\\
\le&Ck_n^{-3/2}.
\end{aligned}
\end{equation*}
This proves~\eqref{noisy:coupled-skew-sum}; symmetry of the noise law
has not been assumed.

Finally, we prove the point and quadratic transfers. At a fixed localization level let
$Z_i=U_{i,n}-U_{i,n}^G-b_n^\epsilon$.
By the local bounds and conditional mean just proved,
$Z_i$ is centered at $\mathcal G_{t_{(i-2)k_nm_n}}$, measurable by
$t_{(i+1)k_nm_n}$, and
\begin{equation*}
\sum_i\|Z_i\|_2^2\le C\eta_{m_n}^2 n^{-\beta/2}.
\end{equation*}
The three-residue martingale argument of Lemma~\ref{noisy:color},
or Doob's inequality on each residue class, gives
\begin{equation*}
\left\|\max_{M}\left|\sum_{i\le M}Z_i\right|\right\|_2^2
\le C\eta_{m_n}^2 n^{-\beta/2}=o(n^{-\beta/2}).
\end{equation*}
Adding~\eqref{noisy:coupled-skew-sum} and dividing by $a_n^{\mathrm N}$ proves
\eqref{noisy:coupled-point}.

For the quadratic form,~\eqref{eq:q-color} gives directly
\begin{equation*}
\E\mathcal Q_{N_n,B_n}(Z)\le3\sum_i\|Z_i\|_2^2
\le C\eta_{m_n}^2n^{-\beta/2}=o(n^{-\beta/2}).
\end{equation*}
Each completion cell is counted exactly $2B_n$ times and the
defining normalization is $1/(2B_n)$.
For the deterministic mean array,
\begin{equation*}
\mathcal Q_{N_n,B_n}((b_n^\epsilon)_i)
\le\bigl(|\mathcal I_n^{\mathrm N}|\,|b_n^\epsilon|\bigr)^2
\le Ck_n^{-3}n^{-\beta/2}=o(n^{-\beta/2}).
\end{equation*}
The triangle inequality proves the first assertion
in~\eqref{noisy:coupled-quadratic}. The Gaussian reference result gives
$Q_n^G=O_p(n^{-\beta/2})$. Hence
\begin{align*}
|Q_n^{\mathrm N}-Q_n^G|
\le&\frac{2}{a_n^{\mathrm N}}\sqrt{Q_n^G\mathcal Q_{N_n,B_n}(U-U^G)}\\
&+\frac{\mathcal Q_{N_n,B_n}(U-U^G)}{(a_n^{\mathrm N})^2}
=o_p(n^{-\beta/2}).
\end{align*}
The exact index set can be used throughout this argument: the
coupled batch variables may be realized beyond $T$ on the independent
extension, while the latent returns use only the observed times.
Alternatively, the at-most-one final contribution has the local moment
bound proved above and is negligible by the
single-cell bound. Proposition~\ref{prop:localization} removes the
fixed localization level as before.
\end{proof}

\begin{proof}[Proof of Theorem~\ref{noisy:clt} and Proposition~\ref{noisy:variance}]
The Gaussian point and variance limits have been established. Consequently,
\eqref{noisy:coupled-point} is negligible on the CLT scale, and
\eqref{noisy:coupled-quadratic} is negligible on the variance scale.
The coupling preserves the joint laws with every bounded
$\F$-measurable random variable. Stable Slutsky therefore
transfers the Gaussian point limit to the actual observation law,
and the same coupling transfers the variance limit. This proves
Theorem~\ref{noisy:clt} and variance consistency in
Proposition~\ref{noisy:variance} under the stated finite eighth
moment assumption. The limiting variance depends on that law only
through $\nu$.

If $\mathcal V_T^{\mathrm N}>0$ almost surely, the argument used for
Proposition~\ref{prop:variance} gives
$\P(Q_n^{\mathrm N}=0)\to0$ and the studentized stable limit.
The ratio may be set to zero when its denominator vanishes.
The normal quantiles give the stated confidence interval.

\end{proof}

\subsection{Estimated roughness}

We first control the dependence of the exact finite-window correction on
its roughness parameter. The bounds are uniform in both integer window lengths.

\begin{lemma}\label{lem:plugin-correction}
Let $I\Subset(0,1/2)$ be a compact interval, and write
$a^{\mathrm N}(h;k,m)$ for the correction in
\eqref{noisy:correction-factor} with $(H,k_n,m_n)=(h,k,m)$.
There are constants $C_I<\infty$ and $k_I\ge2$ such that, for
$h\in I$, $k\ge2$, $m\ge1$ and $r=0,1,2$,
\begin{equation*}
\left|\partial_h^r\{a^{\mathrm N}(h;k,m)-c_m\}\right|
\le C_I k^{-h-1/2}(1+\log k)^r.
\end{equation*}
In particular, $a^{\mathrm N}(h;k,m)\ge1/2$ whenever
$h\in I$, $k\ge k_I$ and $m\ge1$.
\end{lemma}

\begin{proof}
Set $\alpha=h+1/2$ and
$J=I+1/2=[\alpha_0,\alpha_1]\Subset(0,1)$.
The kernel identities in \eqref{noisy:kernel-identities} give
\begin{align*}
&0\le p_k\le c_m\le\sqrt{3/2},\\
&\operatorname{supp}g_m\subset[0,2],\\
&\int_{\mathbb R}g_m(x)^2\,dx=1,\\
&\sum_{j\in\mathbb Z}g_m(x-j)=c_m\quad\text{for almost every }x.
\end{align*}
In particular, $p_k=c_m$ on $(1,k)$ and
\begin{equation*}
\operatorname{supp}\{p_k-c_m\mathbf1_{(0,k)}\}
\subset[0,1]\cup[k,k+1].
\end{equation*}
Define
\begin{align*}
&q_+(t)=\sum_{j=0}^{k-1}g_m(t-j)^2,\\
&q_-(t)=\sum_{j=-k}^{-1}g_m(t-j)^2,\\
&A_\pm(\alpha)=\iint_{s<t}
      p_k(s)q_\pm(t)(t-s)^{\alpha-1}\,dt\,ds.
\end{align*}
\begin{align*}
D(\alpha)=&\sum_{j=-k}^{k-1}(2\mathbf1_{\{j\ge0\}}-1)\\
&\times\iint_{s<t}p_k(t)g_m(s-j)g_m(t-j)(t-s)^{\alpha-1}\,dt\,ds.
\end{align*}
The correction is exactly
\begin{equation*}
a^{\mathrm N}=\alpha(\alpha+1)k^{-\alpha-1}(A_+-A_-+2D).
\end{equation*}
Also, $0\le q_\pm\le c_m^2$, since the sum of the nonnegative translates
of $g_m$ equals $c_m$. For $r=0,1,2$,
\begin{equation*}
\sup_{\alpha\in J}\int_0^2
 x^{\alpha-1}|\log x|^r\,dx<\infty.
\end{equation*}
Thus, for each fixed $k,m$, all the integrals below can be differentiated
twice under the integral sign. The resulting integrands contain
$(t-s)^{\alpha-1}\log^r(t-s)$; the same local integral bounds justify
all differentiations uniformly for $\alpha\in J$.

Replace $p_k$ by $c_m\mathbf1_{(0,k)}$ in $A_+$, and denote the error by
\begin{equation*}
E_p(\alpha)=\iint_{s<t}
 \{p_k(s)-c_m\mathbf1_{(0,k)}(s)\}q_+(t)
 (t-s)^{\alpha-1}\,dt\,ds.
\end{equation*}
The left endpoint strip has $0<s<1$ and $s<t<k+1$; the right
strip has $k<s<t<k+1$. Direct differentiation therefore gives
\begin{align*}
|\partial_\alpha^r E_p(\alpha)|
\le& C_J\left\{1+\int_0^{k+1}
             x^{\alpha-1}|\log x|^r\,dx\right\}\\
\le& C_J\{1+k^\alpha(1+\log^r k)\}
\le C_J k.
\end{align*}
The last bound uses $\alpha\le\alpha_1<1$.

For $t\ge0$, put
\begin{equation*}
F_\alpha(t)=\int_0^{k\wedge t}(t-s)^{\alpha-1}\,ds
 =\frac{t^\alpha-(t-k)_+^\alpha}{\alpha}.
\end{equation*}
We will use, for $r=0,1,2$, the bound
\begin{equation*}
\sup_{\substack{\alpha\in J,\ x,y\ge0\\|x-y|\le2}}
 \left|\partial_\alpha^r(x^\alpha-y^\alpha)\right|
 \le C_J.
\end{equation*}
Indeed, if $\max(x,y)\le4$, it follows from the boundedness of
$x^\alpha|\log x|^r$ on this set, with value zero at $x=0$.
Otherwise $\min(x,y)>2$, and the mean value theorem applies to
\begin{equation*}
\partial_x\partial_\alpha^r x^\alpha
 =x^{\alpha-1}
   \{\alpha(\log x)^r+r(\log x)^{r-1}\},
\end{equation*}
where the second term is omitted for $r=0$.
This expression is uniformly bounded on $x\ge2$, $\alpha\in J$,
since $\alpha_1<1$. Taking positive parts does not increase distances.
Consequently,
\begin{equation*}
\sup_{\substack{\alpha\in J,\ u,v\ge0\\|u-v|\le2}}
 \left|\partial_\alpha^r
       \{F_\alpha(u)-F_\alpha(v)\}\right|\le C_J.
\end{equation*}

Each measure $g_m(t-j)^2\,dt$ has mass one and is supported on
$[j,j+2]$. Couple it with Lebesgue measure on $[j,j+1]$ using their
product measure. This gives the exact identity
\begin{align*}
E_q(\alpha)
:=&c_m\int_0^\infty q_+(t)F_\alpha(t)\,dt
       -c_m\int_0^k F_\alpha(t)\,dt\\
=&c_m\sum_{j=0}^{k-1}\int_0^2\int_0^1g_m(u)^2
   \{F_\alpha(j+u)-F_\alpha(j+v)\}\,dv\,du.
\end{align*}
Differentiating this expression and using the preceding bound yields
\begin{equation*}
|\partial_\alpha^rE_q(\alpha)|\le C_J k,
\qquad r=0,1,2.
\end{equation*}
Moreover,
\begin{align*}
&c_m\int_0^kF_\alpha(t)\,dt
    =\frac{c_m\,k^{\alpha+1}}{\alpha(\alpha+1)},\\
&A_+(\alpha)
    =\frac{c_m\,k^{\alpha+1}}{\alpha(\alpha+1)}+E_p(\alpha)+E_q(\alpha).
\end{align*}

For $A_-$, the conditions $p_k(s)>0$, $q_-(t)>0$ and $s<t$ imply
$0<s<t<1$. Hence the local logarithmic integral bound gives
$|\partial_\alpha^rA_-|\le C_J$.
Each summand in $D$ is supported on $j<s<t<j+2$.
Its $r$th derivative is bounded in absolute value by
\begin{equation*}
C_J\int_0^2\int_u^2
       (v-u)^{\alpha-1}|\log(v-u)|^r\,dv\,du\le C_J.
\end{equation*}
There are $2k$ summands, so $|\partial_\alpha^rD|\le C_Jk$.
Therefore, with
\begin{equation*}
E(\alpha)=E_p(\alpha)+E_q(\alpha)-A_-(\alpha)+2D(\alpha),
\end{equation*}
we have proved directly that
\begin{align*}
&a^{\mathrm N}-c_m=k^{-\alpha-1}\alpha(\alpha+1)E(\alpha),\\
&|\partial_\alpha^rE(\alpha)|\le C_Jk,
       \qquad r=0,1,2.
\end{align*}
The product rule now gives
\begin{equation*}
\partial_\alpha^r(a^{\mathrm N}-c_m)
 =k^{-\alpha-1}\sum_{\ell=0}^r\binom r\ell
       (-\log k)^{r-\ell}\partial_\alpha^\ell\{\alpha(\alpha+1)E(\alpha)\}.
\end{equation*}
This proves the asserted derivative estimates because
$\partial_h=\partial_\alpha$. Finally, $c_m\ge1$, so the estimate for
$r=0$ gives $a^{\mathrm N}\ge1-C_Ik^{-\alpha_0}\ge1/2$ for
$k\ge k_I$, uniformly in $h,m$.
\end{proof}

\begin{proof}[Proof of Remark~\ref{rem:noisy-plugin}]
Write $b_\star=b(H+\delta)$ and $\beta_\delta=1-b_\star$.
Since $\delta\ge0$, $b_\star\ge2H/(2H+1)$, so the exponent in
\eqref{noisy:bandwidth} is exactly $\beta_\delta$.
For this proof, $\widehat\theta_n^{\mathrm N}(h;k)$ and
$Q_n^{\mathrm N}(h;k,B)$ denote the Section~4 statistics evaluated at
roughness $h$ and bandwidths $k,B$, on the same observed path.

\textit{(a) Grid matching and deterministic bandwidths.}
Let $U$ be the independent uniform random variable in the remark.
On its product extension, define the oracle grid value
\begin{equation*}
\overline H_n^o(u)=q_n\{\lfloor H/q_n+u\rfloor-u\},
\qquad 0\le u\le1,
\end{equation*}
with the same truncation to $[0,1/2]$ as in the remark.
Conditional on all variables used to construct $\widehat H_n$, a uniformly
shifted grid separates $H$ and $\widehat H_n$ with probability at most
$\min\{1,|\widehat H_n-H|/q_n\}$. Truncation cannot increase the
mismatch probability. Hence
\begin{align*}
\P\{\overline H_n\ne\overline H_n^o(U)\}
\le&\E\min\{1,|\widehat H_n-H|/q_n\}
\longrightarrow0,
\end{align*}
since $n^{-C}/q_n\to0$.
Put
\begin{equation*}
k_n^o(u)=\left\lceil c(n/m_n)^{b(\overline H_n^o(u)+\delta)}\right\rceil,
\qquad B_n^o(u)=B_n(k_n^o(u)).
\end{equation*}
For each fixed $u$ these are deterministic sequences. Moreover,
$\sup_u|\overline H_n^o(u)-H|\le q_n$ and $b$ is Lipschitz on
$[0,1/2+\delta]$. Thus
\begin{align*}
&\sup_u\left|
 \log\frac{(n/m_n)^{b(\overline H_n^o(u)+\delta)}}
               {(n/m_n)^{b_\star}}\right|
 \le Cq_n\log(n/m_n)\longrightarrow0,\\
&\sup_u\left|\frac{k_n^o(u)}{(n/m_n)^{b_\star}}-c\right|
 \longrightarrow0,\\
&k_n^o(u)m_n\Delta_n\asymp n^{-\beta_\delta/2}
 \quad\text{uniformly in }u.
\end{align*}
By the bandwidth rule in the remark, $B_n^o(u)$ satisfies
\eqref{noisy:Q-bandwidth} for each fixed $u$.
For example, the deterministic rule
\begin{equation*}
B_n(k)=\min\{\lceil k\log n\rceil,\lfloor N_n/2\rfloor\},
\qquad N_n=\lceil n/m_n\rceil,
\end{equation*}
is admissible. Its cap is inactive for all large $n$, uniformly over the
oracle sequences, because $k_n^o(u)\log n/N_n\to0$ uniformly.
For every localization exponent $H_N>0$ in LR,
\begin{align*}
&\frac{k_n^o(u)}{B_n^o(u)}=O((\log n)^{-1}),\\
&B_n^o(u)m_n\Delta_n=O(n^{-\beta_\delta/2}\log n),\\
&n^{\beta_\delta/2}\{B_n^o(u)m_n\Delta_n\}^{1+2H_N}
 =O\!\left(n^{-\beta_\delta H_N}(\log n)^{1+2H_N}\right)
 \longrightarrow0.
\end{align*}

\textit{(b) Transfer of the known-$H$ limits.}
For every fixed $u$, Theorem~\ref{noisy:clt} and
Proposition~\ref{noisy:variance} give
\begin{align*}
&A_n(u):=n^{\beta_\delta/4}
 \{\widehat\theta_n^{\mathrm N}(H;k_n^o(u))-\theta_T\}
 \convst\sqrt{\mathcal V_T^{\mathrm N}}\,\mathcal Z,\\
&n^{\beta_\delta/2}
 Q_n^{\mathrm N}(H;k_n^o(u),B_n^o(u))
 \convp\mathcal V_T^{\mathrm N},
\end{align*}
where the variance is evaluated at the fixed exponent $b=b_\star$,
with the original constants $c,\lambda$. It does not depend on $u$.
Let $Y$ be bounded and $\F$-measurable, let $\eta$ be bounded
and measurable on $[0,1]$, and let $f$ be bounded and continuous.
Independence of $U$ and dominated convergence yield
\begin{align*}
\E[Y\eta(U)f(A_n(U))]
=&\int_0^1\eta(u)\E[Yf(A_n(u))]\,du\\
\longrightarrow&\int_0^1\eta(u)
 \E[Yf(\sqrt{\mathcal V_T^{\mathrm N}}\mathcal Z)]\,du.
\end{align*}
A monotone-class argument for products $Y\eta(U)$ proves stable
convergence relative to $\F\vee\sigma(U)$, with $\mathcal Z$
independent of that sigma-field. Integrating the bounded error probabilities
in the variance-consistency statement gives the corresponding convergence
in probability for $Q$ at $k_n^o(U),B_n^o(U)$.
On the grid-matching event from part~(a), these bandwidths equal
$\widehat k_n,B_n(\widehat k_n)$ exactly. Consequently,
\begin{align*}
&n^{\beta_\delta/4}
 \{\widehat\theta_n^{\mathrm N}(H;\widehat k_n)-\theta_T\}
 \convst\sqrt{\mathcal V_T^{\mathrm N}}\,\mathcal Z,\\
&n^{\beta_\delta/2}
 Q_n^{\mathrm N}(H;\widehat k_n,B_n)
 \convp\mathcal V_T^{\mathrm N}.
\end{align*}
This transfer requires no independence between the pilot and the price data.

\textit{(c) Substitution in the normalization.}
For $h$ in a compact neighborhood $I\Subset(0,1/2)$ of $H$, set
\begin{equation*}
g_n(h;k)=\frac{h+3/2}{a^{\mathrm N}(h;k,m_n)}
                  (km_n\Delta_n)^{1/2-h}.
\end{equation*}
Lemma~\ref{lem:plugin-correction} gives, uniformly for $h\in I$,
\begin{align*}
&\partial_h\log g_n(h;k)
 =-\log(km_n\Delta_n)+\frac1{h+3/2}
       -\frac{\partial_h a^{\mathrm N}(h;k,m_n)}
                   {a^{\mathrm N}(h;k,m_n)},\\
&\left|\partial_h\log a^{\mathrm N}(h;k,m_n)\right|
 \le C_Ik^{-h-1/2}(1+\log k),\\
&\partial_h\log g_n(h;k)=-\log(km_n\Delta_n)+O_I(1).
\end{align*}
The pilot belongs to $I$ with probability tending to one. On this event,
apply the mean value theorem at the same realized block length
$k=\widehat k_n$. Part~(a) gives
$-\log(\widehat k_nm_n\Delta_n)=O_p(\log n)$, so
\begin{align*}
&d_n:=\frac{g_n(\widehat H_n;\widehat k_n)}
                 {g_n(H;\widehat k_n)},\\
&\log d_n=O_p(n^{-C}\log n),\qquad
 d_n-1=O_p(n^{-C}\log n),\\
&n^{\beta_\delta/4}(d_n-1)
 =O_p(n^{\beta_\delta/4-C}\log n)=o_p(1).
\end{align*}
The products $P_iD_i$ and their completion cells do not depend on $h$
when $k,m_n,B$ are fixed. Hence the cumulative-estimator definition gives
exactly
\begin{align*}
&\widehat\theta_n^{\mathrm N}(\widehat H_n;\widehat k_n)
 =d_n\widehat\theta_n^{\mathrm N}(H;\widehat k_n),\\
&Q_n^{\mathrm N}(\widehat H_n;\widehat k_n,B_n)
 =d_n^2Q_n^{\mathrm N}(H;\widehat k_n,B_n).
\end{align*}
It follows that
\begin{align*}
n^{\beta_\delta/4}
 \{\widehat\theta_n^{\mathrm N}(\widehat H_n;\widehat k_n)-\theta_T\}
=&d_n n^{\beta_\delta/4}
 \{\widehat\theta_n^{\mathrm N}(H;\widehat k_n)-\theta_T\}\\
&+n^{\beta_\delta/4}(d_n-1)\theta_T.
\end{align*}
Since $\theta_T$ is finite almost surely, part~(b) and stable Slutsky give
\begin{align*}
&n^{\beta_\delta/4}
 \{\widehat\theta_n^{\mathrm N}(\widehat H_n;\widehat k_n)-\theta_T\}
 \convst\sqrt{\mathcal V_T^{\mathrm N}}\,\mathcal Z,\\
&n^{\beta_\delta/2}
 Q_n^{\mathrm N}(\widehat H_n;\widehat k_n,B_n)
 \convp\mathcal V_T^{\mathrm N}.
\end{align*}
Under the stated variance positivity condition, a final application of
stable Slutsky proves
\begin{equation*}
\frac{\widehat\theta_n^{\mathrm N}(\widehat H_n;\widehat k_n)-\theta_T}
 {\sqrt{Q_n^{\mathrm N}(\widehat H_n;\widehat k_n,B_n)}}
 \convst N(0,1).
\end{equation*}
Finally, $\beta_\delta=(1+2H+2\delta)^{-1}$, so
$C>\beta_\delta/4$ is equivalent to the inequality for $\delta$ in the
remark. If $C=1/[4(2H+1)]$, every fixed $\delta>0$ makes this inequality
strict.
\end{proof}

%% file: appendices/F_sim.tex
\section{Supplementary simulation results}
\setcounter{table}{0}
\renewcommand{\thetable}{F.\arabic{table}}
\setcounter{figure}{0}
\renewcommand{\thefigure}{F.\arabic{figure}}

We report noise-free and noisy results for the rough cubic (RC) and rough constant (RCO) specifications. The nine values of $H$, the $5000$ paths per cell, the observation grid, and the statistics are the same as in the main text. The noisy design uses the same pre-averaging bandwidth rule and single-scale variance estimator. All tables report empirical means and variances of $Z$ and 95\% coverage.

Under the simulation normalization of the variance equation~\eqref{eq:rough-heston},
the Volterra loadings are
\[
\begin{aligned}
&v_t^{\mathrm{sim}}=\tfrac12\Delta_{\mathrm{ref}}^{1/2-H}\sigma_t^3
\qquad\text{(RC)},\\
&v_t^{\mathrm{sim}}=\frac{K_H}{2\Gamma(H+1/2)}\Delta_{\mathrm{ref}}^{1/2-H}
\qquad\text{(RCO)}.
\end{aligned}
\]
We set $\sigma_0^2=\vartheta=0.02$ for RC and $\sigma_0^2=\vartheta=0.10$ for RCO,
with the remaining drift and correlation parameters as in the main-text design
and $\Delta_{\mathrm{ref}}=1/(252\cdot2340)$ held fixed.
The targets use the actual simulation loading and
$v_t^{\mathrm{th}}=K_Hv_t^{\mathrm{sim}}/\Gamma(H+1/2)$.
Both specifications use variance-protected simulation paths and provide
supplementary numerical diagnostics for the two loading forms.

\subsection{Rough cubic}

\begin{table}[!htbp]
\centering
\normalsize
\linespread{1.6}\selectfont
\setlength{\tabcolsep}{4pt}
\caption{Rough cubic: empirical means, variances, and 95\% coverage without noise and with observation noise.}
\begin{tabular}{c|rrr|rrr}
\toprule
$H$ & \multicolumn{3}{c|}{Noise-free} & \multicolumn{3}{c}{Noisy} \\
\cmidrule(lr){2-4}\cmidrule(lr){5-7}
 & Mean $Z$ & Var. $Z$ & Cov. 95\% & Mean $Z$ & Var. $Z$ & Cov. 95\% \\
\midrule
0.05 & -0.012 & 1.036 & 0.9440 & -0.000 & 0.995 & 0.9520 \\
0.10 & 0.027 & 1.009 & 0.9476 & 0.005 & 1.021 & 0.9494 \\
0.15 & 0.003 & 1.003 & 0.9492 & 0.021 & 1.002 & 0.9520 \\
0.20 & -0.009 & 1.005 & 0.9502 & 0.001 & 0.984 & 0.9510 \\
0.25 & -0.005 & 0.988 & 0.9566 & 0.025 & 1.024 & 0.9502 \\
0.30 & 0.002 & 1.011 & 0.9474 & 0.000 & 1.021 & 0.9510 \\
0.35 & 0.006 & 1.010 & 0.9500 & 0.002 & 1.054 & 0.9486 \\
0.40 & 0.003 & 1.013 & 0.9492 & -0.005 & 1.102 & 0.9420 \\
0.45 & 0.005 & 0.995 & 0.9536 & 0.035 & 1.078 & 0.9454 \\
\bottomrule
\end{tabular}
\end{table}

The RC coverage ranges are $94.40\%$--$95.66\%$ without noise and $94.20\%$--$95.20\%$ with observation noise.

\begin{figure}[!htbp]
\centering
\begin{subfigure}[t]{0.32\textwidth}
\includegraphics[width=\textwidth]{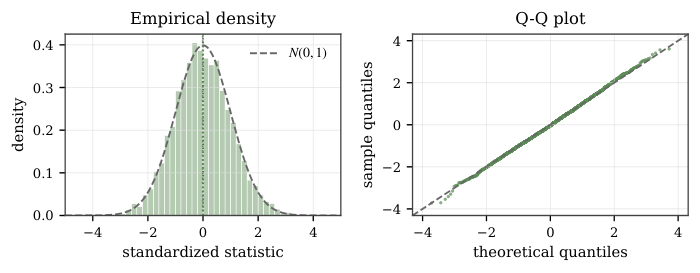}
\caption{$H=0.05$}
\end{subfigure}
\hfill
\begin{subfigure}[t]{0.32\textwidth}
\includegraphics[width=\textwidth]{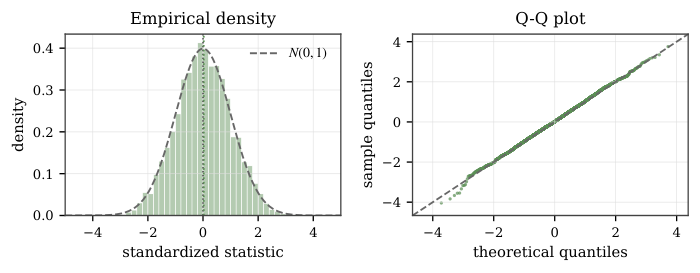}
\caption{$H=0.10$}
\end{subfigure}
\hfill
\begin{subfigure}[t]{0.32\textwidth}
\includegraphics[width=\textwidth]{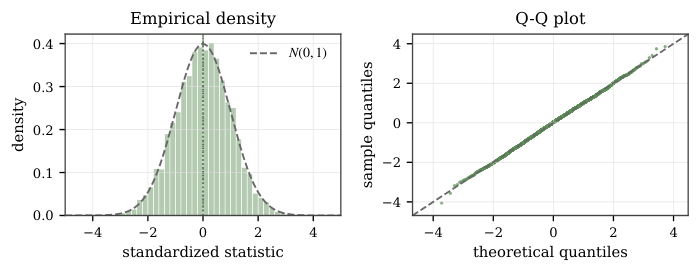}
\caption{$H=0.15$}
\end{subfigure}

\begin{subfigure}[t]{0.32\textwidth}
\includegraphics[width=\textwidth]{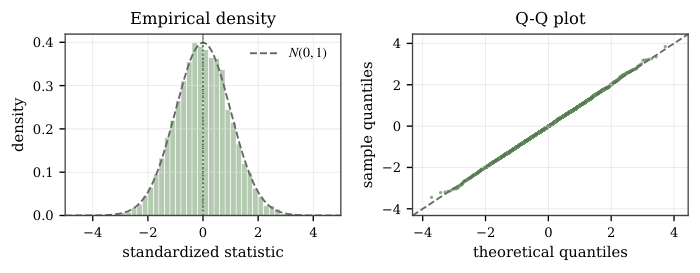}
\caption{$H=0.20$}
\end{subfigure}
\hfill
\begin{subfigure}[t]{0.32\textwidth}
\includegraphics[width=\textwidth]{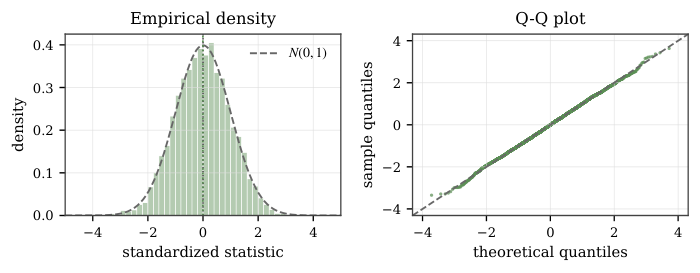}
\caption{$H=0.25$}
\end{subfigure}
\hfill
\begin{subfigure}[t]{0.32\textwidth}
\includegraphics[width=\textwidth]{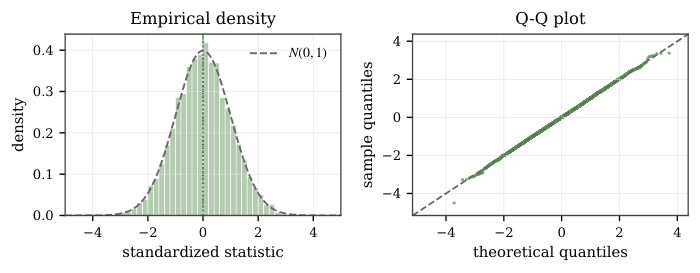}
\caption{$H=0.30$}
\end{subfigure}

\begin{subfigure}[t]{0.32\textwidth}
\includegraphics[width=\textwidth]{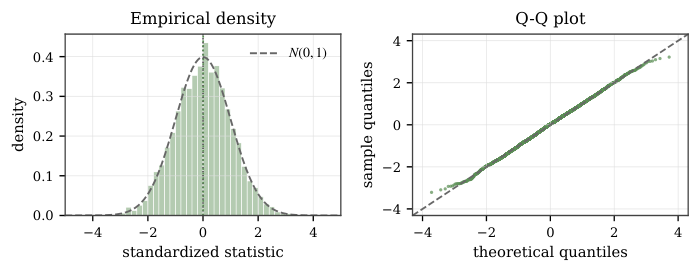}
\caption{$H=0.35$}
\end{subfigure}
\hfill
\begin{subfigure}[t]{0.32\textwidth}
\includegraphics[width=\textwidth]{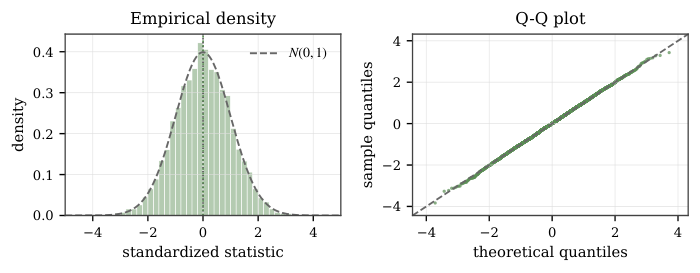}
\caption{$H=0.40$}
\end{subfigure}
\hfill
\begin{subfigure}[t]{0.32\textwidth}
\includegraphics[width=\textwidth]{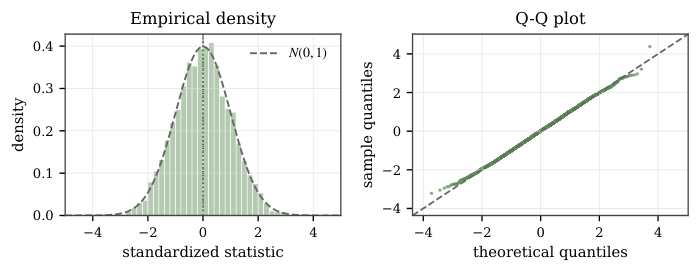}
\caption{$H=0.45$}
\end{subfigure}
\caption{Rough cubic without observation noise: densities of $Z$ and normal QQ plots at the nine values of $H$.}
\end{figure}

\begin{figure}[!htbp]
\centering
\begin{subfigure}[t]{0.32\textwidth}
\includegraphics[width=\textwidth]{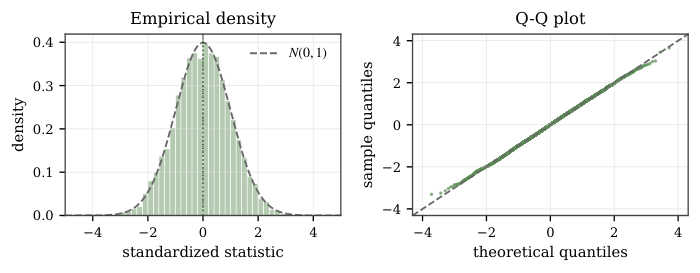}
\caption{$H=0.05$}
\end{subfigure}
\hfill
\begin{subfigure}[t]{0.32\textwidth}
\includegraphics[width=\textwidth]{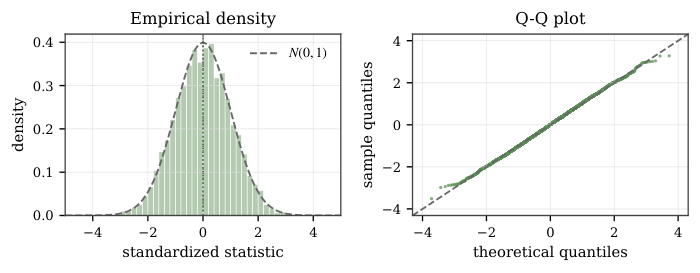}
\caption{$H=0.10$}
\end{subfigure}
\hfill
\begin{subfigure}[t]{0.32\textwidth}
\includegraphics[width=\textwidth]{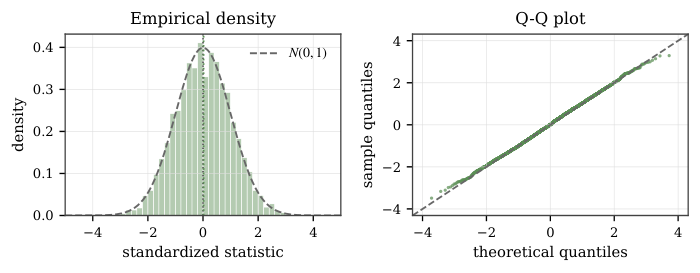}
\caption{$H=0.15$}
\end{subfigure}

\begin{subfigure}[t]{0.32\textwidth}
\includegraphics[width=\textwidth]{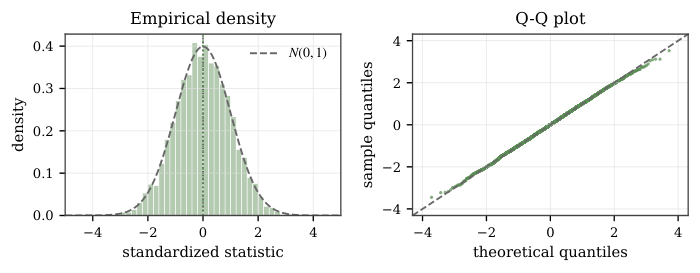}
\caption{$H=0.20$}
\end{subfigure}
\hfill
\begin{subfigure}[t]{0.32\textwidth}
\includegraphics[width=\textwidth]{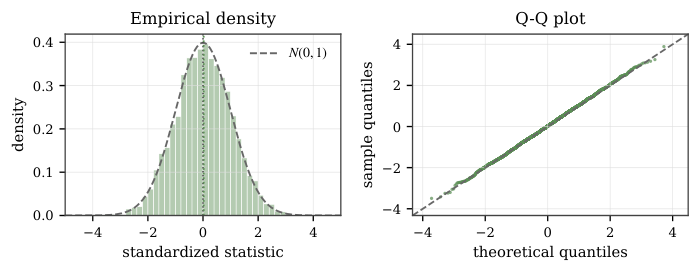}
\caption{$H=0.25$}
\end{subfigure}
\hfill
\begin{subfigure}[t]{0.32\textwidth}
\includegraphics[width=\textwidth]{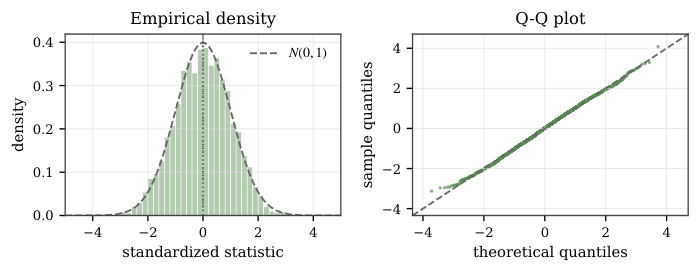}
\caption{$H=0.30$}
\end{subfigure}

\begin{subfigure}[t]{0.32\textwidth}
\includegraphics[width=\textwidth]{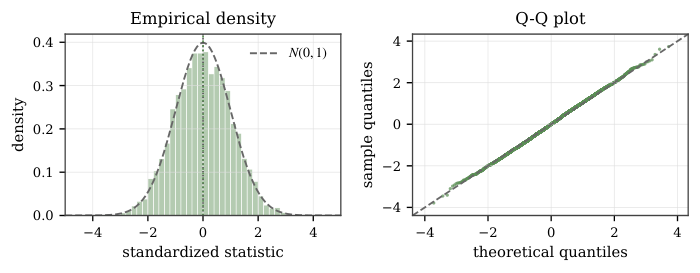}
\caption{$H=0.35$}
\end{subfigure}
\hfill
\begin{subfigure}[t]{0.32\textwidth}
\includegraphics[width=\textwidth]{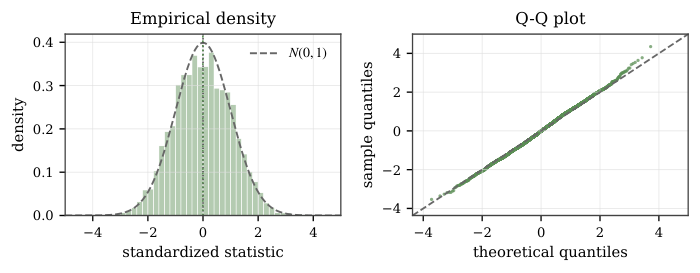}
\caption{$H=0.40$}
\end{subfigure}
\hfill
\begin{subfigure}[t]{0.32\textwidth}
\includegraphics[width=\textwidth]{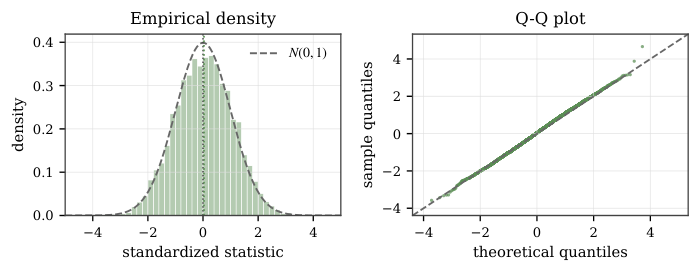}
\caption{$H=0.45$}
\end{subfigure}
\caption{Rough cubic with observation noise: densities of $Z$ and normal QQ plots at the nine values of $H$.}
\end{figure}

\subsection{Rough constant}

\begin{table}[!htbp]
\centering
\normalsize
\linespread{1.6}\selectfont
\setlength{\tabcolsep}{4pt}
\caption{Rough constant: empirical means, variances, and 95\% coverage without noise and with observation noise.}
\begin{tabular}{c|rrr|rrr}
\toprule
$H$ & \multicolumn{3}{c|}{Noise-free} & \multicolumn{3}{c}{Noisy} \\
\cmidrule(lr){2-4}\cmidrule(lr){5-7}
 & Mean $Z$ & Var. $Z$ & Cov. 95\% & Mean $Z$ & Var. $Z$ & Cov. 95\% \\
\midrule
0.05 & 0.069 & 0.976 & 0.9490 & 0.009 & 0.952 & 0.9526 \\
0.10 & 0.040 & 1.008 & 0.9494 & 0.024 & 1.017 & 0.9516 \\
0.15 & 0.001 & 0.950 & 0.9532 & 0.028 & 0.994 & 0.9496 \\
0.20 & 0.006 & 1.031 & 0.9438 & 0.004 & 0.982 & 0.9560 \\
0.25 & 0.020 & 1.004 & 0.9488 & 0.014 & 1.035 & 0.9486 \\
0.30 & 0.018 & 1.021 & 0.9436 & -0.022 & 1.014 & 0.9498 \\
0.35 & 0.027 & 1.006 & 0.9526 & 0.037 & 1.039 & 0.9492 \\
0.40 & -0.010 & 0.997 & 0.9544 & 0.030 & 1.031 & 0.9498 \\
0.45 & 0.021 & 1.020 & 0.9466 & 0.083 & 1.038 & 0.9528 \\
\bottomrule
\end{tabular}
\end{table}

The RCO coverage ranges are $94.36\%$--$95.44\%$ without noise and $94.86\%$--$95.60\%$ with observation noise.

\begin{figure}[!htbp]
\centering
\begin{subfigure}[t]{0.32\textwidth}
\includegraphics[width=\textwidth]{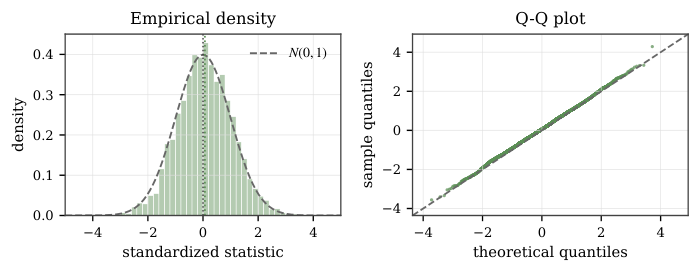}
\caption{$H=0.05$}
\end{subfigure}
\hfill
\begin{subfigure}[t]{0.32\textwidth}
\includegraphics[width=\textwidth]{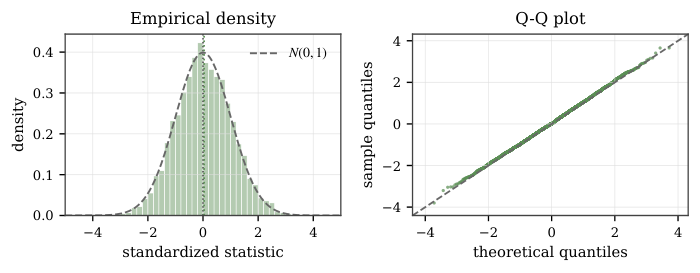}
\caption{$H=0.10$}
\end{subfigure}
\hfill
\begin{subfigure}[t]{0.32\textwidth}
\includegraphics[width=\textwidth]{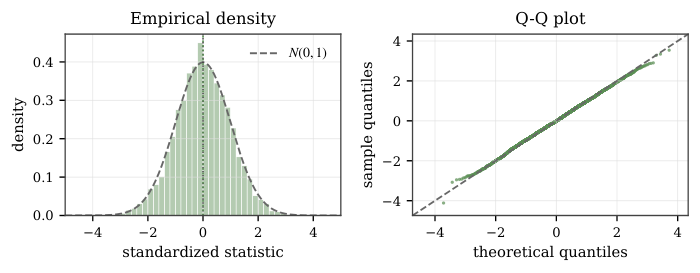}
\caption{$H=0.15$}
\end{subfigure}

\begin{subfigure}[t]{0.32\textwidth}
\includegraphics[width=\textwidth]{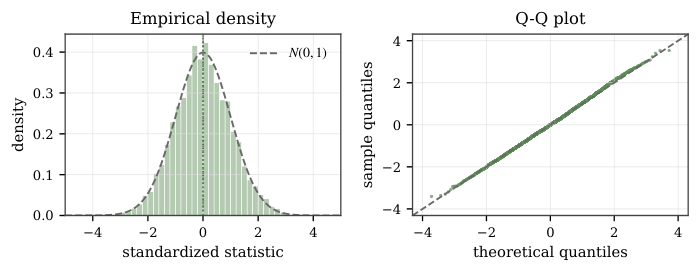}
\caption{$H=0.20$}
\end{subfigure}
\hfill
\begin{subfigure}[t]{0.32\textwidth}
\includegraphics[width=\textwidth]{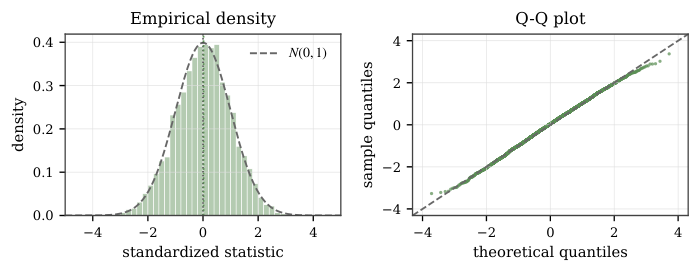}
\caption{$H=0.25$}
\end{subfigure}
\hfill
\begin{subfigure}[t]{0.32\textwidth}
\includegraphics[width=\textwidth]{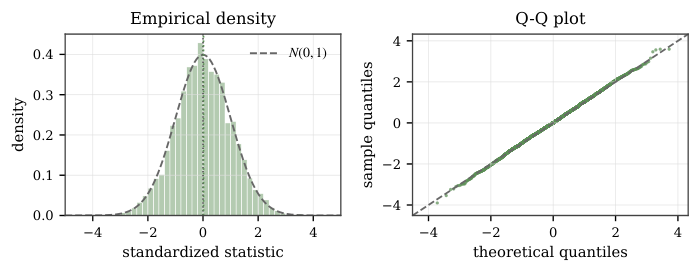}
\caption{$H=0.30$}
\end{subfigure}

\begin{subfigure}[t]{0.32\textwidth}
\includegraphics[width=\textwidth]{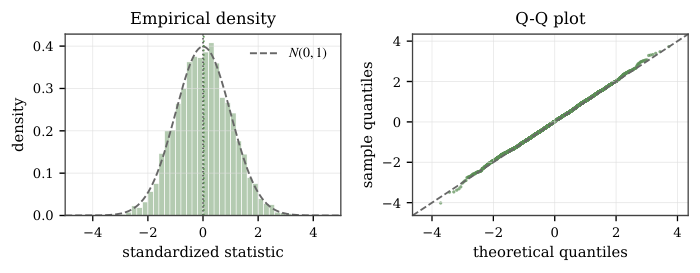}
\caption{$H=0.35$}
\end{subfigure}
\hfill
\begin{subfigure}[t]{0.32\textwidth}
\includegraphics[width=\textwidth]{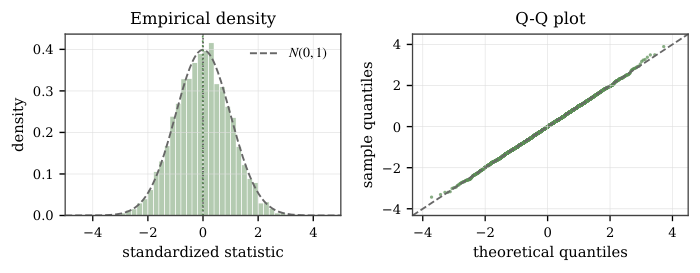}
\caption{$H=0.40$}
\end{subfigure}
\hfill
\begin{subfigure}[t]{0.32\textwidth}
\includegraphics[width=\textwidth]{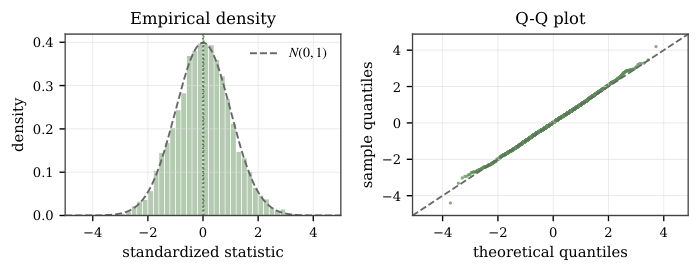}
\caption{$H=0.45$}
\end{subfigure}
\caption{Rough constant without observation noise: densities of $Z$ and normal QQ plots at the nine values of $H$.}
\end{figure}

\begin{figure}[!htbp]
\centering
\begin{subfigure}[t]{0.32\textwidth}
\includegraphics[width=\textwidth]{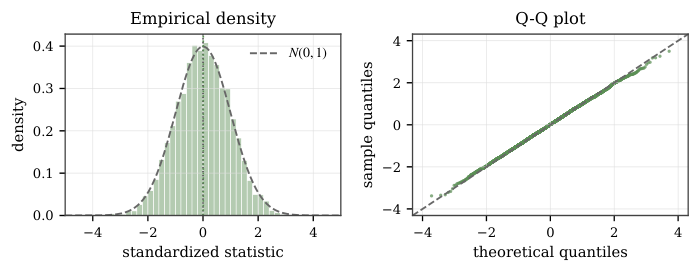}
\caption{$H=0.05$}
\end{subfigure}
\hfill
\begin{subfigure}[t]{0.32\textwidth}
\includegraphics[width=\textwidth]{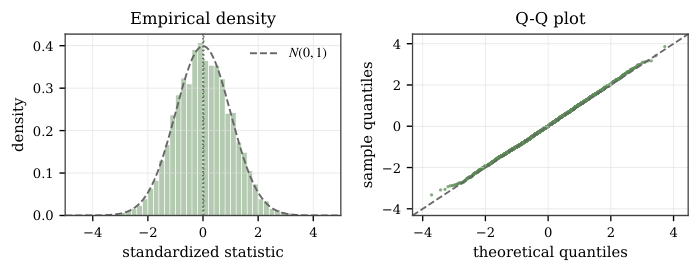}
\caption{$H=0.10$}
\end{subfigure}
\hfill
\begin{subfigure}[t]{0.32\textwidth}
\includegraphics[width=\textwidth]{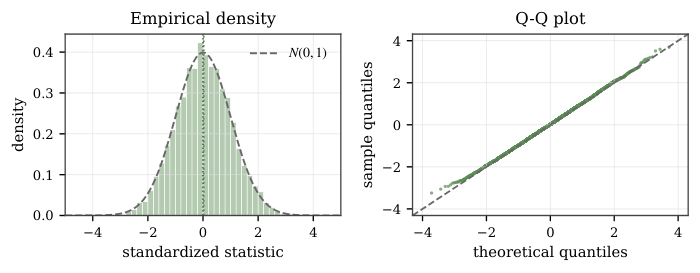}
\caption{$H=0.15$}
\end{subfigure}

\begin{subfigure}[t]{0.32\textwidth}
\includegraphics[width=\textwidth]{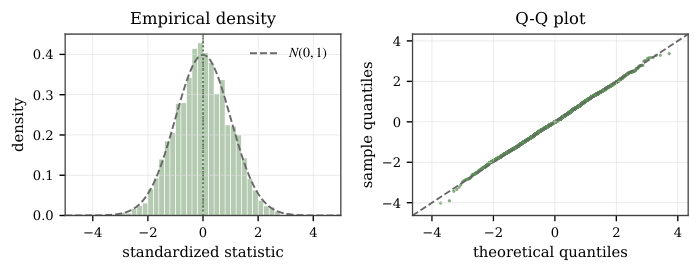}
\caption{$H=0.20$}
\end{subfigure}
\hfill
\begin{subfigure}[t]{0.32\textwidth}
\includegraphics[width=\textwidth]{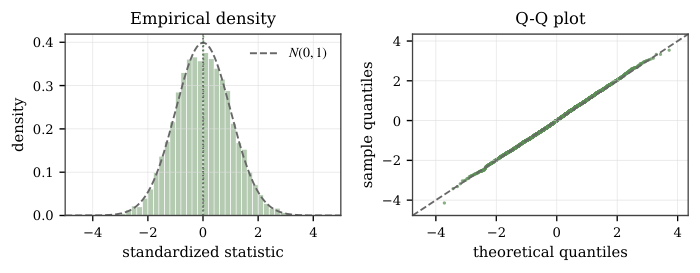}
\caption{$H=0.25$}
\end{subfigure}
\hfill
\begin{subfigure}[t]{0.32\textwidth}
\includegraphics[width=\textwidth]{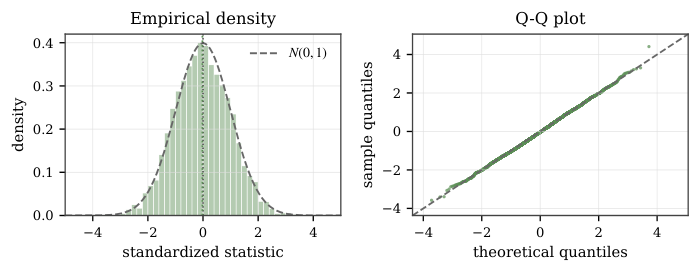}
\caption{$H=0.30$}
\end{subfigure}

\begin{subfigure}[t]{0.32\textwidth}
\includegraphics[width=\textwidth]{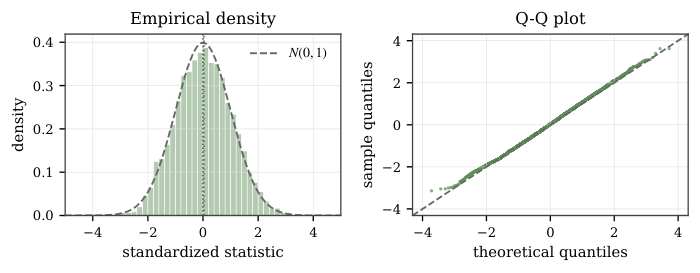}
\caption{$H=0.35$}
\end{subfigure}
\hfill
\begin{subfigure}[t]{0.32\textwidth}
\includegraphics[width=\textwidth]{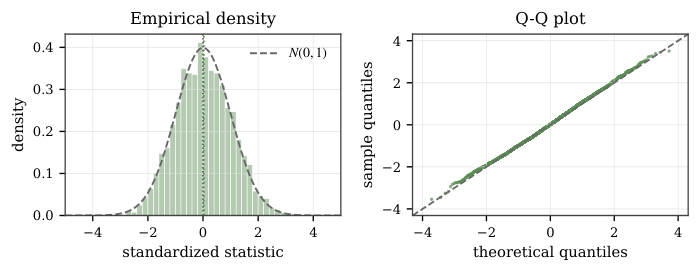}
\caption{$H=0.40$}
\end{subfigure}
\hfill
\begin{subfigure}[t]{0.32\textwidth}
\includegraphics[width=\textwidth]{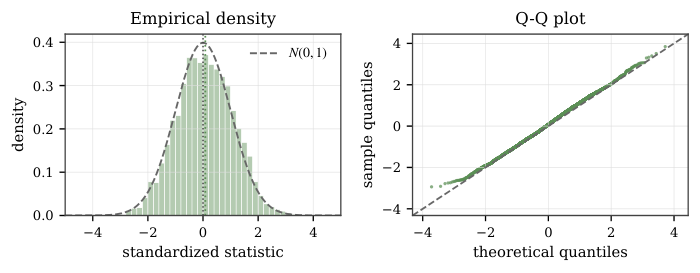}
\caption{$H=0.45$}
\end{subfigure}
\caption{Rough constant with observation noise: densities of $Z$ and normal QQ plots at the nine values of $H$.}
\end{figure}

%% file: appendices/G_emp.tex
\section{Empirical sample and additional results}

\setcounter{table}{0}
\renewcommand{\thetable}{G.\arabic{table}}
\setcounter{figure}{0}
\renewcommand{\thefigure}{G.\arabic{figure}}

To assess sensitivity to the fitted $H$, we perturb it upward and downward by approximately $0.016$, hold the adopted bandwidths fixed, and recompute the exact normalization. The median absolute change divided by the original standard error is $0.140$ for weekly estimates and $0.145$ for monthly estimates. The point estimates are relatively robust to these perturbations under the fixed-bandwidth diagnostic.

The $N$ columns report the estimation-window counts for Table~\ref{tab:empirical}.

\begin{table}[!htbp]
\centering
\normalsize
\linespread{1.6}\selectfont
\setlength{\tabcolsep}{4pt}
\caption{Retained trading days and estimation windows.}
\begin{tabular}{lr|rr|rr}
\toprule
Ticker & \shortstack{Jump days\\removed} & \multicolumn{2}{c|}{Monthly windows} & \multicolumn{2}{c}{Weekly windows} \\
\cmidrule(lr){3-4}\cmidrule(lr){5-6}
 & & Days & $N$ & Days & $N$ \\
\midrule
AAPL & 32 & 3698 & 178 & 3730 & 781 \\
AMZN & 37 & 3695 & 178 & 3725 & 781 \\
DIA & 15 & 3715 & 178 & 3747 & 781 \\
IWM & 9 & 3720 & 178 & 3753 & 781 \\
MSFT & 32 & 3698 & 178 & 3730 & 781 \\
NVDA & 30 & 3699 & 178 & 3732 & 781 \\
QQQ & 7 & 3668 & 176 & 3700 & 770 \\
SPY & 8 & 3721 & 178 & 3754 & 781 \\
TSLA & 51 & 3681 & 178 & 3711 & 781 \\
\bottomrule
\end{tabular}
\end{table}
\begin{table}[!htbp]
\centering
\normalsize
\linespread{1.6}\selectfont
\setlength{\tabcolsep}{4pt}
\caption{Additional summary statistics of fixed-five-minute pre-averaging rough leverage estimates.}
\begin{tabular}{@{}lrrrrrrrrr@{}}
\toprule
Statistic & AAPL & AMZN & DIA & IWM & MSFT & NVDA & QQQ & SPY & TSLA \\
\midrule
$\widehat H_n$ & 0.153 & 0.181 & 0.204 & 0.206 & 0.179 & 0.147 & 0.184 & 0.200 & 0.150 \\
\midrule
\multicolumn{10}{l}{\textbf{A: Monthly windows}} \\
Median & -0.09 & -1.15 & -0.29 & -0.92 & -0.53 & -2.05 & -0.83 & -0.36 & -3.14 \\
Mean & -0.70 & -5.63 & -0.76 & -1.56 & -1.75 & -4.70 & -2.14 & -0.65 & -27.77 \\
SD & 13.71 & 30.27 & 3.42 & 8.64 & 14.37 & 42.17 & 11.79 & 4.18 & 140.37 \\
Neg. & 51.7 & 53.9 & 71.9 & 69.1 & 55.6 & 57.9 & 76.7 & 79.8 & 55.6 \\
IQR & 6.65 & 11.65 & 0.87 & 3.75 & 7.15 & 30.10 & 2.10 & 0.97 & 59.37 \\
SD ratio & 20.94 & 8.53 & 11.11 & 10.01 & 32.86 & 18.00 & 11.18 & 7.92 & 14.25 \\
Sig. neg. & 3.9 & 2.2 & 8.4 & 7.9 & 2.2 & 4.5 & 15.9 & 12.4 & 4.5 \\
\midrule
\multicolumn{10}{l}{\textbf{B: Weekly windows}} \\
Median & -0.05 & -0.12 & -0.05 & -0.22 & -0.11 & -0.44 & -0.15 & -0.06 & -0.39 \\
Mean & -1.23 & -0.84 & -0.16 & -0.63 & -0.45 & -1.64 & -0.91 & -0.22 & -4.76 \\
SD & 15.56 & 12.36 & 3.41 & 4.83 & 7.92 & 24.18 & 12.91 & 2.92 & 60.91 \\
Neg. & 52.2 & 52.5 & 64.3 & 67.0 & 53.9 & 53.8 & 71.7 & 71.7 & 52.9 \\
IQR & 1.53 & 3.18 & 0.26 & 0.90 & 1.69 & 7.76 & 0.52 & 0.26 & 14.64 \\
SD ratio & 12.72 & 9.26 & 12.85 & 20.71 & 14.36 & 15.63 & 8.91 & 10.18 & 11.12 \\
Sig. neg. & 2.7 & 1.7 & 4.6 & 5.1 & 1.9 & 3.2 & 7.1 & 7.2 & 2.3 \\
\bottomrule
\end{tabular}

\begin{minipage}{\textwidth}
\normalsize
\linespread{1}\selectfont
\raggedright
Note: Median, mean, standard deviation (SD), and interquartile range (IQR) are in units of $10^{-4}$. Neg. is the percentage of negative estimates. Sig. neg. is the percentage of 95\% confidence intervals entirely below zero.
SD ratio is the sample standard deviation of the $H=1/2$ estimates divided by that of the $\widehat H_n$ estimates; both use fixed five-minute pre-averaging and exact finite-window correction, with time measured in years.
\end{minipage}
\end{table}

The following plots show selected weekly and corresponding monthly windows for the remaining five assets: TSLA, IWM, DIA, QQQ, and SPY. Returns are aggregated within each trading day, discarding incomplete end-of-day groups, then concatenated; the resulting return count and retained-day horizon in years determine $\Delta_n=T/n$. Five-minute pre-averaging, finite-window correction, and confidence intervals based on the single-scale $Q_n^{\mathrm N}$ follow the main signature plots. The top and bottom rows show weekly and monthly windows; the left and right columns use $\widehat H_n$ and $H=1/2$, respectively.

\begin{figure}[!htbp]
	\centering
	\begin{subfigure}[t]{0.48\textwidth}
		\centering
		\includegraphics[width=\linewidth]{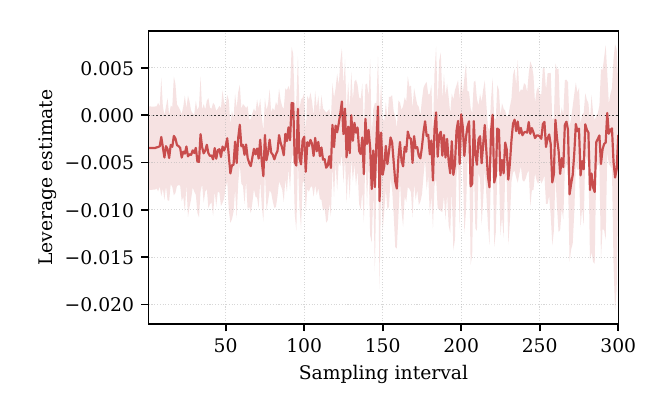}
		\caption{Weekly, $H=\widehat H_n$}
	\end{subfigure}
	\hfill
	\begin{subfigure}[t]{0.48\textwidth}
		\centering
		\includegraphics[width=\linewidth]{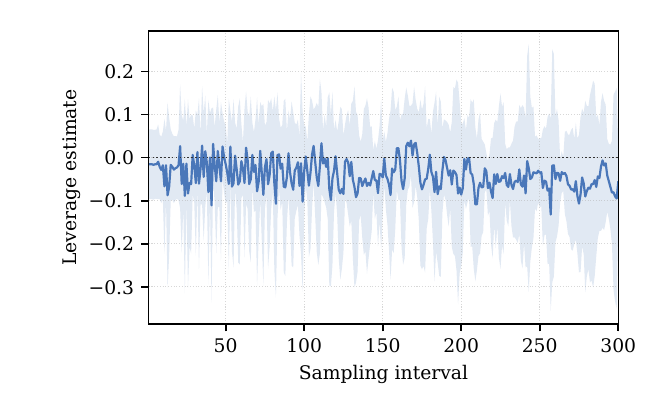}
		\caption{Weekly, $H=1/2$}
	\end{subfigure}

	\begin{subfigure}[t]{0.48\textwidth}
		\centering
		\includegraphics[width=\linewidth]{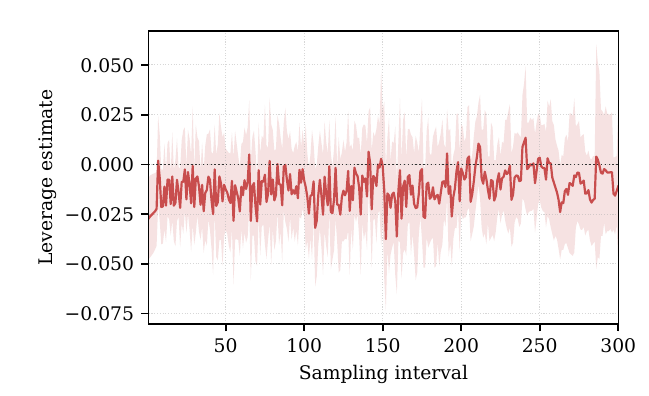}
		\caption{Monthly, $H=\widehat H_n$}
	\end{subfigure}
	\hfill
	\begin{subfigure}[t]{0.48\textwidth}
		\centering
		\includegraphics[width=\linewidth]{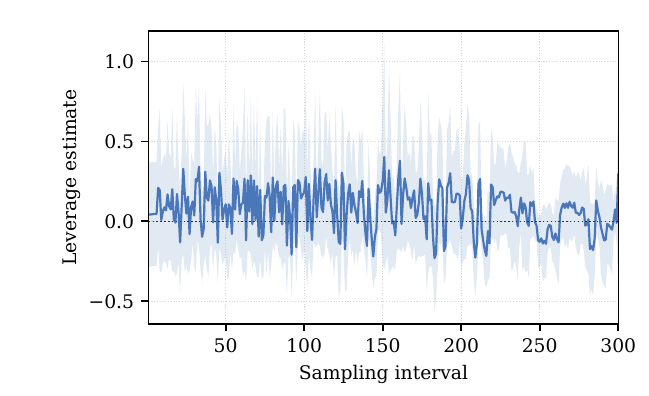}
		\caption{Monthly, $H=1/2$}
	\end{subfigure}
	\caption{TSLA: week 2022-05-16; month 2022-05.}
\end{figure}

\begin{figure}[!htbp]
	\centering
	\begin{subfigure}[t]{0.48\textwidth}
		\centering
		\includegraphics[width=\linewidth]{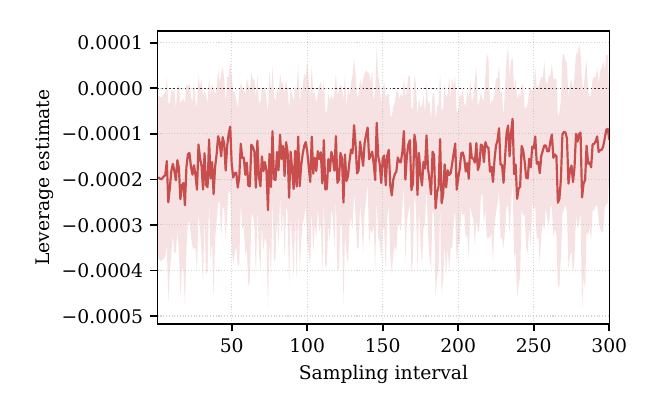}
		\caption{Weekly, $H=\widehat H_n$}
	\end{subfigure}
	\hfill
	\begin{subfigure}[t]{0.48\textwidth}
		\centering
		\includegraphics[width=\linewidth]{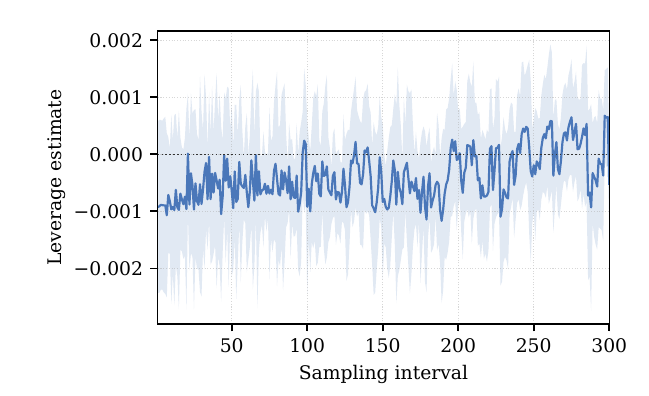}
		\caption{Weekly, $H=1/2$}
	\end{subfigure}

	\begin{subfigure}[t]{0.48\textwidth}
		\centering
		\includegraphics[width=\linewidth]{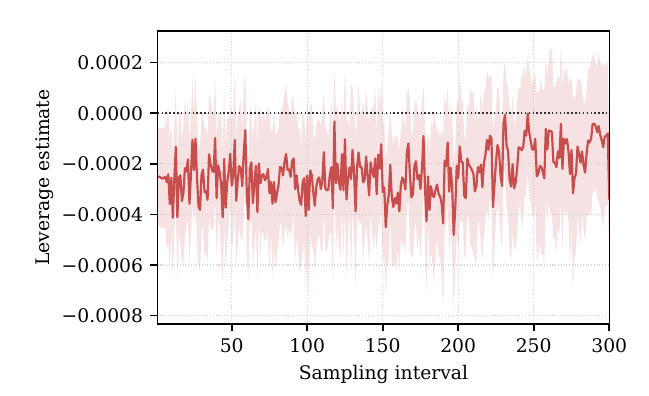}
		\caption{Monthly, $H=\widehat H_n$}
	\end{subfigure}
	\hfill
	\begin{subfigure}[t]{0.48\textwidth}
		\centering
		\includegraphics[width=\linewidth]{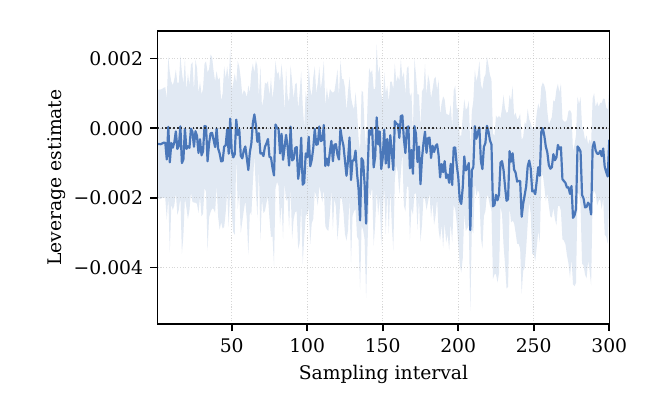}
		\caption{Monthly, $H=1/2$}
	\end{subfigure}
	\caption{IWM: week 2025-06-02; month 2025-06.}
\end{figure}

\begin{figure}[!htbp]
	\centering
	\begin{subfigure}[t]{0.48\textwidth}
		\centering
		\includegraphics[width=\linewidth]{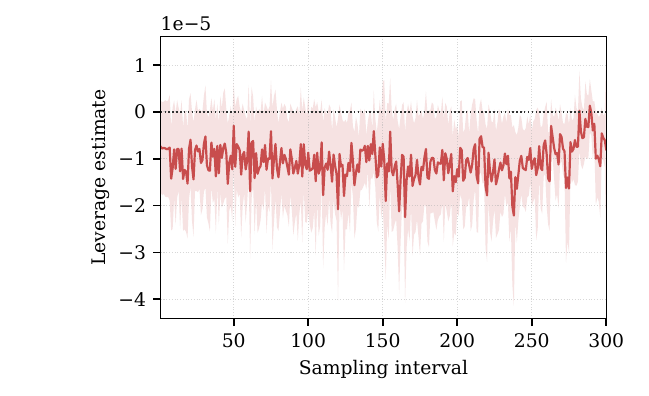}
		\caption{Weekly, $H=\widehat H_n$}
	\end{subfigure}
	\hfill
	\begin{subfigure}[t]{0.48\textwidth}
		\centering
		\includegraphics[width=\linewidth]{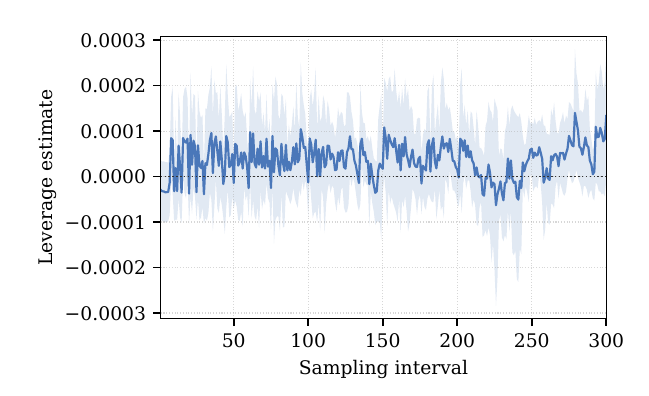}
		\caption{Weekly, $H=1/2$}
	\end{subfigure}

	\begin{subfigure}[t]{0.48\textwidth}
		\centering
		\includegraphics[width=\linewidth]{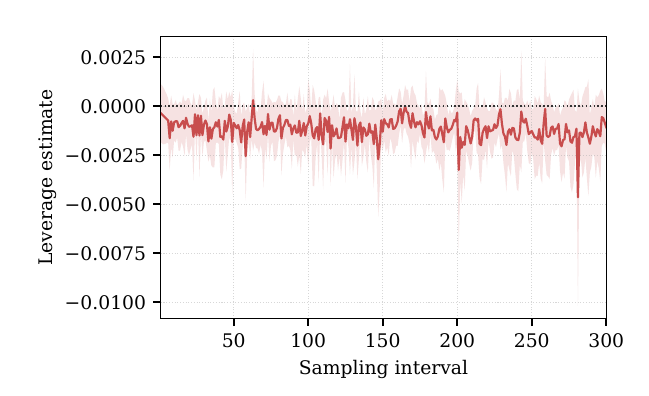}
		\caption{Monthly, $H=\widehat H_n$}
	\end{subfigure}
	\hfill
	\begin{subfigure}[t]{0.48\textwidth}
		\centering
		\includegraphics[width=\linewidth]{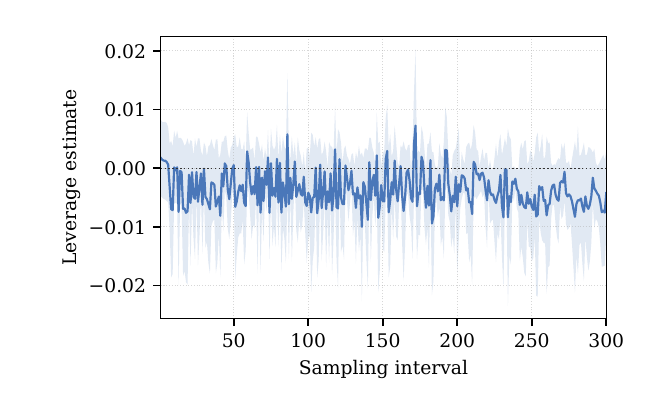}
		\caption{Monthly, $H=1/2$}
	\end{subfigure}
	\caption{DIA: week 2020-02-10; month 2020-02.}
\end{figure}

\begin{figure}[!htbp]
	\centering
	\begin{subfigure}[t]{0.48\textwidth}
		\centering
		\includegraphics[width=\linewidth]{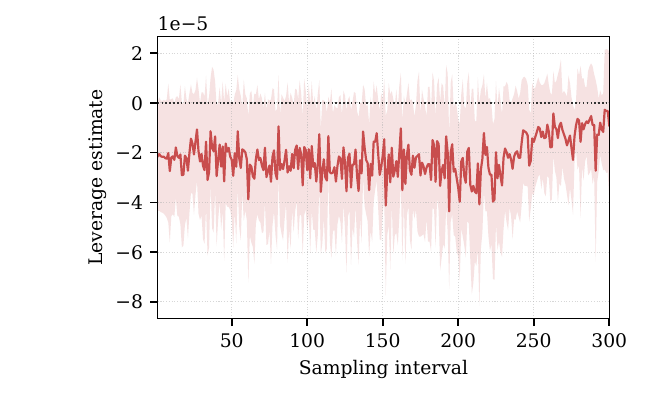}
		\caption{Weekly, $H=\widehat H_n$}
	\end{subfigure}
	\hfill
	\begin{subfigure}[t]{0.48\textwidth}
		\centering
		\includegraphics[width=\linewidth]{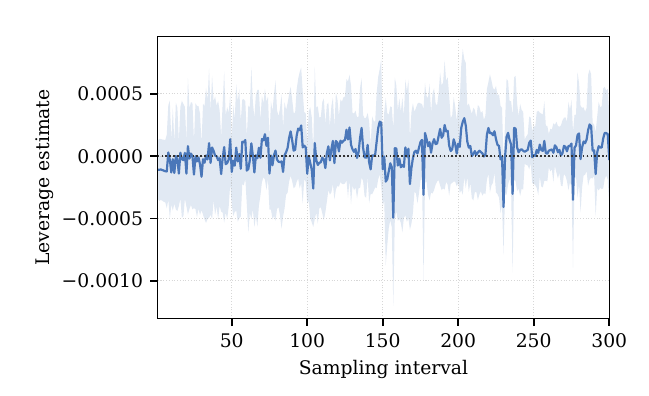}
		\caption{Weekly, $H=1/2$}
	\end{subfigure}

	\begin{subfigure}[t]{0.48\textwidth}
		\centering
		\includegraphics[width=\linewidth]{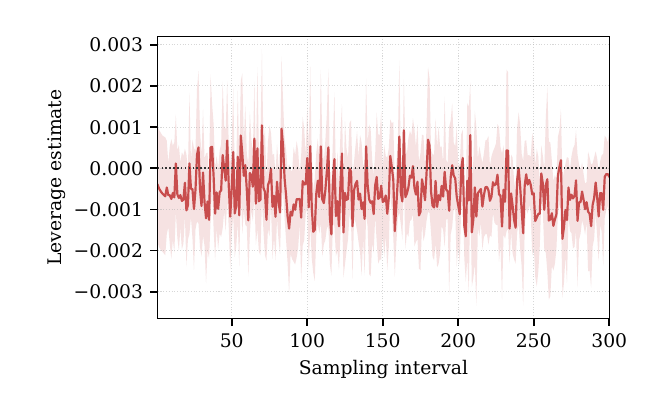}
		\caption{Monthly, $H=\widehat H_n$}
	\end{subfigure}
	\hfill
	\begin{subfigure}[t]{0.48\textwidth}
		\centering
		\includegraphics[width=\linewidth]{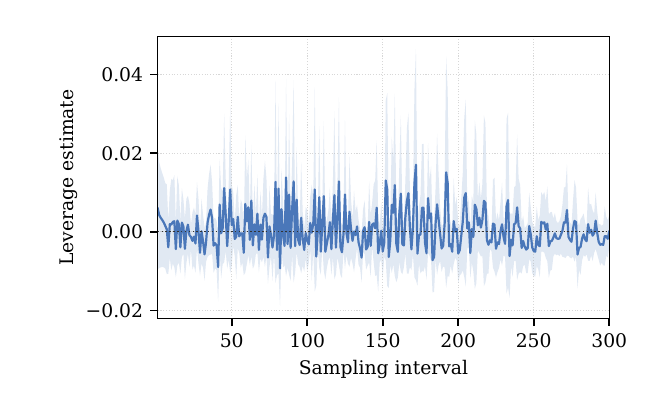}
		\caption{Monthly, $H=1/2$}
	\end{subfigure}
	\caption{QQQ: week 2020-02-10; month 2020-02.}
\end{figure}

\begin{figure}[!htbp]
	\centering
	\begin{subfigure}[t]{0.48\textwidth}
		\centering
		\includegraphics[width=\linewidth]{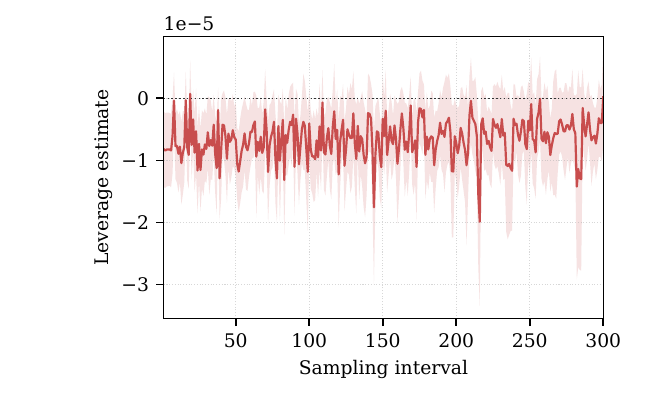}
		\caption{Weekly, $H=\widehat H_n$}
	\end{subfigure}
	\hfill
	\begin{subfigure}[t]{0.48\textwidth}
		\centering
		\includegraphics[width=\linewidth]{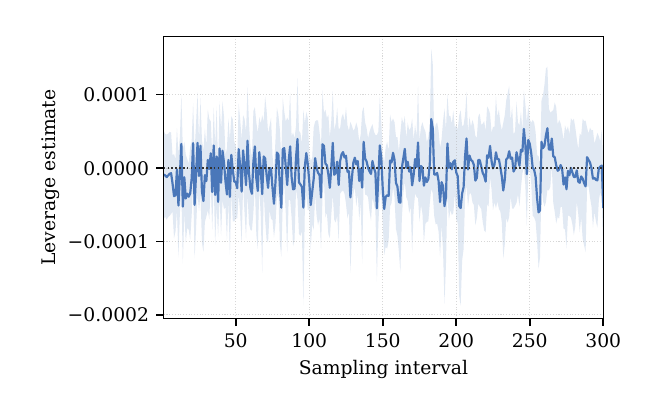}
		\caption{Weekly, $H=1/2$}
	\end{subfigure}

	\begin{subfigure}[t]{0.48\textwidth}
		\centering
		\includegraphics[width=\linewidth]{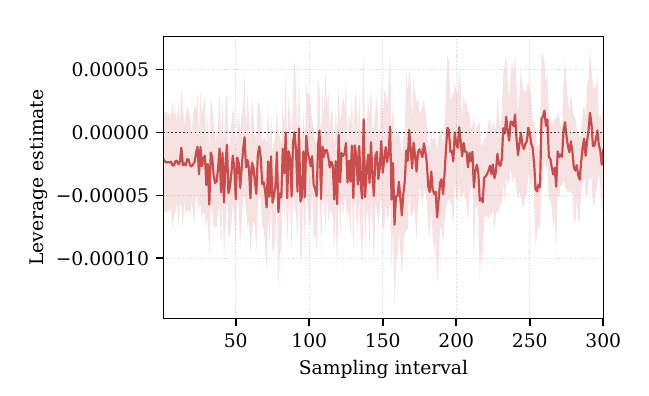}
		\caption{Monthly, $H=\widehat H_n$}
	\end{subfigure}
	\hfill
	\begin{subfigure}[t]{0.48\textwidth}
		\centering
		\includegraphics[width=\linewidth]{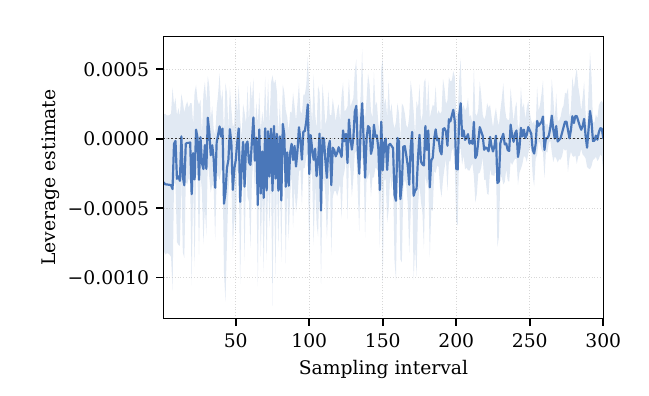}
		\caption{Monthly, $H=1/2$}
	\end{subfigure}
	\caption{SPY: week 2023-07-17; month 2023-07.}
\end{figure}